\documentclass[fleqn,usenatbib]{mnras}
\usepackage{newtxtext,newtxmath}
\usepackage[T1]{fontenc}
\DeclareRobustCommand{\VAN}[3]{#2}
\let\VANthebibliography\thebibliography
\def\thebibliography{\DeclareRobustCommand{\VAN}[3]{##3}\VANthebibliography}

\usepackage{booktabs}
\usepackage{ae,aecompl}
\usepackage{xcolor}

\usepackage{graphicx}
\usepackage{subfigure}
\usepackage{amsmath}	
\usepackage{amssymb}	

\title[]{On the Precision of Full-spectrum Fitting of Simple Stellar Populations. I. Well-sampled Populations.}
\author[Asa'd \& Goudfrooij]{Randa Asa'd$^{1,2}$ and Paul Goudfrooij$^2$ \smallskip \\ 
$^1$Physics Department, American University of Sharjah, P.O.\ Box 26666, Sharjah, UAE \\
$^2$Space Telescope Science Institute, 3700 San Martin Drive, Baltimore, MD 21218, USA
}
\date{Accepted XXX. Received XXX; in original form 2020 XXX}

\begin{document}

\pagerange{\pageref{firstpage}\,--\,44} \pubyear{2020} 

\maketitle

\label{firstpage}

\begin{abstract}
We investigate the precision of the ages and
metallicities of 21,000 mock simple stellar populations (SSPs)
determined through full-spectrum fitting. The mock SSPs cover 
an age range of 6.8 $<$ log\,(age/yr) $<$ 10.2, for three wavelength
ranges in the optical regime, using both Padova and MIST
isochrone models. Random noise is added to the model spectra to
  achieve S/N ratios between 10 to 100 per wavelength pixel. 
We find that for S/N $\geq$ 50, this technique can yield ages of
SSPs to an overall precision of $\Delta\,\mbox{log(age/yr)} \sim 0.1$ for ages in
the ranges 7.0 $\leq$ log\,(age/yr) $\leq$ 8.3 and 8.9 $\leq$ log\,(age/yr)
$\leq$ 9.4. For the age ranges of 8.3 $\leq$ log\,(age/yr) $\leq$ 8.9 and
log\,(age/yr) $\geq$ 9.5, which have significant flux contributions from
asymptotic giant branch (AGB) and red giant branch (RGB) stars, respectively,
the age uncertainty rises to about $\pm 0.3$ dex. 
The precision of age and metallicity estimation using this method depends
significantly on the S/N and the wavelength range used in the fitting. 
We quantify the systematic differences in age predicted by the MIST and
Padova isochrone models, due to their different assumptions about
stellar physics in various important (i.e., luminous) phases of stellar evolution,
which needs to be taken in consideration when comparing ages of star clusters
obtained using these popular models. 
Knowing the strengths and limitations of this technique is crucial in
interpreting the results obtained for real star clusters and for deciding the
optimal instrument setup before performing the observations.   

\end{abstract}


\section{Introduction}
\label{Introduction}

Determining accurate ages and metallicities of star clusters has been an
important goal for both galactic and extragalactic astrophysics, because it
makes it possible to study the formation and evolution of stars, stellar
populations, galaxies and the universe as a whole.
Star clusters are generally considered to be the best known examples of simple
  stellar populations (SSPs), thus providing valuable snapshots of the chemical
enrichment history of the region of the galaxy where it resided at the time of
its formation. While most old globular clusters are now known to host
variations of abundances of several light elements \citep[mainly He, C, N, O,
  Na,  and Al;  see][for details]{BastianLardo18}, such variations have not yet
been found in clusters younger than $\sim$\,2 Gyr \citep[see][]{Martocchia18}. 
    In this paper, we simplify this situation by considering star clusters to
    be well described by SSP models  
(i.e., models constructed of stars of the same age and metallicity), and we
    assume solar element abundance ratios.  

The most common and direct method to estimate the age of a star cluster is
to obtain a colour-magnitude diagram (CMD) of the constituent stars in the
cluster and perform isochrone fitting \citep[e.g.,][]{Goudfrooij11a,
  Goudfrooij2014, Goudfrooij17, Correnti14, Niederhofer15a, Bastian16, Milone16, 
  Milone17a}. However, for distant galaxies where star clusters cannot be
resolved, we have to rely on integrated-light measurements to derive ages and metallicities, e.g., using spectra \citep{AA02, Puzia05,
  Puzia06, Santos06, Palma08, Talavera10, Fernandes10, Asad13, Asad14,
  Chilingarian18}. Some of these studies examined the accuracy of the 
  integrated-light method by comparing the results with results obtained from
CMDs for a small number of clusters. 
Several SSP 
models \citep[e.g.,][]{Bruzual03, Maraston05, Vazdekis10, Maraston2011}, and different full-spectrum fitting programs \citep[e.g.,][]{Fernandes10, Asad14, Wilkinson17} are available for analyzing integrated-light spectra of star clusters.
In this first paper of a series aimed at providing useful insights on the
accuracy and precision of the method of full-spectrum fitting of integrated-light
  spectra, we examine the intrinsic precision of this fitting technique in 
determining the age and metallicity of simple stellar populations (SSPs) of
  ``pseudo-infinite'' mass, for which the effects of stochastic fluctuations of
  the number of stars near the maximum stellar mass at a given SSP age are
  negligible. In practice this is equivalent to cluster masses $M_{\rm cl} \ga
  10^6\;M_{\odot}$  \citep[see, e.g.,][]{CL04,Pessev08}.
For this work, we create 21,000 mock clusters of known age and
metallicity, in the range 6.8 $<$ log\,(age/yr) $<$ 10.2, with S/N ratios
between 10 to 100 for three wavelength ranges in the optical regime, using both
Padova and MIST isochrone models.
The main question that we want to address is:
How does the accuracy and precision of ages and metallicities determined from full-spectrum fitting of integrated-light spectra depend on S$/$N, the
wavelength range, and the model used?  
Answering this question will not only help astronomers interpret the results
they obtain with this method, but also provide useful information for deciding
the optimal instrument set-up before performing such observations (like
the optimal wavelength range to be observed and the minimum S$/$N required to
extract reliable information from the data).   

\section{Data and Method}
\label{Data}

\subsection{Mock Star Clusters}

We use the flexible stellar population synthesis (FSPS) code
\citep{Conroy09, Conroy10} operated through the Python package python-FSPS
\citep{python-fsps} to create two sets of integrated-light spectra (units of L$_{\odot}$/\AA) of SSPs with
\citet{Kroupa01} IMF and metallicity [Z$/$H] = $-$0.4 for the age range 6.8
  $<$ log age $<$ 10.2 in steps of 0.1 dex. The first set uses Padova isochrones 
  \citep{Padova}, while the second set uses MIST isochrones \citep{Choi17}. In both cases we use the MILES spectral library \citep{MILESI, Vazdekis10, Vazdekis16}, which has a spectral resolution of 2.5 \AA.
  We used the metallicity [Z$/$H] = $-$0.4 to mimic the average metallicity of the Large Magellanic Cloud (LMC)\footnote{Later papers in this series will compare observed spectra of LMC clusters with SSP model predictions.}.
  We use the default settings of python-FSPS, including the absence of nebular emission for the youngest ages, and a zero fraction of blue horizontal branch stars at old ages.
For the wavelength range we choose 3700\,--\,6200 \AA,
encompassing the range typically used in full spectrum fitting analysis in
  the literature \citep[e.g.,][]{Cappellari11,Asad13,Asad16,McDermid15}. 
We also perform the same analysis after splitting this wavelength range in
two roughly equal parts, $3700-5000$\,\AA\ and  $5000-6200$\,\AA, in order to
allow one to judge both the relative sensitivity of those two wavelength regions to
variations in age and/or metallicity, and the impact of using a wavelength
range that is roughly twice as small, corresponding to the use of a spectral
grating with roughly twice the resolving power $R$.

Mock cluster spectra are created by adding 30 different realizations
of random noise, corresponding to S/N values between 10
to 100 (in steps of 10) for each SSP, thus producing a grand total of 21,000
mock clusters.   

\subsection{Method}

Using FSPS, we produce a model grid that varies in both age ((6.8 $<$ log\,(age) $<$ 10.2, in steps of 0.1 dex) and metallicity ($-$1.0 $<$ [Z$/$H] $<$ 0.2,  in steps of 0.2 dex). Our two sets of model grids are created using FSPS in order to ensure that the only difference between them is the isochrone family (Padova vs.\  MILES). We use the full-spectrum fitting program {\sc ASAD}$_{\rm 2}$
 \citep[for a full description, see][]{Asad13,Asad16,Asad14} to obtain the
best-fitting age and metallicity using the following equation:  

\begin{equation} 
\sum_{\lambda=\lambda_{\rm initial }}^{\lambda_{\rm final}}
\frac{[(OF)_{\lambda} - (MF)_{\lambda}]^{2}}{(OF)_{\lambda_{\rm norm}}}.
\label{Eq_1SSP}
\end{equation} 

where OF is the mock cluster flux, MF is the SSP model flux, and
$\lambda_{\rm norm}$ is the wavelength at which the model and the mock cluster spectra are normalized.

\section{Results and Discussion}
\label{Results} 

\subsection{Padova Models}

\subsubsection{Full Wavelength Range: $3700 \leqslant \lambda/\mbox{\AA} \leqslant 6200$}

\subparagraph{A. Age Determinations:}

Table \ref{T1} shows the mean  value and its uncertainty (calculated as the
standard deviation $\sigma$) of the difference between the derived and the true
age for each input age and S/N in this wavelength region. 

\begin{figure*}
\resizebox{85mm}{!}{\includegraphics[angle=270]{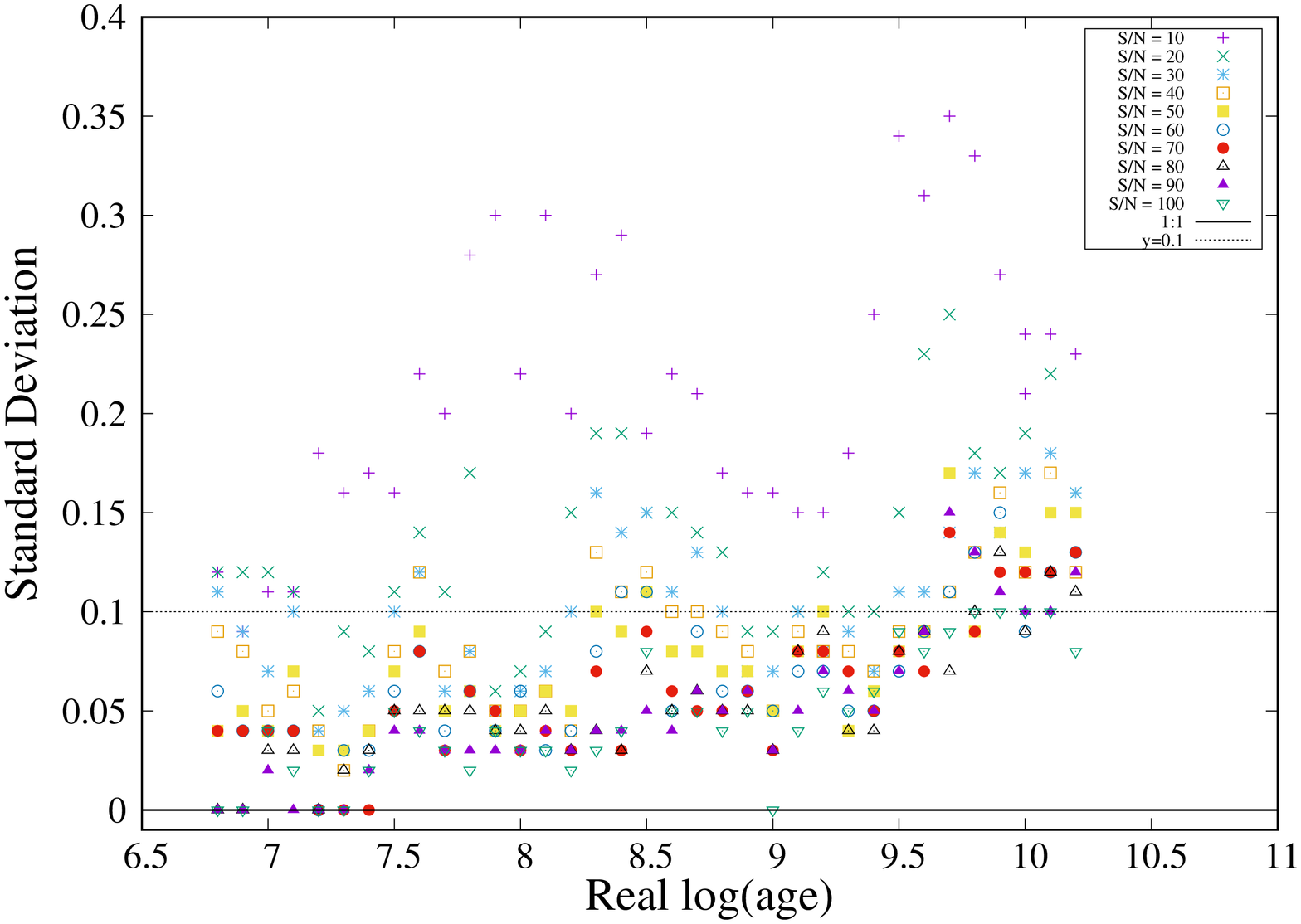}}
\resizebox{85mm}{!}{\includegraphics[angle=270]{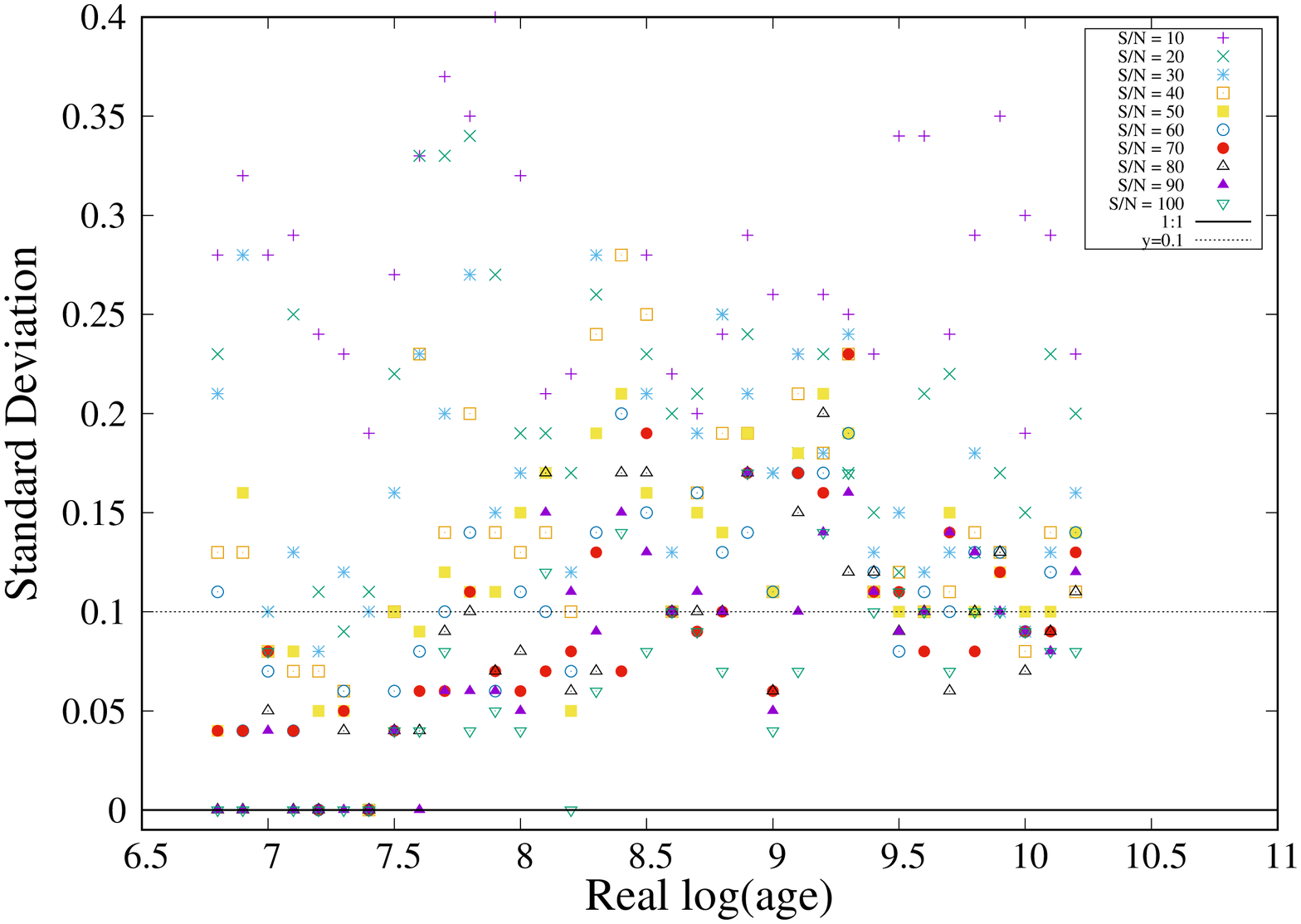}}

\caption{The standard deviation ($\sigma$) of the differences of derived log\,(age) (left panel) and [Z/H] (right panel) relative to their true (input) values versus log\,(age) for the full mock clusters
  sample for all values of S/N (see boxed-in legend). The horizontal line at $\sigma = 0.1$ is plotted to guide the eye.}
\label{SV}
\end{figure*}

To visualize the uncertainties associated with the mean derived ages, we
plot the standard deviation (uncertainty) versus log\,(age) for all S/N
ratios on the left panel of Figure \ref{SV}.
The uncertainties are generally largest in the age intervals
  $8.3 \la \mbox{log\,(age/yr)} \la 8.9$ (especially for S/N $\la 40$) and
  $\mbox{log\,(age/yr)} \ga 9.6$. For the latter age range,  we find
  $\sigma(\mbox{log\,(age/yr)}) \ga 0.1$  even for the highest S/N values.

To shed more light on the accuracy of the derived ages, the top row of
  panels in Figure~\ref{Fig1} shows the number of correctly recovered ages i.e., the number of times the derived age matches exactly the input age.)
using our full-spectrum fitting method as a function of age. When no point is present for a certain input age it means that specific age was never correctly recovered. As might have
  been expected, we find that the higher the S/N, the greater the number of the
correctly derived results for a given age. However, the number of
correctly derived ages also depends on the age: 
a significant drop in the number of correctly recovered ages is seen around log
(age) = 8.5 which corresponds to the asymptotic giant branch (AGB) phase transition in stellar evolution, and around log (age) 9.0$-$9.2 which corresponds to the red giant branch (RGB) phase transition, which is confirmed by the CMDs discussed in section \ref{CMDs}.

The second row of panels of Figure \ref{Fig1} shows the difference between
the derived and real log\,(age) versus real  log\,(age) for the
entire sample. The left panels are for S/N values from 10 to 50, and the
right panels are for S/N values from 60 to 100. 

 In addition to illustrating and quantifying the benefit of higher S/N in age
  determinations from full-spectrum fitting, this figure also shows that certain
  age ranges are generally more susceptible to erroneous age determination than
  others. Specifically, concentrating on S/N values $> 50$, relatively large
  errors are found in the age ranges $7.5 \la \mbox{log\,(age)} \la 7.8$,
  $8.5 \la \mbox{log\,(age)} \la 8.9$ and $\mbox{log\,(age)} \ga 9.1$,
  corresponding to the periods during which the integrated light of SSPs is
  dominated by supergiant stars, AGB stars, and RGB and horizontal branch (HB) stars, respectively.

\subparagraph{B. Metallicity Predictions:}
All our mock clusters have a metallicity of [Z$/$H] $= -$0.4, while our
model clusters cover the metallicity range $-$1.0 $<$ [Z$/$H] $<$ 0.2 in
steps of 0.2 dex.

The right panel of Figure \ref{SV} shows that the uncertainties associated
  with the mean metallicity value are large (relative to the uncertainties for
  log\,(age)), especially for the lower S/N values.

The third row of panels of Figure \ref{Fig1} shows the fraction of
correctly recovered [Z/H] as function of age and S/N.
   Similar to the age predictions, the fraction of correctly recovered
  [Z/H] generally increases with increased S/N, 
 while the age ranges with the lowest fraction of correctly recovered [Z/H]
  are $8.4 \la \mbox{log\,(age)} \la 8.9$ and log\,(age) $\ga 9.1$, even at the
  higher values of S$/$N. 
The bottom row of panels of Figure \ref{Fig1} shows the metallicities
derived for our sample as a function of age. 
 Note that the deviation from the real [Z/H] spans the full range
  considered here (i.e., $\pm 0.6$ dex) for spectra with S$/$N $\la 50$. 

Table \ref{T13} lists the mean values of [Z/H] for each log\,(age)
and S$/$N ratio along with their uncertainties.    
\subsubsection{Wavelength Ranges   $3700 \leqslant \lambda/\mbox{\AA} \leqslant
  5000$ and $5000 \leqslant \lambda/\mbox{\AA} \leqslant 6200$} 

When repeating the analysis using the shorter wavelength range 
$3700 \leqslant \lambda/\mbox{\AA} \leqslant 5000$, Figure \ref{Fig2} shows that
the age determinations are worse than the full wavelength range for log\,(age)
$>$ 9.5, but they are actually {\it better} for the younger ages with log\,(age)
$\la 8.5$. This likely reflects the power of the Balmer jump at
  $\sim$\,4000\AA\ for age determination using the full-spectrum fitting method,
  which is weakened when including a larger wavelength interval beyond
  5000\,\AA\ \citep[see also][for a similar result]{Wilkinson17}.  
Similarly, the overall number of correctly recovered age and metallicity shown
in Figure \ref{Fig2} are improved in the age range 6.8 $\leqslant$ log (age) $<$
8.2 for S/N $=$ 100.
Table \ref{T2} shows that the mean age is always correctly recovered for S$/$N
$\geq$ 50. 

For the wavelength range $5000 \leqslant \lambda/\mbox{\AA} \leqslant 6200$, the
results are generally significantly worse than those for the
  3700\,--\,5000\,\AA\ range, except around input log\,(age) $\sim 9.0$ for 
age determination, and for log\,(age) $\ga 9.5$ for metallicity determination
(See Figure \ref{Fig3}).
This is likely due to the dominance of RGB stars in the energy production of
SSPs with ages above $\sim$\,1 Gyr, since RGB stars mainly produce light at
the longer wavelengths, and many of the stronger metallicity-sensitive
spectral features (e.g., MgH, Mg$_2$, several Fe lines; see, e.g., \citealt{Worthey94}) are located
beyond 5000\,\AA.

\subsection{MIST Models}

We repeat the analysis described in Section~3.1, now using the MIST isochrone
models. Looking at Figure ~\ref{Fig4} to \ref{Fig6}, the general results
remain the same as for the Padova models:\ results for S$/$N = 10 are not 
reliable regardless of the wavelength range, the wavelength range
$5000 - 6200$~\AA\ should never be used for age determination, and metallicity
determination from simple integrated-light spectrum fitting is not as powerful
as age determination, especially for low values of S/N. 

In the wavelength range $3700 - 6200$~\AA, the fraction of correctly recovered
metallicities drops beyond log\,(age) = 8.2 for Padova models, and at log\,(age)
= 8.4 for MIST models.
This difference in the exact age drop for MIST versus Padova is because the two
models make different assumptions about stellar physics in various important
(i.e., luminous) phases of stellar evolution. 

In the wavelength range $3700 - 5000$~\AA, the
recovery of ages is very good with MIST models for all ages $\leqslant$ 7.5; 
however, the drop in recovery fraction for the older ages is steeper
relative to the full wavelength
  range.
Good recovery of metallicity in this wavelength range is achieved up to
log\,(age) = 8.2 but it has a steep gap at log\,(age) = 7.0 (this gap is located
at log\,(age) = 7.3 for Padova models).

\subsection{A Closer Look into the Differences between the Models}
\label{Two_Models}

\subsubsection{Shifts in derived ages}
\label{ageshifts}

For the analysis discussed above, the full fitting technique was applied by
fitting SSP models with mock clusters built with the same SSP models, which means that we started with SSPs of known ages and metallicities and added various amounts of Poisson noise, then compared those artificial 
clusters with the full SSP model grid. This was done for both Padova and MIST models. 

An interesting question arises: how would the results change if we perform the
fitting by comparing Padova-based mock clusters with MIST models, and
vice versa (i.e., MIST-based mock clusters with Padova models)?
The results of this investigation is shown in Figures
  \ref{Fig7}\,--\,\ref{Fig12} and Tables \ref{T7}\,--\,\ref{T12}.  
In the figures showing the number of correctly recovered ages as function
of real age, when no point is present for a certain input age it means that
specific age was never correctly recovered.

The results show clear differences between some derived ages and the
real ones. Just to list a few such differences for the full
wavelength range $3700 \leqslant \lambda/\mbox{\AA} \leqslant 6200$, when
using MIST models with Padova-based mock clusters for log\,(age) = 6.8  we
  find the mean derived age to be log\,(age) = 7.2, for log\,(age) = 7.0 the
mean derived age is log\,(age) = 6.9, for log\,(age) = 7.1 the mean derived age is
log\,(age) = 7.0, and for log\,(age) = 7.2 the mean derived age is log\,(age) = 7.0.

Overall, this shift effect is less significant for analysis based on
the wavelength range $3700 \leqslant \lambda/\mbox{\AA} \leqslant 5000$.
We find a shift of $\Delta$\,log\,(age/yr) $\sim$ +0.1 within the age range
$7.1 \leqslant \mbox{log\,(age/yr)} \leqslant 7.6$, except for log\,(age/yr) = 7.3 
and 7.5 (that is, the real age is greater than the mean derived age).
This shift is +0.2 for log\,(age) = 7.7, 7.8, and 7.9, it is +0.1 for
log(age) = 8.1, 8.2, and 8.3, and it is $-0.1$ for the range
$9.4 \leqslant \mbox{log\,(age)} \leqslant 9.7$. 
This shift is evident for the highest S/N ratios, confirming that the shift is not
due to the bad quality of the data. The shift is present in all wavelength
ranges. This effect will be discussed in details in section \ref{CMDs}.

When using Padova models with MIST-based mock clusters, the shift observed in
the range $3700 \leqslant \lambda/\mbox{\AA} \leqslant 5000$ is again than that in the range
$3700 \leqslant \lambda/\mbox{\AA} \leqslant 6200$, especially in 
the age range $7.5 \leqslant \mbox{log\,(age)} \leqslant 8.5$.
 In fact, 
in the range $3700 \leqslant \lambda/\mbox{\AA} \leqslant 6200$, none of the ages
in the range 7.5 $<$ log\,(age) $<$ 8.5 were correctly recovered.
 As shown in the plots of (derived\,--\,real) versus real age, the shifts
 are very large for 7.5 $<$ log (age) $<$ 8.5, with values ranging up 
to $-$0.7 to $-$0.9. An interesting observation is that there is
almost a linear correlation between the shift and the age in the range 7.5
$\leqslant$ log\,(age) $\leqslant$ 8.8, with the shift decreasing almost
linearly with increasing age. The slope of this linear correlation is about 0.7.
 However, when looking at the results in the wavelength range
  $3700 \leqslant \lambda/\mbox{\AA} \leqslant 5000$, the typical shift within
  this age range is only $-$0.3 except for shifts of $-$0.5 and $-$0.6 for
  log\,(age) = 8.3 and 8.4, respectively.

In the age range 9.3 $\leqslant$ log\,(age) $\leqslant$ 9.7, we find that
  the shift is almost never less than +0.2 when using the wavelength range
$3700 \leqslant \lambda/\mbox{\AA} \leqslant 5000$.
However, the shifts fluctuate when using the full wavelength range
$3700 \leqslant \lambda/\mbox{\AA} \leqslant 6200$. 

The main conclusion of this comparison exercise is that

 for a given observed integrated-light spectrum of an SSP, different
models can predict significantly different ages, especially in some age ranges
as detailed above. This point will be discussed and elaborated on
real clusters in an upcoming paper.

\subsubsection{CMDs of Padova and MIST isochrones}
\label{CMDs}

Figure \ref{CMD_P_M} compares the Padova and MIST isochrones in $V$ vs.\ $B-R$ CMDs from log age = 7.0 to 10.1, using [Z/H] = -0.4.

In each of these sets we note that there are certain age ranges where not much
is changing in the shape of the isochrone from one age to the next, which
makes it challenging to obtain an accurate age determination in those age
ranges. Good examples of such age ranges for the Padova isochrones are 7.5 $<$
log\,(age) $<$ 7.9, just before the AGB starts growing. Another example is the
range 8.5 $<$ log(age) $<$ 8.9, just before the RGB starts growing.

This is also observed in MIST isochrones, except that the exact ages differ
somewhat from one model to the other, because two models make different
assumptions about stellar physics in various important (luminous) phases of
stellar evolution.
The main differences between the MIST and Padova isochrones in the age range studied here are: \textit{(i)} differences in the onset and extent of the blue loop of core helium burning stars at young ages (log\,(age/yr) $\la 7.8$); \textit{(ii)} the
growth of the AGB and then RGB occur a bit earlier in the MIST isochrones than in the  Padova ones; and \textit{(iii)} the luminosities of carbon stars in the thermally pulsing AGB, which are significantly higher for the Padova isochrones \citep[for details, see][]{Padova,Choi16}.  We also note that the MIST isochrones include post-AGB
evolution and the Padova ones do not, but this is not relevant for a typical star cluster given the
  very small numbers of post-AGB stars and the speed with which such stars move
  through the optically luminous part of the CMD \citep[see, e.g.,][]{brown00}.

 The differences mentioned above explain the shifts seen between ages
  of mock clusters built using one SSP model and fit by another
  one.
For example, when fitting Padova-based clusters  with MIST models,
the model predicts log\,(age) = 7.2 for the real log\,(age) = 6.8 which is
  likely caused by the ``blue loop'' of supergiants which starts significantly
earlier in the Padova models than in the MIST models.

\section{Can fixing the metallicity improve the derived age?}
\label{Fixed_Z}

So far all analysis was done by varying both age and metallicity. Solving
for fewer unknowns would certainly improve the fitting, however, when
applying this technique to typical observations, neither the age nor the
metallicity is known. In the previous sections we saw that age determination
using the full-spectrum fitting technique can be successful to an accuracy
  of 0.1 dex (up to log\,(age/yr) $\sim$\,9.5), however, the metallicity
determination is not as good, especially for lower S$/$N. It is then interesting
to study the case of using models of fixed metallicity, and explore
  whether the age determination is improved. 

As mentioned in section \ref{Data}, all our mock clusters have metallicity
  [Z/H] = $-$0.4. Assuming that the metallicity is unknown, we created Padova
models for the same age range, but for a fixed solar metallicity. We applied the
full spectrum fitting and analyzed the results. We then repeated the analysis
with Padova models of fixed metallicity [Z/H] = $-$0.8.

The results are shown in figures \ref{Z8}. When solar metallicity
is used, over all ages are underestimated. The strongest underestimation
is for log\,(age) $>$ 9.5.
Using the wavelength range $3700  \leqslant \lambda/\mbox{\AA} \leqslant 5000$
with high S/N reduces the underestimation of age.

The opposite effect is seen when using [Z/H] = $-$0.8. Most ages are
overestimated, however this is slightly improved for high S/N when using the
wavelength range $3700 \leqslant \lambda/\mbox{\AA} \leqslant 6200$.
See \citet{Beasley02} and \citet{Asad16} for more discussion on the effect of
metallicity. 

\begin{figure}
\resizebox{85mm}{!}{\includegraphics[angle=270]{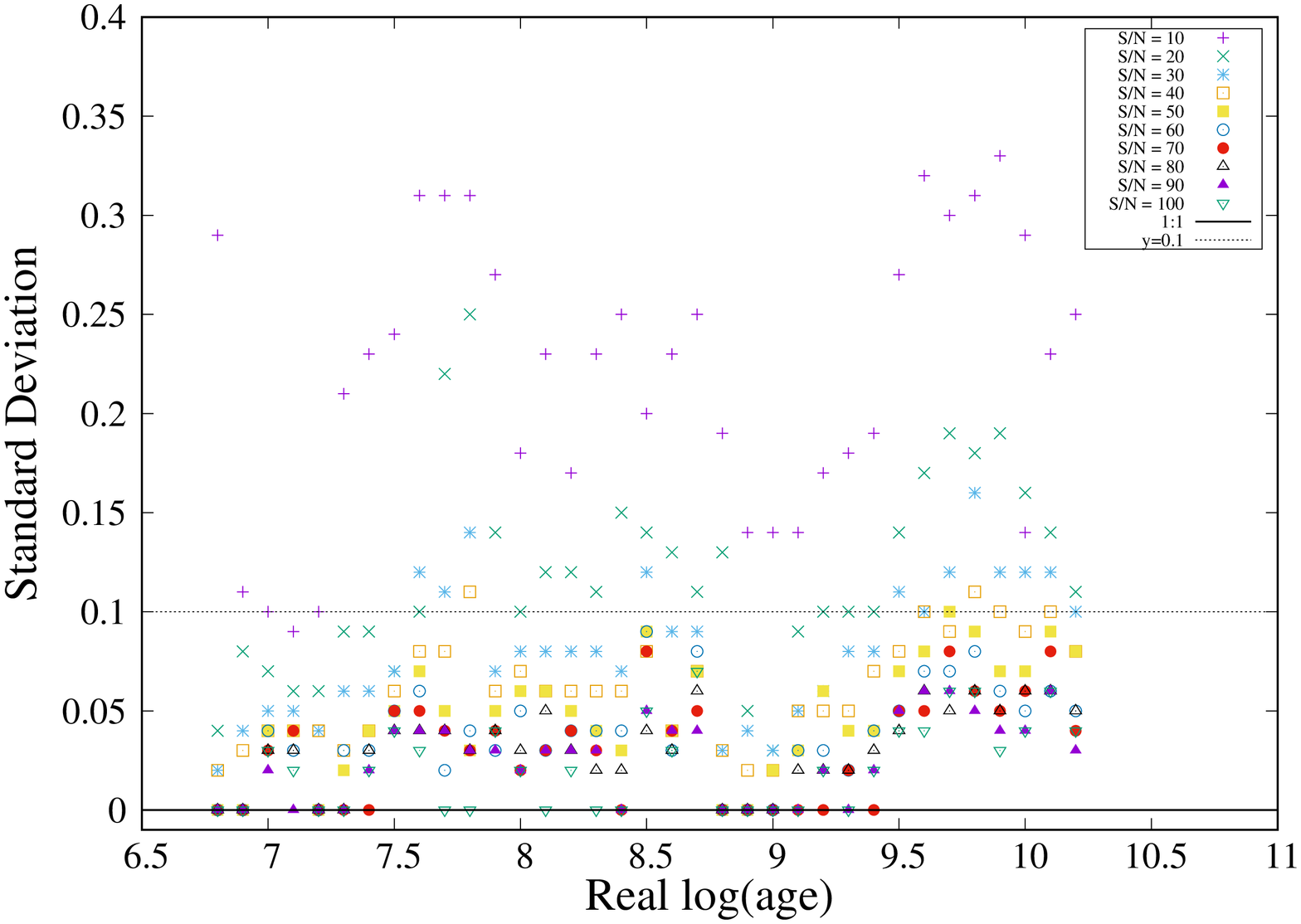}}
\caption{The standard deviation ($\sigma$) for derived log\,(age) minus real log\,(age), at a fixed [Z/H] = -0.4 (see discussion in Section~\ref{Fixed_Z}). The horizontal line at $\sigma = 0.1$ is plotted to guide the eye.}
\label{SV_fixedZ}
\end{figure}

\begin{figure*}
\resizebox{75mm}{!}{\includegraphics[angle=270]{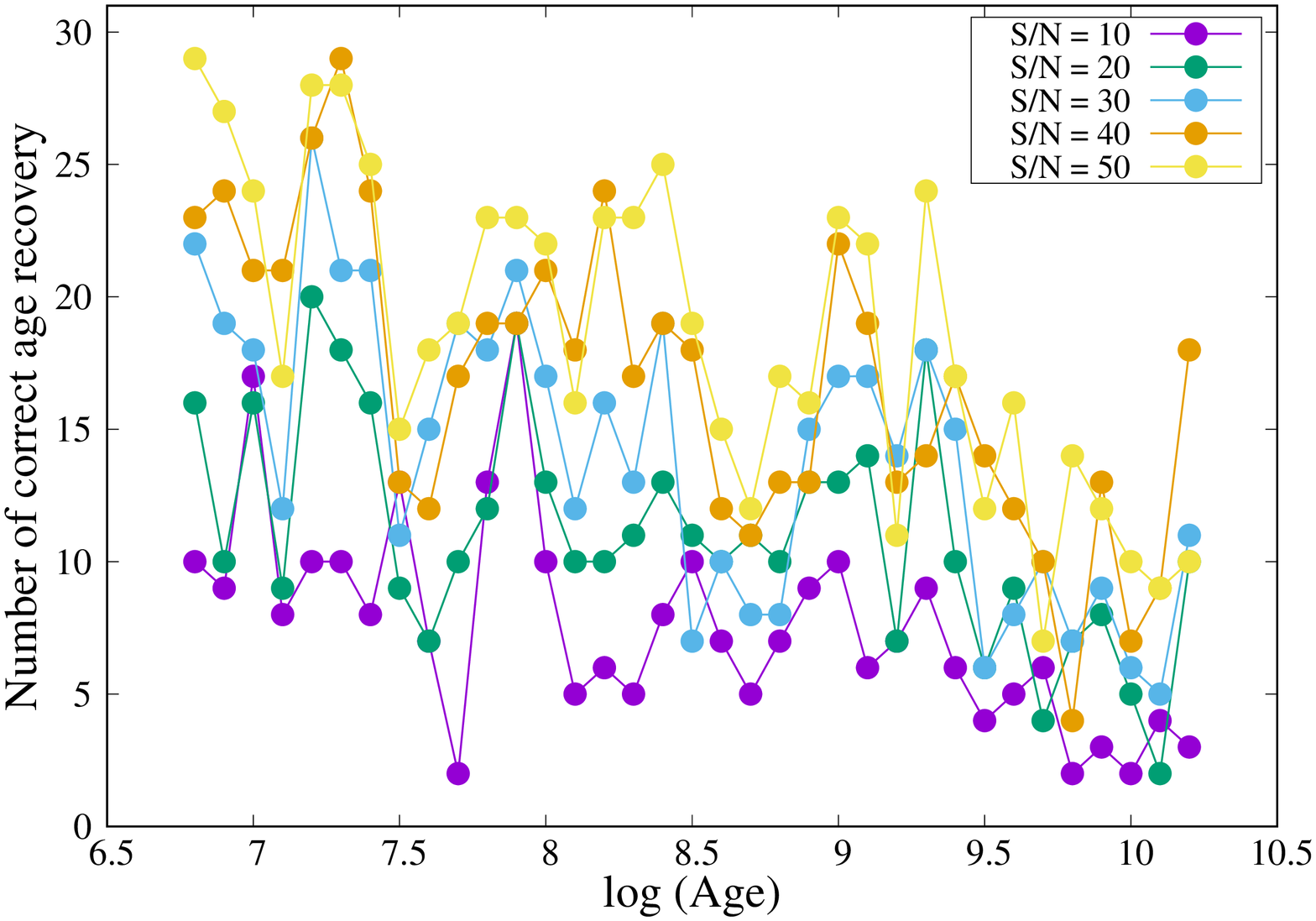}}
\resizebox{75mm}{!}{\includegraphics[angle=270]{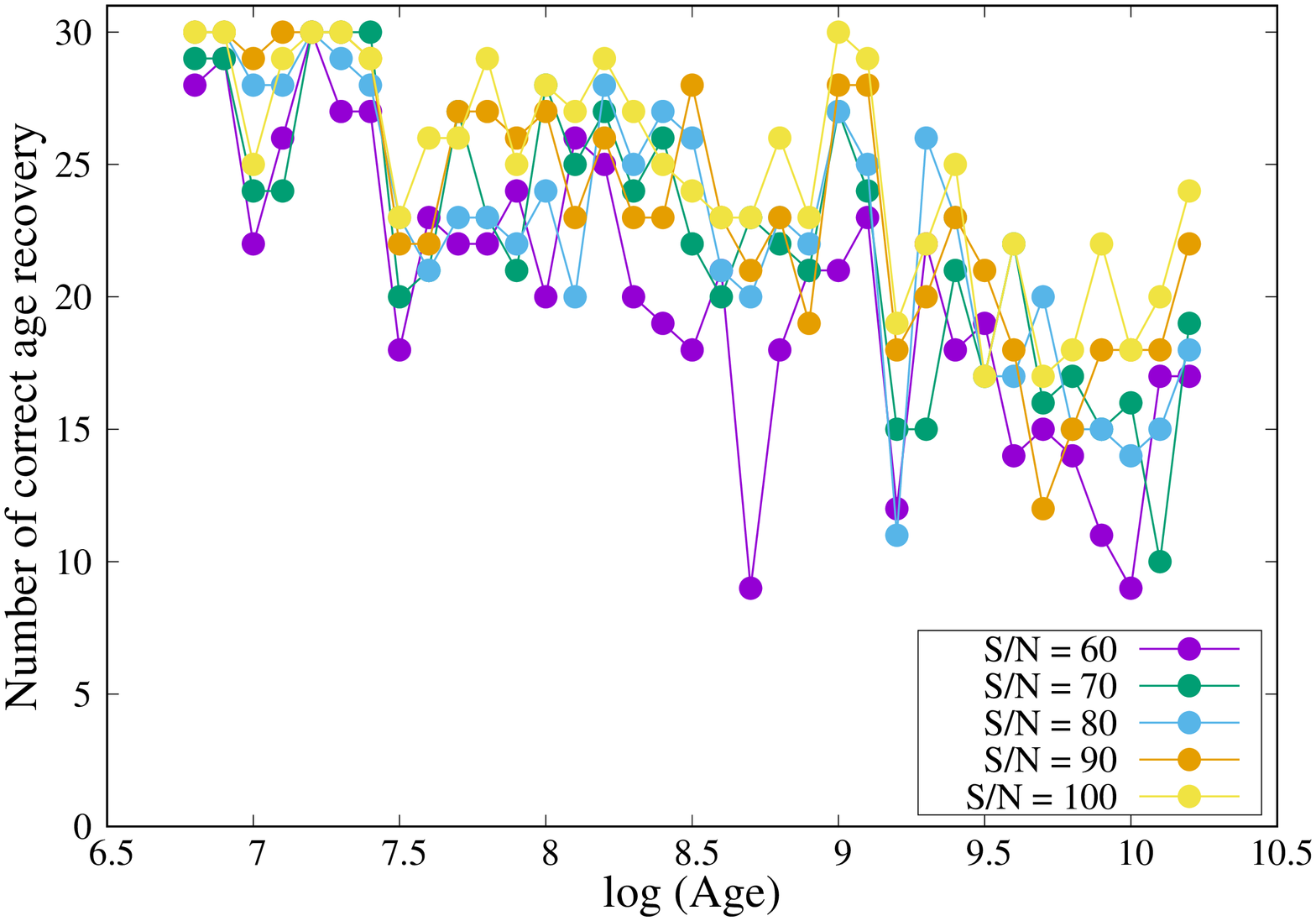}}
\resizebox{75mm}{!}{\includegraphics[angle=270]{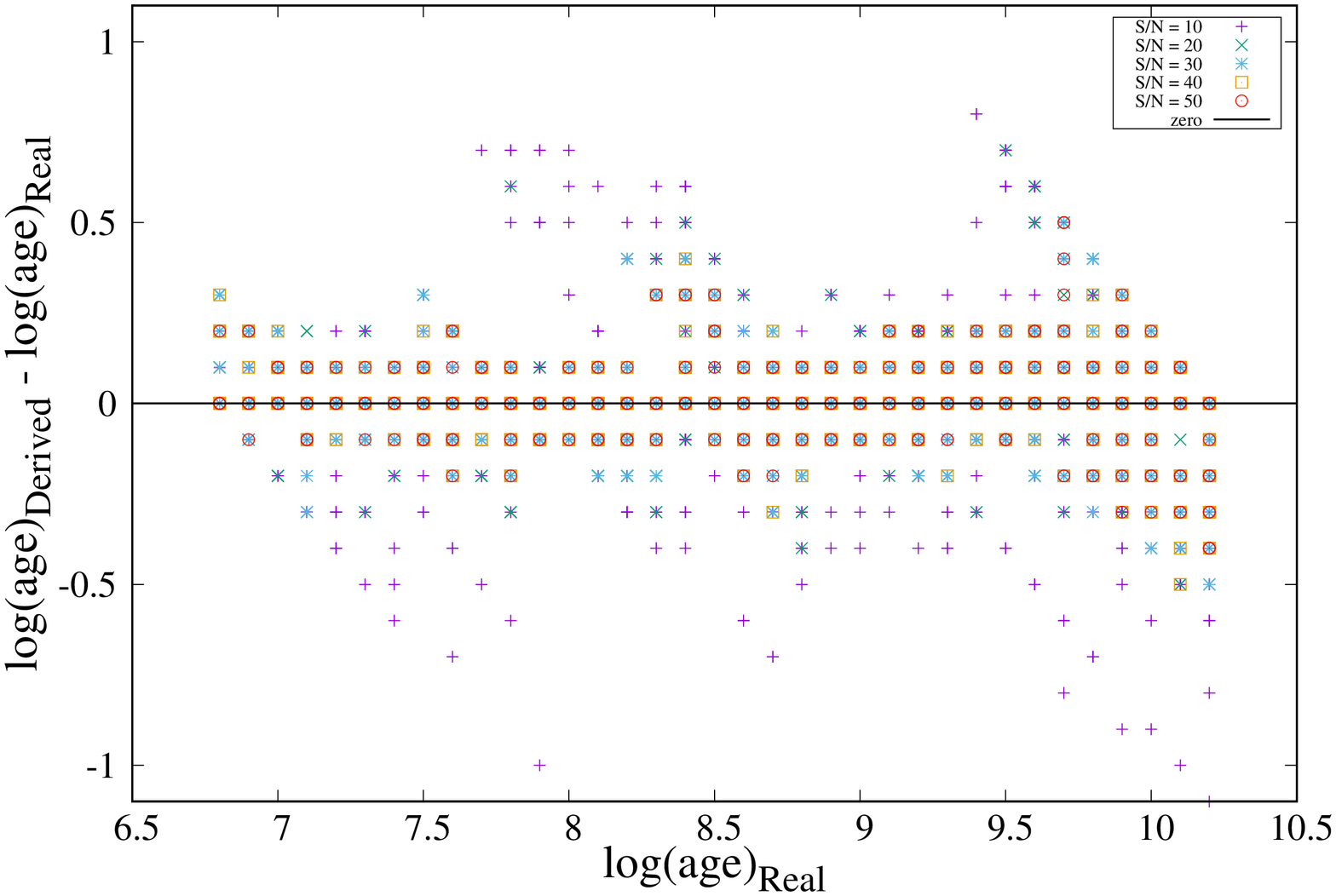}}
\resizebox{75mm}{!}{\includegraphics[angle=270]{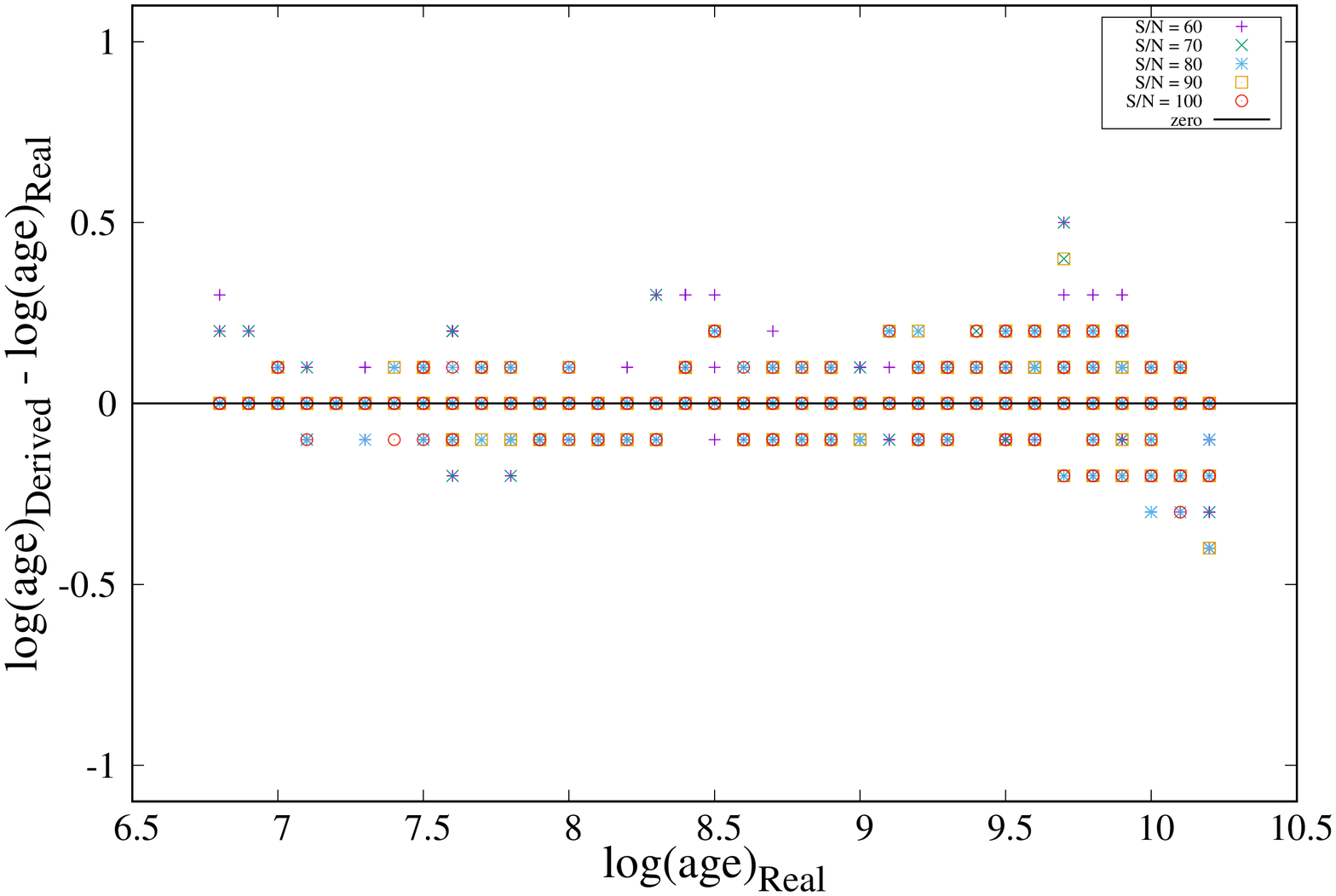}}
\resizebox{75mm}{!}{\includegraphics[angle=270]{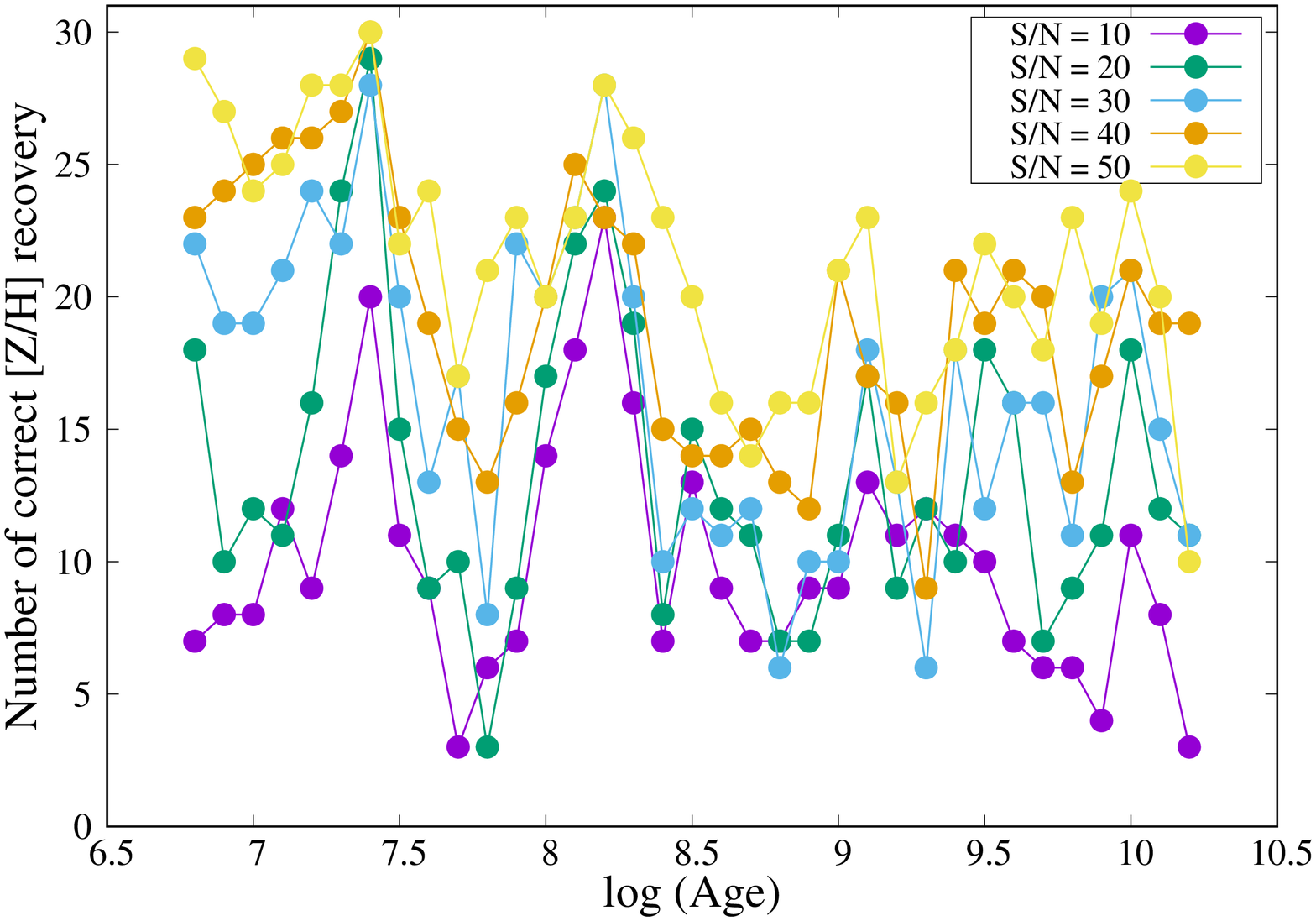}}
\resizebox{75mm}{!}{\includegraphics[angle=270]{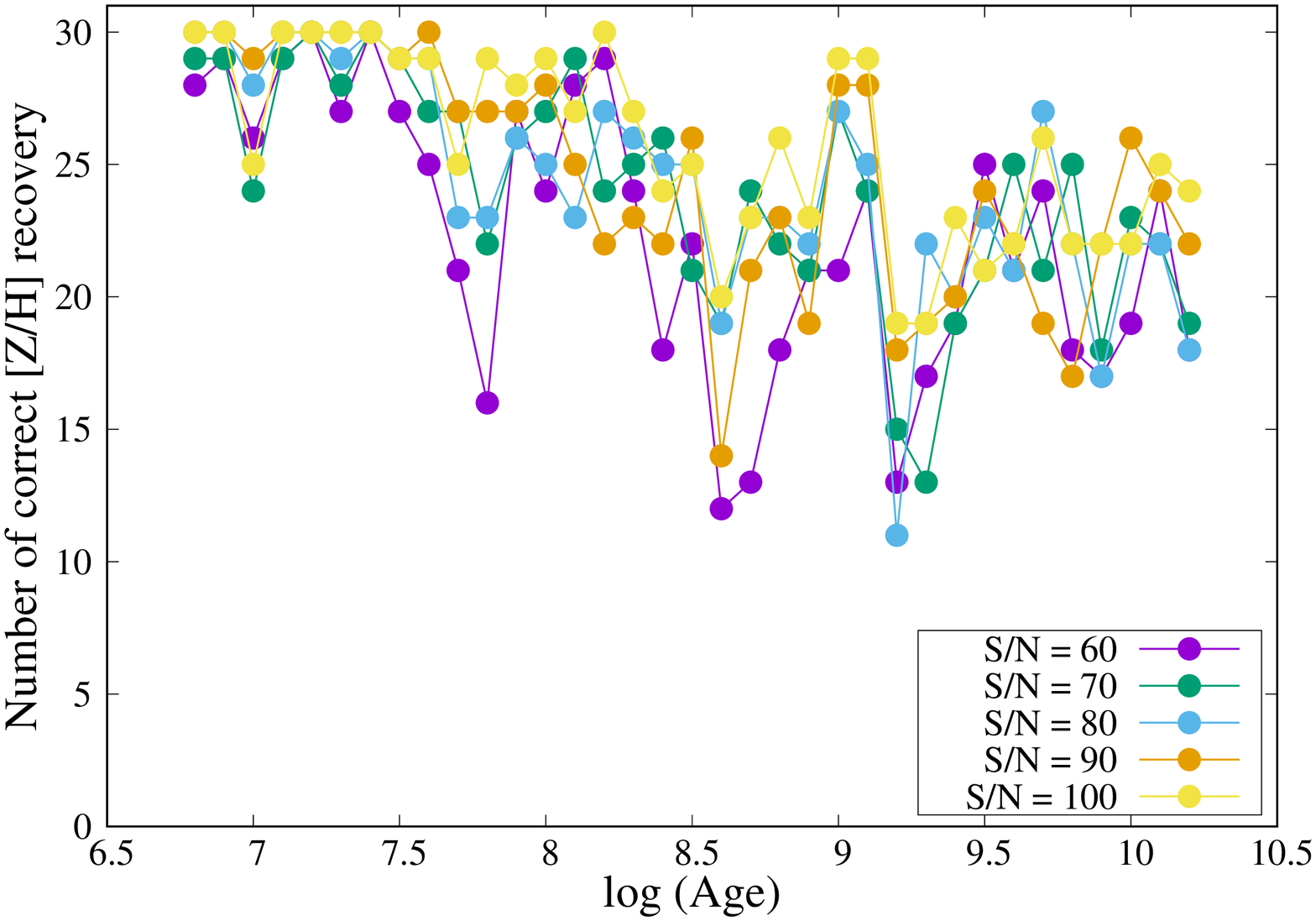}}
\resizebox{75mm}{!}{\includegraphics[width=\columnwidth, angle=270]{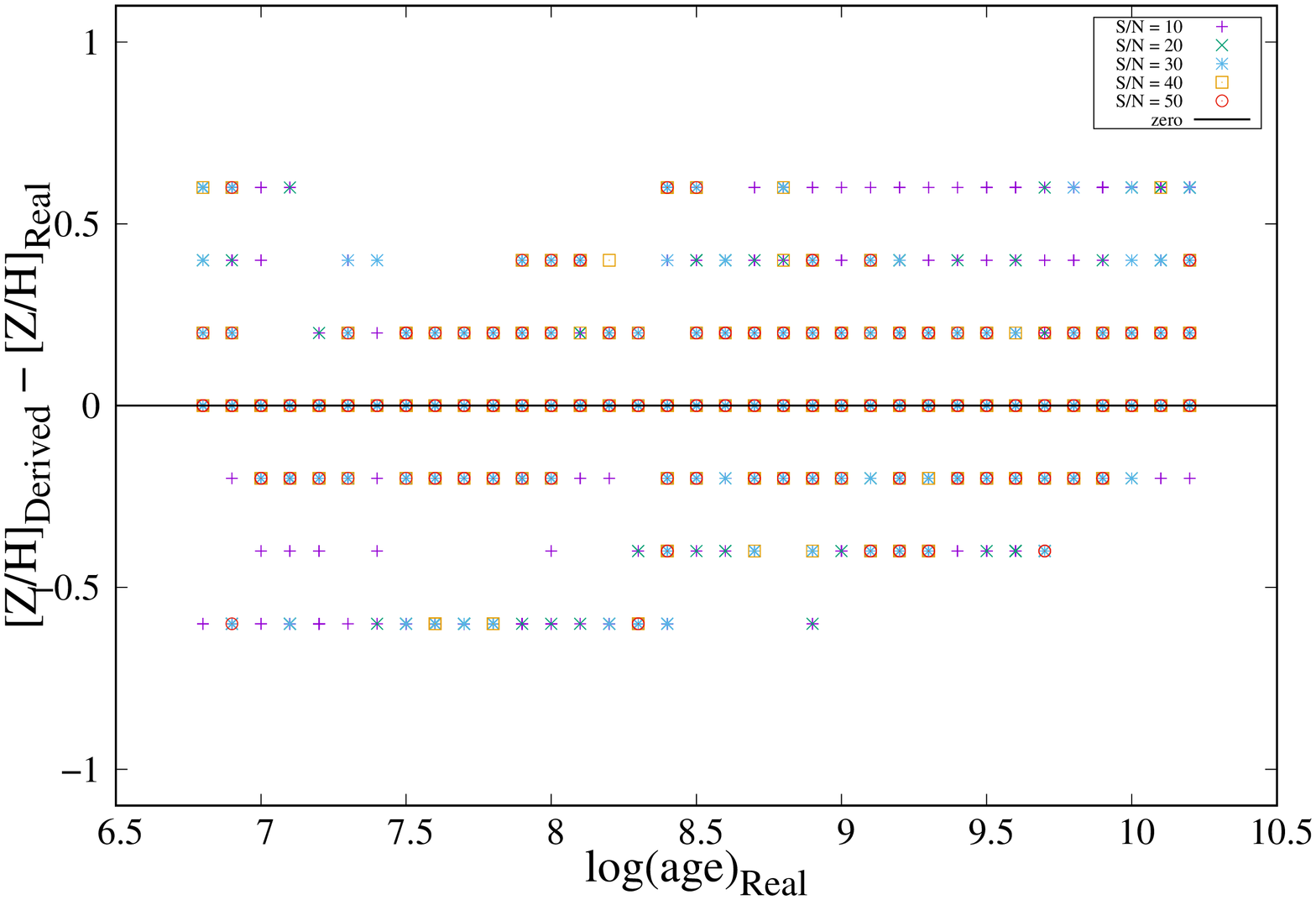}}
\resizebox{75mm}{!}{\includegraphics[width=\columnwidth, angle=270]{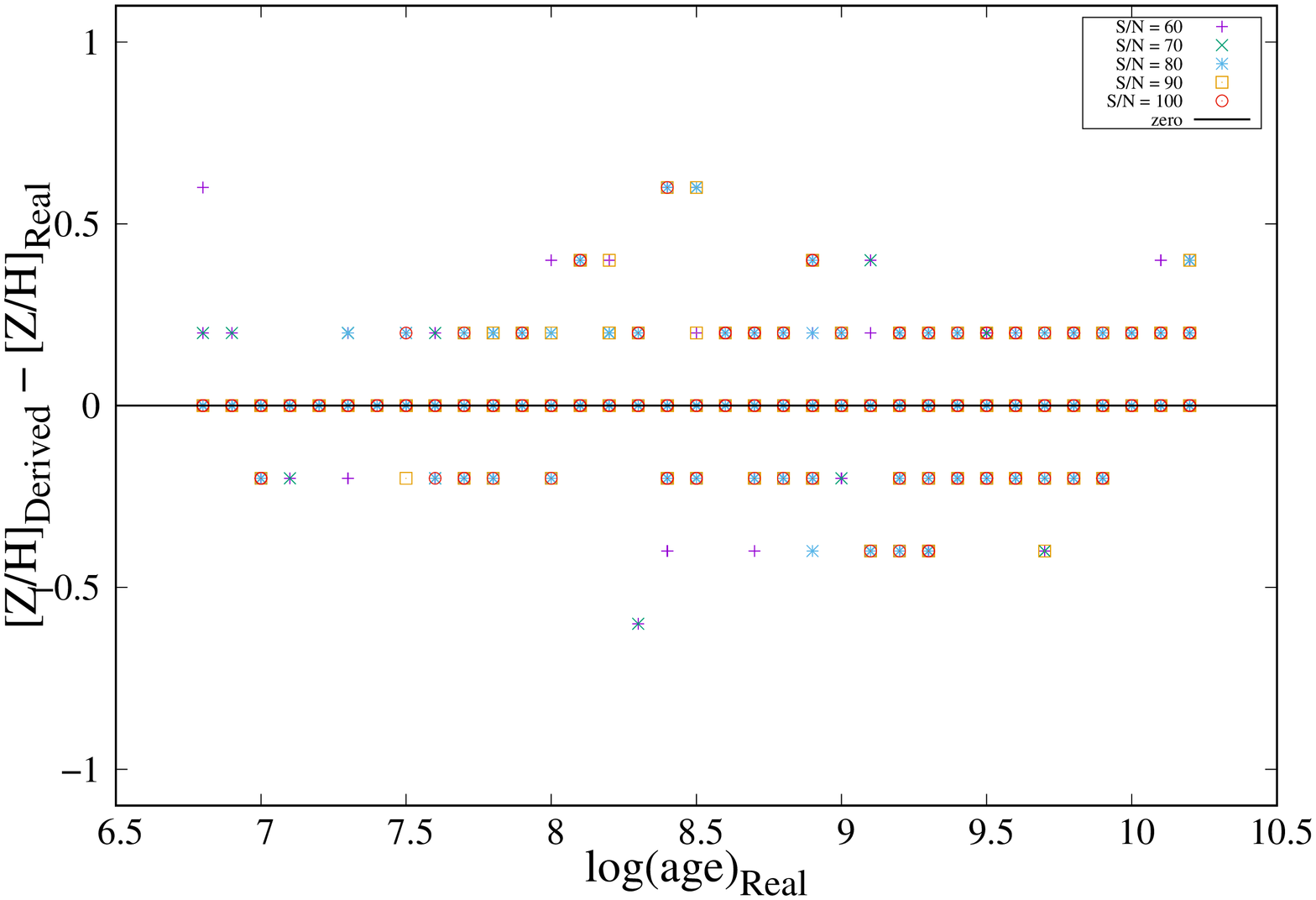}}

\caption{ The results obtained with the full-spectrum fitting
  technique in the range $3700 \leqslant \lambda/\mbox{\AA} \leqslant
  6200$ using Padova isochrone models. The left panels show the
  results for 10 $\leqslant$ S/N $\leqslant$ 50 and the right panels
  do so for 60 $\leqslant$ S/N $\leqslant$ 100. 
The first panel shows the number of correctly recovered age as a function of
input age. The second panel shows (Derived log(age) $-$ Real log(age))
versus Real log(age).  The third panel shows the number of correctly
recovered metallicity as a function of log(age). The fourth panel
shows (Derived metallicity  - Real metallicity ) versus Real
log(age). }

\label{Fig1}
\end{figure*}

\begin{figure*}

\resizebox{75mm}{!}{\includegraphics[width=\columnwidth, angle=270]{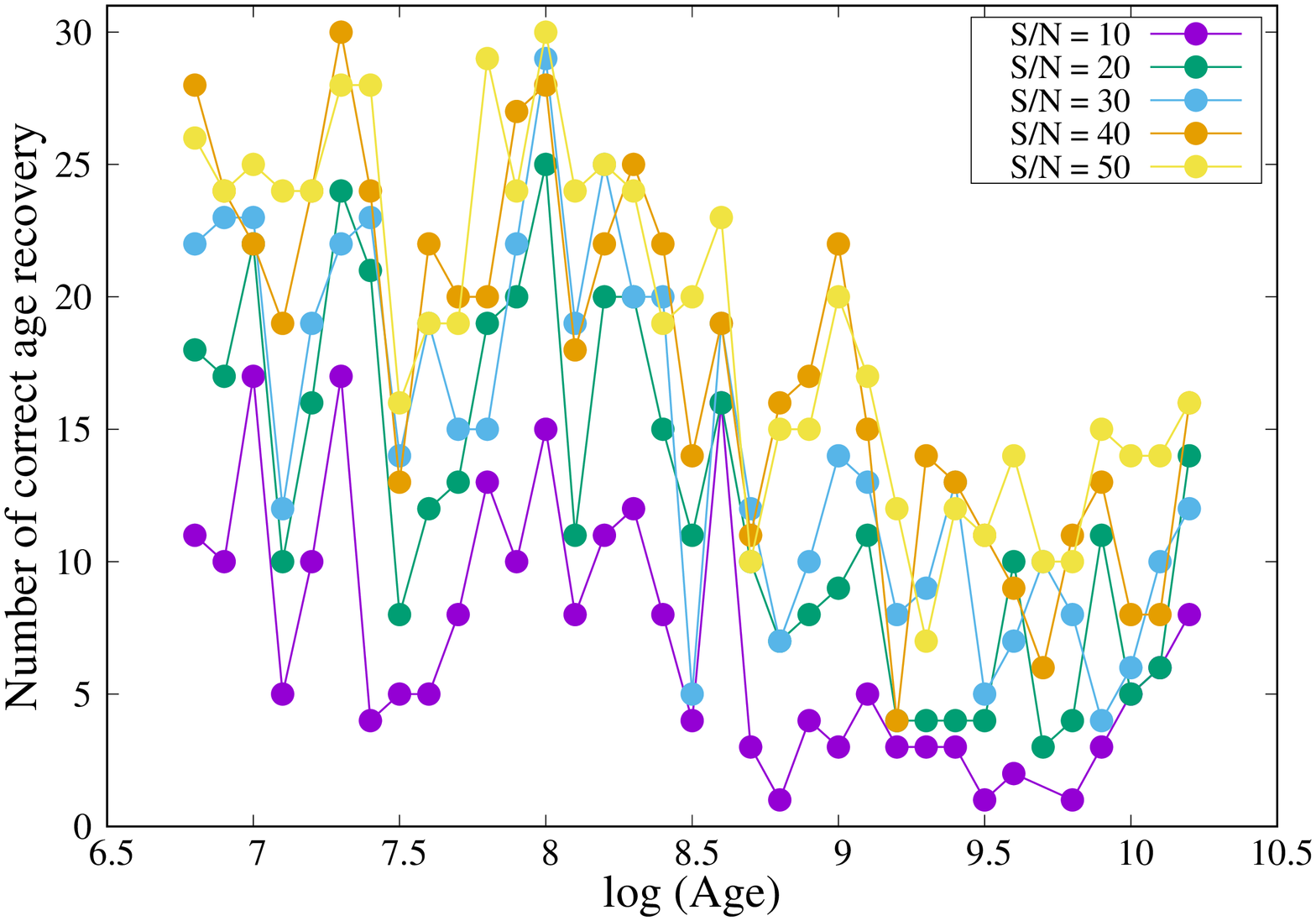}}
\resizebox{75mm}{!}{\includegraphics[width=\columnwidth, angle=270]{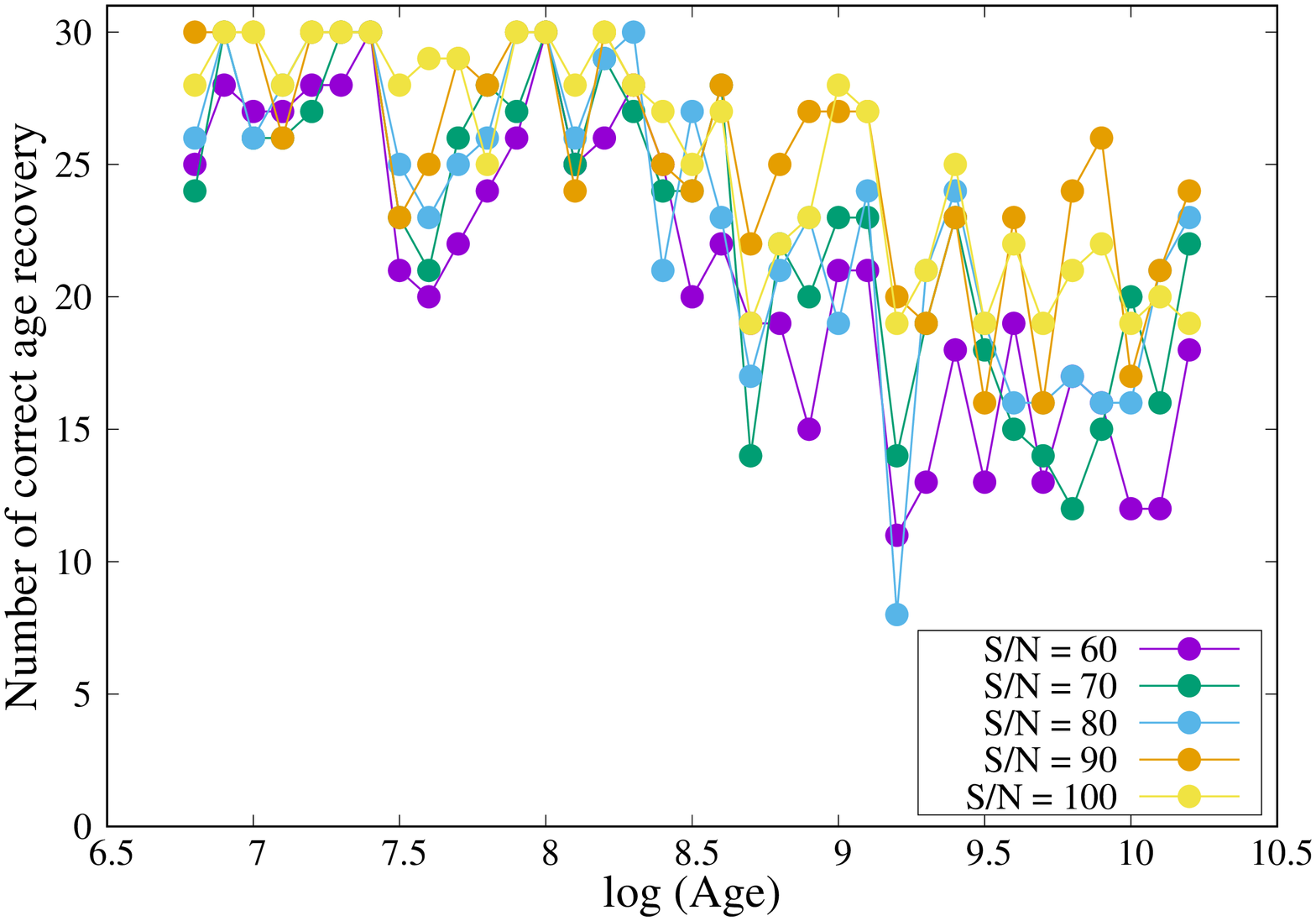}}

\resizebox{75mm}{!}{\includegraphics[width=\columnwidth, angle=270]{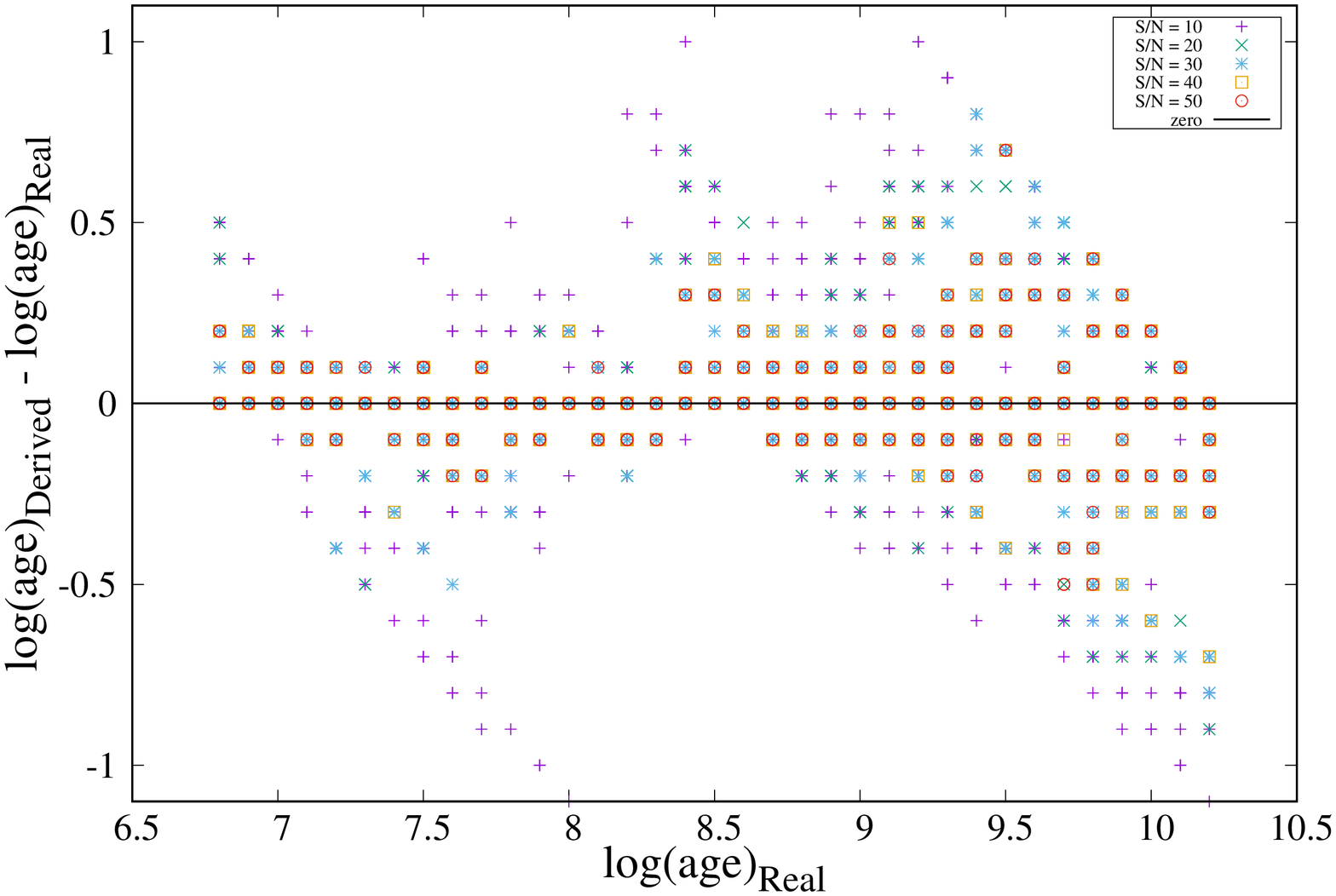}}
\resizebox{75mm}{!}{\includegraphics[width=\columnwidth, angle=270]{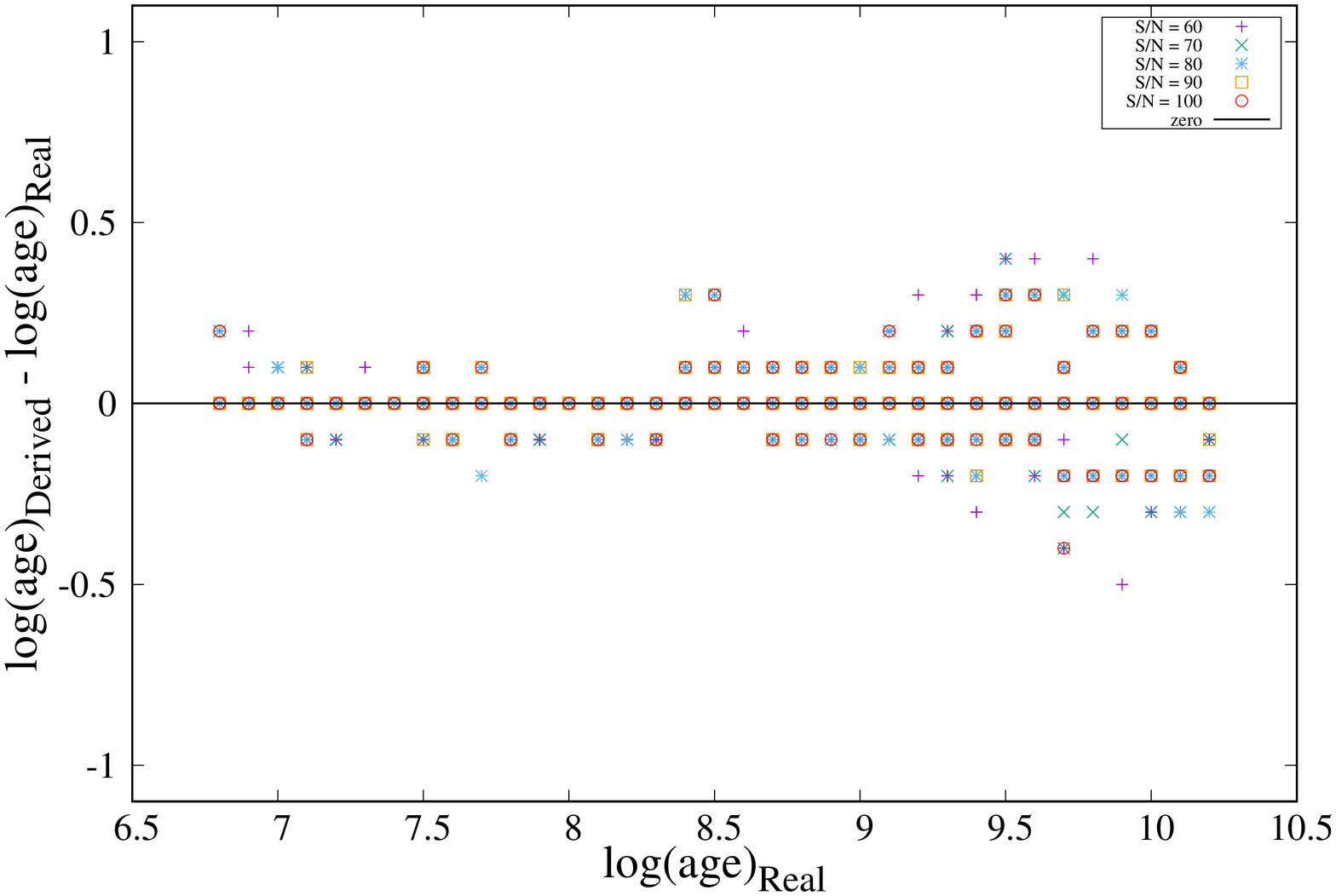}}
\resizebox{75mm}{!}{\includegraphics[angle=270]{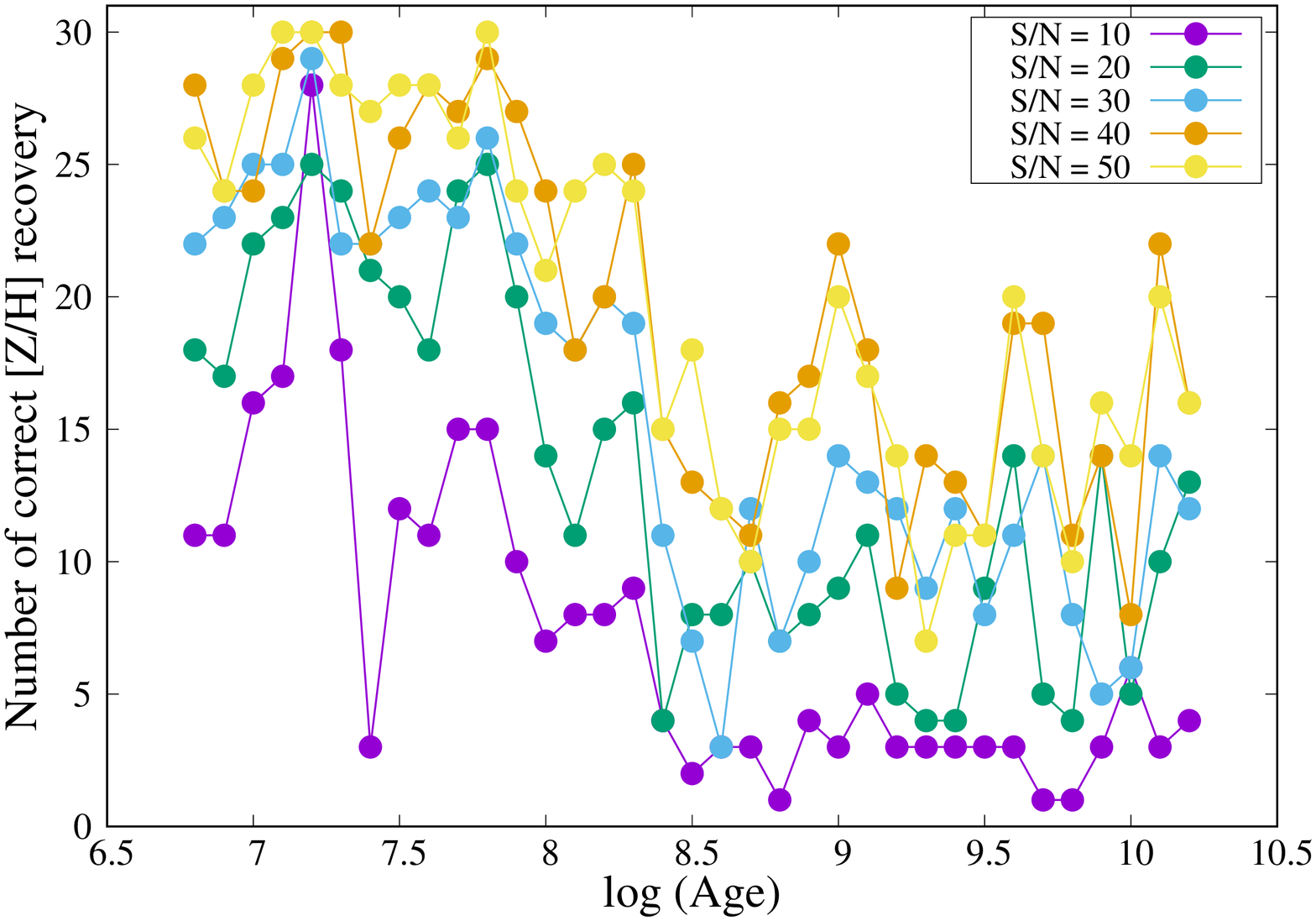}}
\resizebox{75mm}{!}{\includegraphics[angle=270]{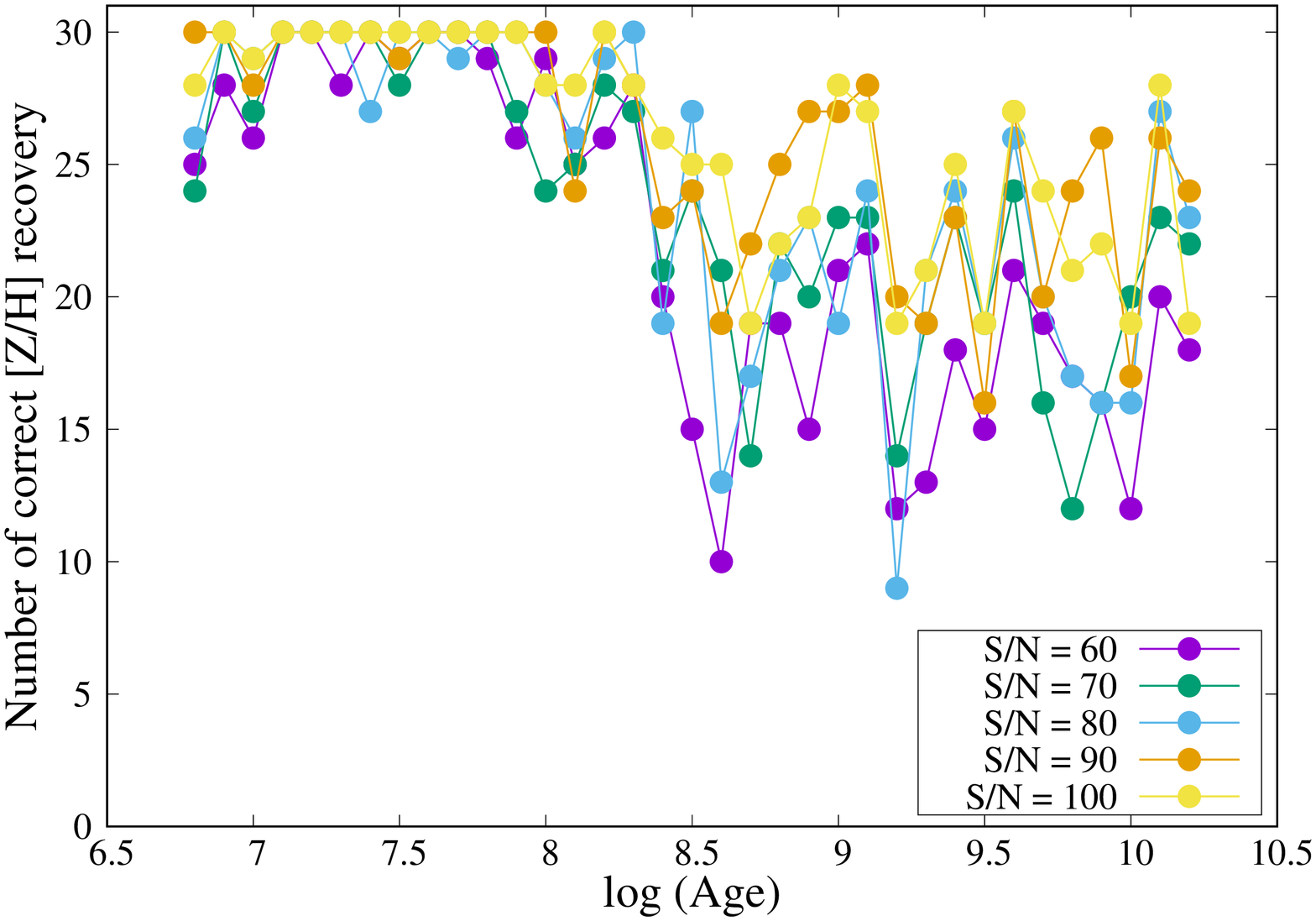}}
\resizebox{75mm}{!}{\includegraphics[width=\columnwidth, angle=270]{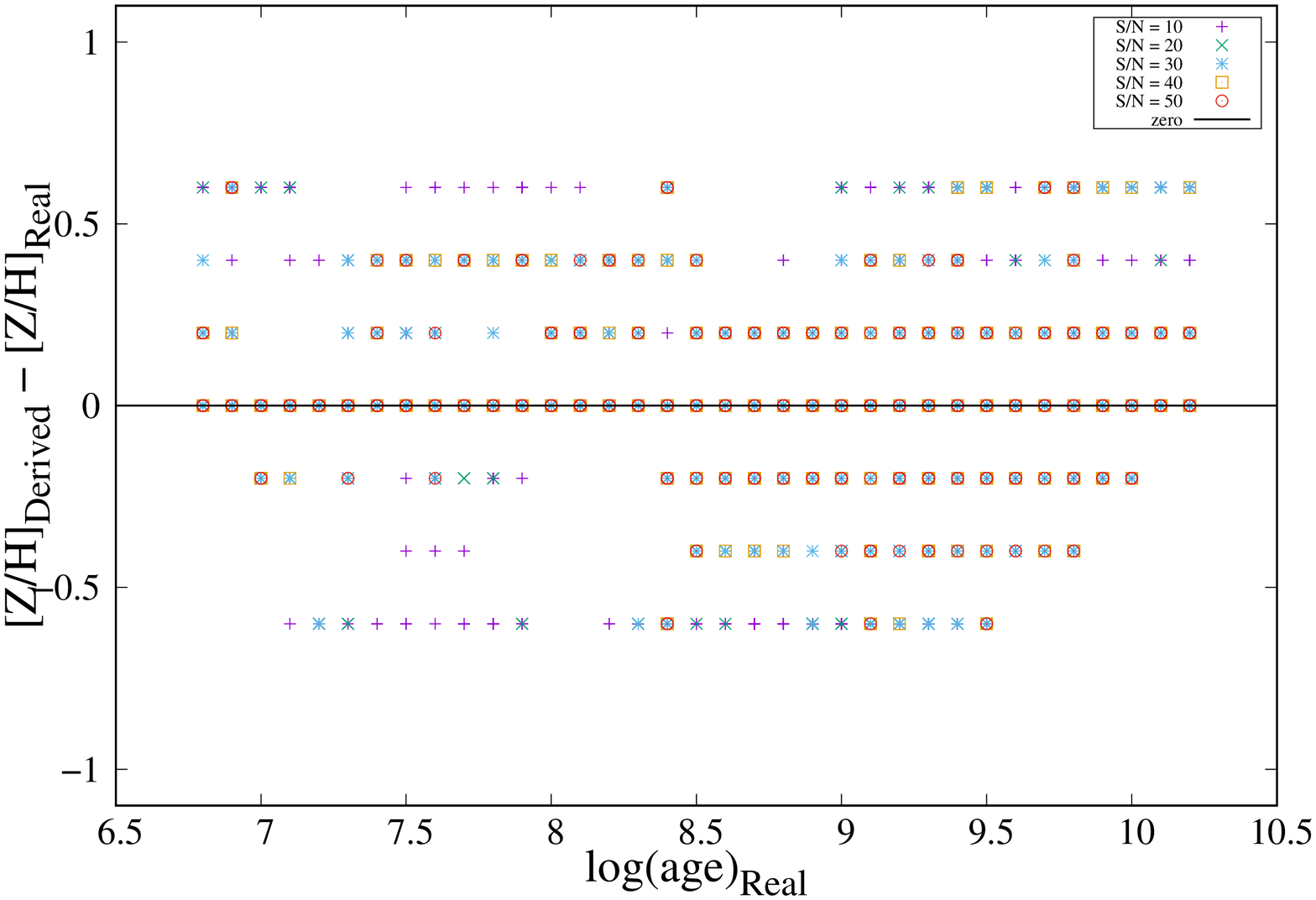}}
\resizebox{75mm}{!}{\includegraphics[width=\columnwidth, angle=270]{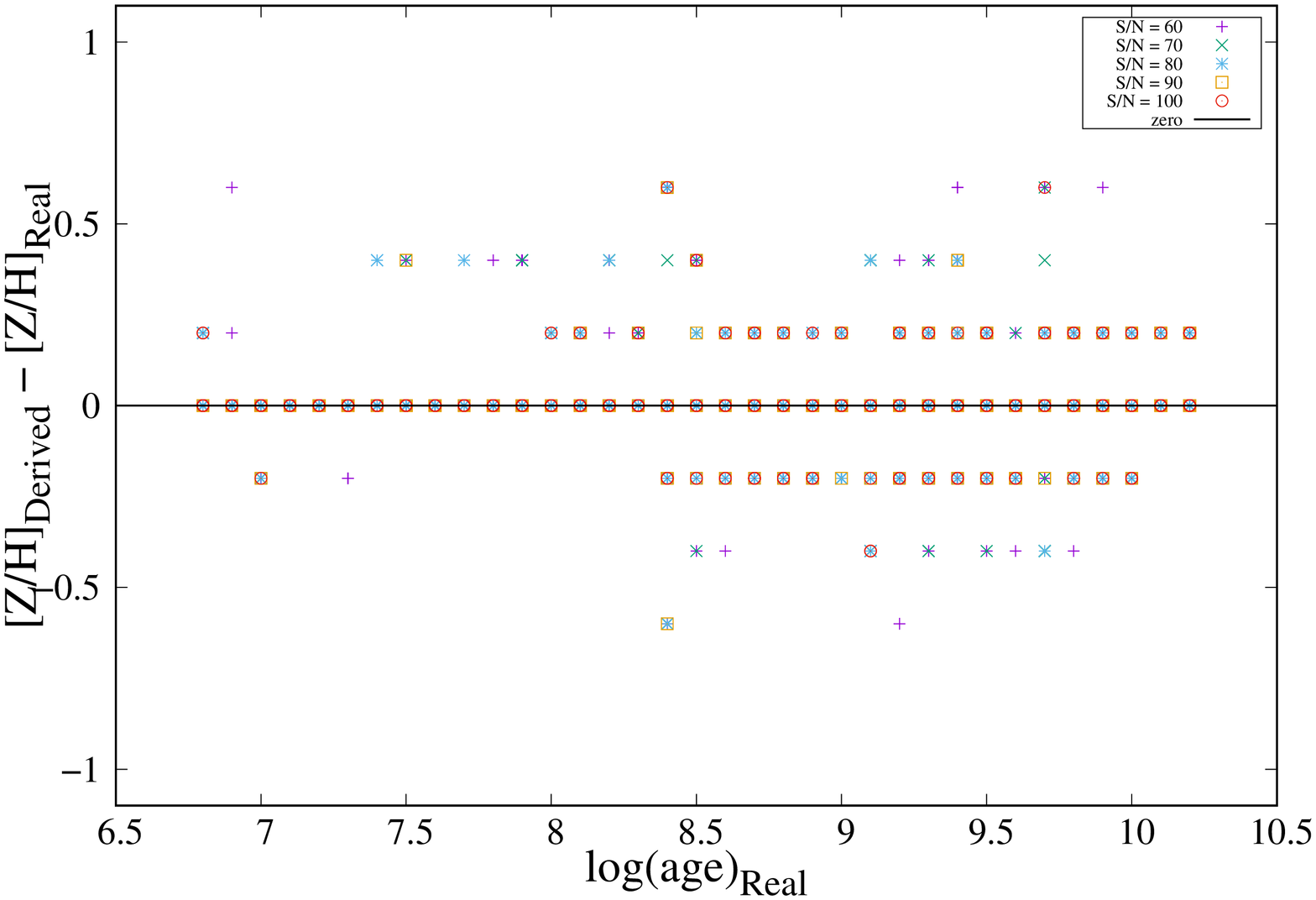}}

\caption{-PADOVA- The same as figure \ref{Fig1} for the range $3700 \leqslant \lambda/\mbox{\AA} \leqslant 5000$ }
\label{Fig2}
\end{figure*}

\begin{figure*}

\resizebox{75mm}{!}{\includegraphics[width=\columnwidth, angle=270]{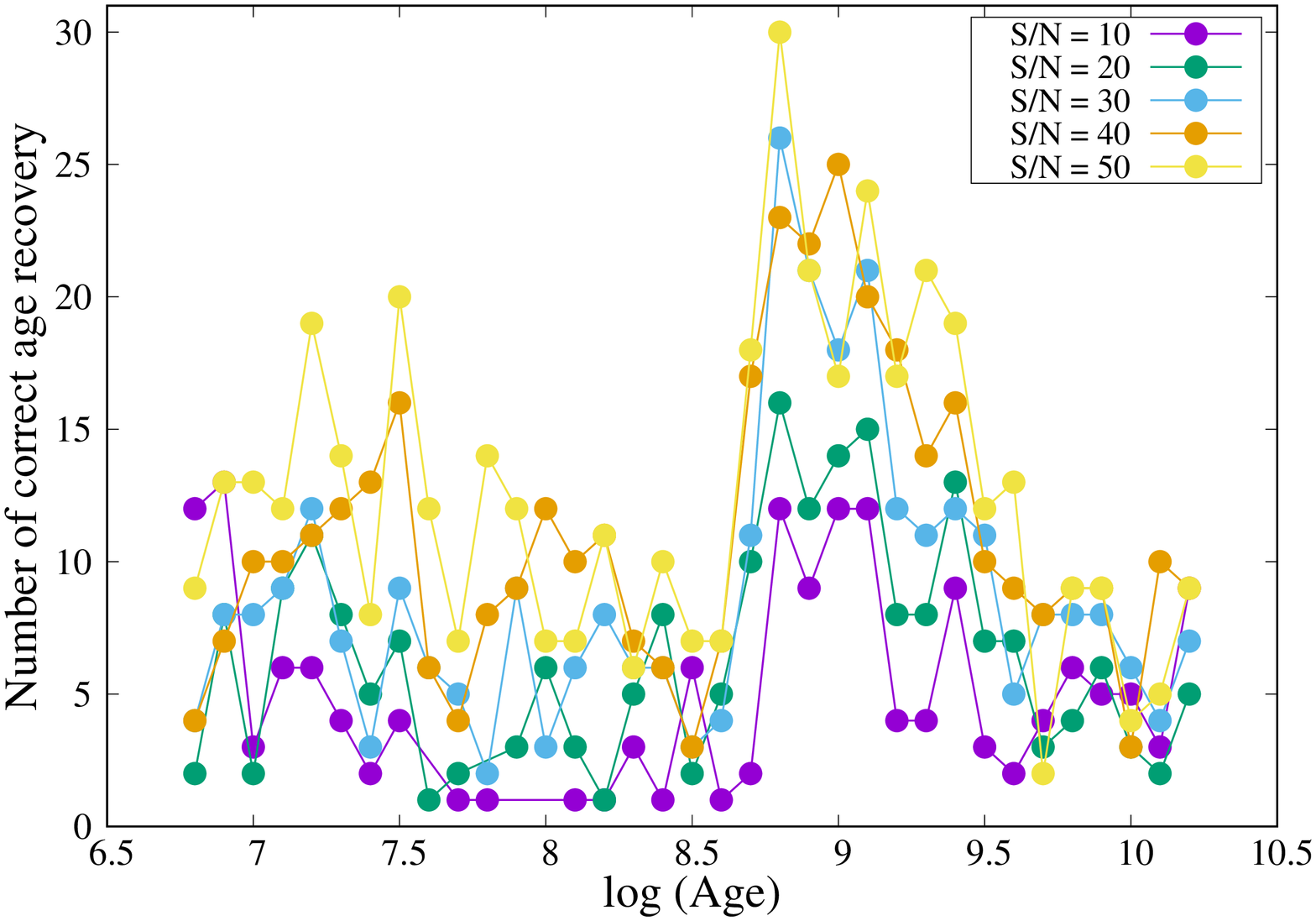}}
\resizebox{75mm}{!}{\includegraphics[width=\columnwidth, angle=270]{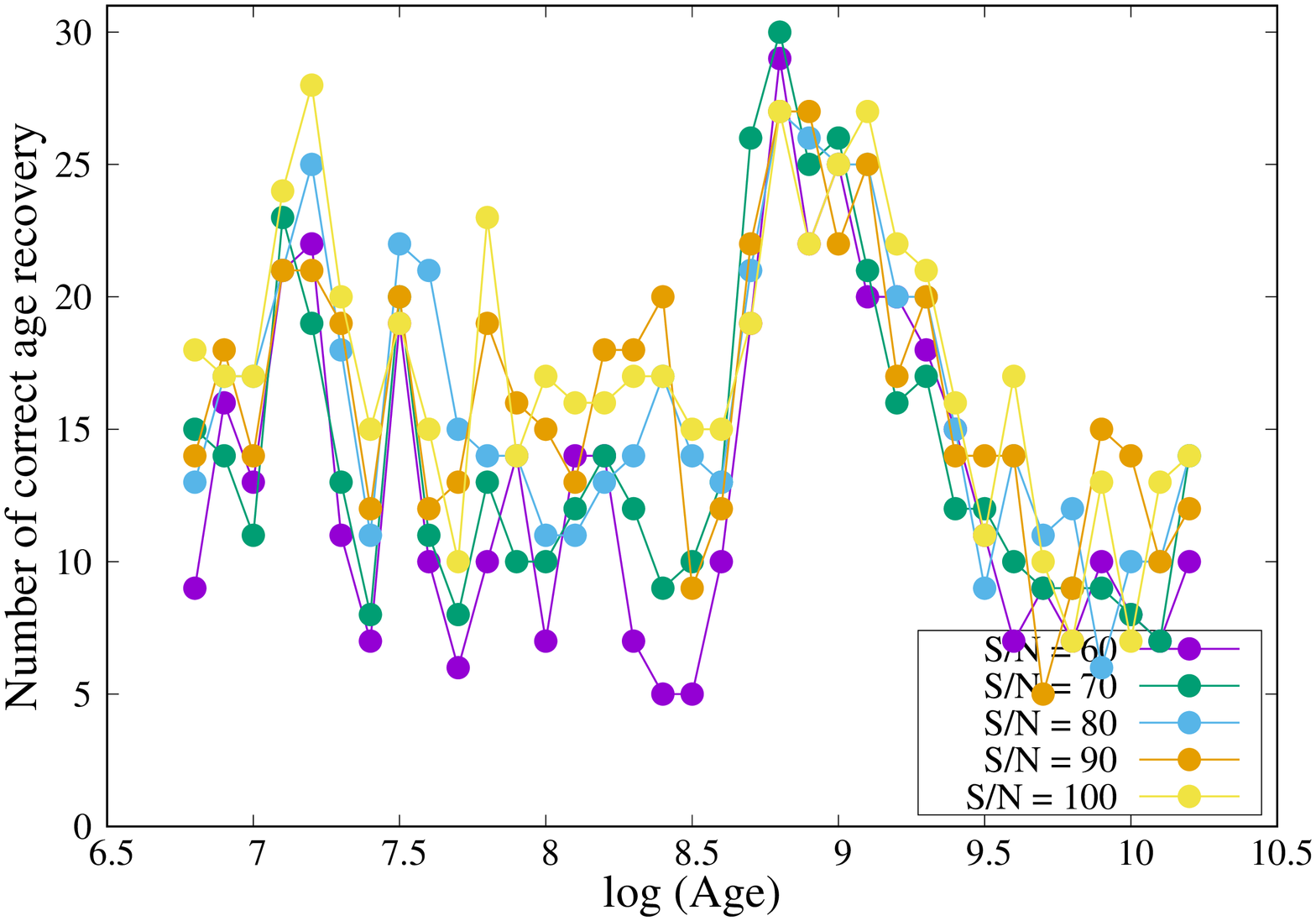}}
\resizebox{75mm}{!}{\includegraphics[angle=270]{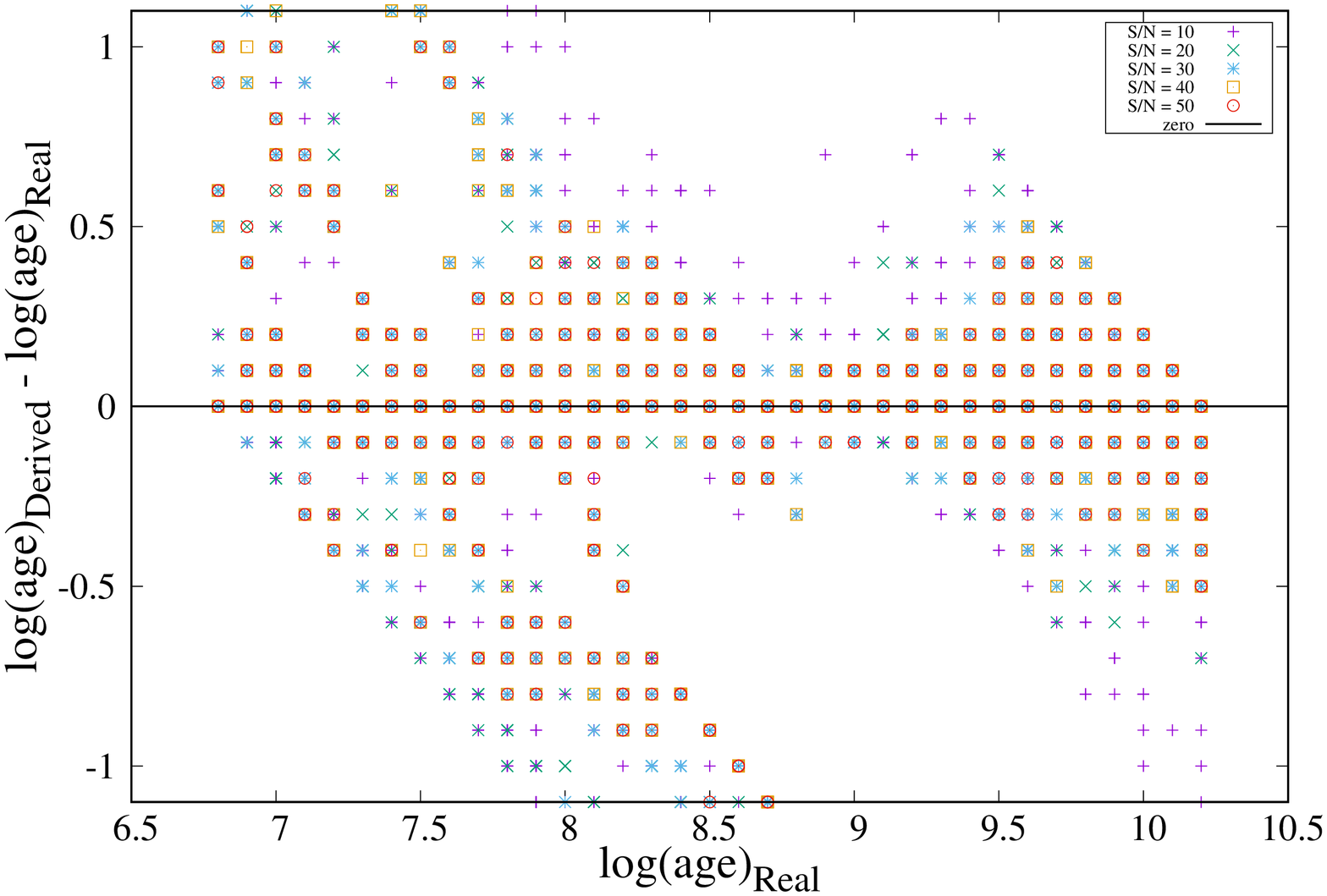}}
\resizebox{75mm}{!}{\includegraphics[angle=270]{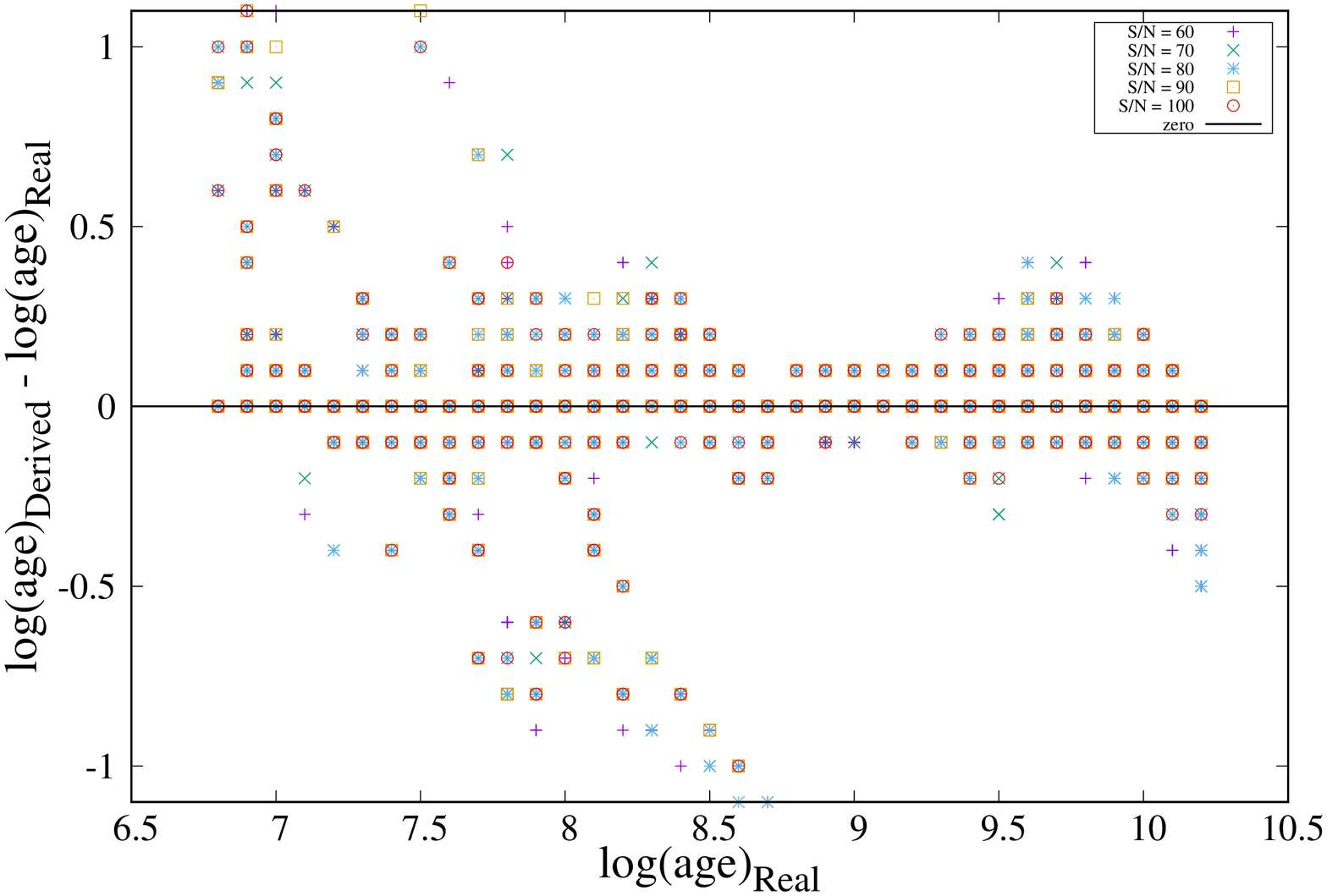}}
\resizebox{75mm}{!}{\includegraphics[angle=270]{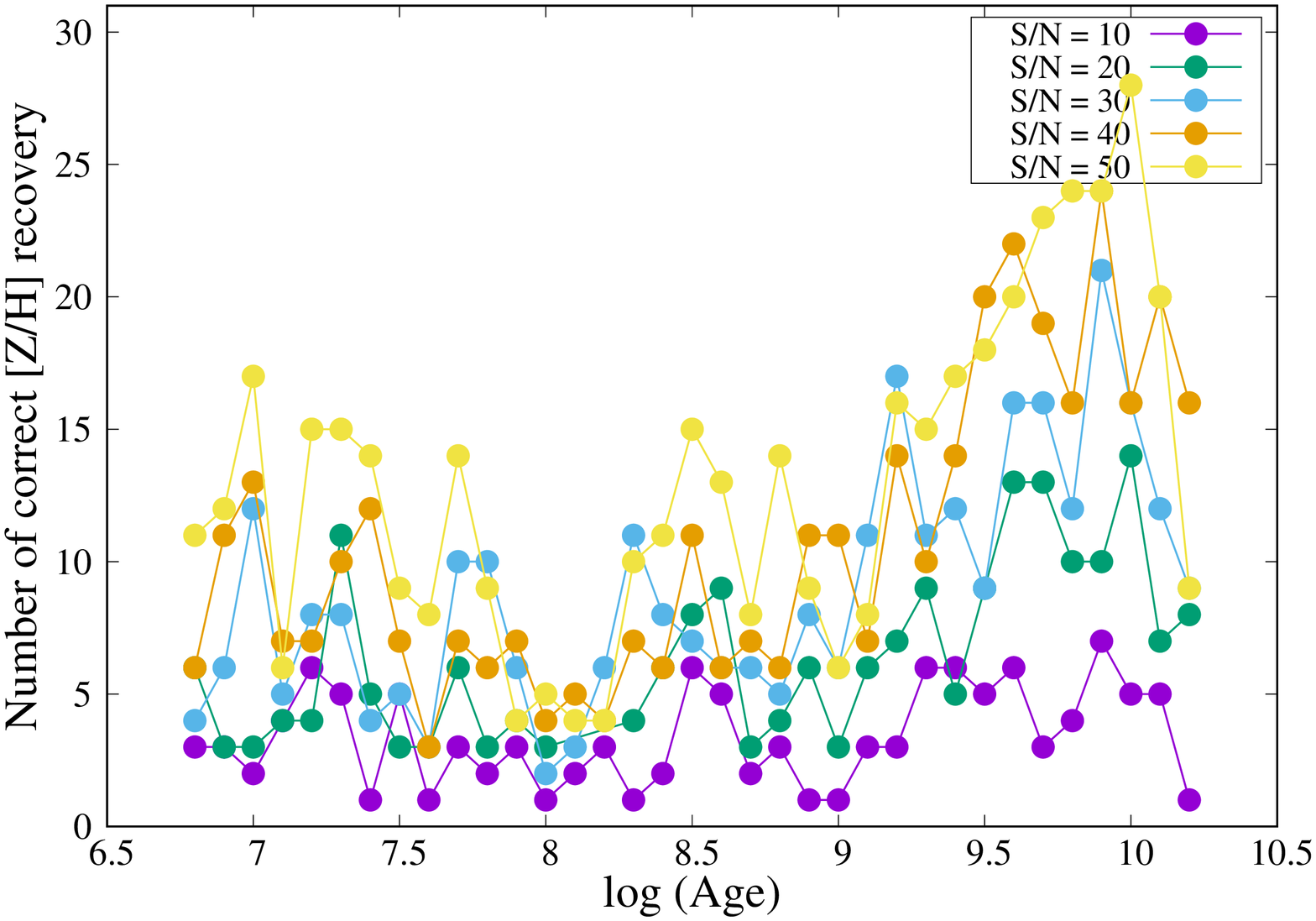}}
\resizebox{75mm}{!}{\includegraphics[angle=270]{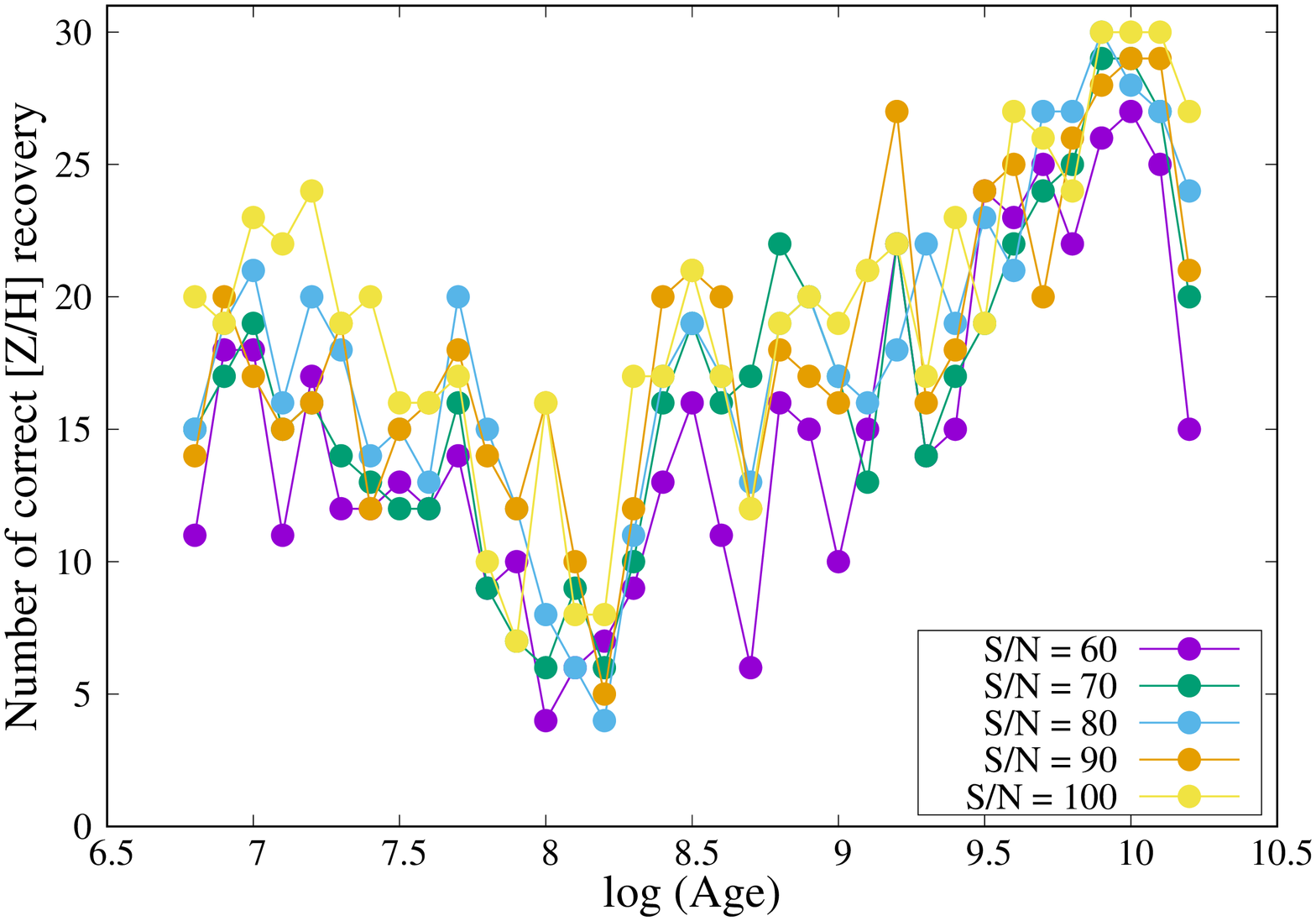}}
\resizebox{75mm}{!}{\includegraphics[width=\columnwidth, angle=270]{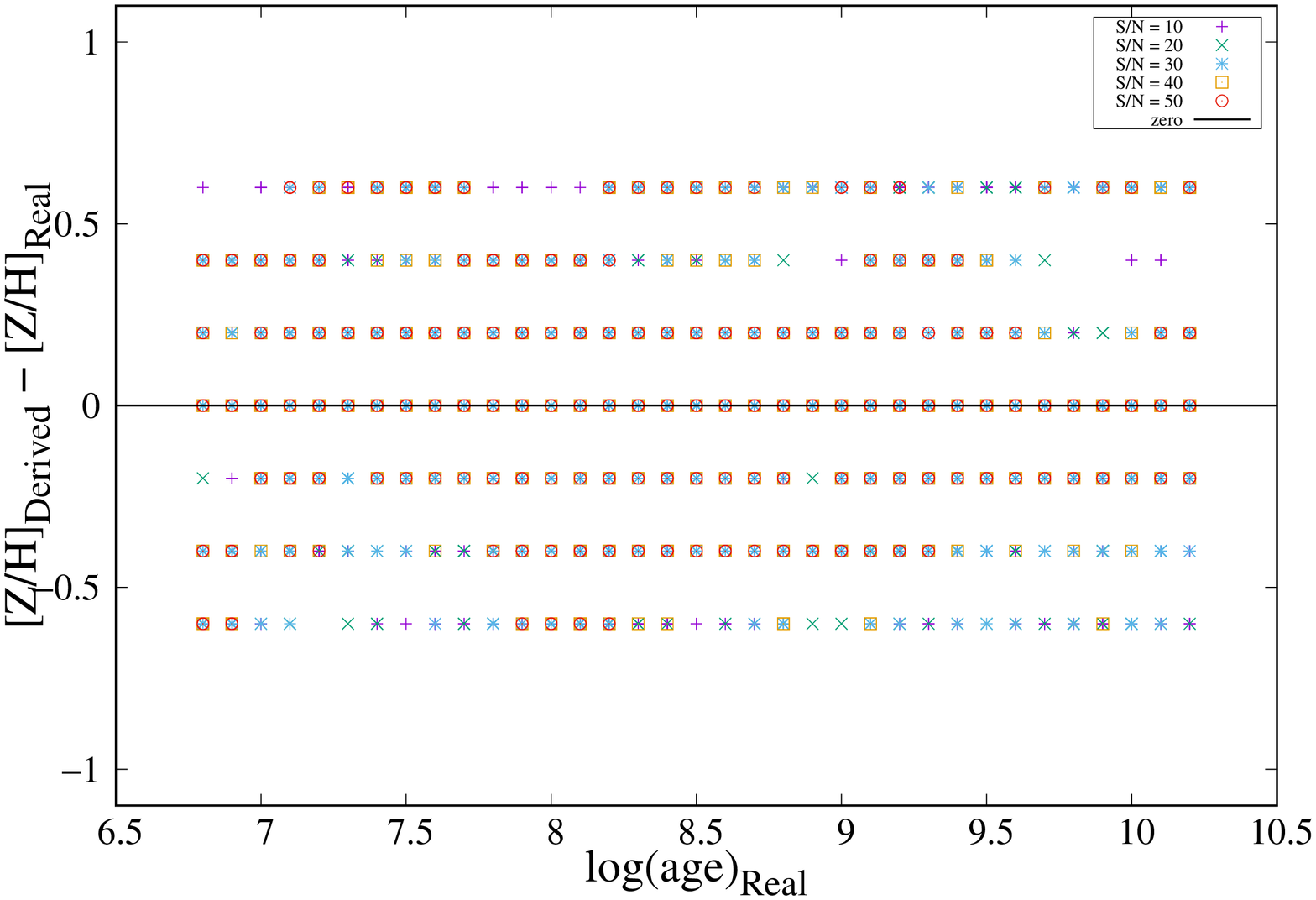}}
\resizebox{75mm}{!}{\includegraphics[width=\columnwidth, angle=270]{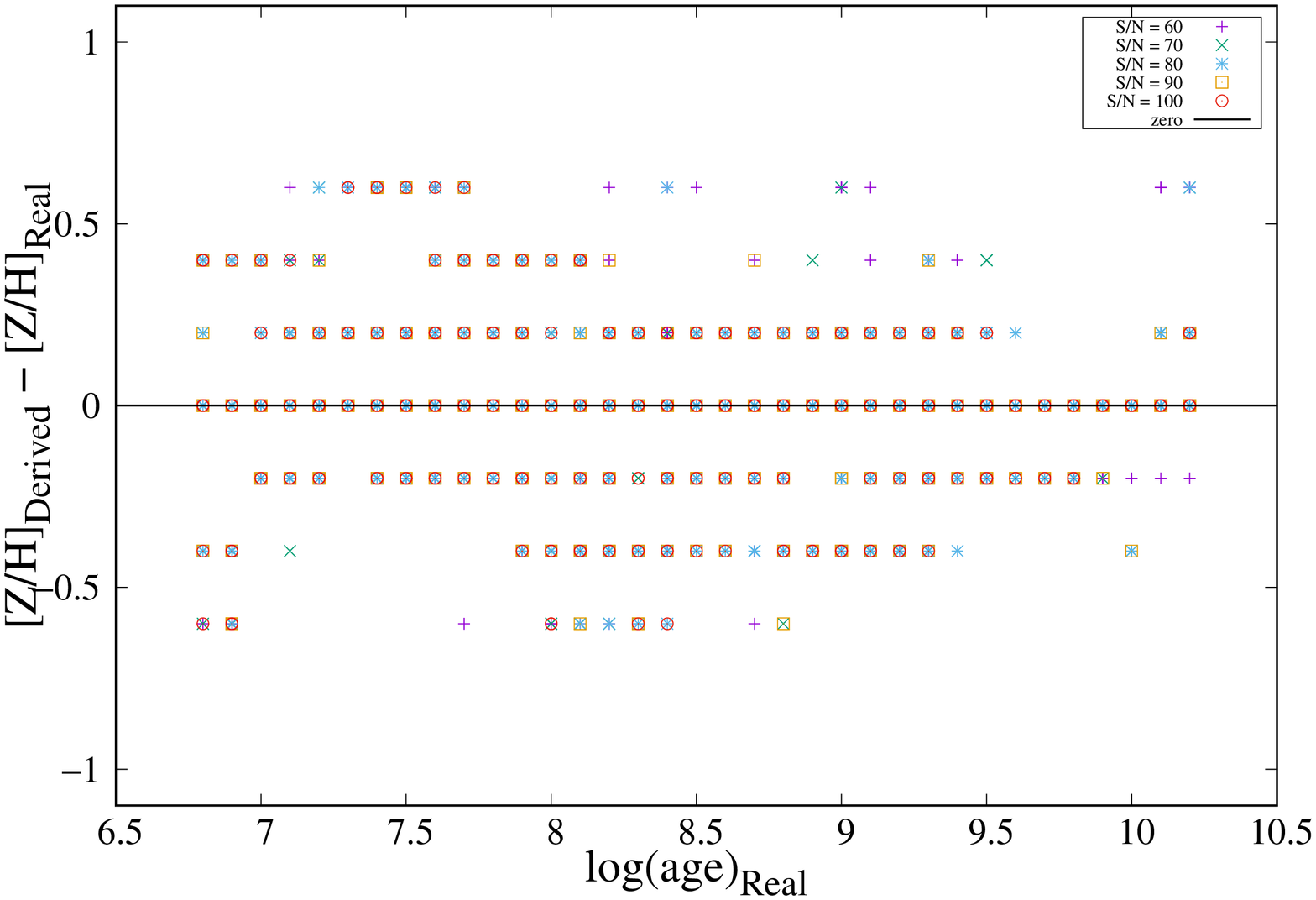}}

\caption{-PADOVA- The same as figure \ref{Fig1} for the range  $5000 \leqslant
  \lambda/\mbox{\AA} \leqslant 6200$ }
\label{Fig3}
\end{figure*}

\begin{figure*}

\resizebox{75mm}{!}{\includegraphics[width=\columnwidth, angle=270]{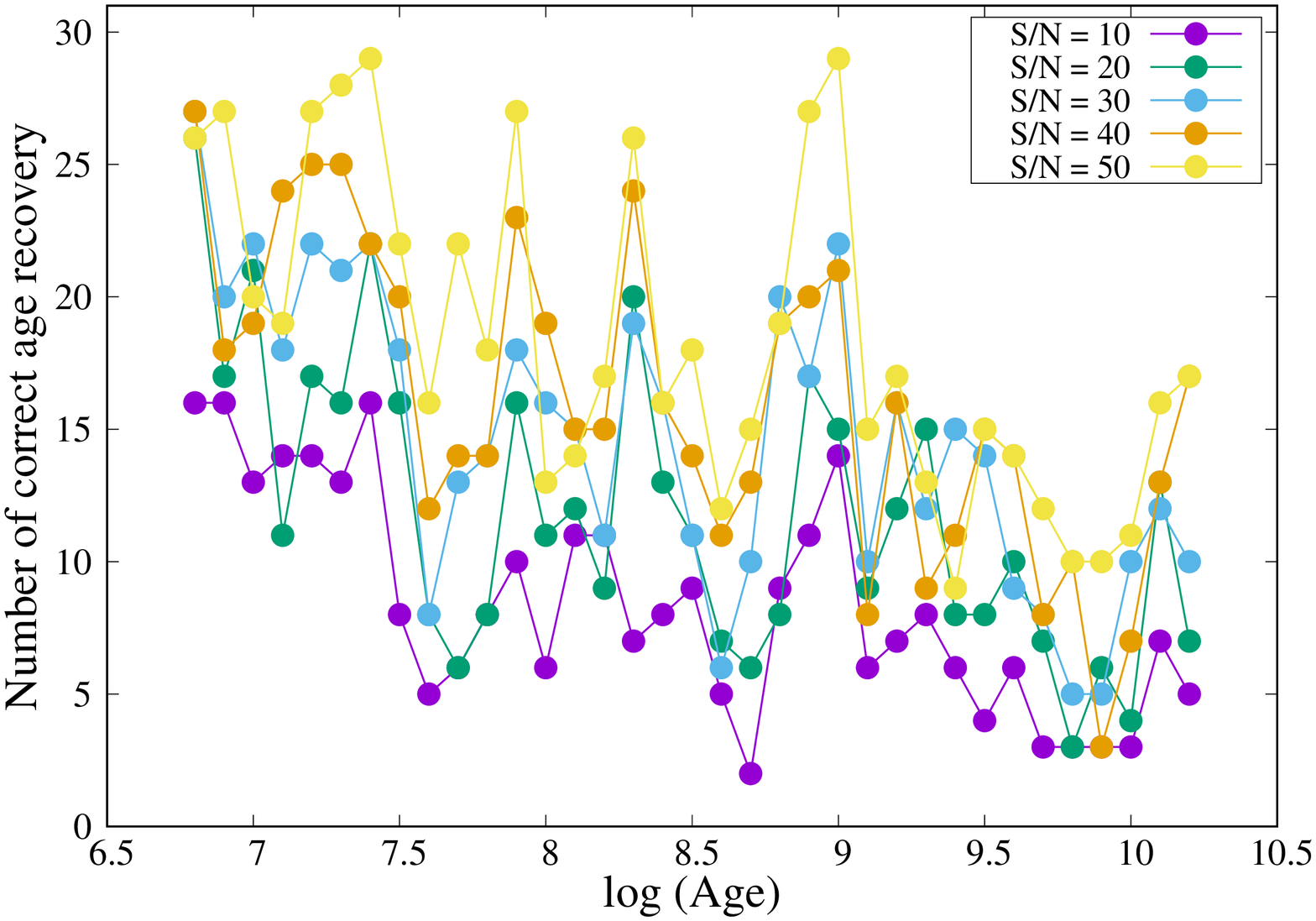}}
\resizebox{75mm}{!}{\includegraphics[width=\columnwidth, angle=270]{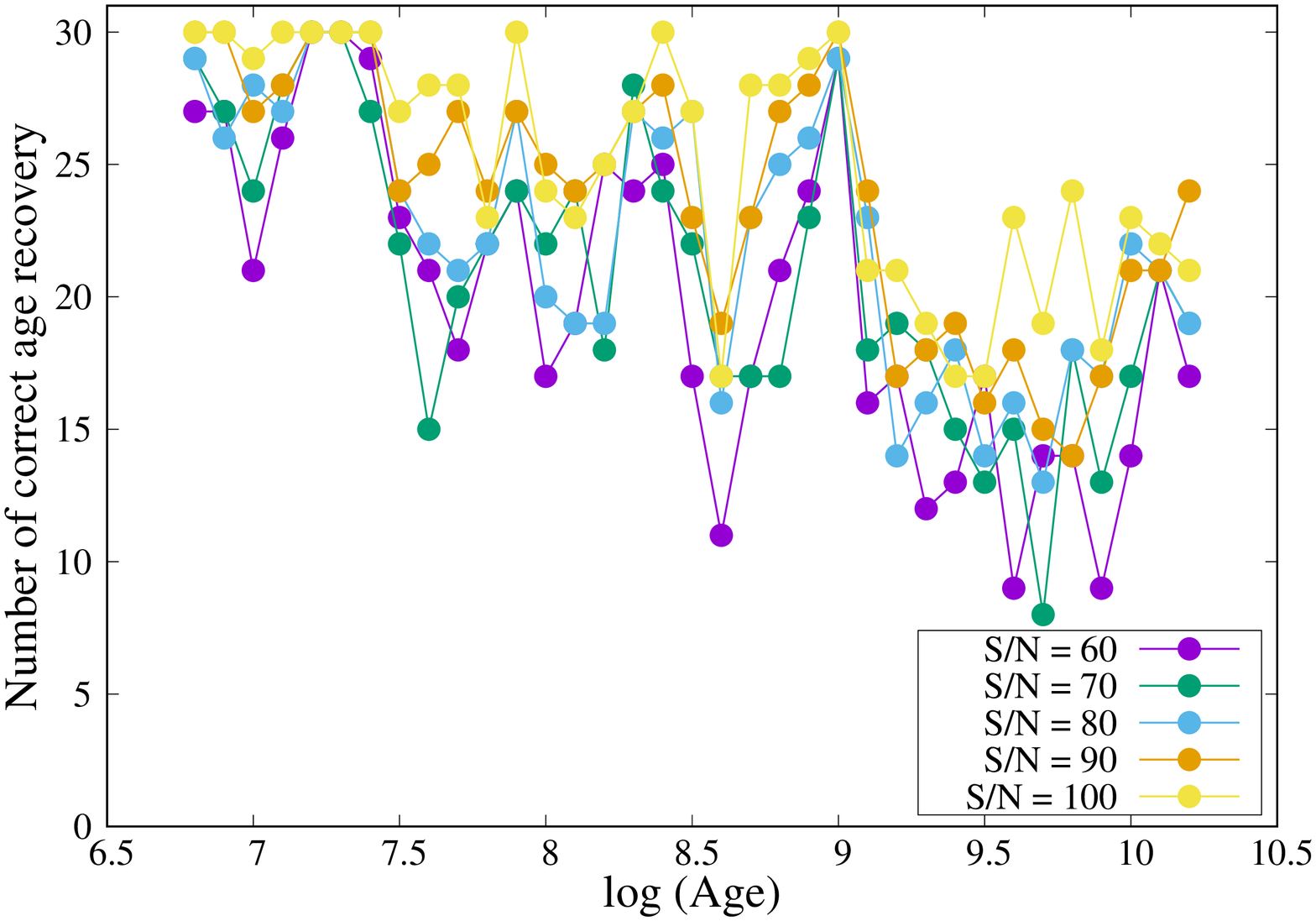}}
\resizebox{75mm}{!}{\includegraphics[width=\columnwidth, angle=270]{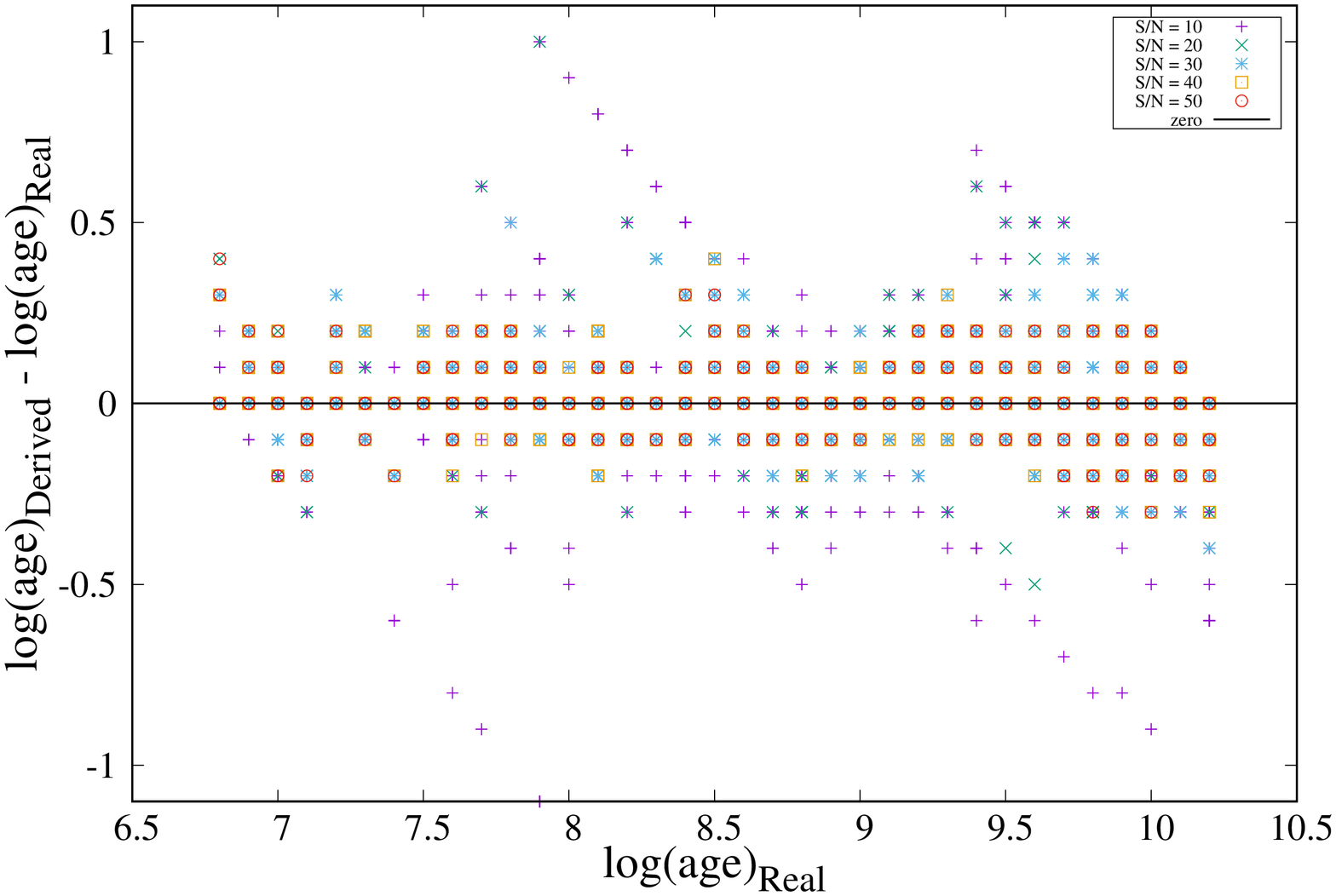}}
\resizebox{75mm}{!}{\includegraphics[width=\columnwidth, angle=270]{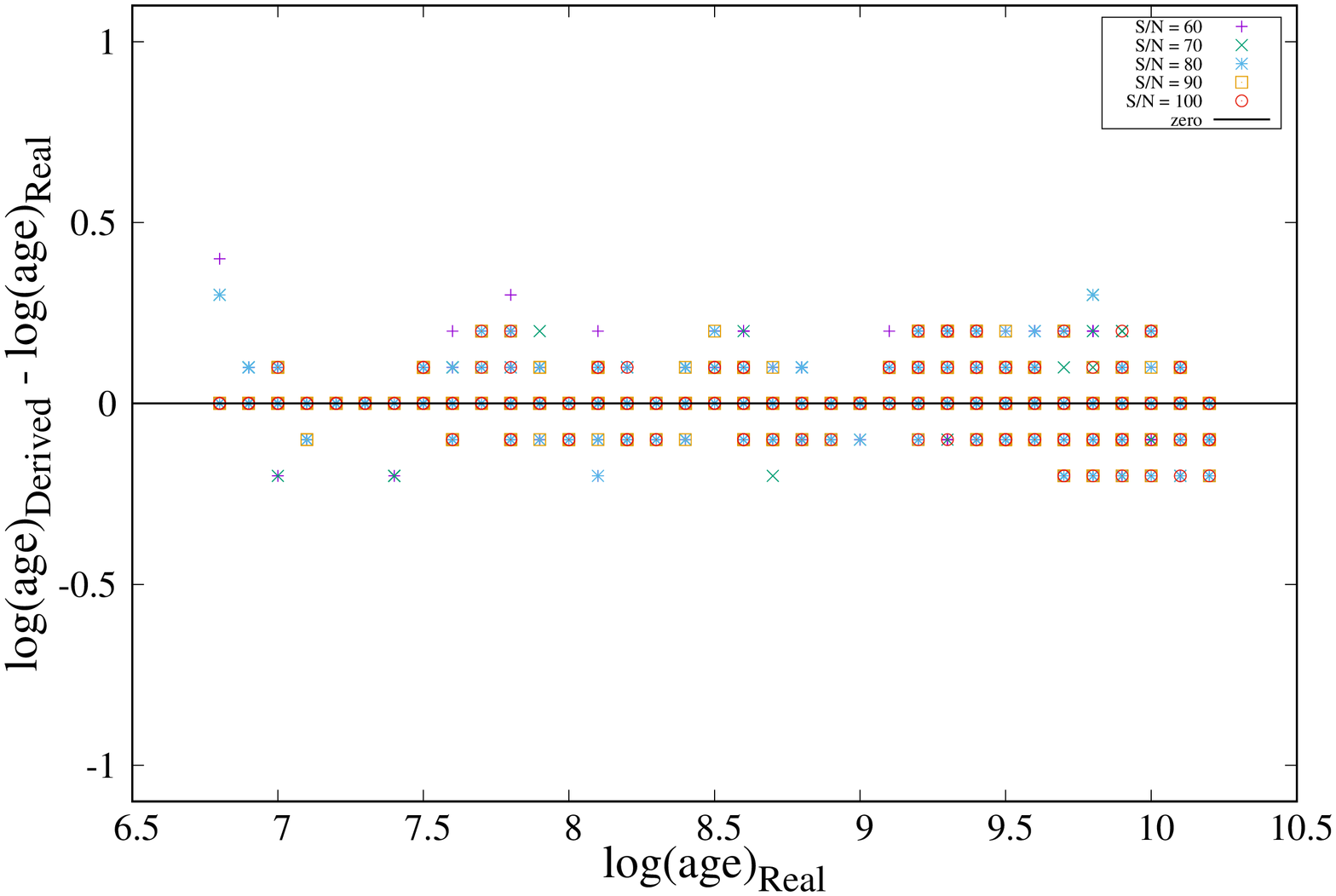}}
\resizebox{75mm}{!}{\includegraphics[angle=270]{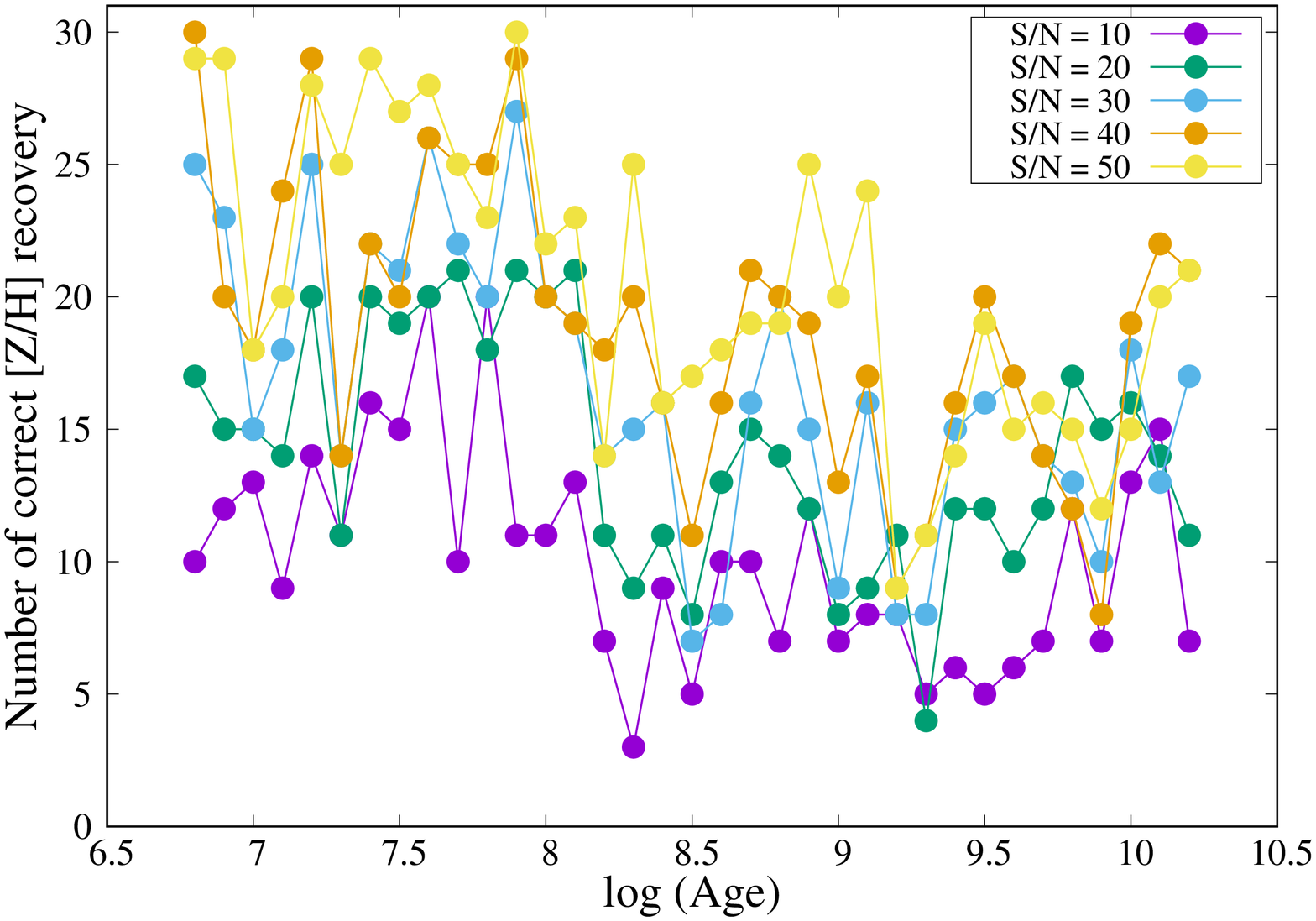}}
\resizebox{75mm}{!}{\includegraphics[angle=270]{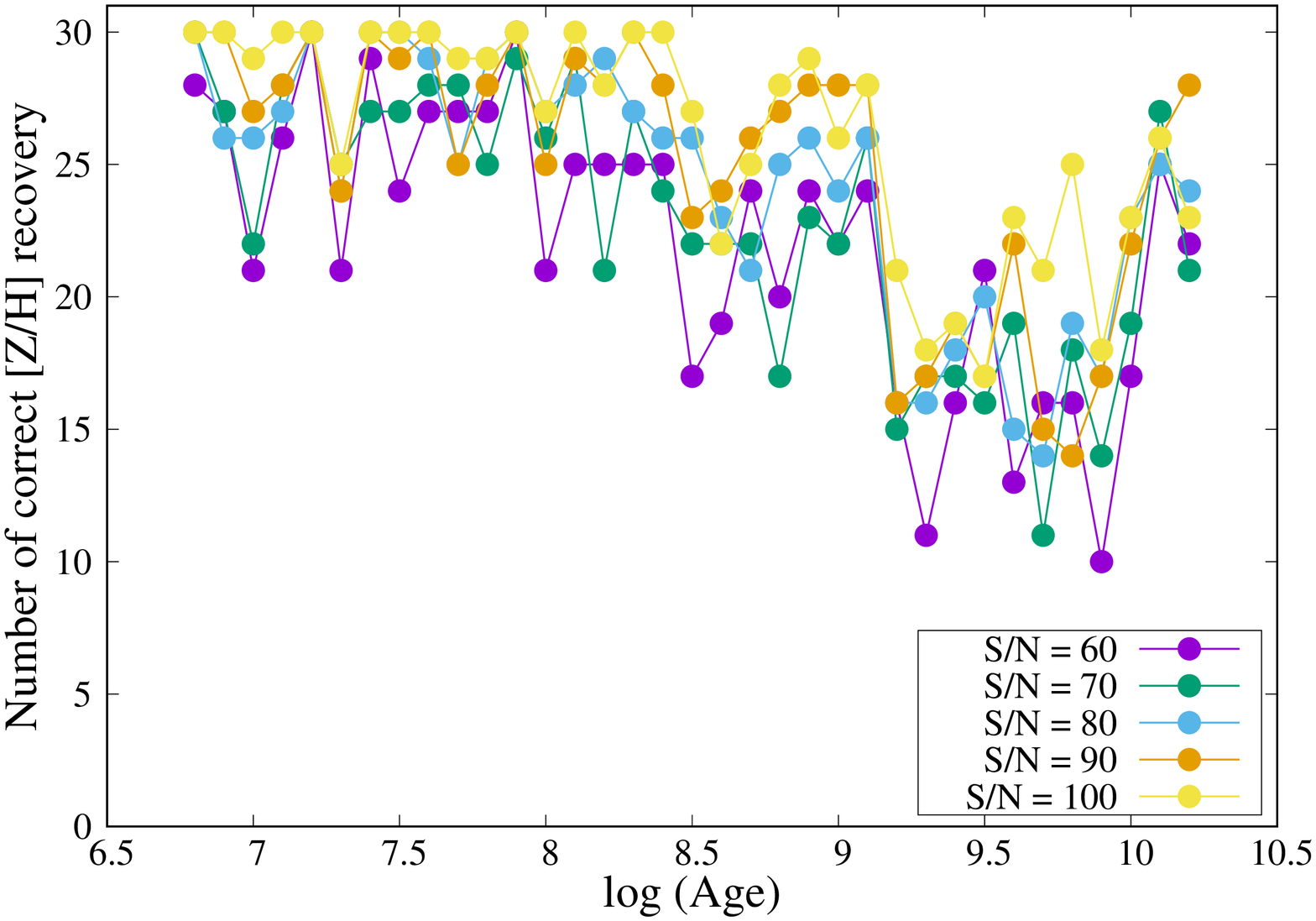}}
\resizebox{75mm}{!}{\includegraphics[width=\columnwidth, angle=270]{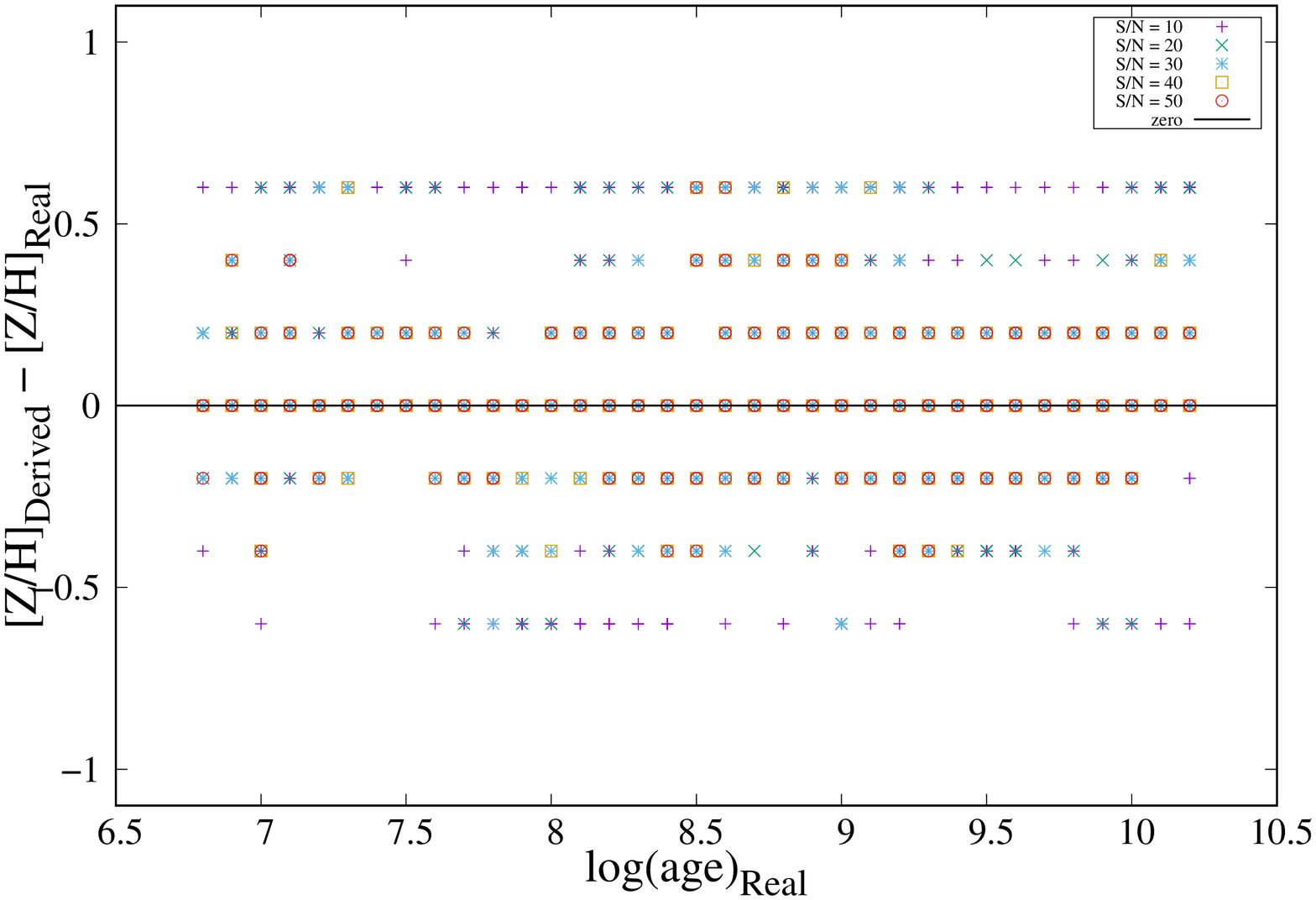}}
\resizebox{75mm}{!}{\includegraphics[width=\columnwidth, angle=270]{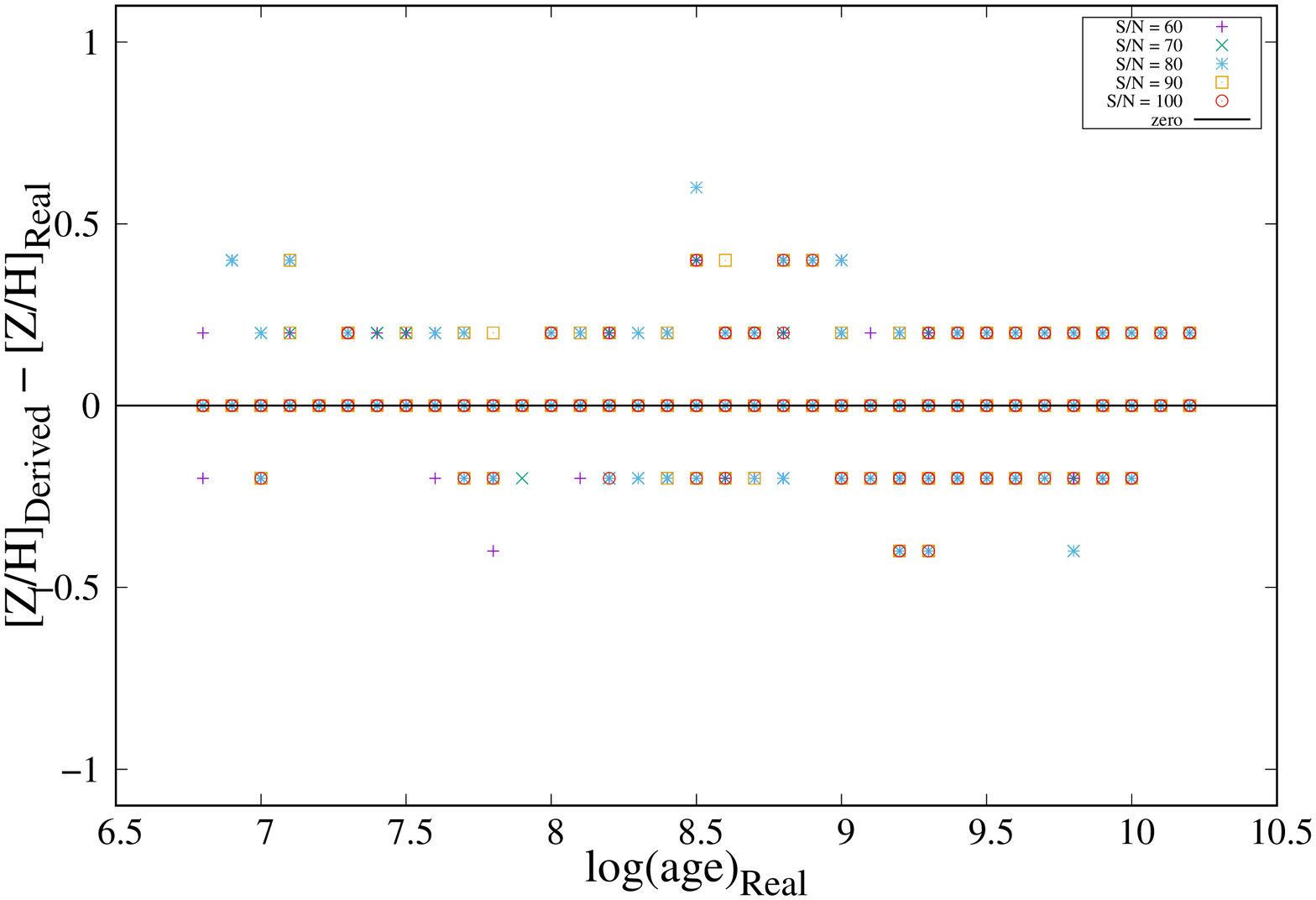}}

\caption{The same as figure \ref{Fig1} using MIST models}
\label{Fig4}
\end{figure*}

\begin{figure*}

\resizebox{75mm}{!}{\includegraphics[width=\columnwidth, angle=270]{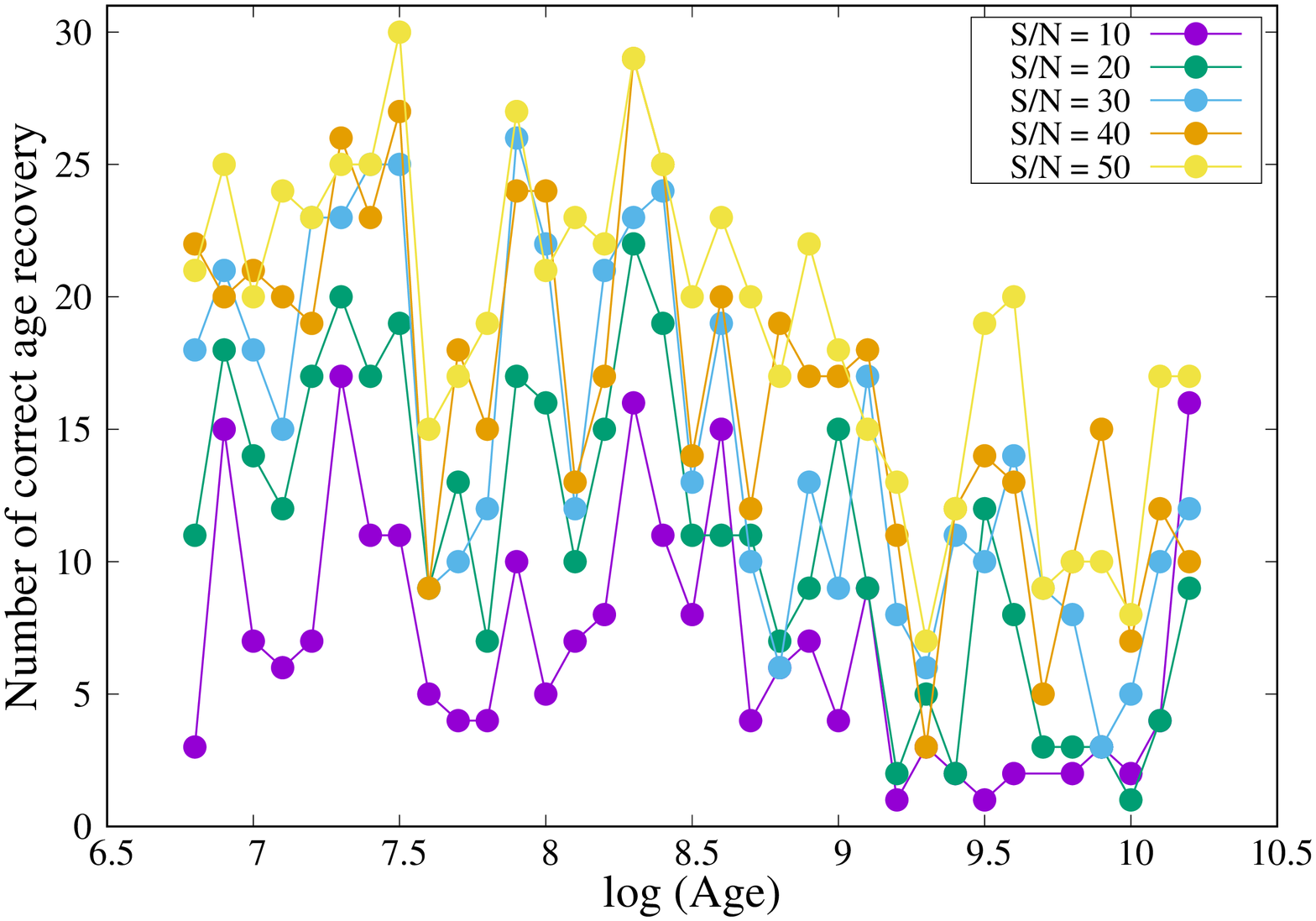}}
\resizebox{75mm}{!}{\includegraphics[width=\columnwidth, angle=270]{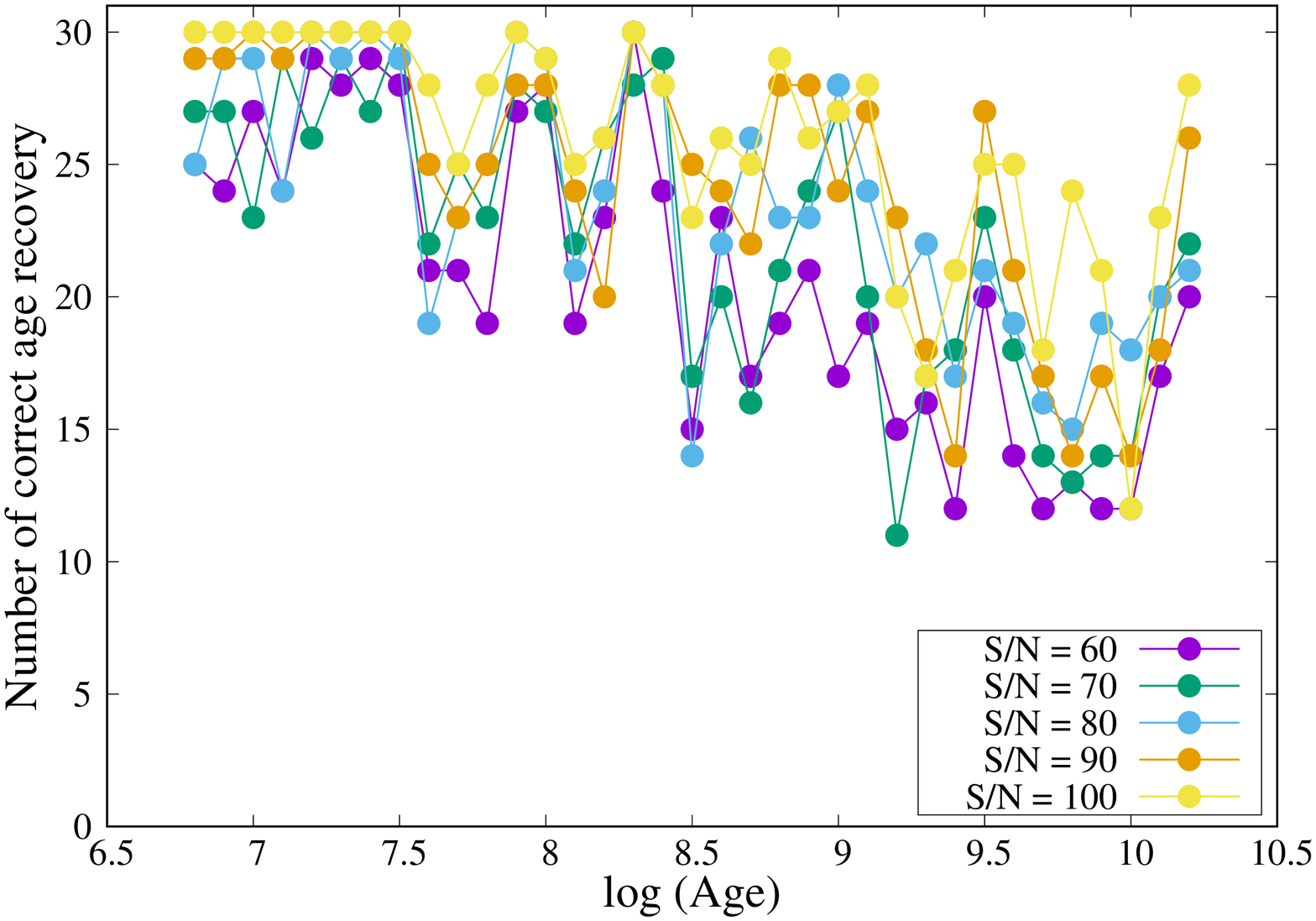}}
\resizebox{75mm}{!}{\includegraphics[width=\columnwidth, angle=270]{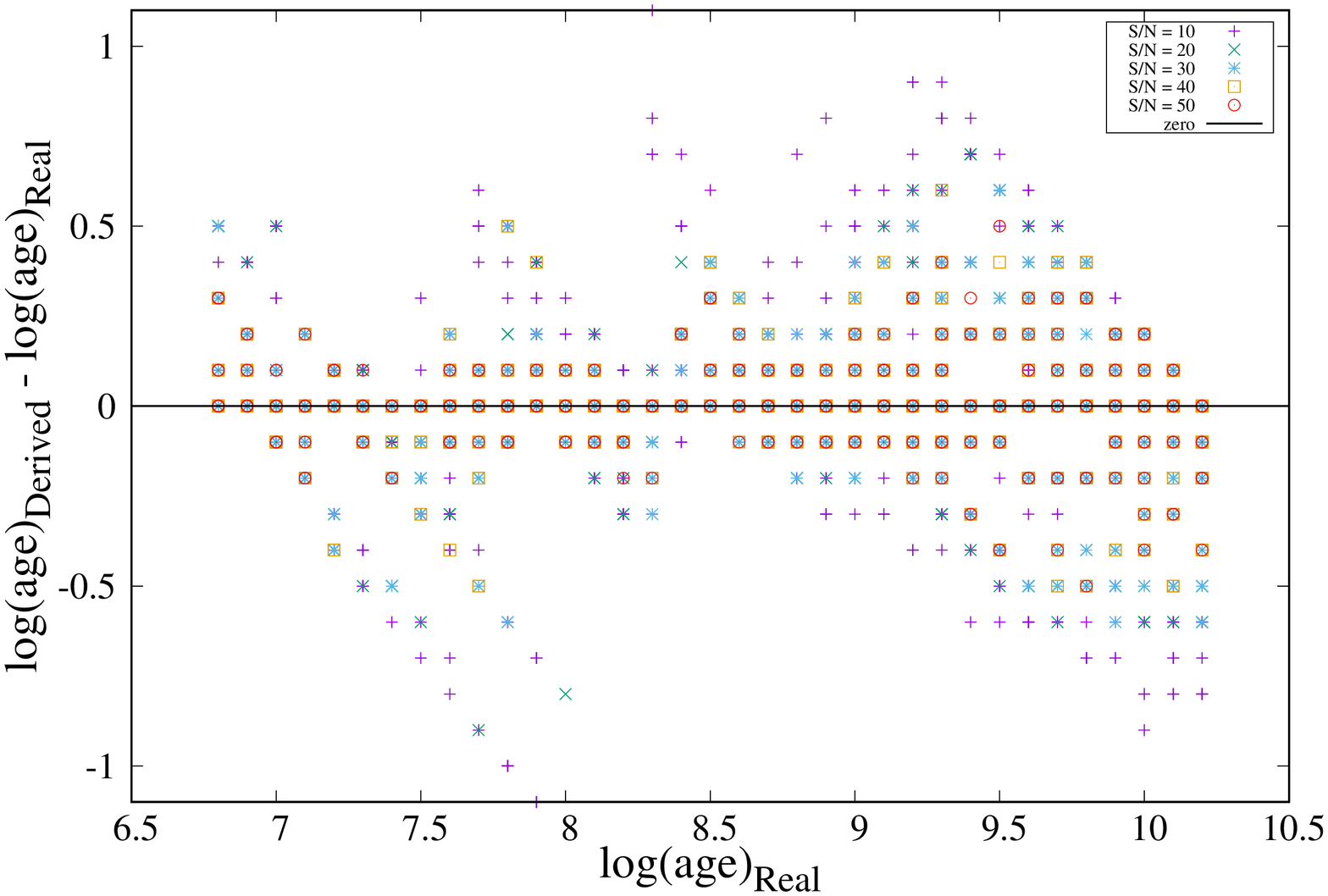}}
\resizebox{75mm}{!}{\includegraphics[width=\columnwidth, angle=270]{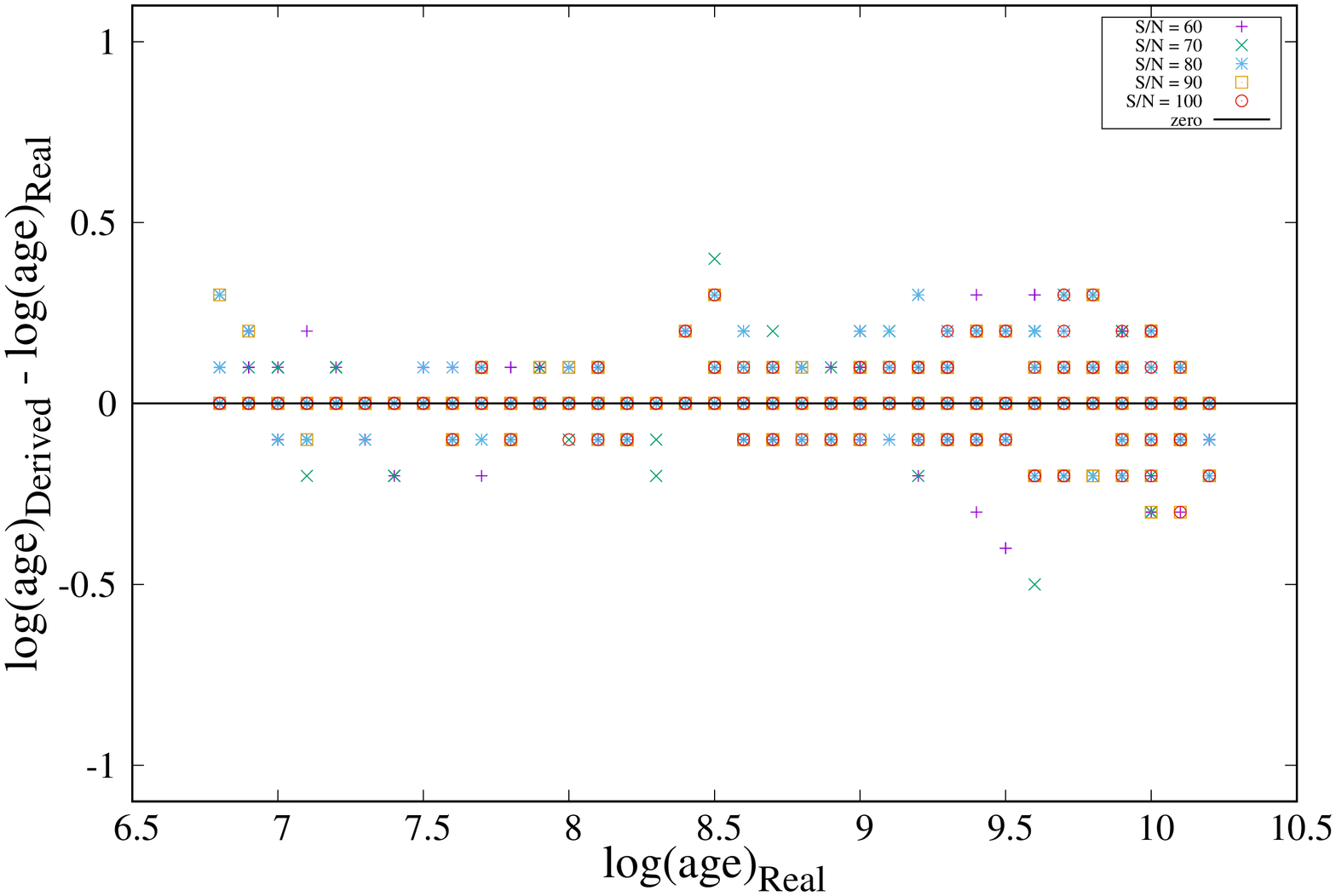}}
\resizebox{75mm}{!}{\includegraphics[width=\columnwidth, angle=270]{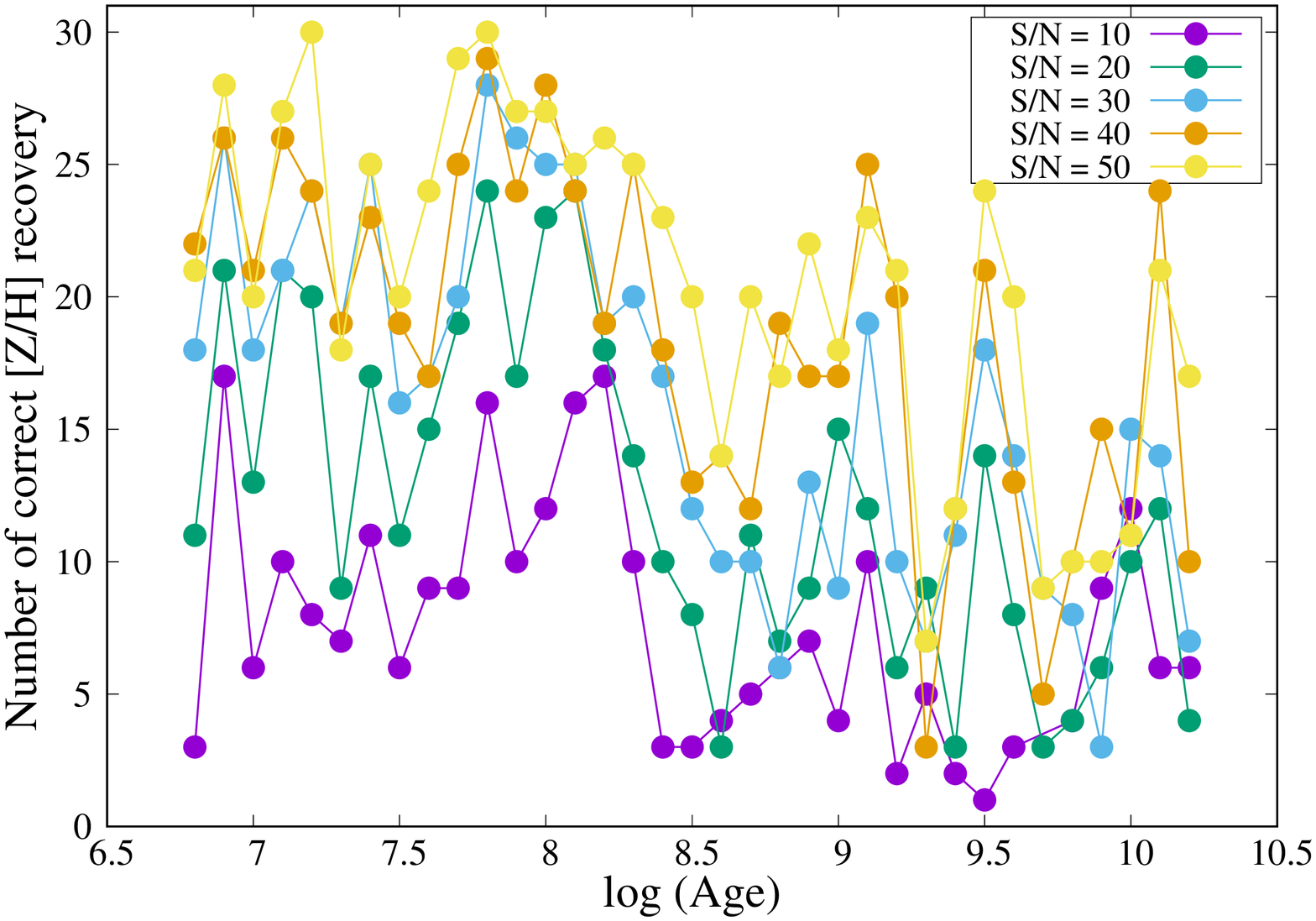}}
\resizebox{75mm}{!}{\includegraphics[width=\columnwidth, angle=270]{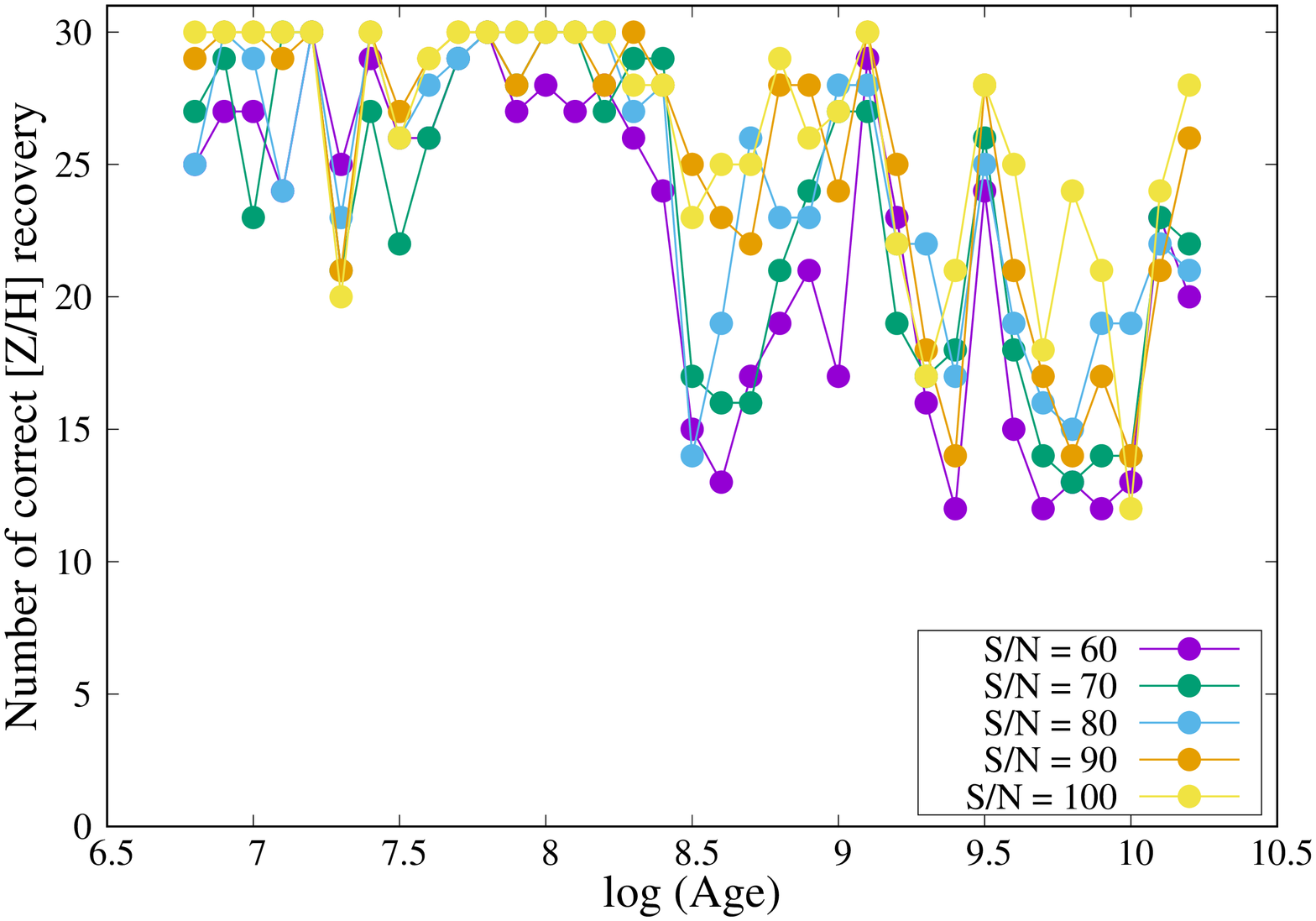}}
\resizebox{75mm}{!}{\includegraphics[width=\columnwidth, angle=270]{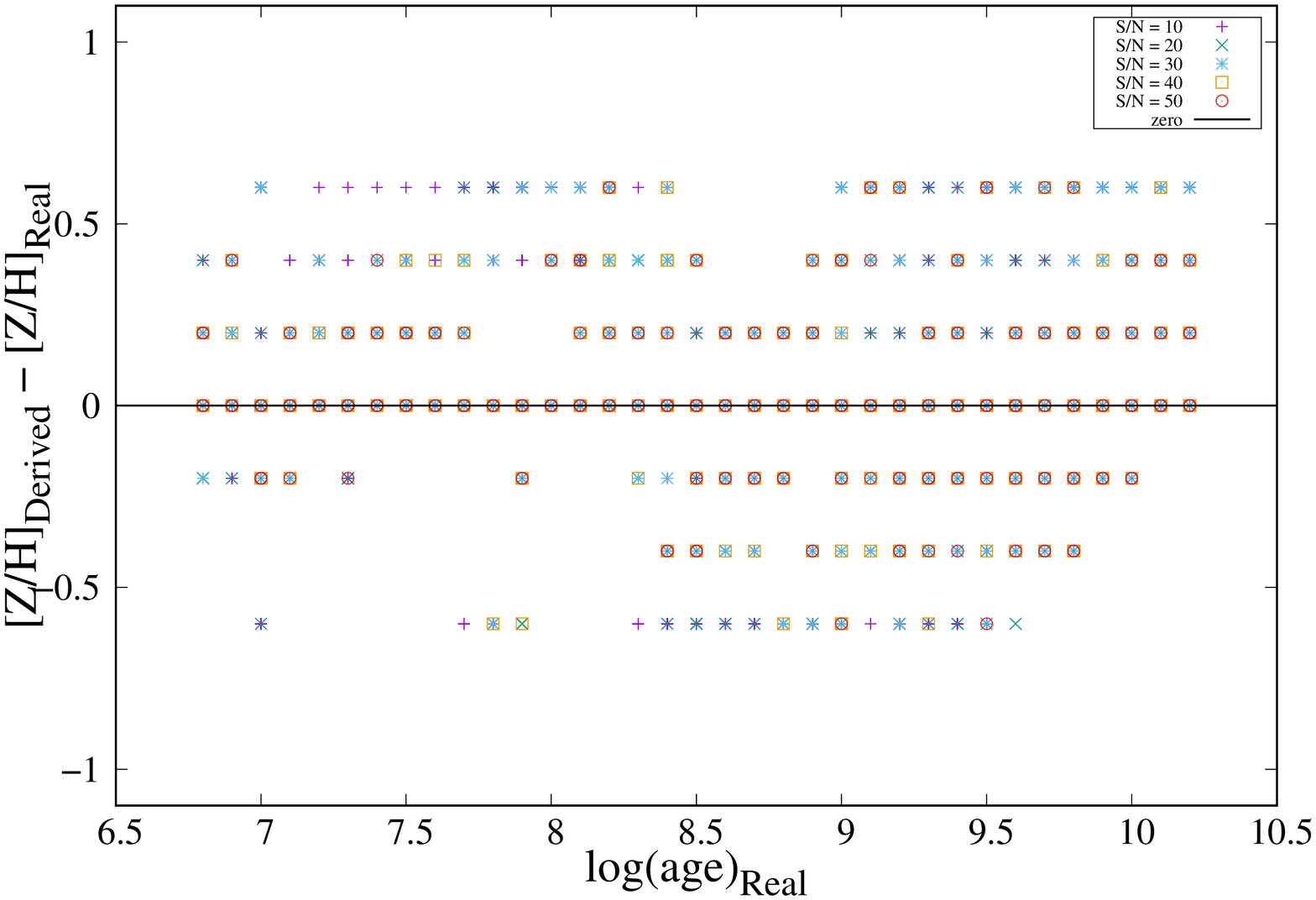}}
\resizebox{75mm}{!}{\includegraphics[width=\columnwidth, angle=270]{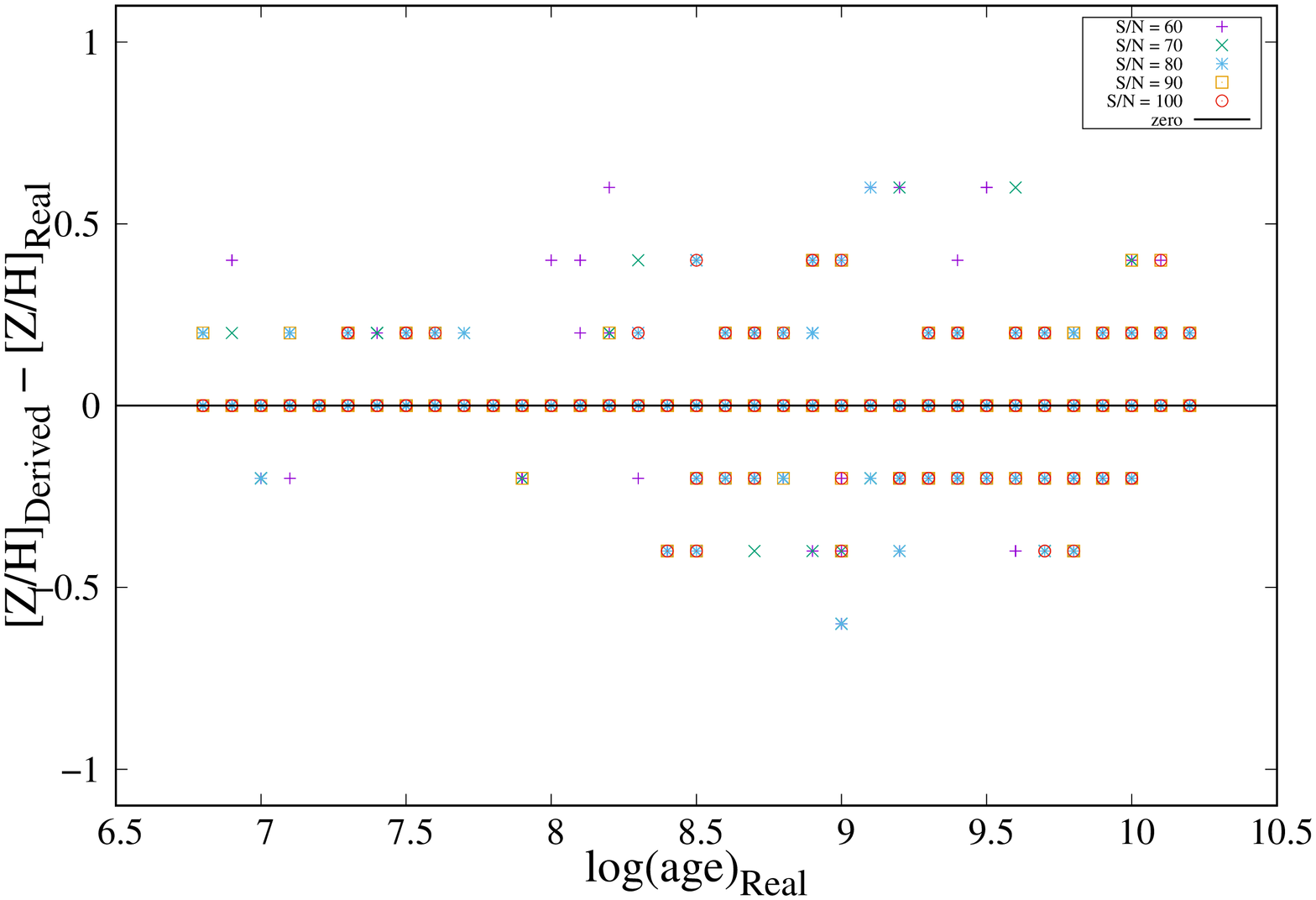}}

\caption{The same as figure \ref{Fig4} for the range $3700 \leqslant \lambda/\mbox{\AA} \leqslant 5000$ }
\label{Fig5}
\end{figure*}

\begin{figure*}

\resizebox{75mm}{!}{\includegraphics[width=\columnwidth, angle=270]{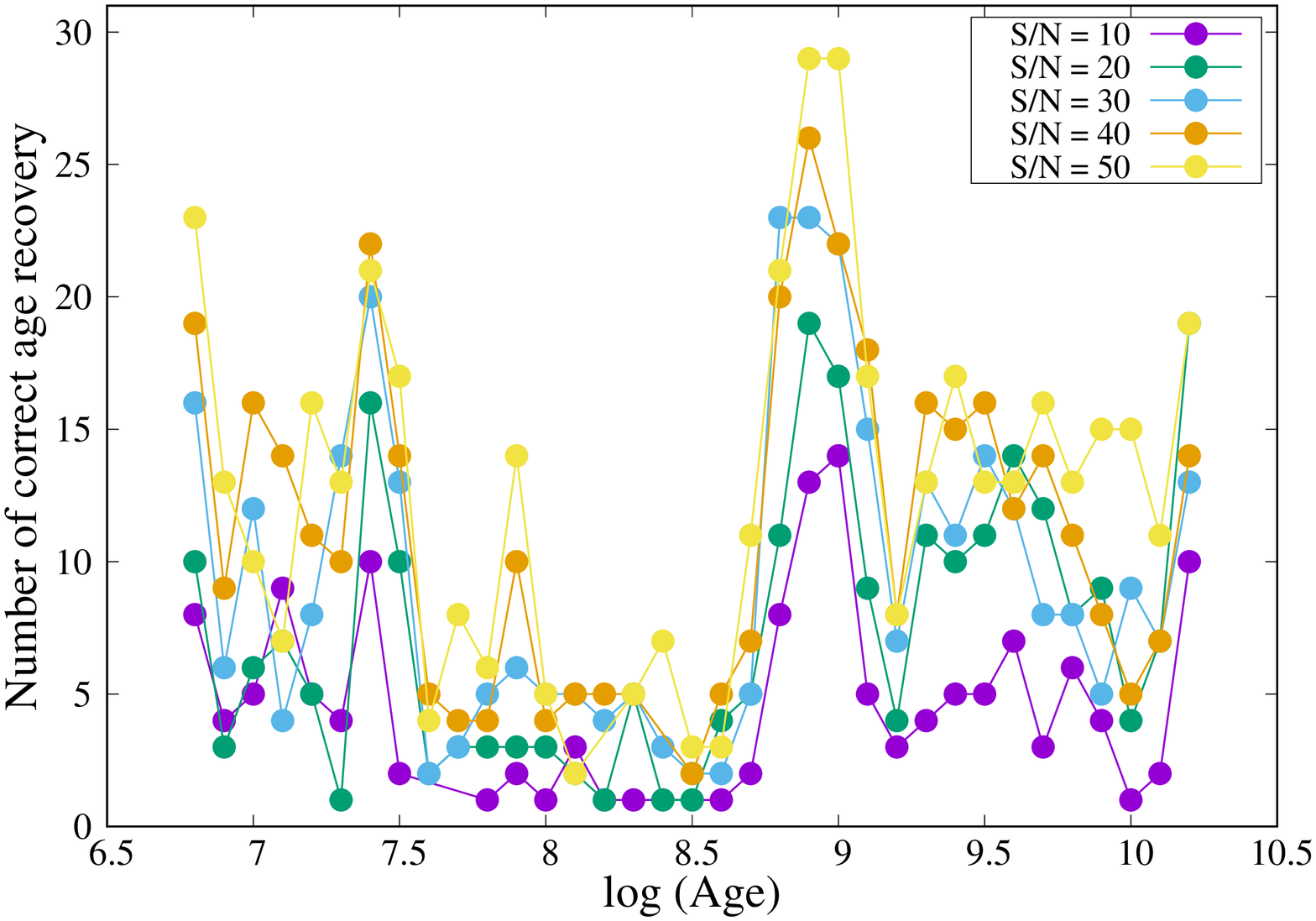}}
\resizebox{75mm}{!}{\includegraphics[width=\columnwidth, angle=270]{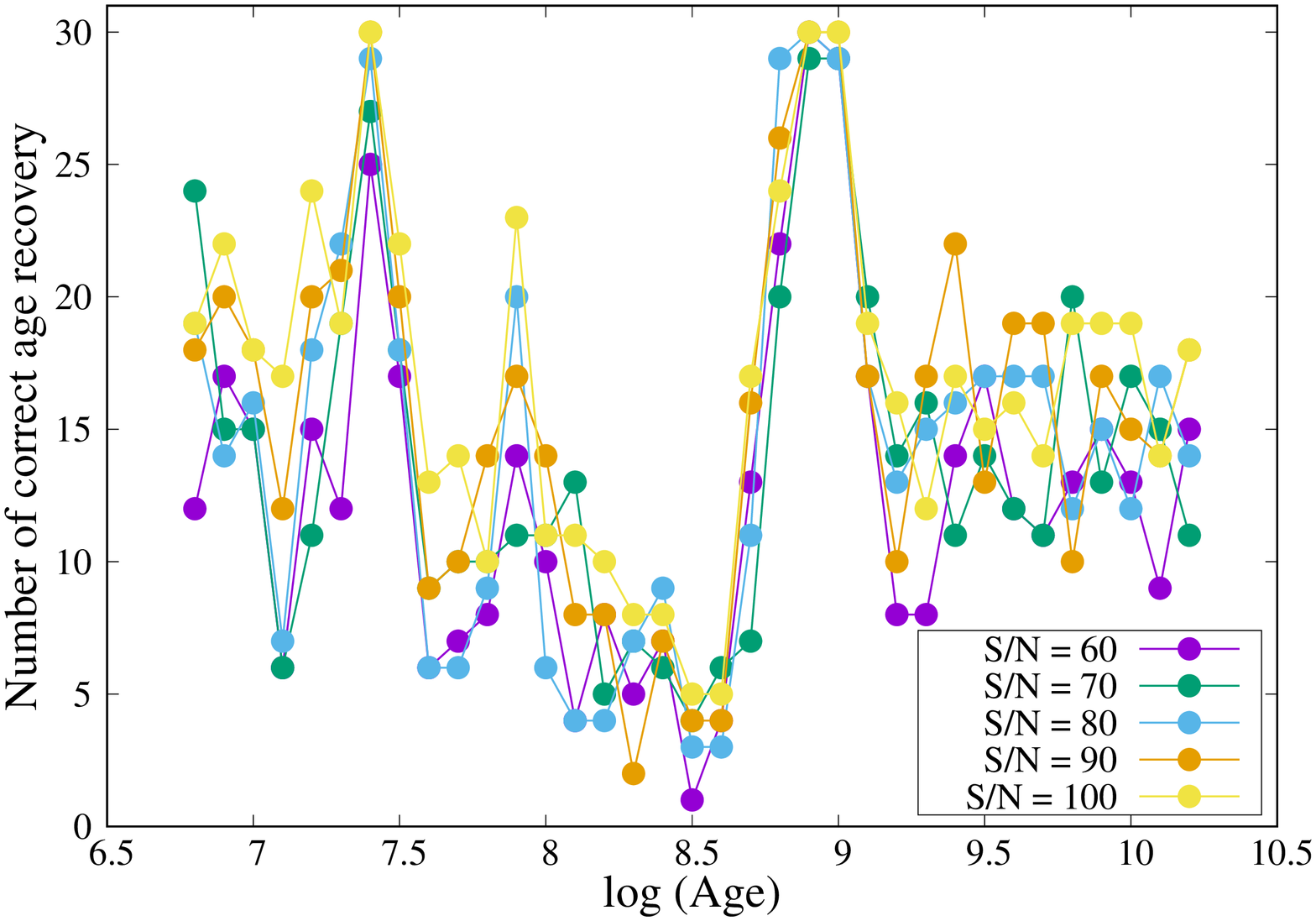}}
\resizebox{75mm}{!}{\includegraphics[width=\columnwidth, angle=270]{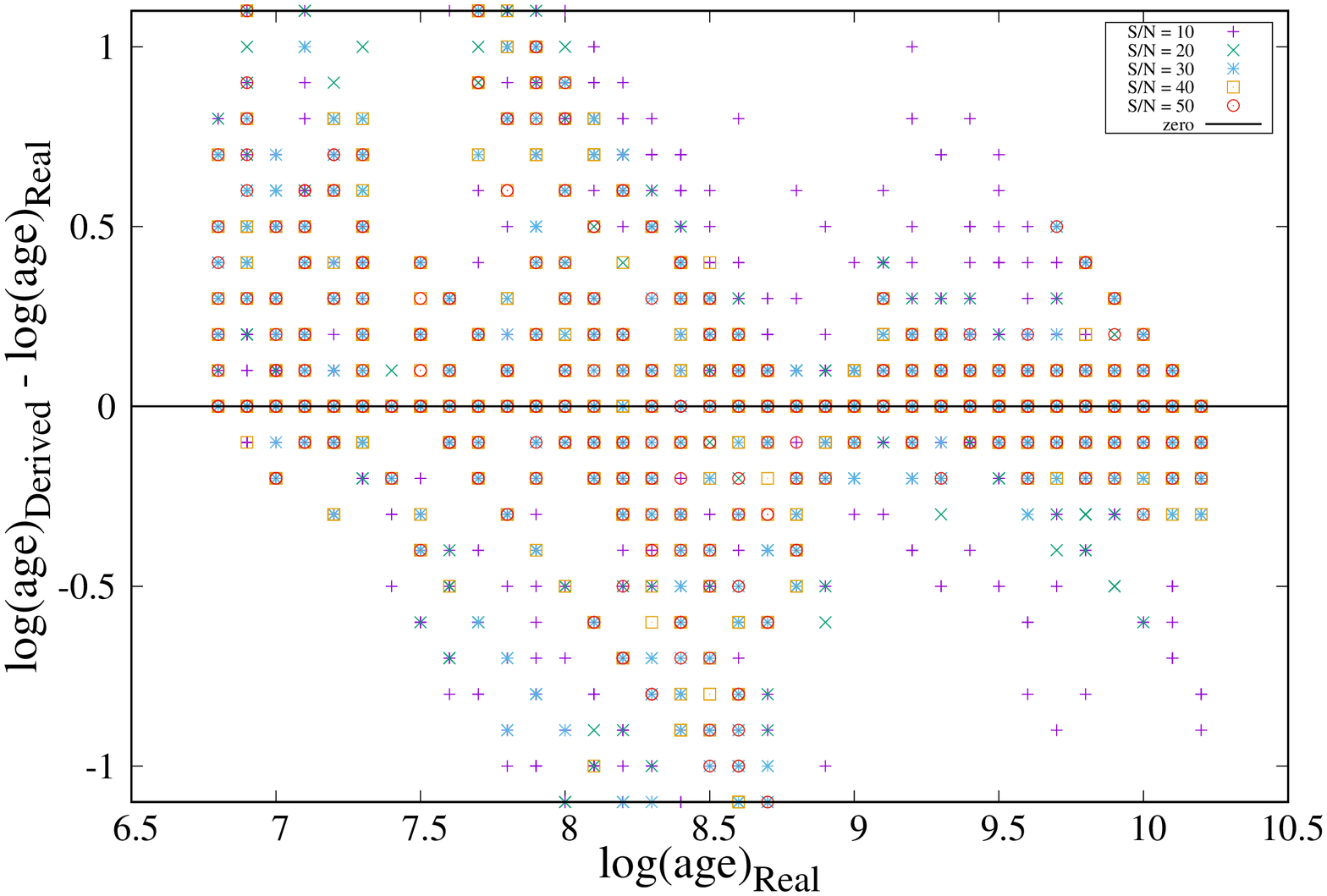}}
\resizebox{75mm}{!}{\includegraphics[width=\columnwidth, angle=270]{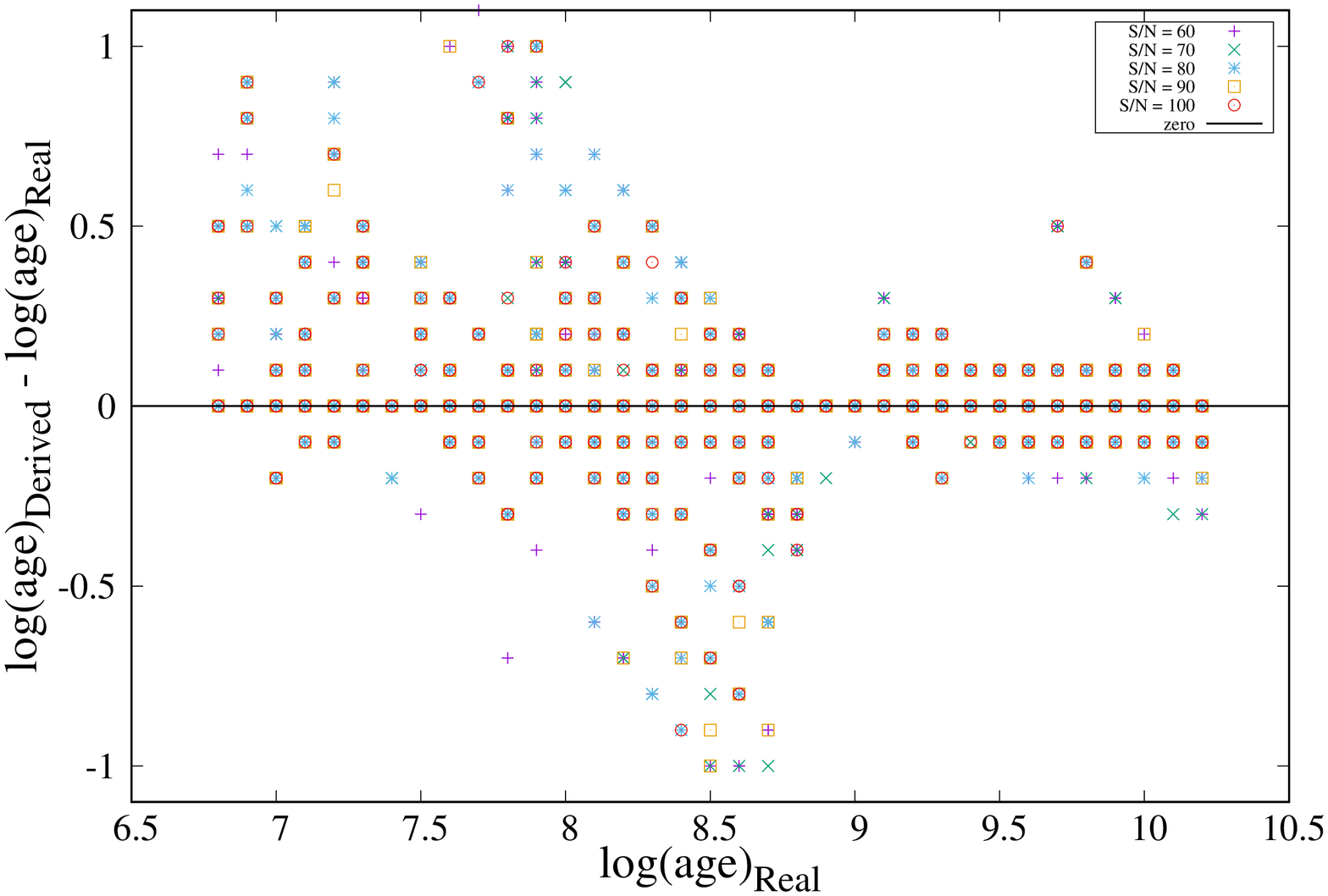}}
\resizebox{75mm}{!}{\includegraphics[width=\columnwidth, angle=270]{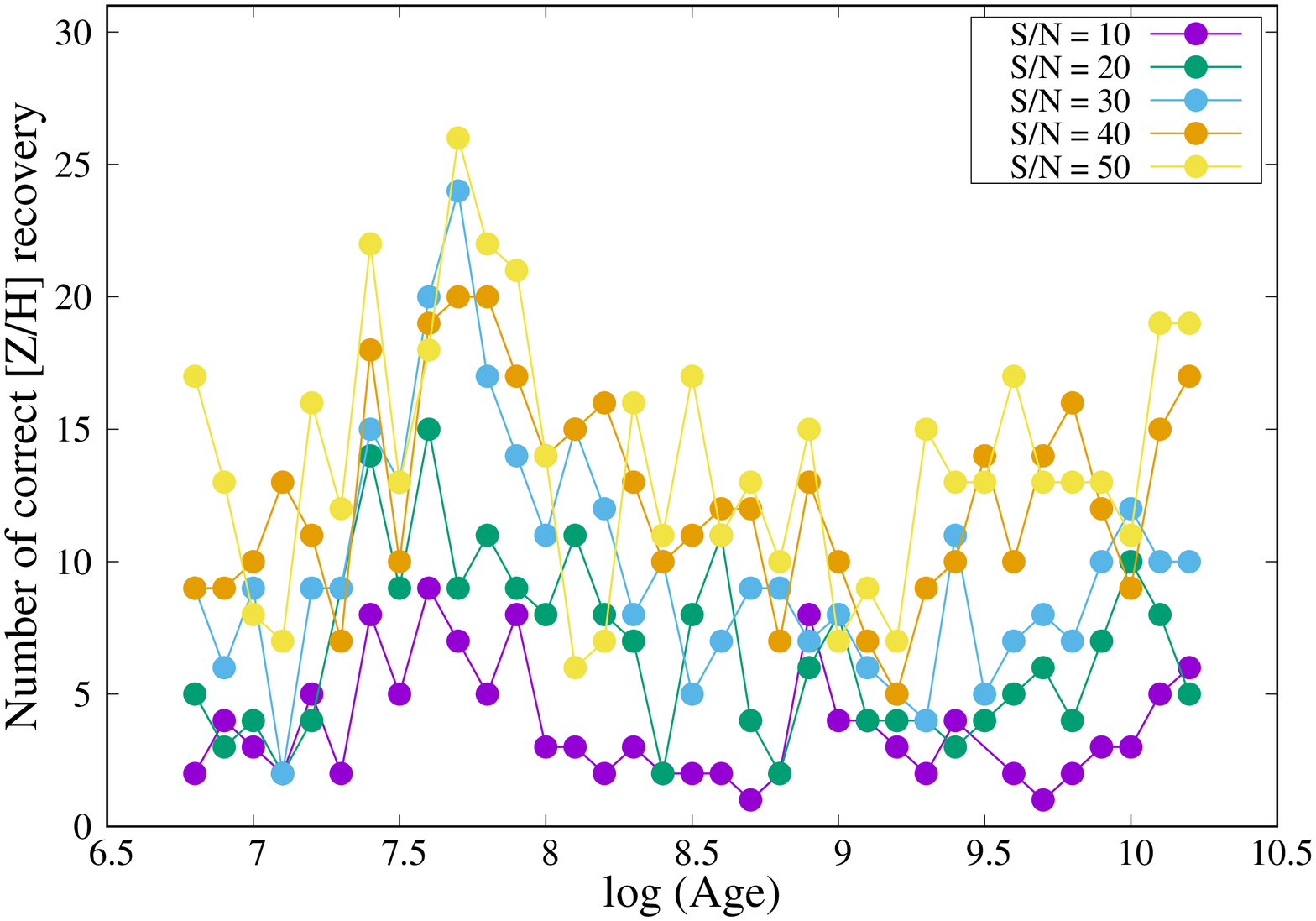}}
\resizebox{75mm}{!}{\includegraphics[width=\columnwidth, angle=270]{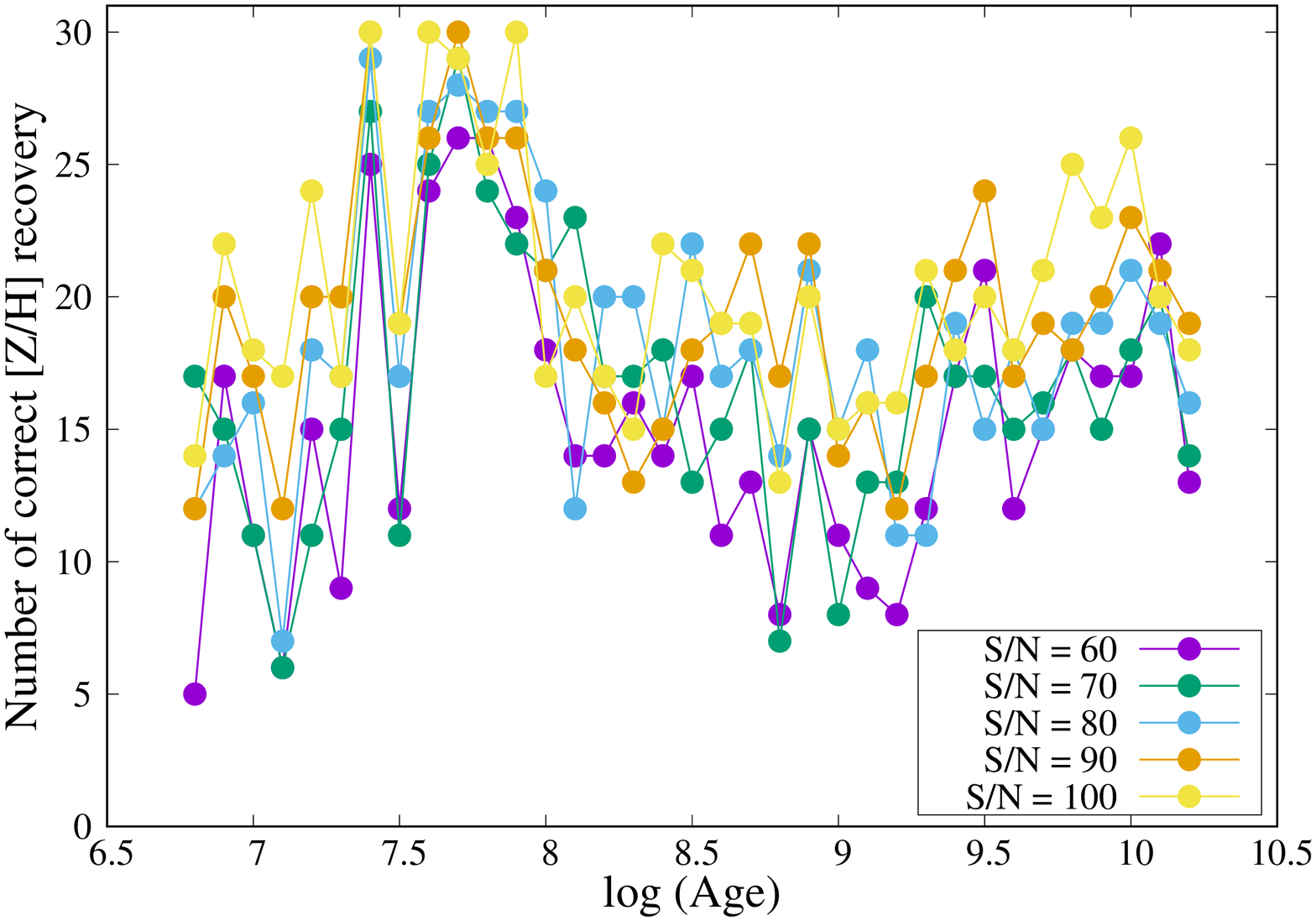}}
\resizebox{75mm}{!}{\includegraphics[width=\columnwidth, angle=270]{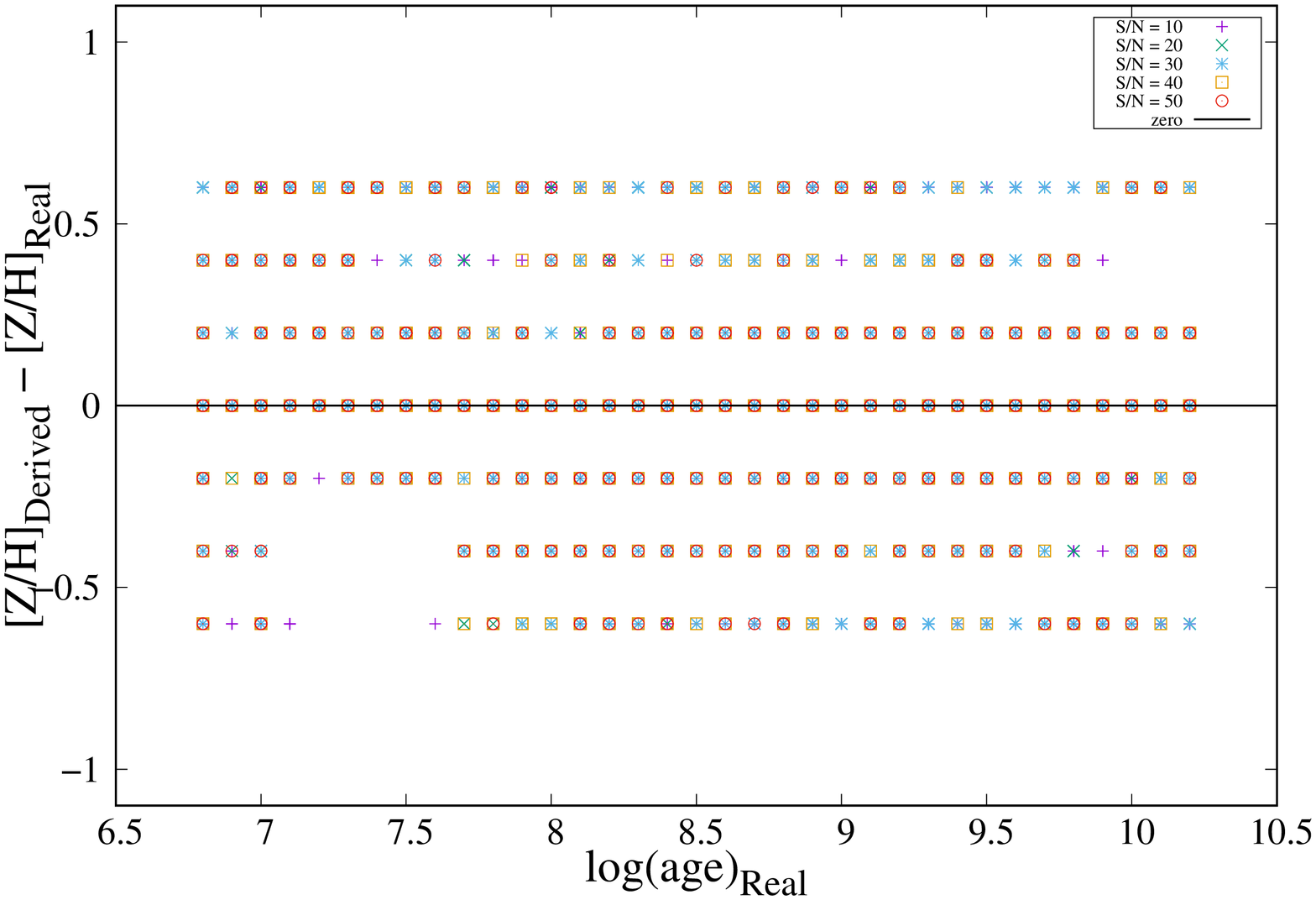}}
\resizebox{75mm}{!}{\includegraphics[width=\columnwidth, angle=270]{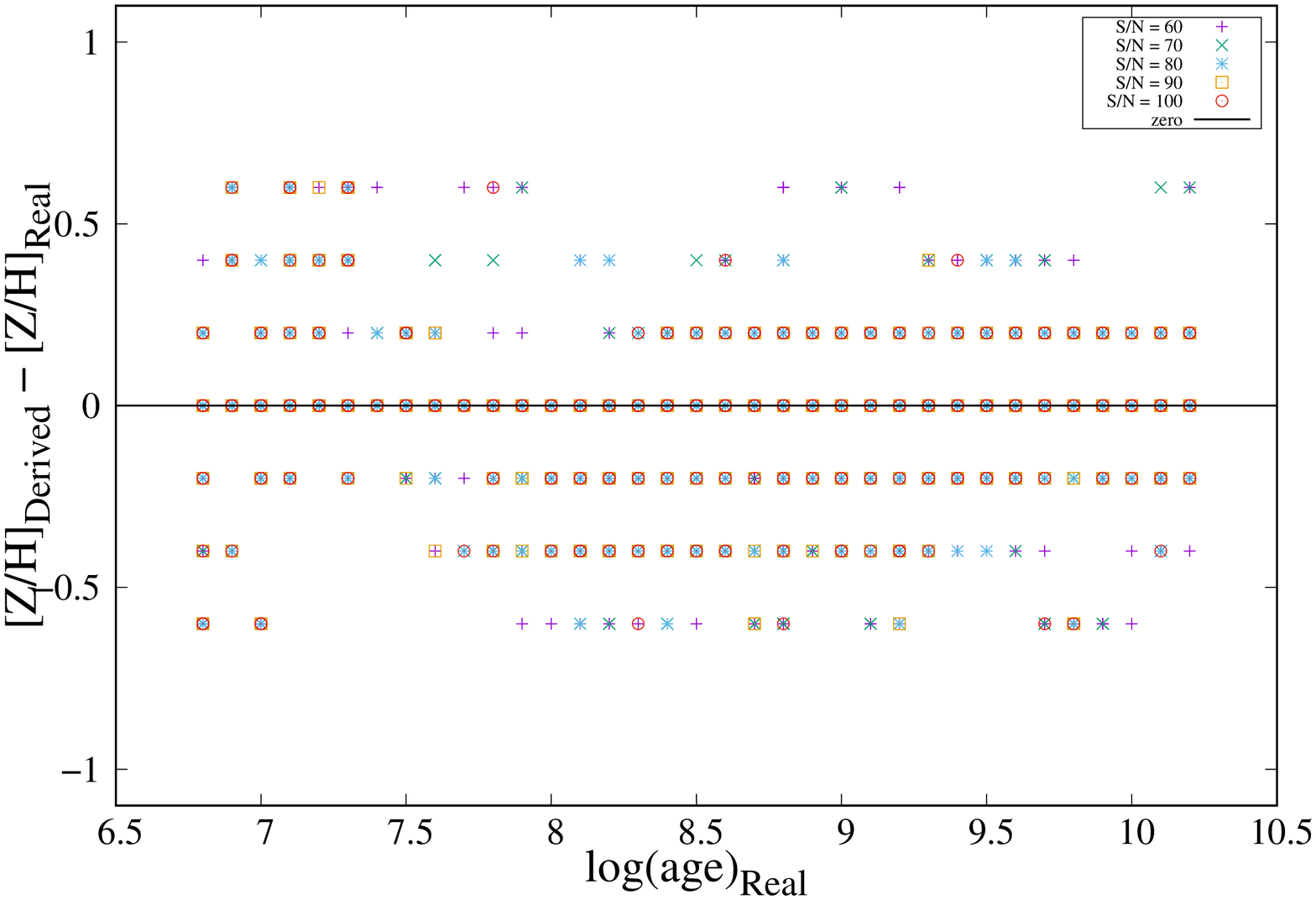}}

\caption{The same as \ref{Fig4} for $5000 \leqslant \lambda/\mbox{\AA} \leqslant 6200$ }

\label{Fig6}
\end{figure*}

\begin{figure*}

\resizebox{75mm}{!}{\includegraphics[width=\columnwidth, angle=270]{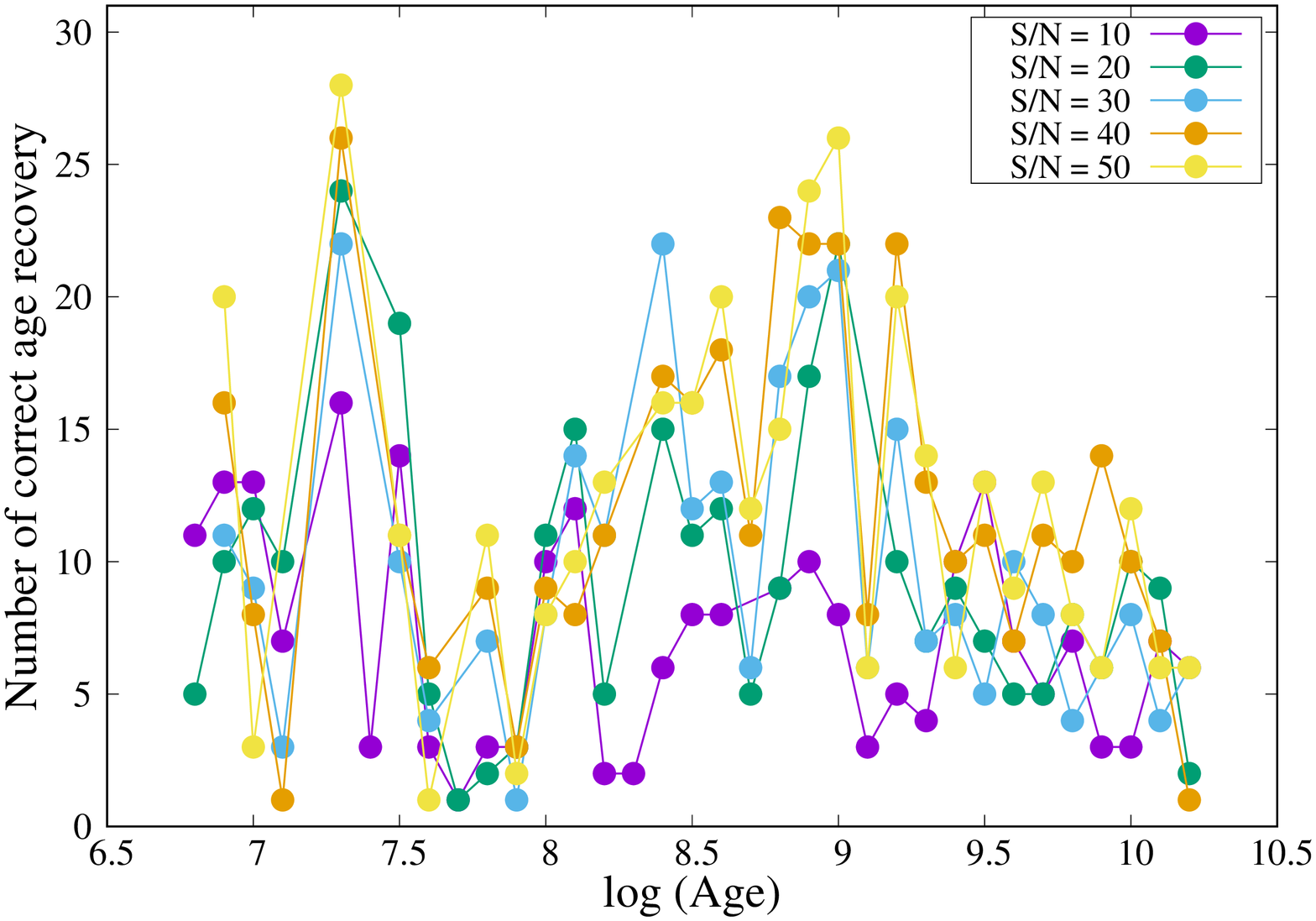}}
\resizebox{75mm}{!}{\includegraphics[width=\columnwidth, angle=270]{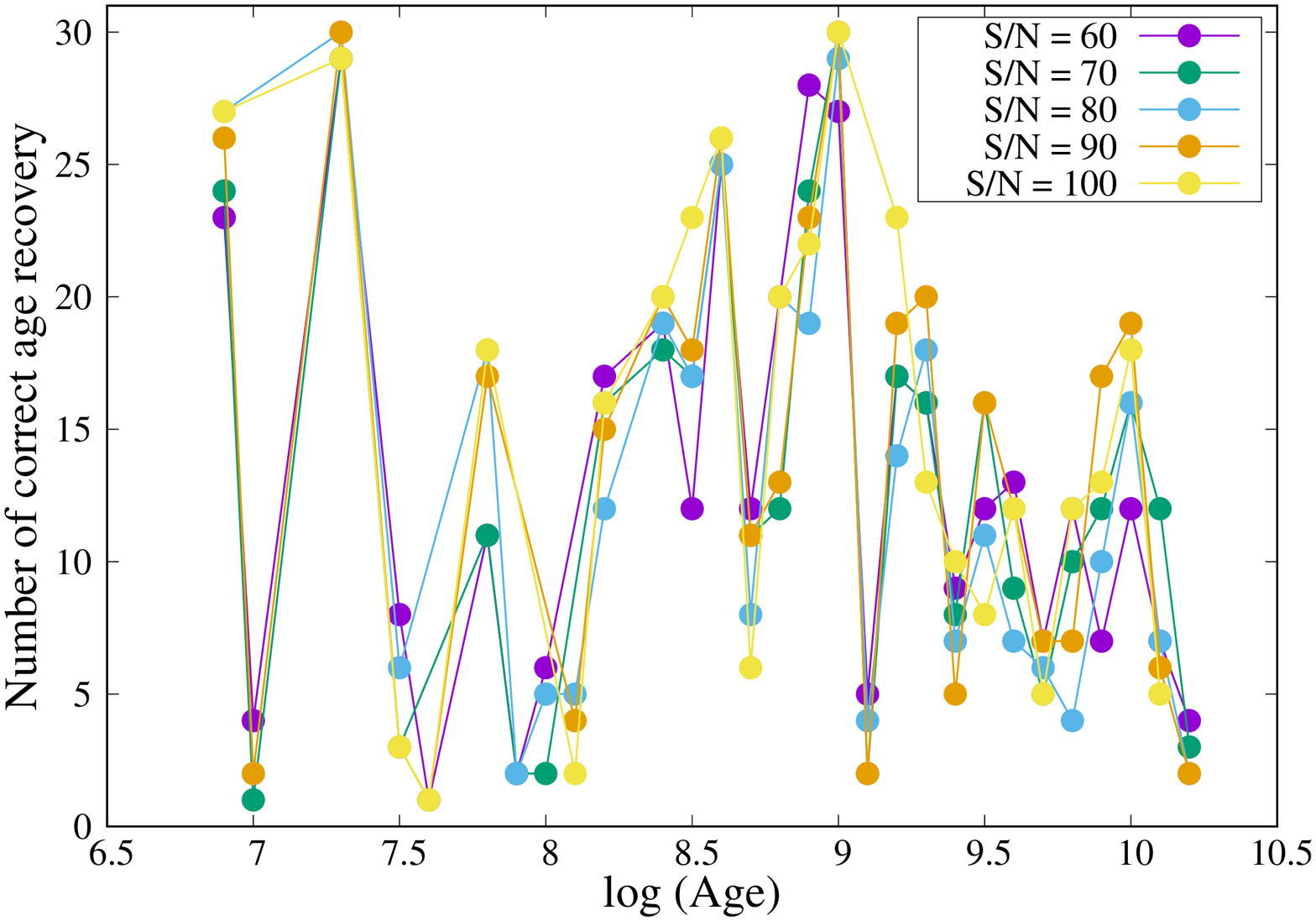}}
\resizebox{75mm}{!}{\includegraphics[width=\columnwidth, angle=270]{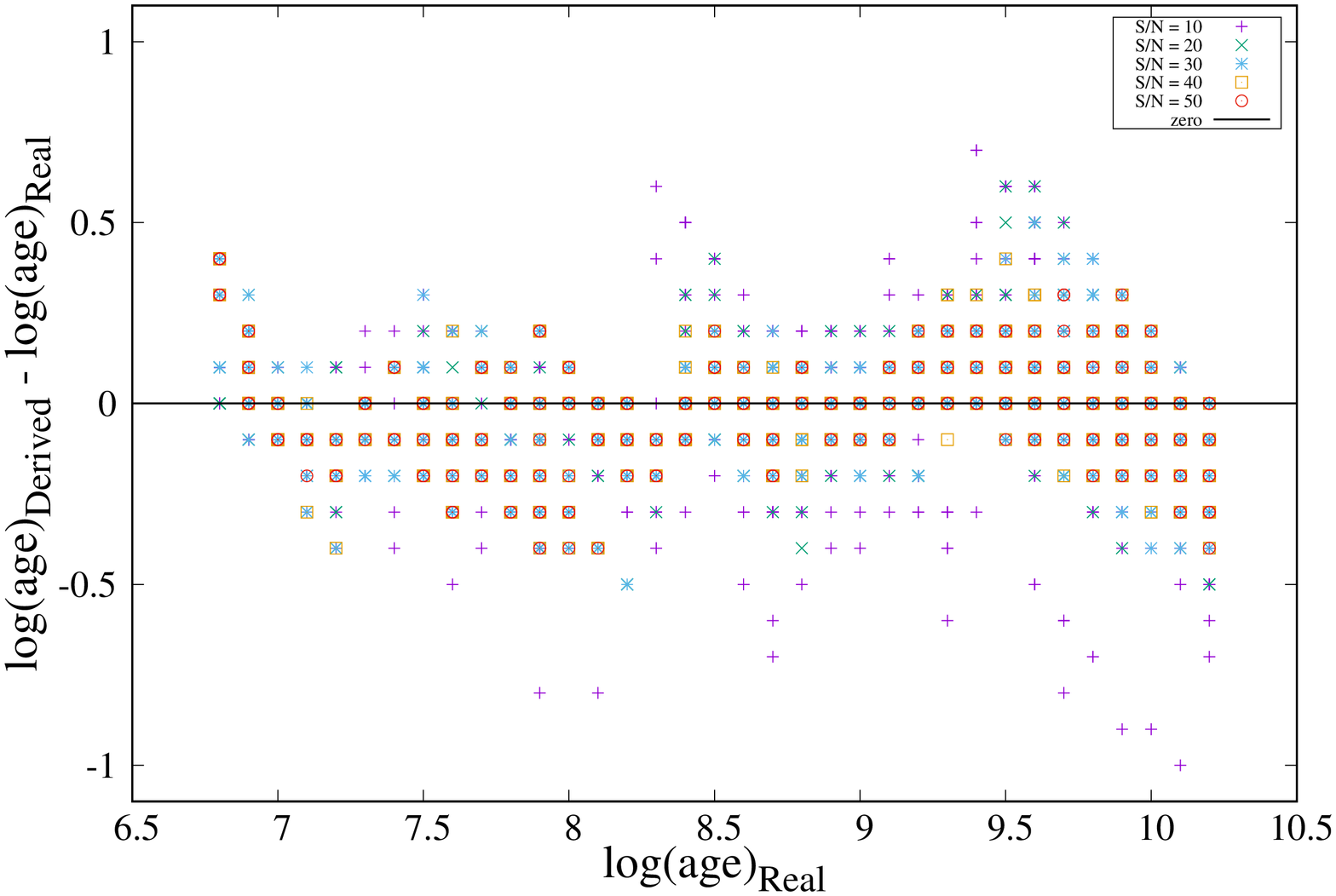}}
\resizebox{75mm}{!}{\includegraphics[width=\columnwidth, angle=270]{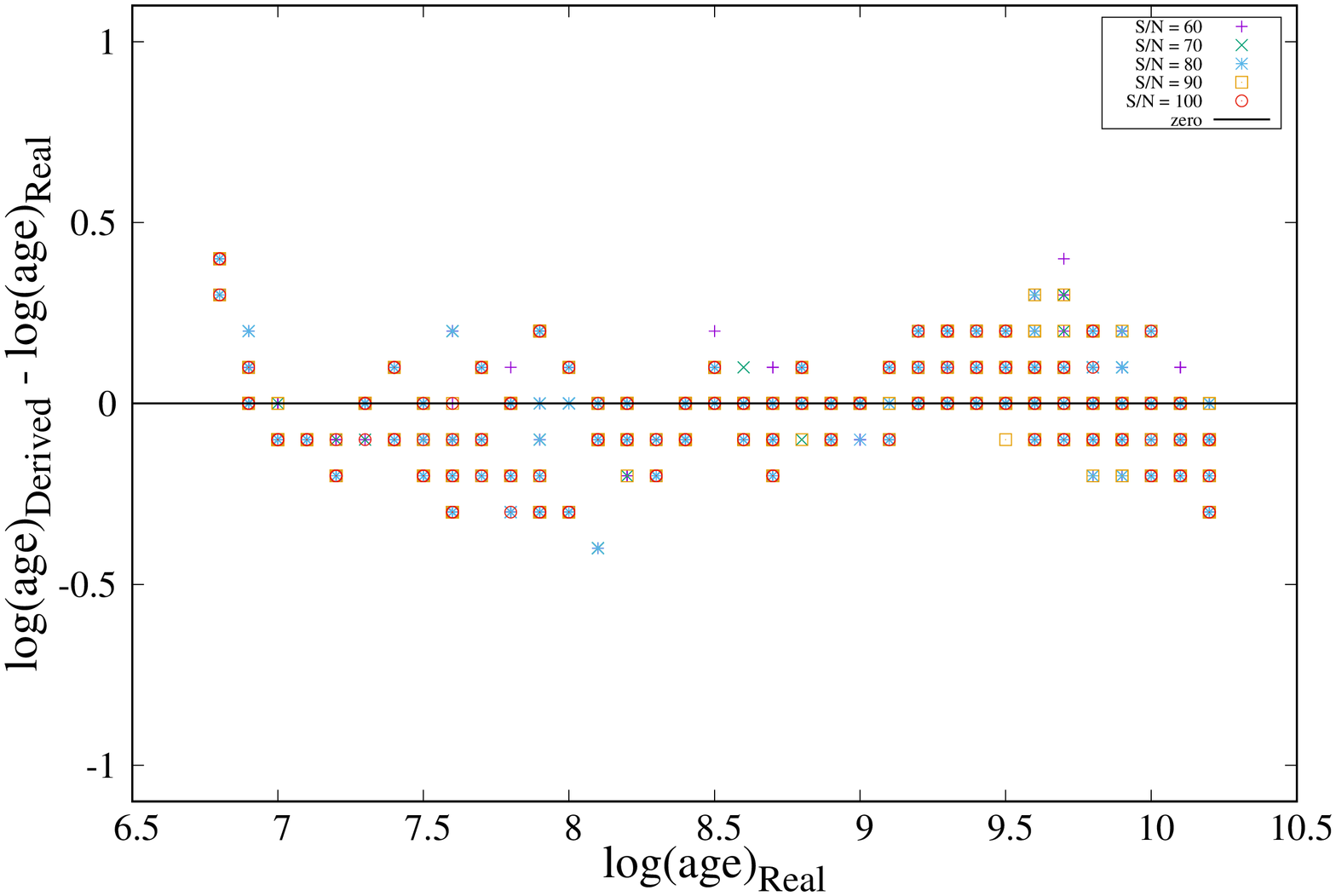}}
\resizebox{75mm}{!}{\includegraphics[width=\columnwidth, angle=270]{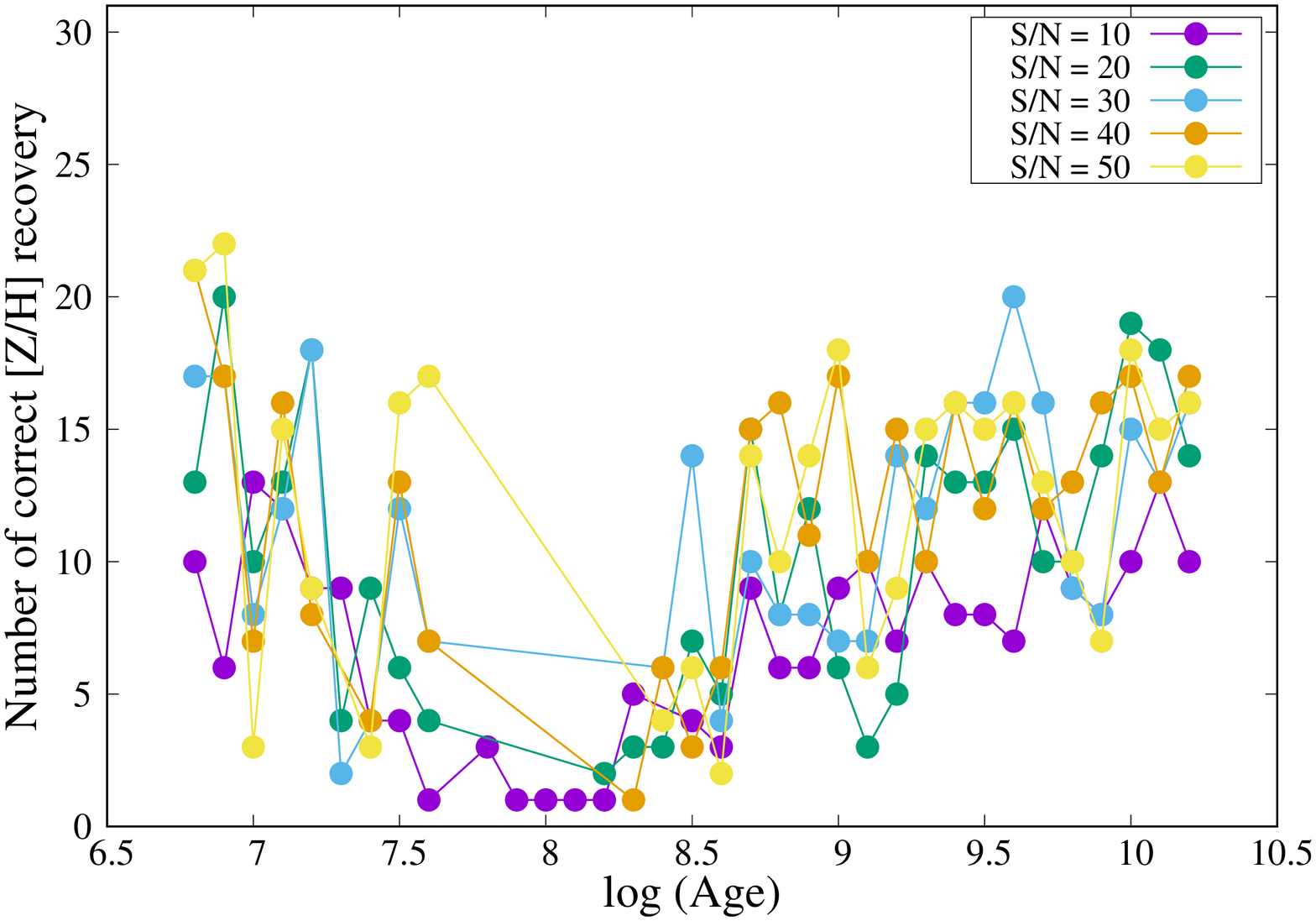}}
\resizebox{75mm}{!}{\includegraphics[width=\columnwidth, angle=270]{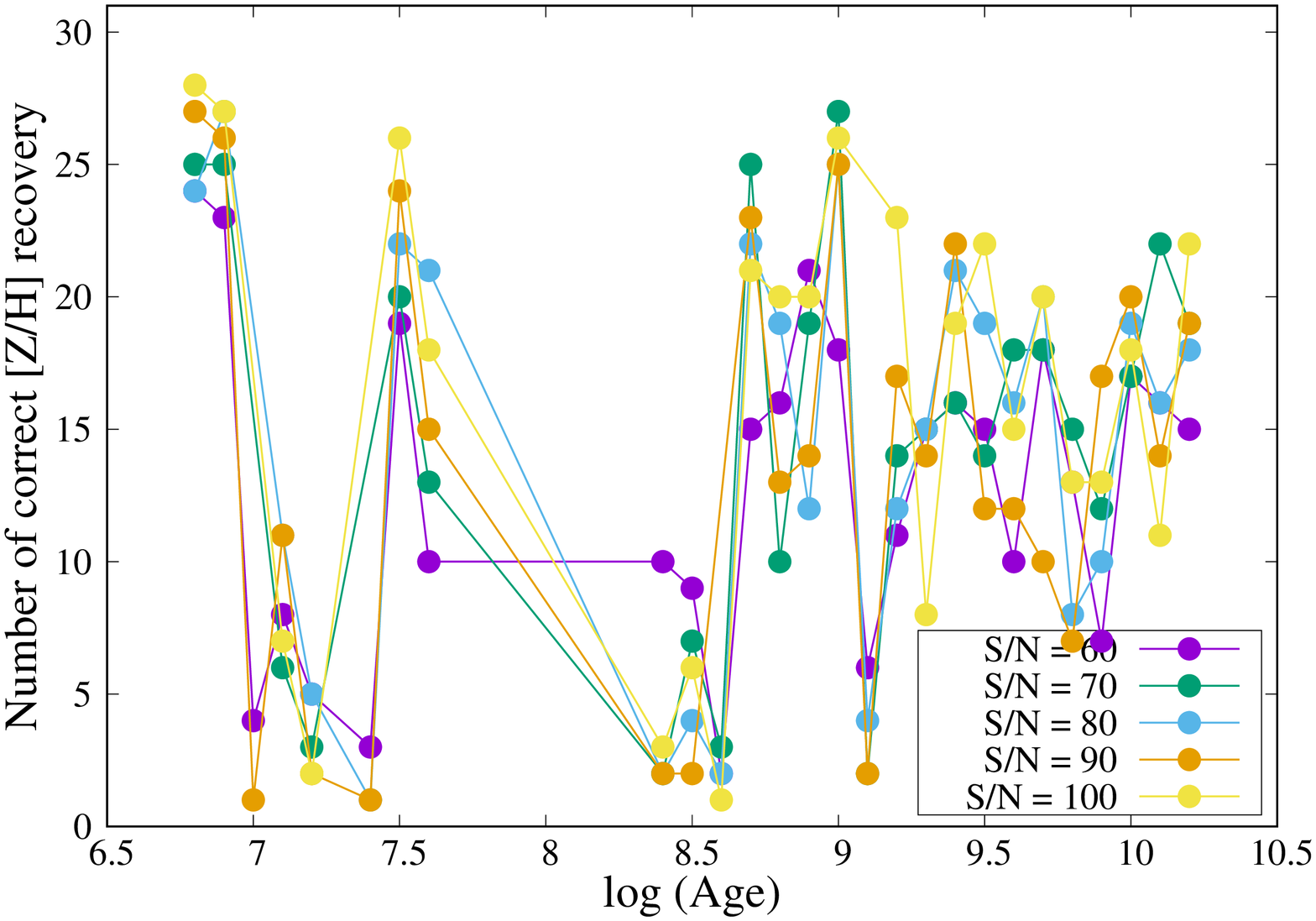}}
\resizebox{75mm}{!}{\includegraphics[width=\columnwidth, angle=270]{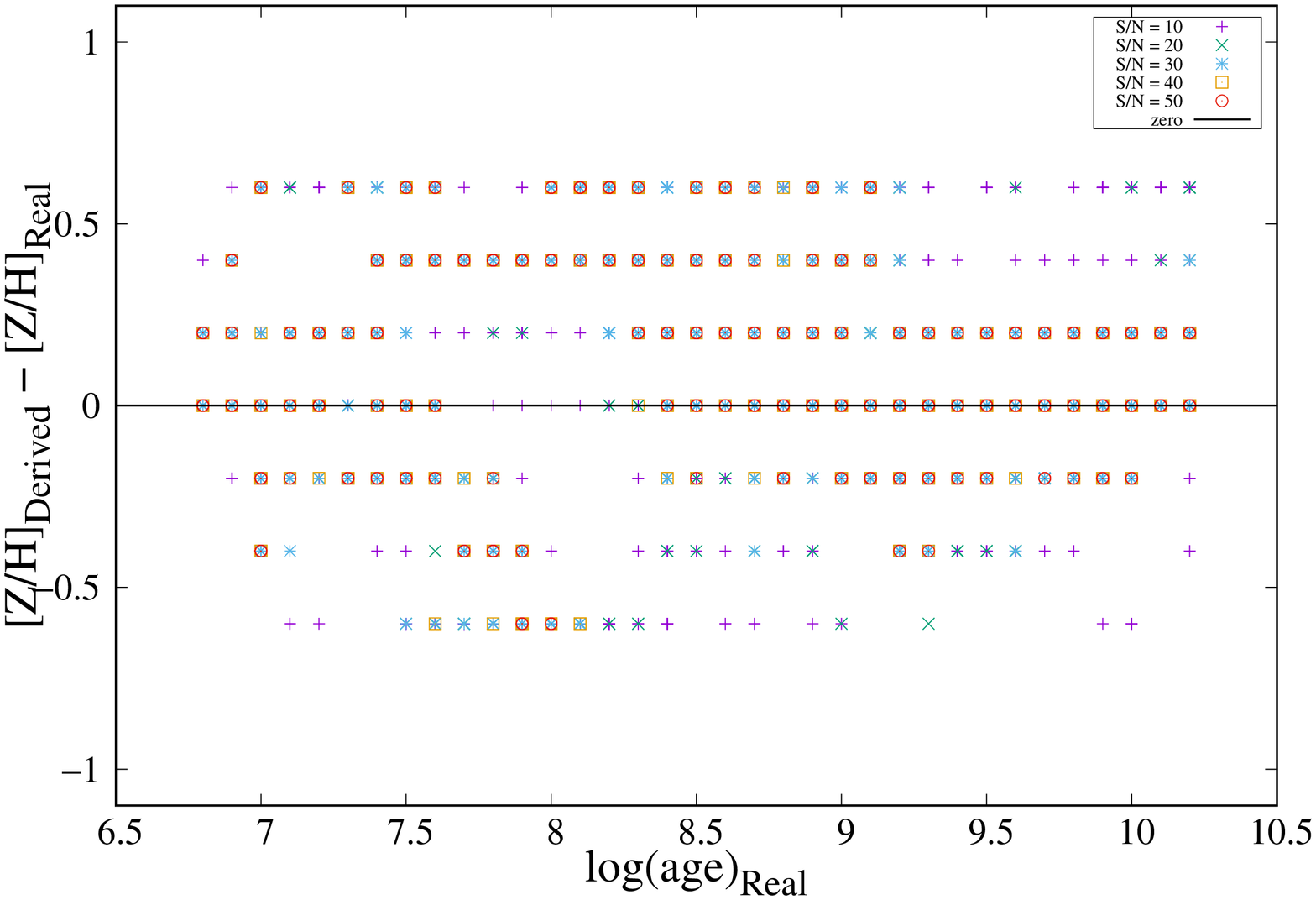}}
\resizebox{75mm}{!}{\includegraphics[width=\columnwidth, angle=270]{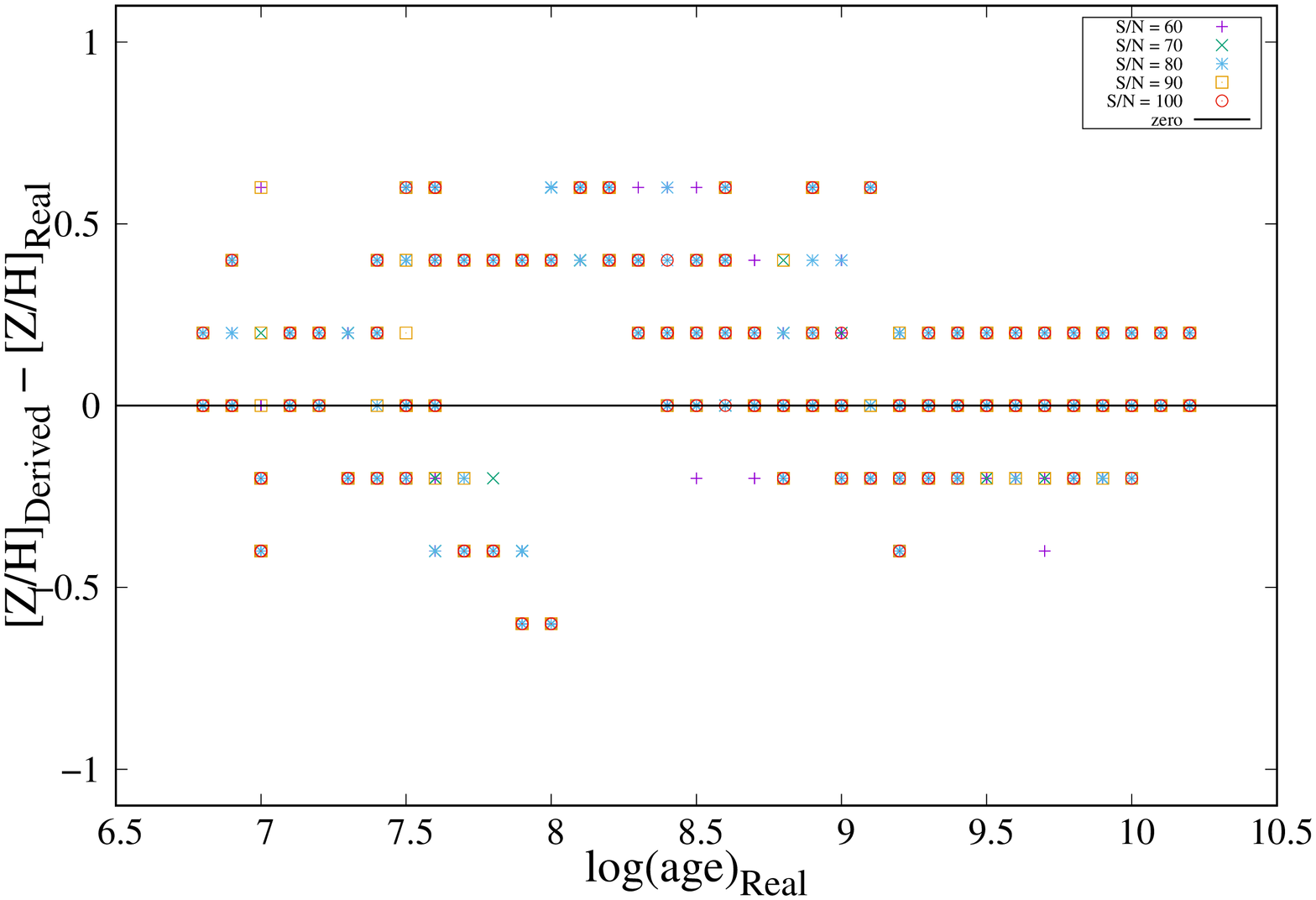}}

\caption{The same as figure \ref{Fig1} for the range $3700 \leqslant \lambda/\mbox{\AA} \leqslant 6200$ using MIST models and Padova mock clusters}
\label{Fig7}
\end{figure*}

When using [Z/H] = $-$0.4, the overall age determination improves compared to the results when the metallicity is unknown a priori. Figure \ref{SV_fixedZ} shows that this is especially the case for log\,(age/yr) $\ga$ 8.0 (compare with Figure~\ref{SV}). In fact, for S/N $>$ 30, the standard deviation ($\sigma$) is found to stay $<$ 0.1 for all ages. The overall improvement in $\sigma$ relative to the case of letting metallicity be a free parameter is of order factors of 1.5\,--\,2 depending on the input age.

\section{The role of extinction}
\label{red}

Although several recently developed full-spectrum fitting techniques do not
directly fit for extinction \citep[e.g.,][]{Chilingarian07, Larsen12,
  Hernandez17}, it is important to note that age and extinction do affect the
shape of spectra in a similar way, especially when the spectral range does not
include a strong break such as the Balmer jump. 

In this section we discuss the effect of including extinction as a
free parameter on the fitting results obtained on our mock
clusters. This is done by allowing our fitting technique to apply the
\cite{Cardelli89} extinction law 
with $R_V = 3.1$, using the range of E(B-V) from 0.0 to 0.5, in steps of 0.01.
We perform this analysis by comparing Padova model clusters with the Padova
 isochrone models for the full wavelength range 
$3700 \leqslant \lambda/\mbox{\AA} \leqslant 6200$.

We start with analyzing the case of fixed metallicity. Here we use the models
with the same metallicity as the mock clusters in order to constrain that
variable in the fitting procedure. The left panel of Figure \ref{red} shows
log\,(age) obtained when accounting for extinction versus log\,(age) obtained
without accounting for extinction.

Overall, we find that ages are correctly retrieved for the higher
S/N ratios (above $\sim$\,50). However,  for the lower S/N ratios, there is a
tendency to fit somewhat younger ages when reddening is a variable in the
fitting. This tendency is strongest in the age range 7.8 $<$ log\,(age) $<$ 8.5,
and is also apparent for log\,(age) $\ga 9.5$. This tendency is however not seen
for the youngest ages (log\,(age) $\la 7.5$), which is probably mainly due to
the prominent Balmer jump at those ages.  

We repeat this analysis by letting all parameters (age, metallicity and
reddening) vary. The results are shown in the right panel of Figure
\ref{red}.  It turns out that this extra degree of freedom does not worsen
the tendency to fit slightly younger ages along with small amounts of
  extinction, nor does it change the age ranges where this tendency is observed
  in any significant way.

\section{Summary}

In this work we used synthetic spectra of 21,000 artificial star clusters
spanning an age range of 6.8 $<$ log age $<$ 10.2, with S/N ratios between 10 to
100 using isochrone models of both Padova and MIST to analyze the accuracy
  and precision of ages and metallicities of SSPs obtained from full-spectrum
fitting in three wavelength intervals. Our results can be summarized in the
following points: 

(A) Overall, this method can be used in obtaining ages of simple
stellar populations to a precision of 0.1 dex (i.e., $\sim$\,25\%) in the
age range $7.0 \leqslant \mbox{log\,(age/yr)} \leqslant 9.5$, provided
  that S/N $\geq$ 50. Metallicity determinations using this technique are
  however generally not as precise. For S/N $\ga 50$, typical standard
  deviations $\sigma$ for [Z/H] range between $\sim$\,0.05 dex and 0.20 dex, depending on   the age, and worse for lower S/N. 

(B) Age determinations using the full wavelength range
$3700 \leqslant \lambda/\mbox{\AA} \leqslant 6200$ are generally not as
  precise as those obtained for the narrower range 
$3700 \leqslant \lambda/\mbox{\AA} \leqslant 5000$. This likely reflects
  the power of the 4000\,\AA\ Balmer jump for age determination, which is
  diluted when including more spectral range on the long-wavelength end of the
  optical range. 

 (C) For optical integrated-light spectra of SSPs with S/N $\leq 50$,
  full-spectrum fitting yields significant age uncertainties in the age ranges
  $8.2 \la \mbox{log\,(age/yr)} \la 8.8$ and log\,(age/yr) $\ga 9.5$ (for the
  Padova models). This is due to the fact that the shape of the CMD does not
  change significantly with age in those intervals.

(D) There is no significant difference in the precision of age and metallicity determination when using MIST models or Padova models
for clusters based on the same models.

(E) MIST and Padova isochrones make different assumptions about stellar physics 
in various important (i.e., luminous) phases of stellar evolution  (e.g.,
  the onsets of the supergiants, AGB, and RGB phases). We quantify the impact of
  these differences to the derived ages as a function of age, finding
  significant differences in the age ranges $7.5 \la \mbox{log\,(age/yr)} \la 8.5$
  and $9.3 \la \mbox{log\,(age/yr)} \la 9.7$. 
  This issue needs to be taken in consideration when comparing  SSP ages 
  obtained by full-spectrum fitting using different models.

(F) When using models of fixed metallicity to predict the age, ages are
generally underestimated when using models of higher metallicities than the mock
clusters' metallicity, and overestimated when using models of lower metallicity
then the mock clusters.  

(G) We assess the effect of including extinction as an additional free
  parameter in full-spectrum fitting to the accuracy of age and metallicity
  determination, using a range $0.0 \leq E(B-V) \leq 0.5$. 
Overall, we find that ages are correctly retrieved to within 0.1 dex for
spectra with S/N $\ga 50$. However, for lower S/N ratios, there is a tendency to
fit younger ages  (by $\Delta$\,log\,(age/yr) $\la$ 0.7, depending on the age) when extinction is a variable in the fitting.

\section*{Acknowledgments}
R.A. thanks the Space Telescope Science Institute for a sabbatical visitorship
including travel and subsistence support as well as access to their science
cluster computer facilities. This work is supported in part by the
EFRG18-SET-CAS-74 and FRG19-M-S77 (Grant P.I., R. Asa'd) from American
University of Sharjah. R.A. is grateful to Benjamin Johnson for useful
discussions on using python-FSPS.  

\bibliographystyle{mnras}
\bibliography{References}


\clearpage


\clearpage



\begin{figure*}

\resizebox{75mm}{!}{\includegraphics[width=\columnwidth, angle=270]{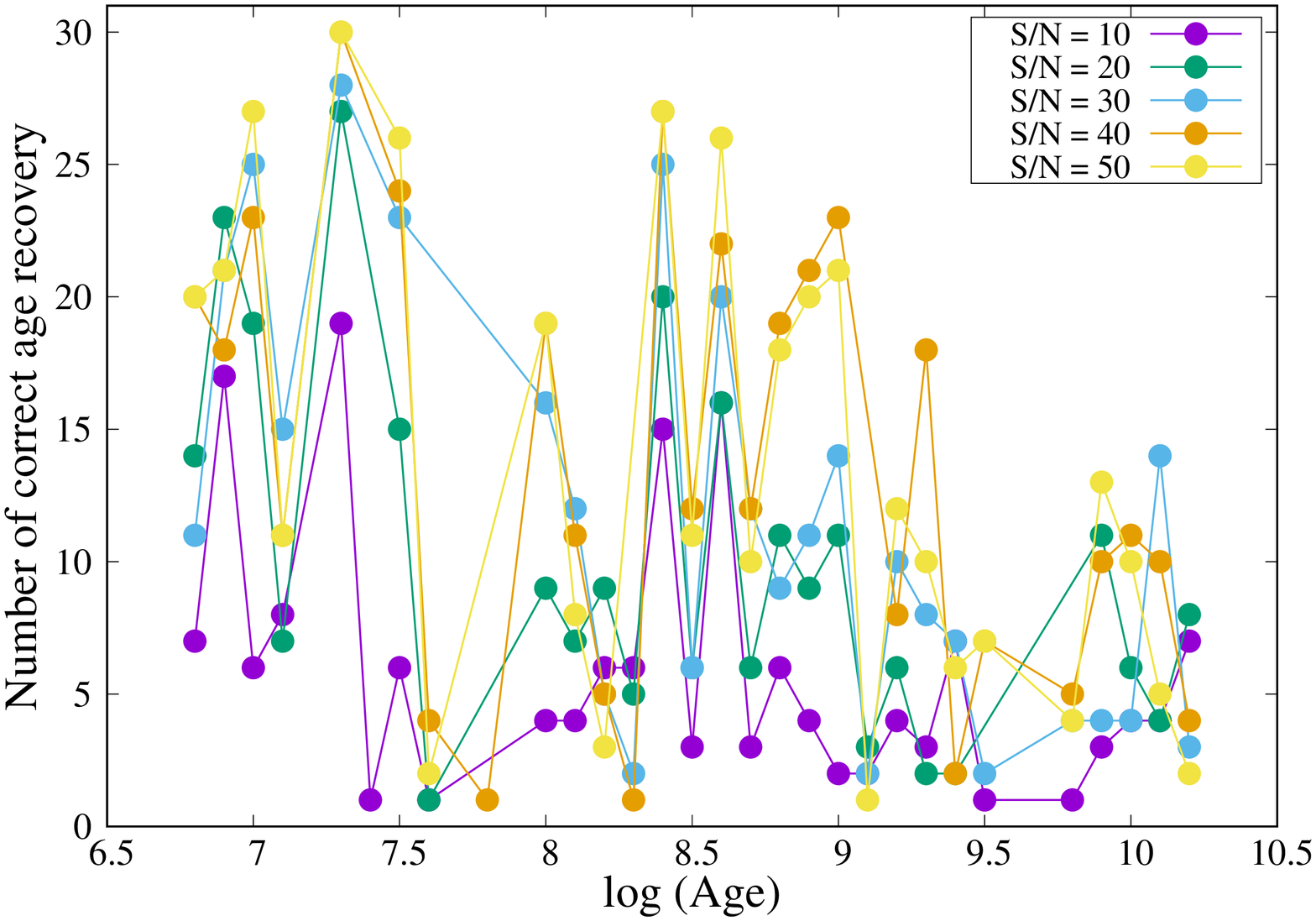}}
\resizebox{75mm}{!}{\includegraphics[width=\columnwidth, angle=270]{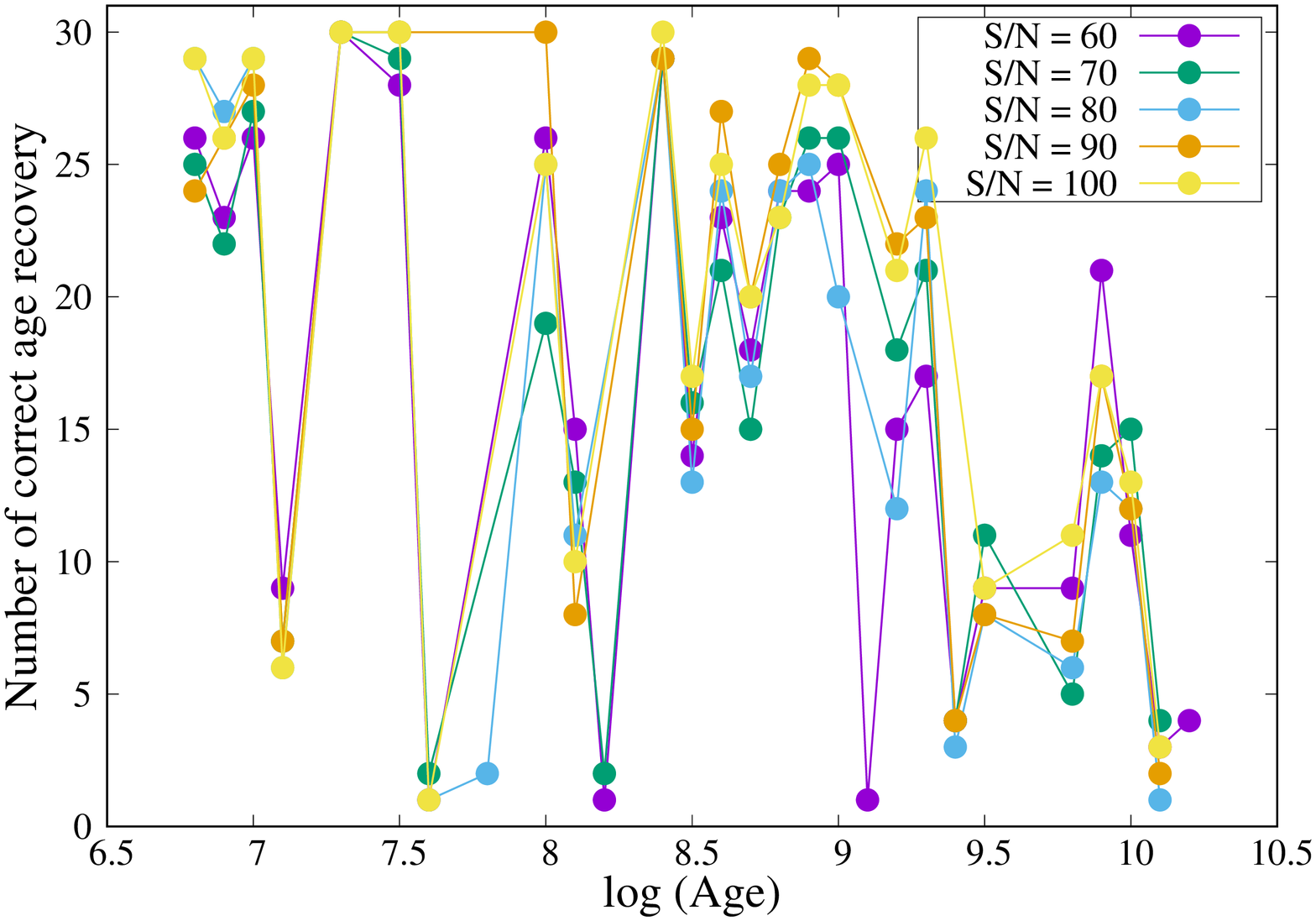}}
\resizebox{75mm}{!}{\includegraphics[width=\columnwidth, angle=270]{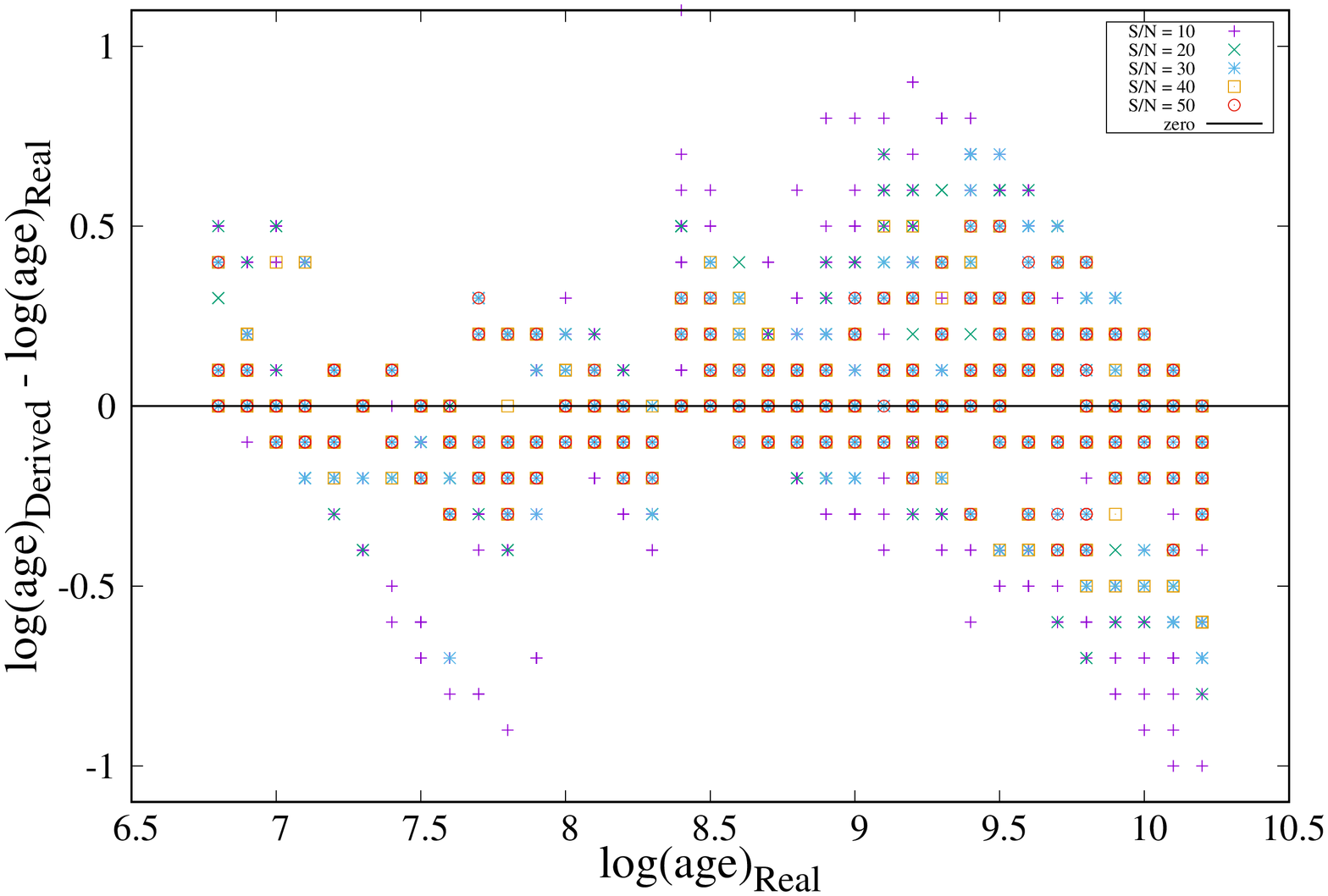}}
\resizebox{75mm}{!}{\includegraphics[width=\columnwidth, angle=270]{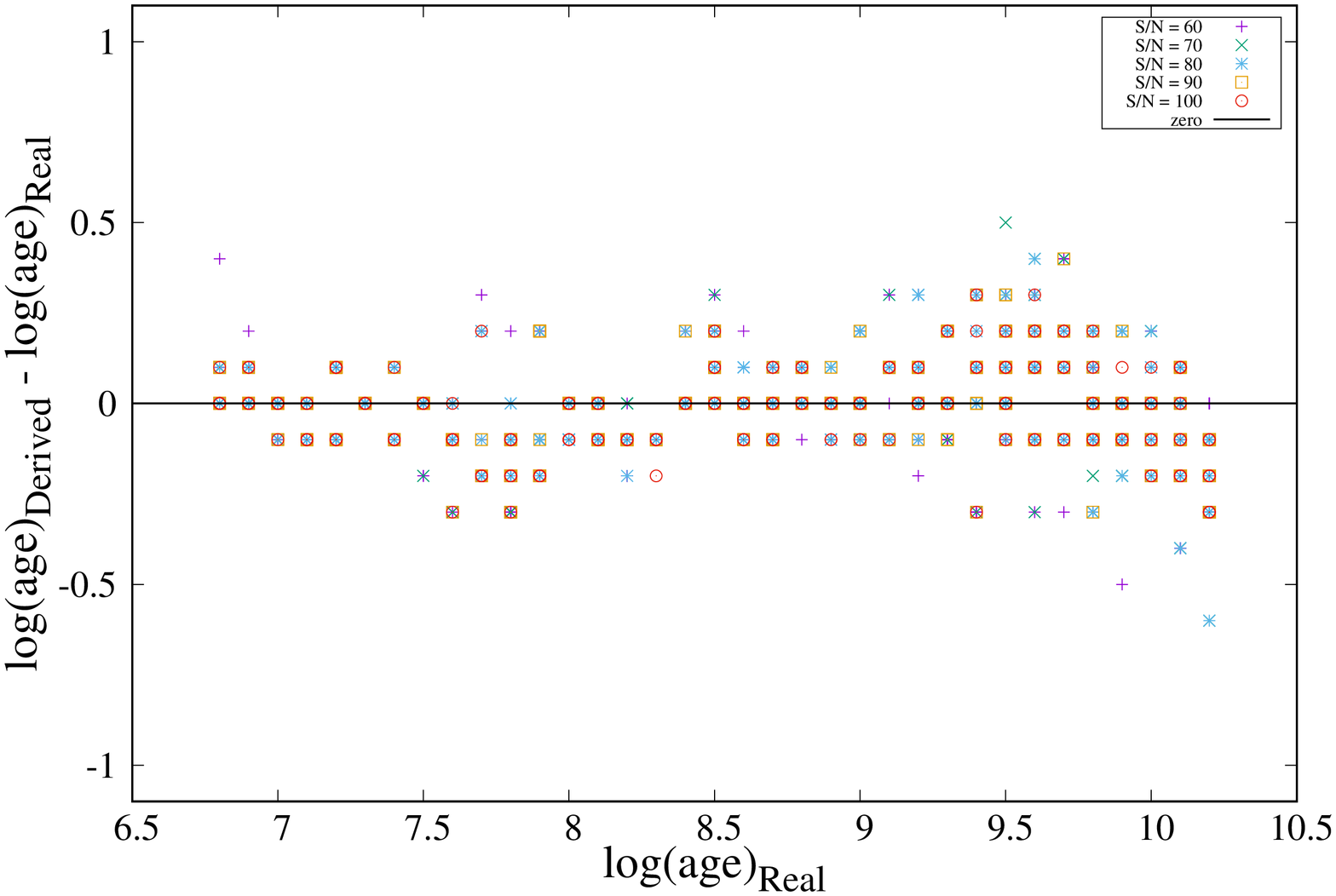}}
\resizebox{75mm}{!}{\includegraphics[width=\columnwidth, angle=270]{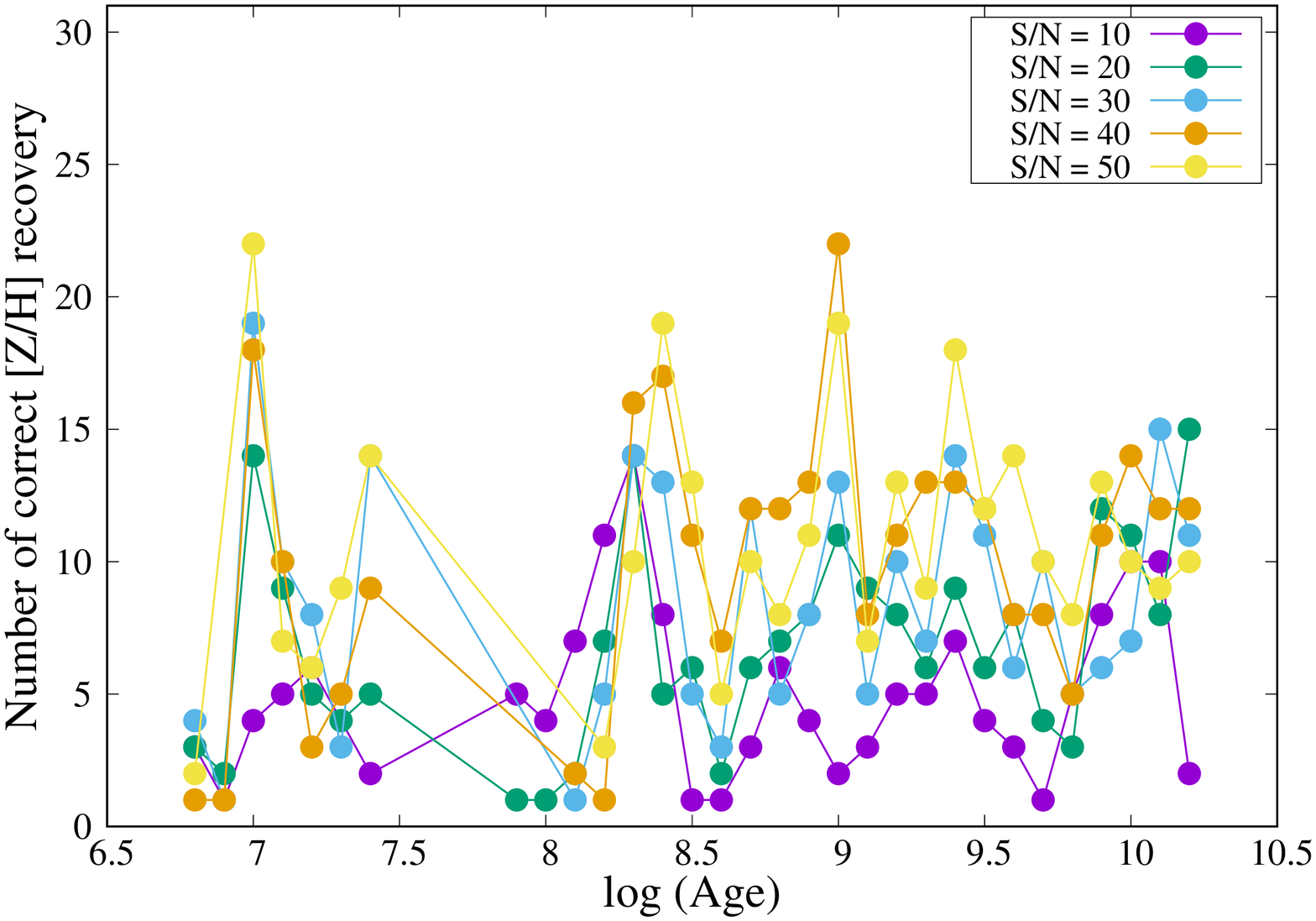}}
\resizebox{75mm}{!}{\includegraphics[width=\columnwidth, angle=270]{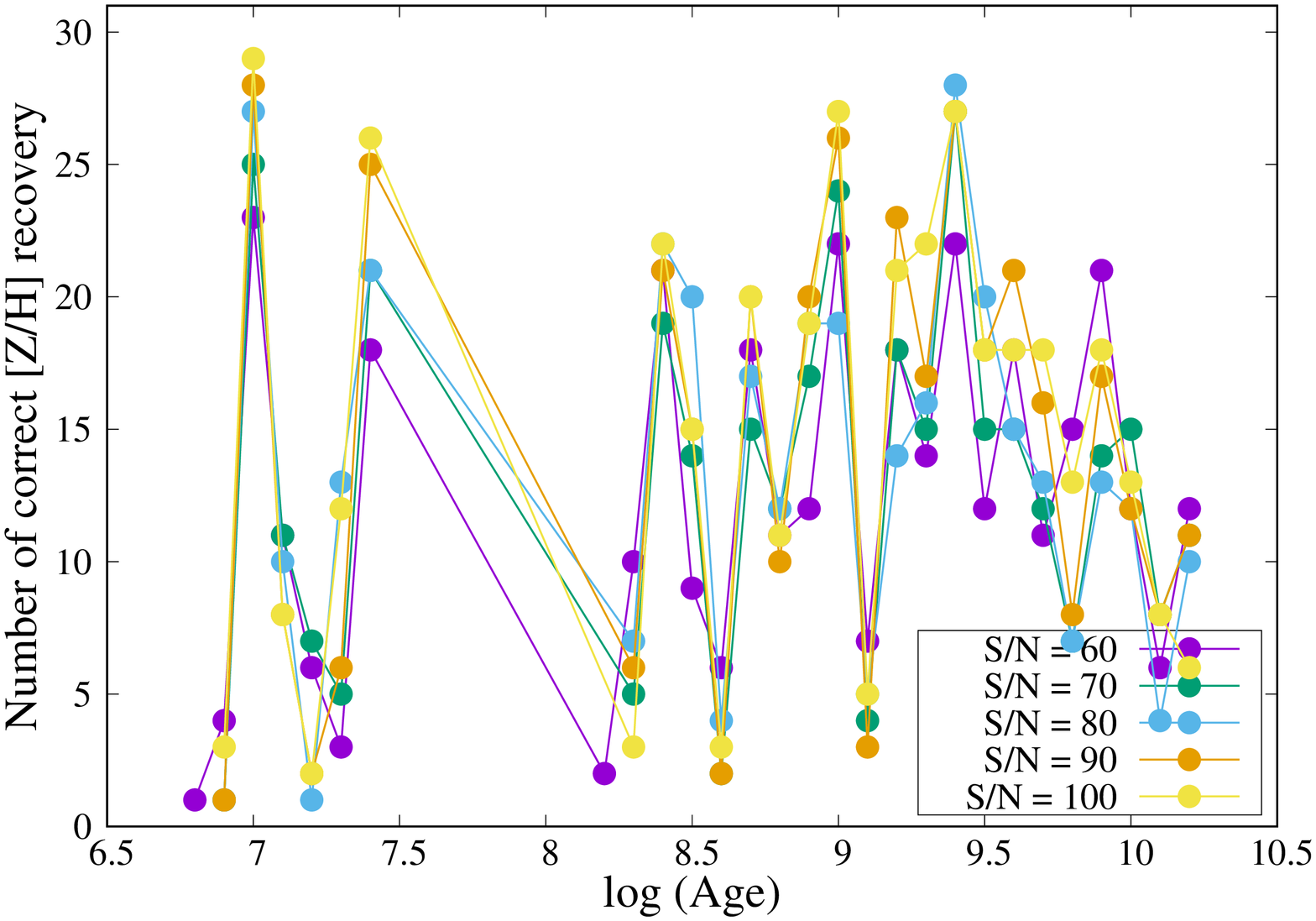}}
\resizebox{75mm}{!}{\includegraphics[width=\columnwidth, angle=270]{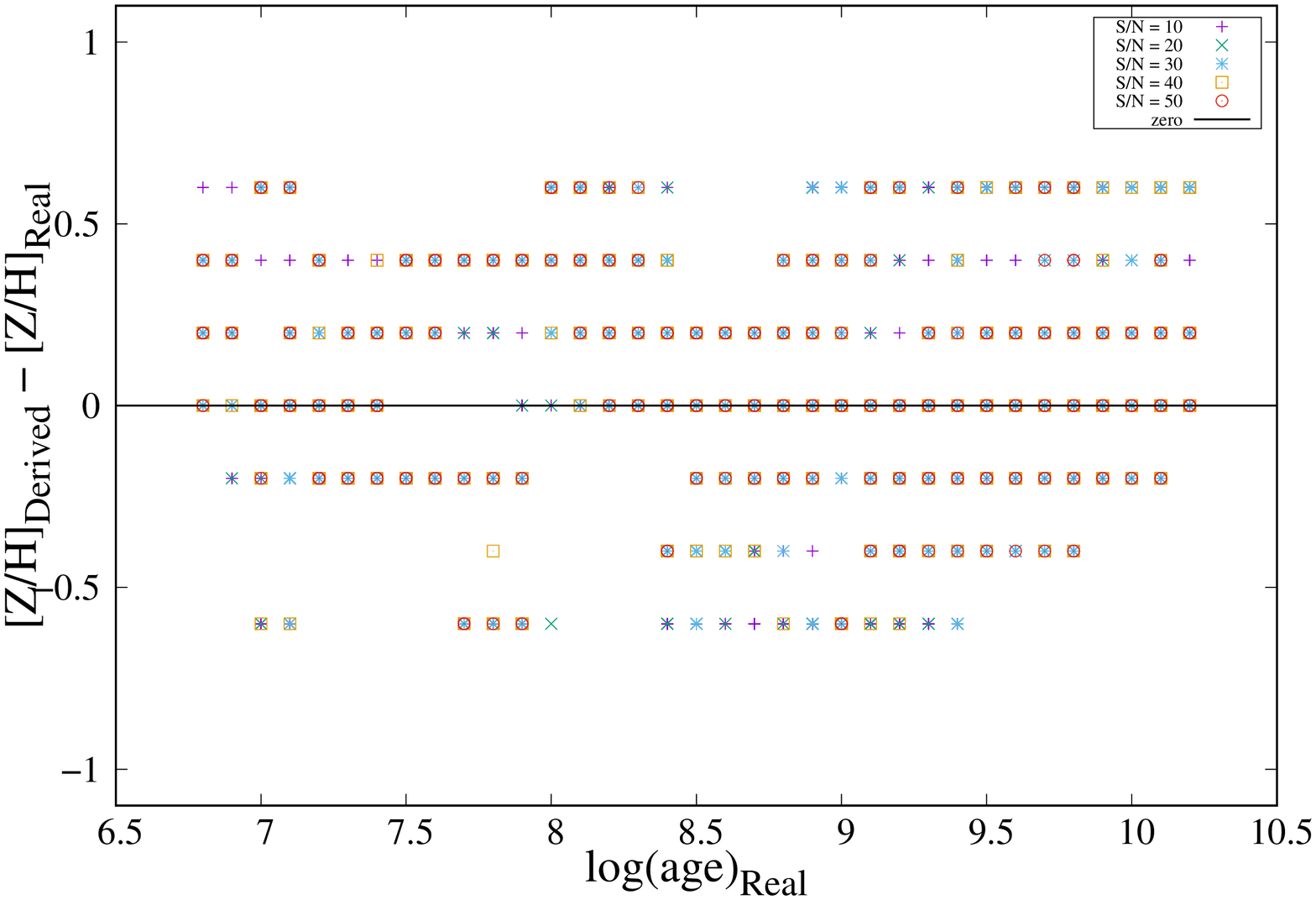}}
\resizebox{75mm}{!}{\includegraphics[width=\columnwidth, angle=270]{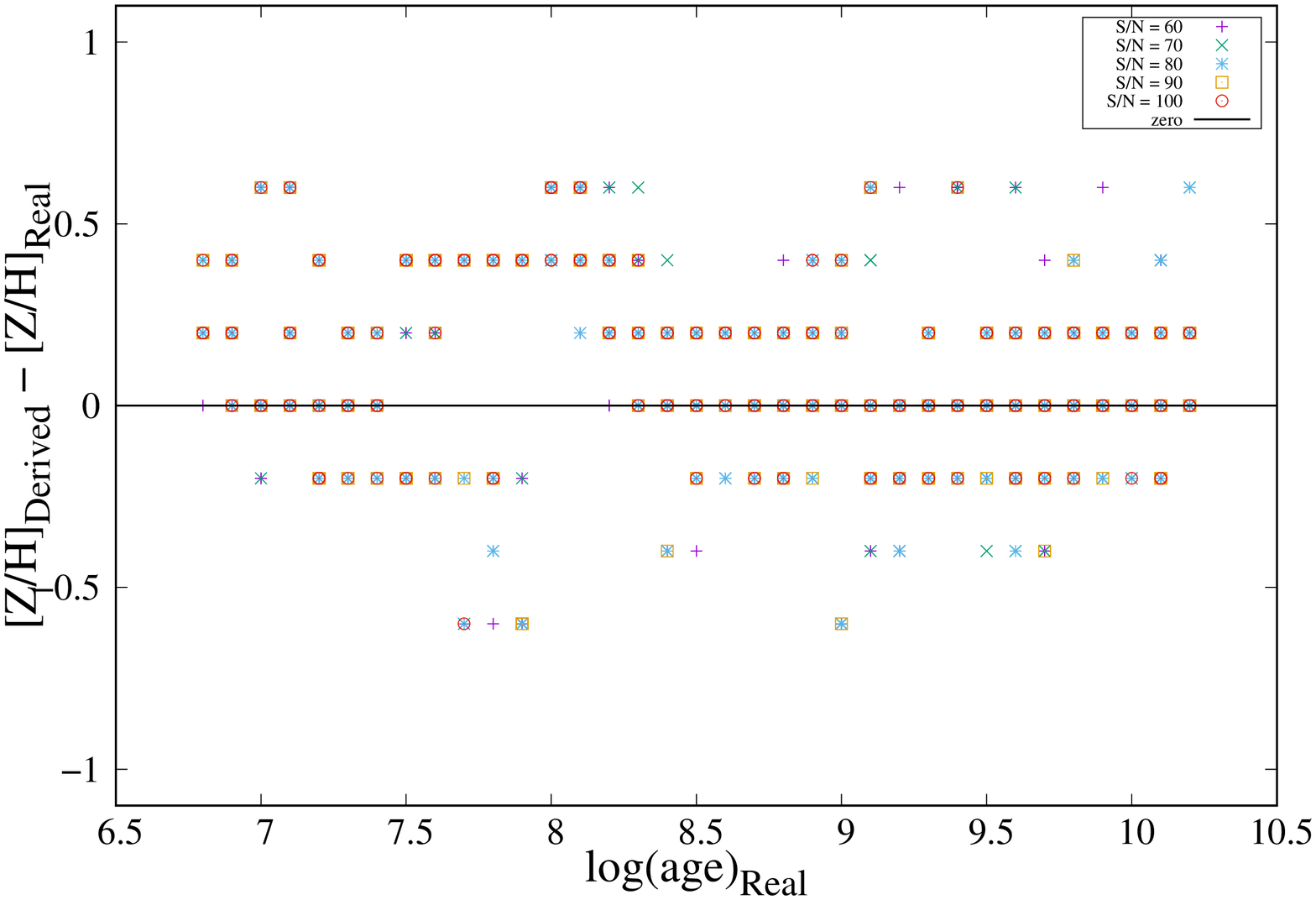}}

\caption{The same as figure \ref{Fig7} for the range $3700 \leqslant \lambda/\mbox{\AA} \leqslant 5000$ }
\label{Fig8}
\end{figure*}

\begin{figure*}

\resizebox{75mm}{!}{\includegraphics[width=\columnwidth, angle=270]{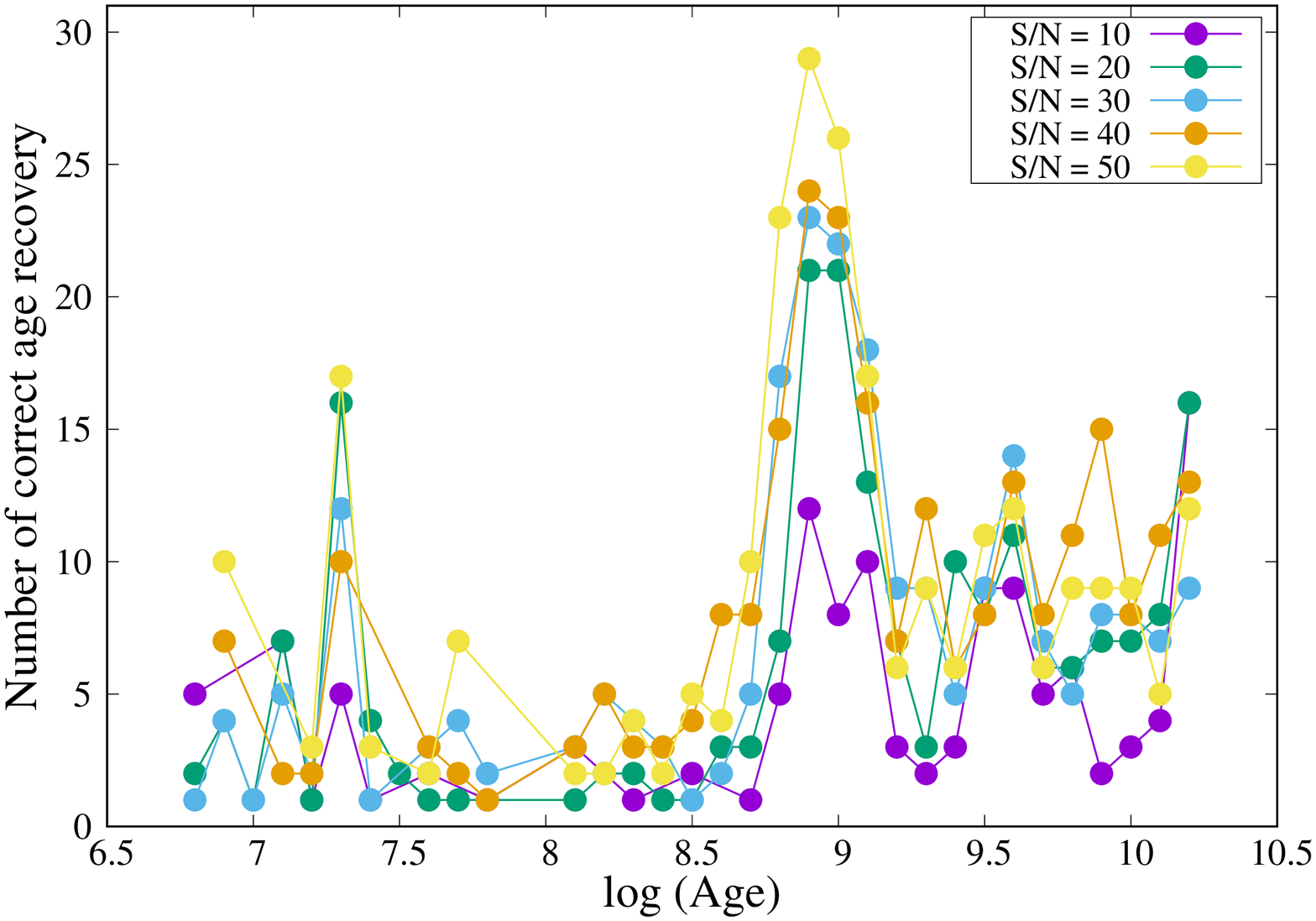}}
\resizebox{75mm}{!}{\includegraphics[width=\columnwidth, angle=270]{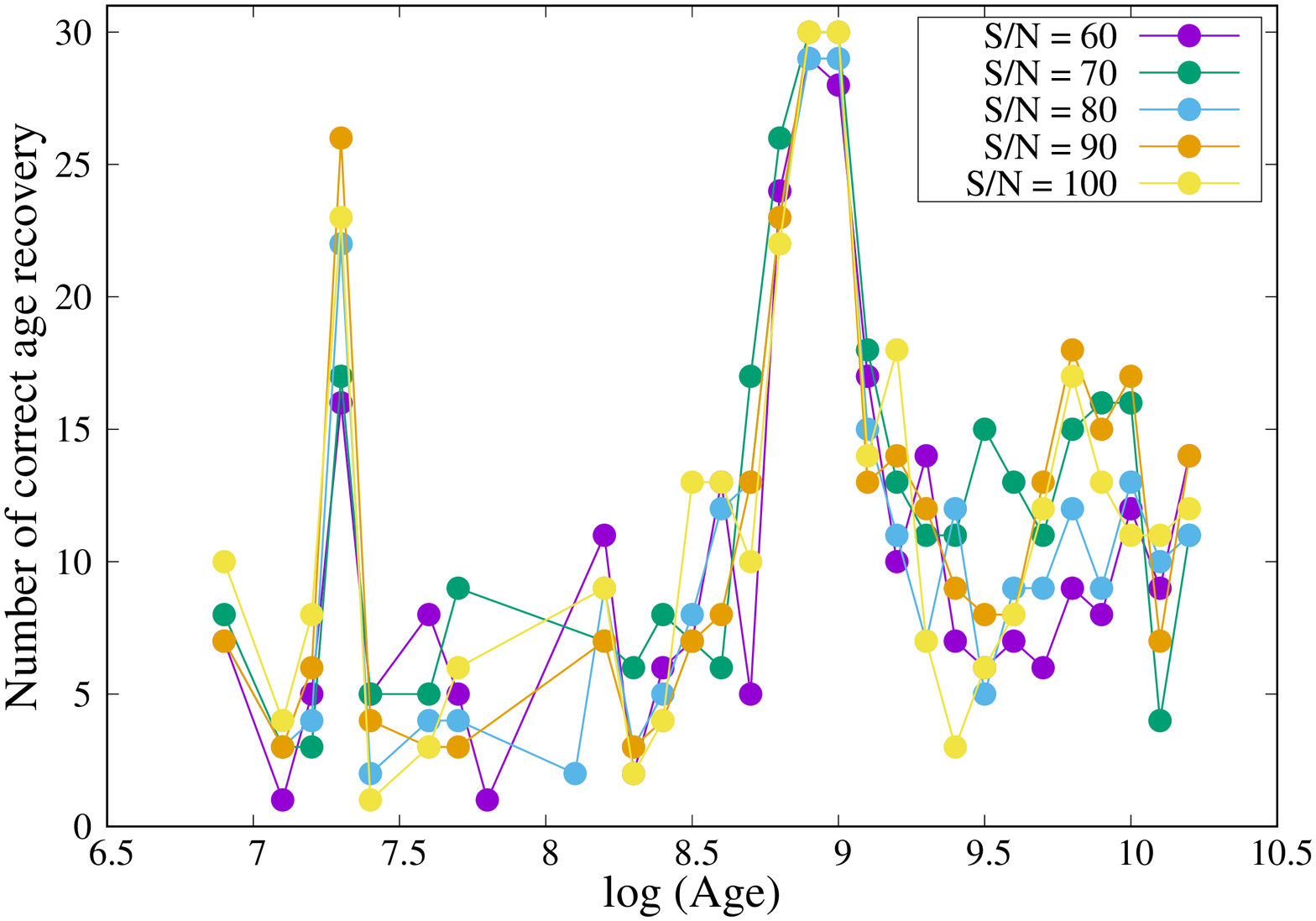}}
\resizebox{75mm}{!}{\includegraphics[width=\columnwidth, angle=270]{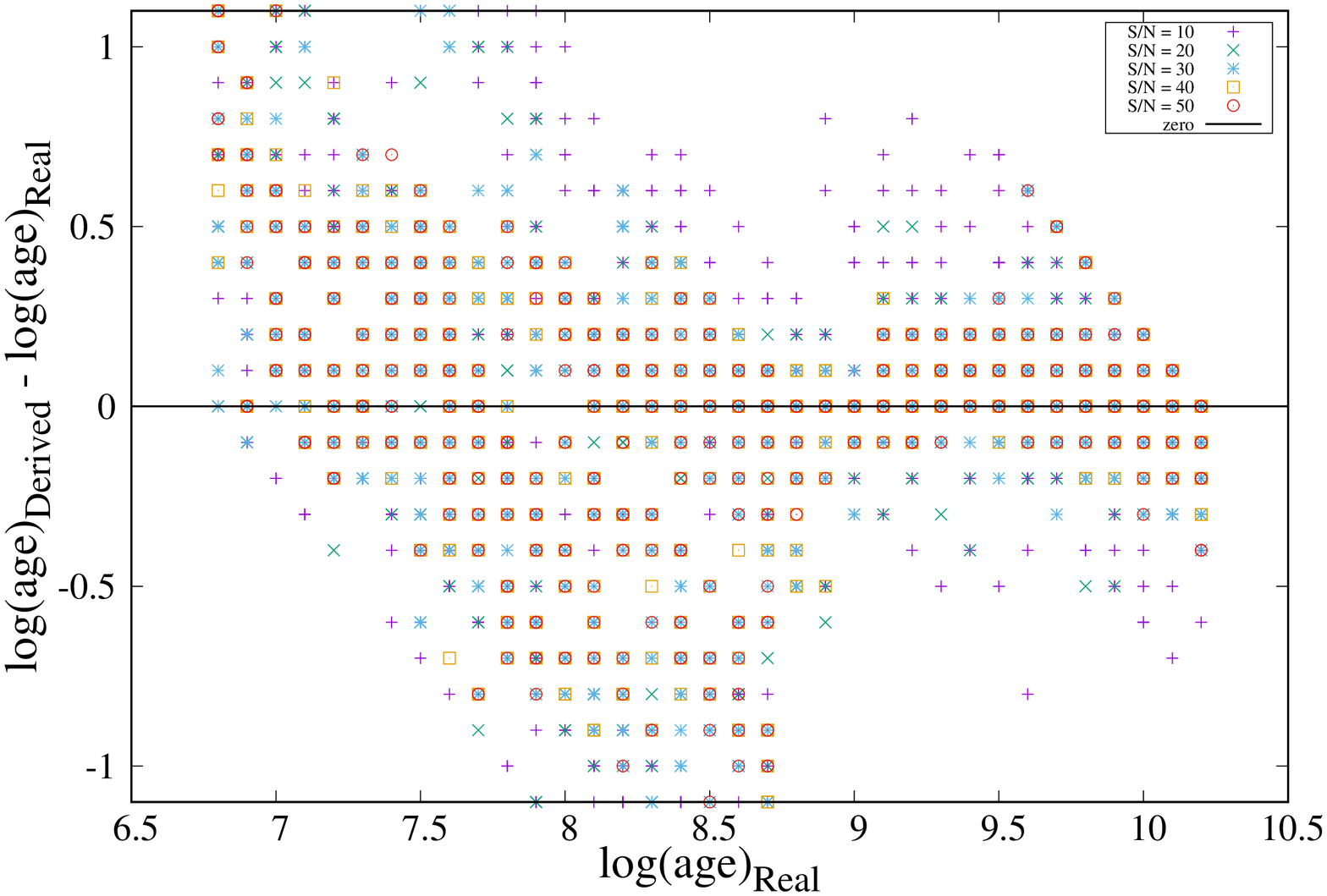}}
\resizebox{75mm}{!}{\includegraphics[width=\columnwidth, angle=270]{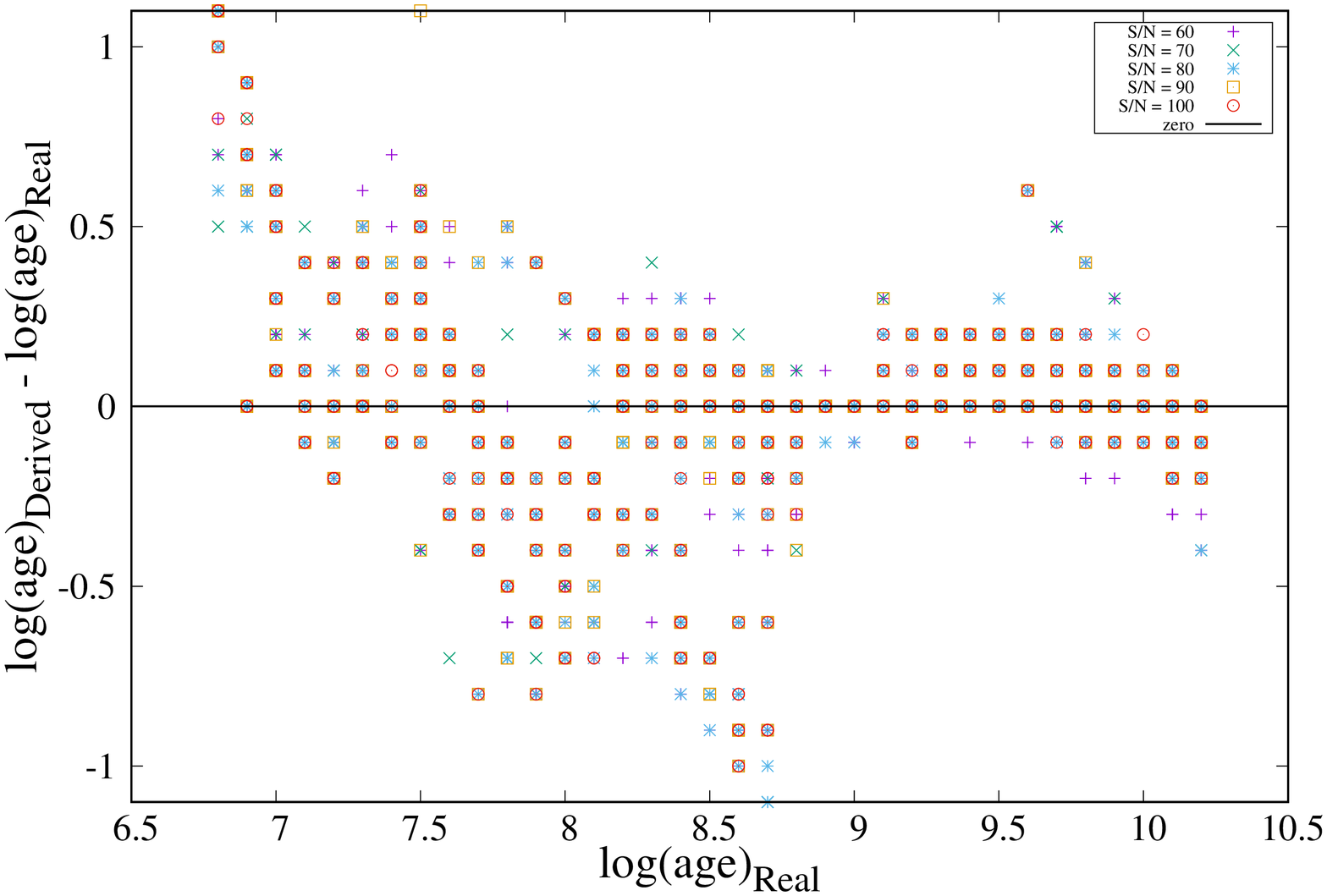}}
\resizebox{75mm}{!}{\includegraphics[width=\columnwidth, angle=270]{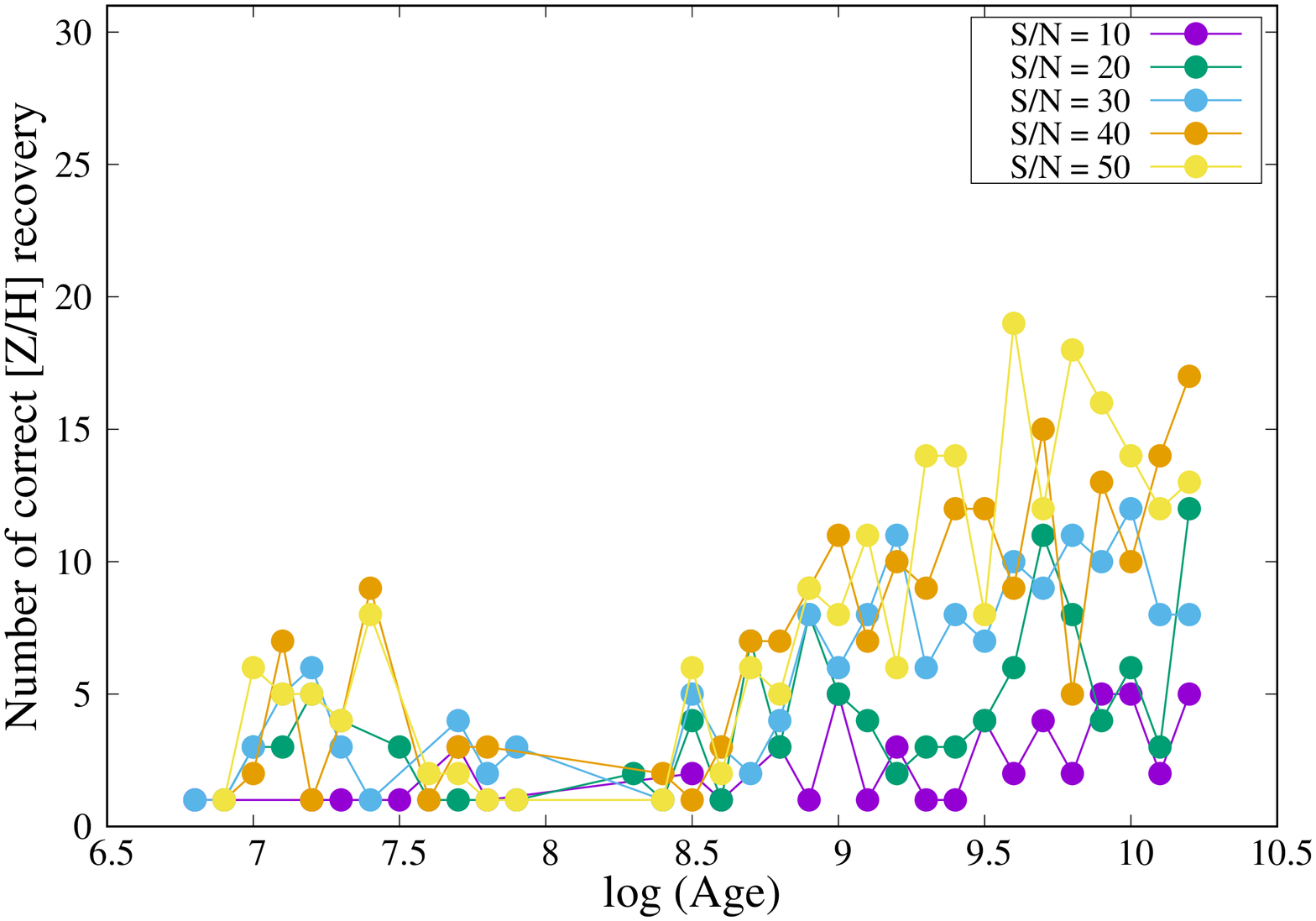}}
\resizebox{75mm}{!}{\includegraphics[width=\columnwidth, angle=270]{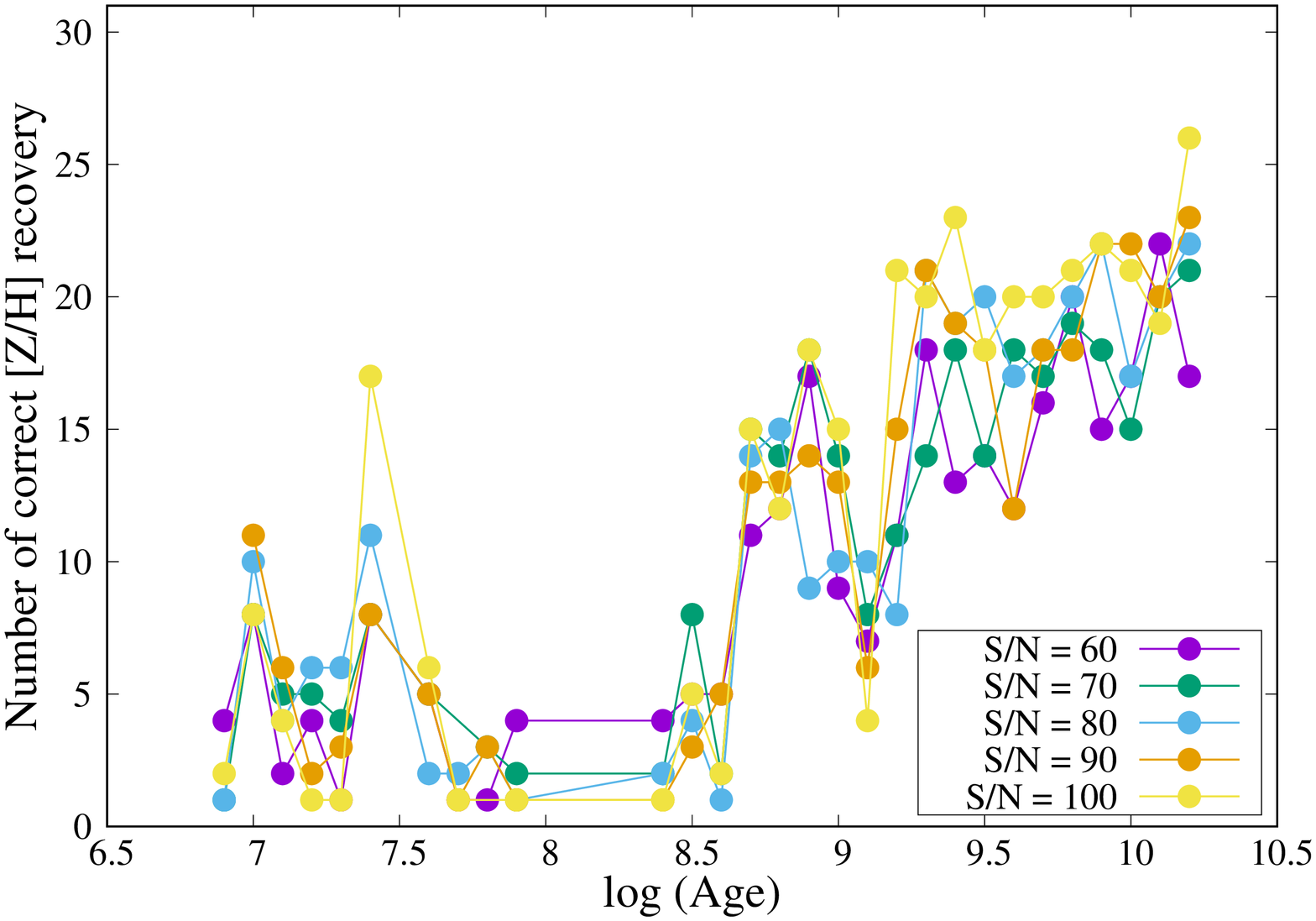}}
\resizebox{75mm}{!}{\includegraphics[width=\columnwidth, angle=270]{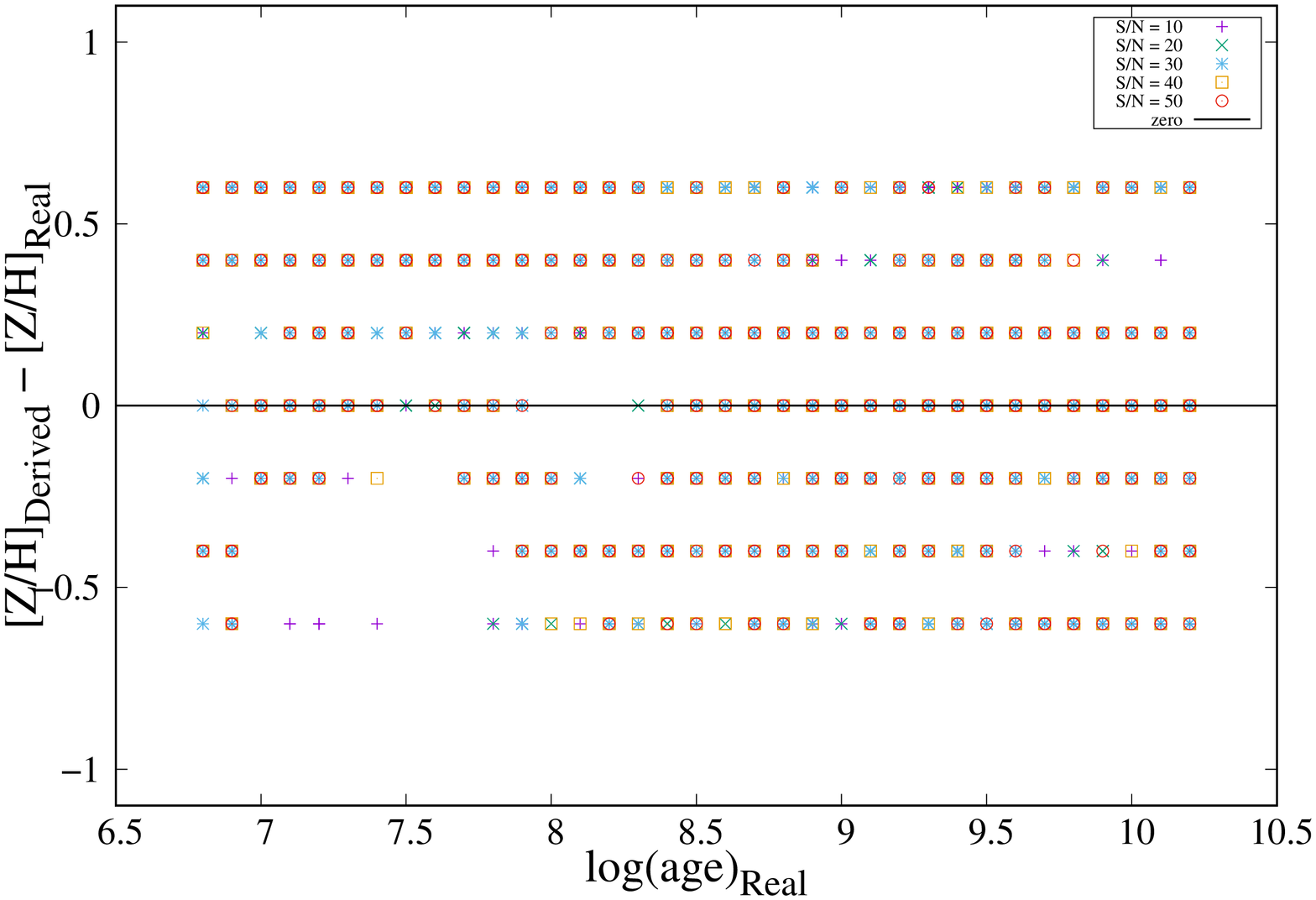}}
\resizebox{75mm}{!}{\includegraphics[width=\columnwidth, angle=270]{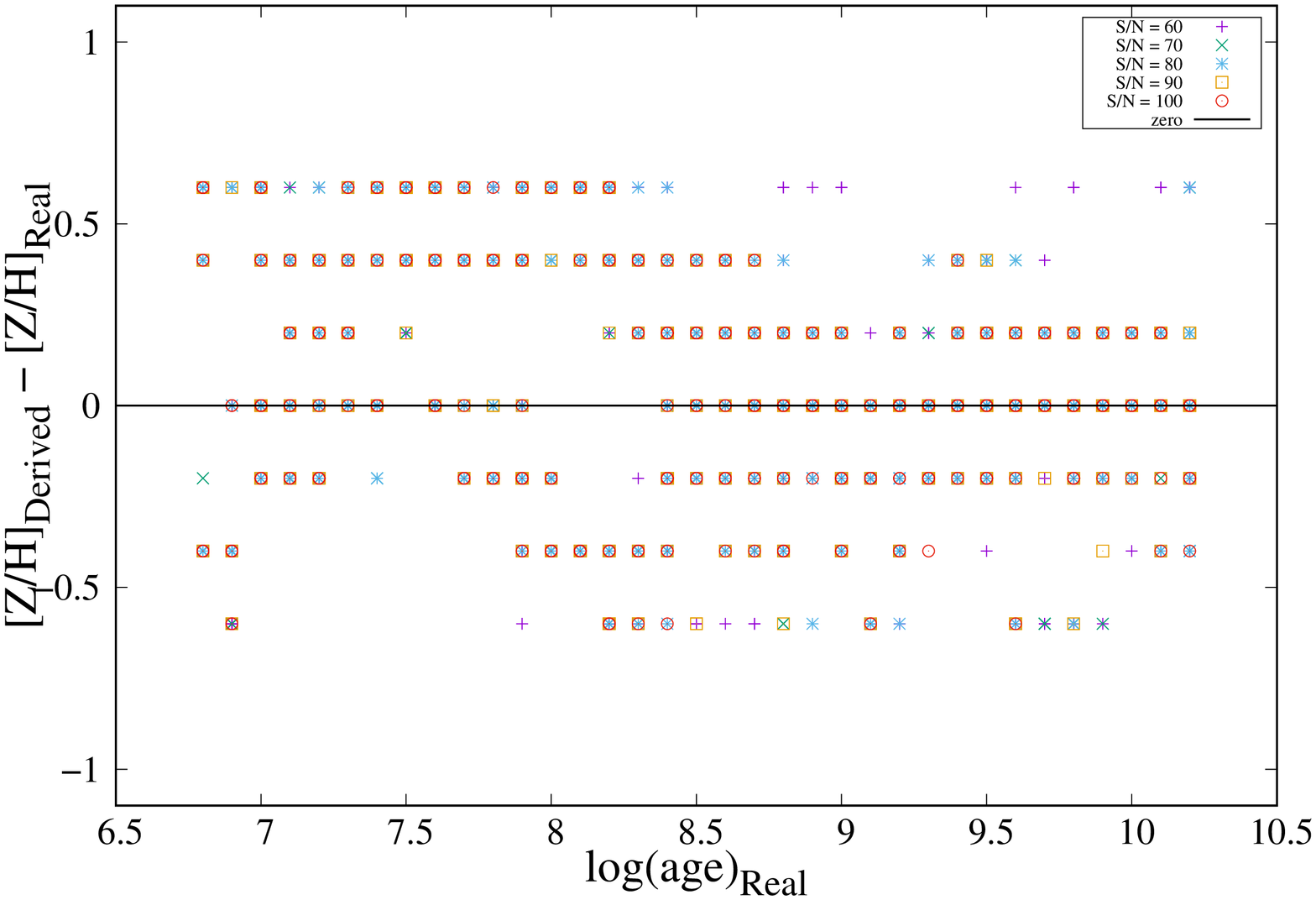}}

\caption{The same as figure \ref{Fig7} for the range $5000 \leqslant \lambda/\mbox{\AA} \leqslant 6200$ }
\label{Fig9}
\end{figure*}

\clearpage
\begin{figure*}

\resizebox{75mm}{!}{\includegraphics[width=\columnwidth, angle=270]{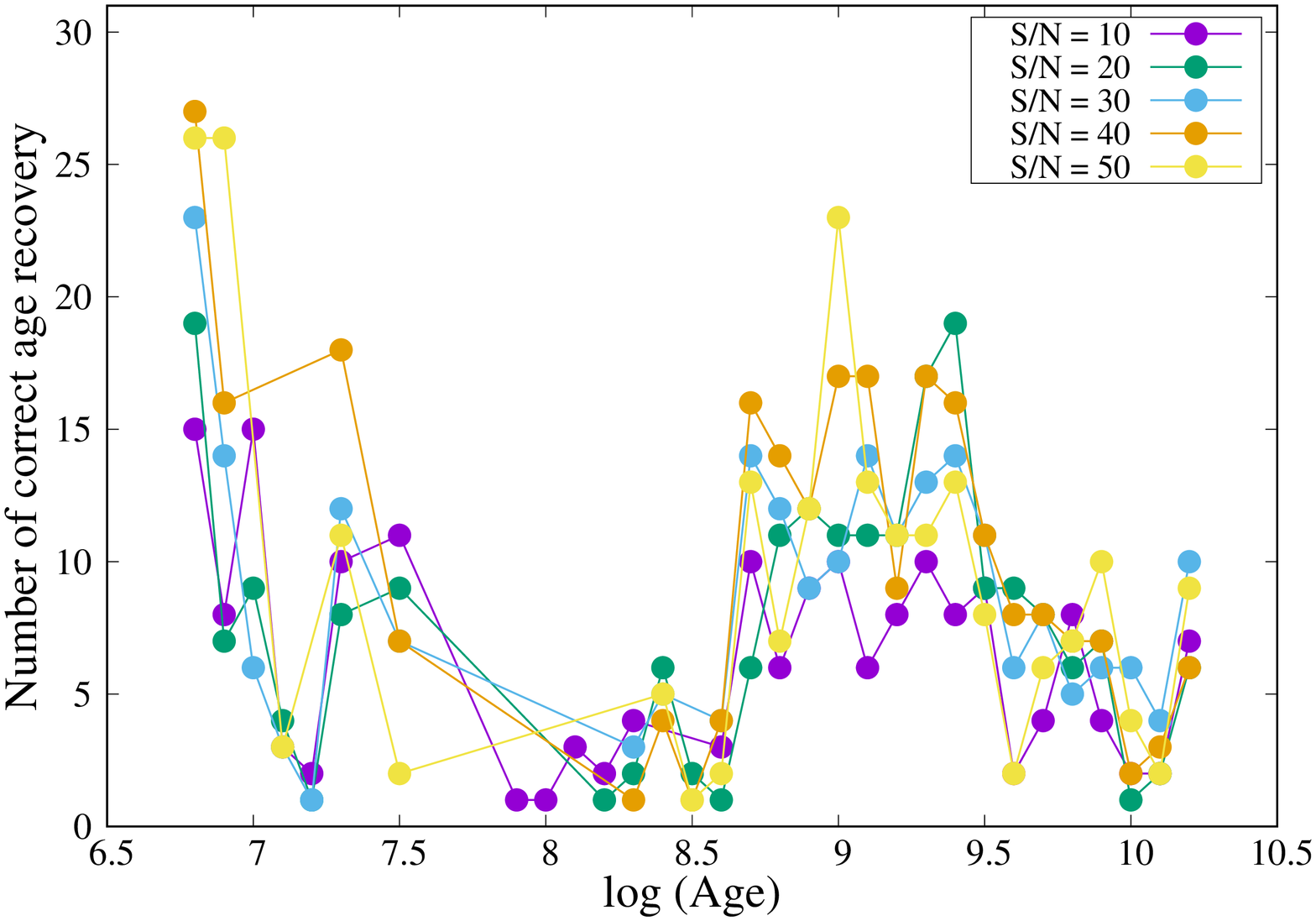}}
\resizebox{75mm}{!}{\includegraphics[width=\columnwidth, angle=270]{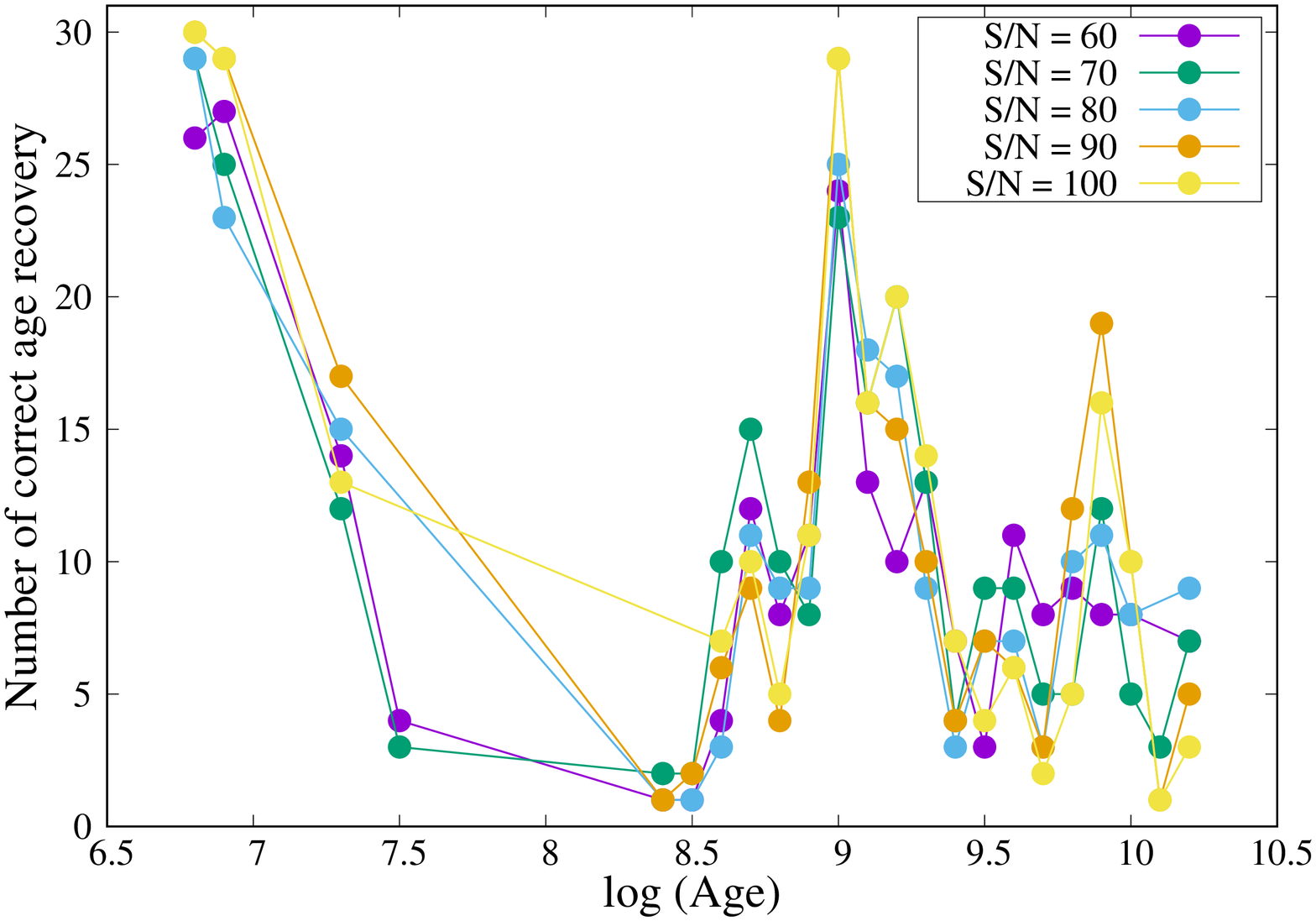}}
\resizebox{75mm}{!}{\includegraphics[width=\columnwidth, angle=270]{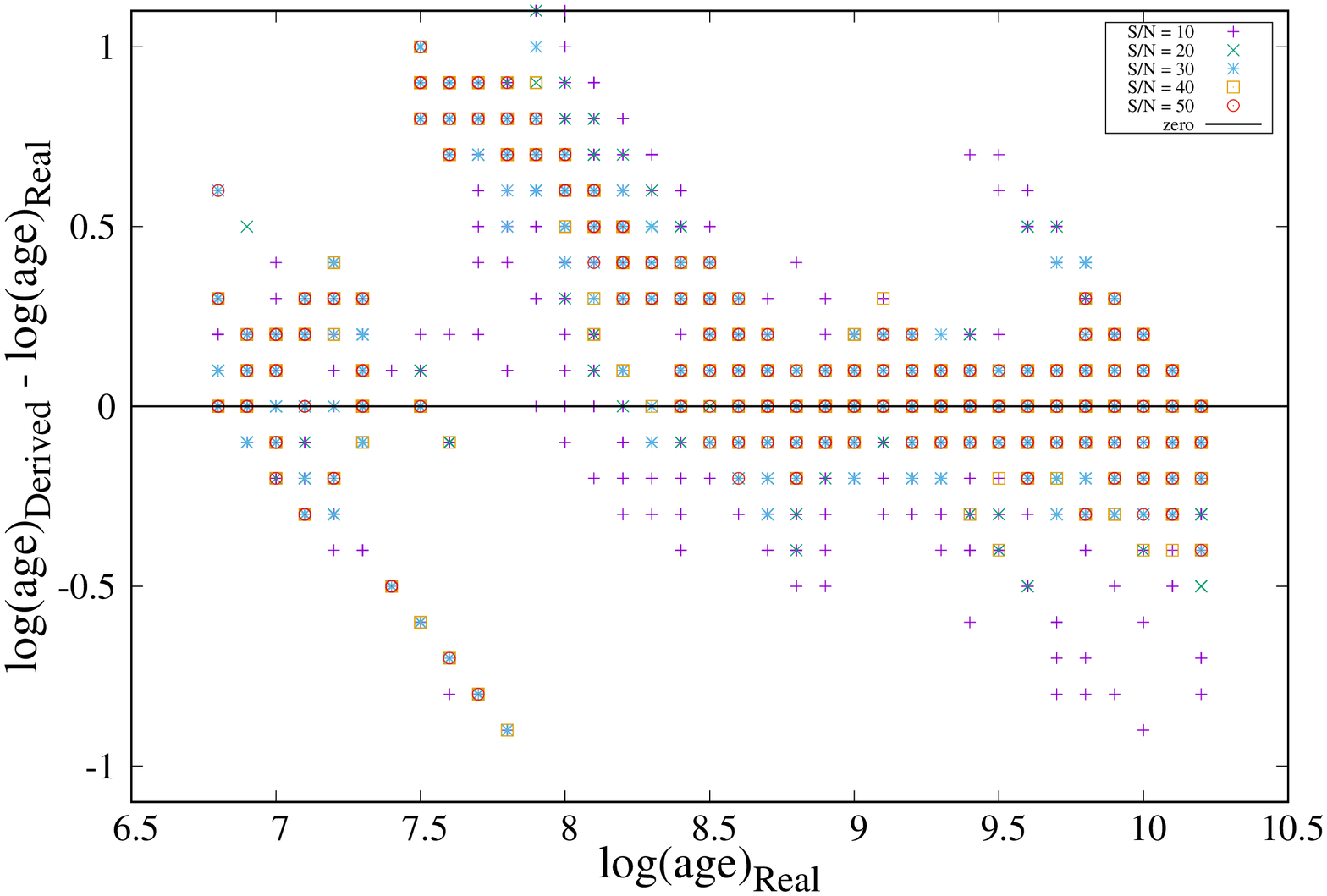}}
\resizebox{75mm}{!}{\includegraphics[width=\columnwidth, angle=270]{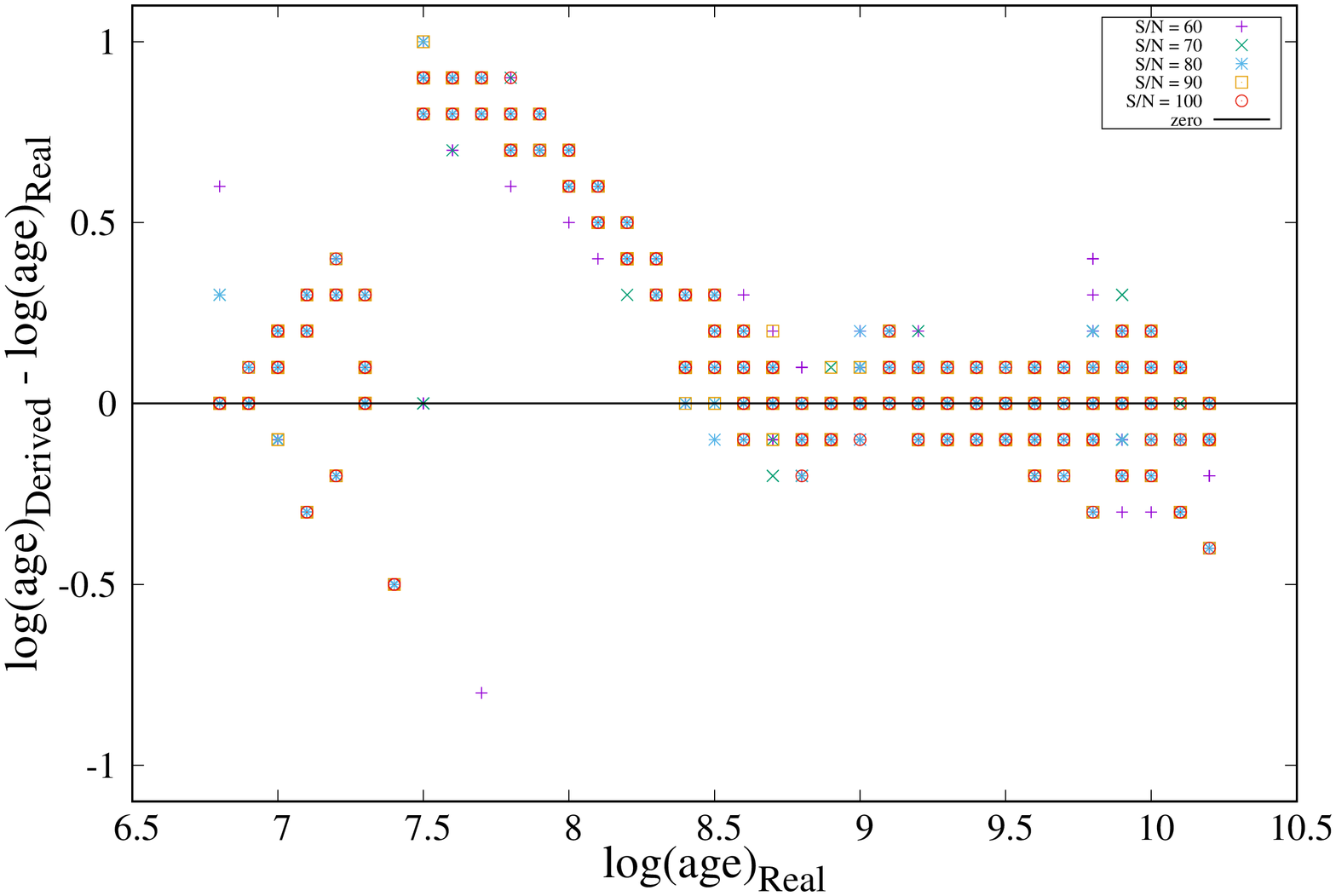}}
\resizebox{75mm}{!}{\includegraphics[width=\columnwidth, angle=270]{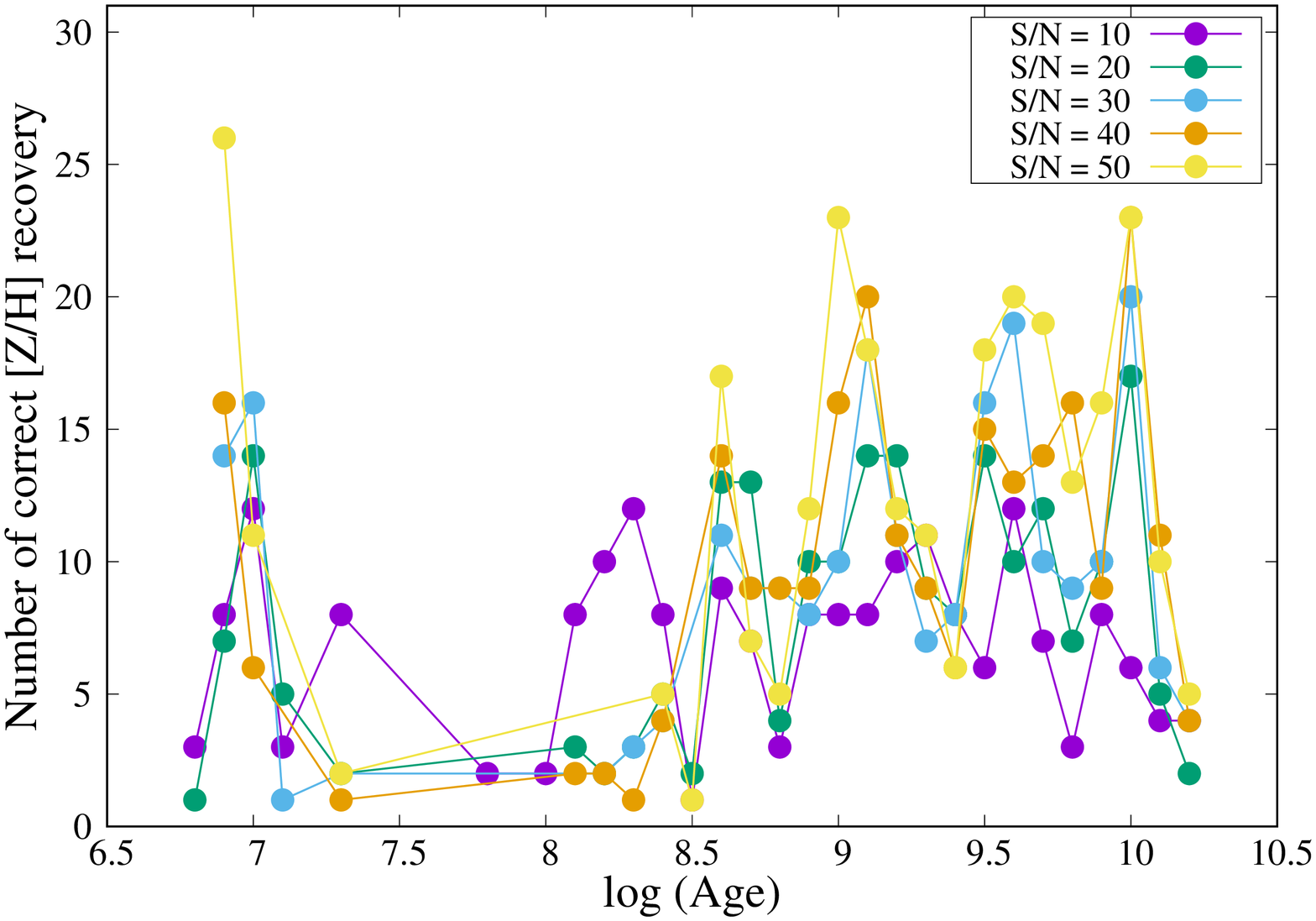}}
\resizebox{75mm}{!}{\includegraphics[width=\columnwidth, angle=270]{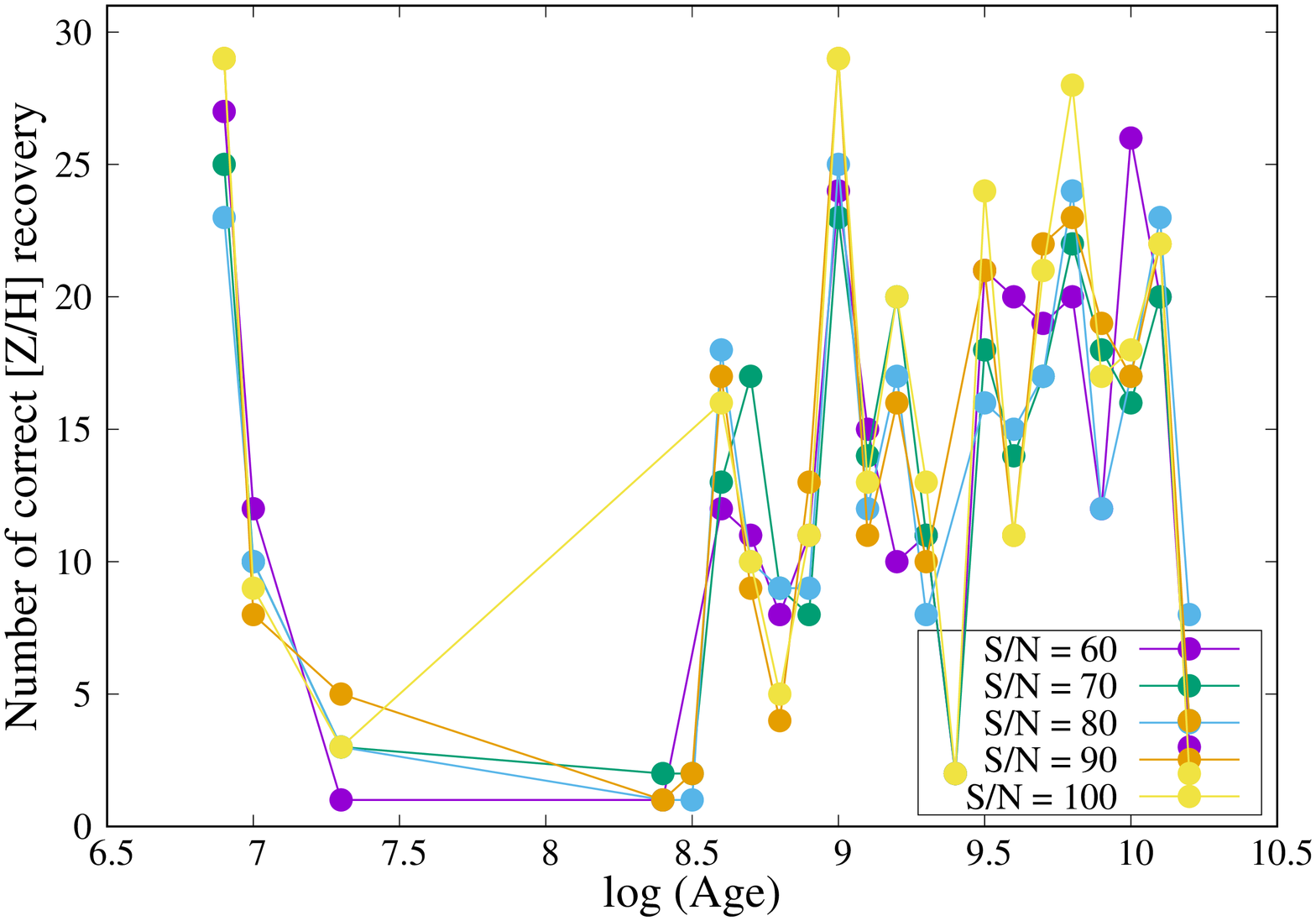}}
\resizebox{75mm}{!}{\includegraphics[width=\columnwidth, angle=270]{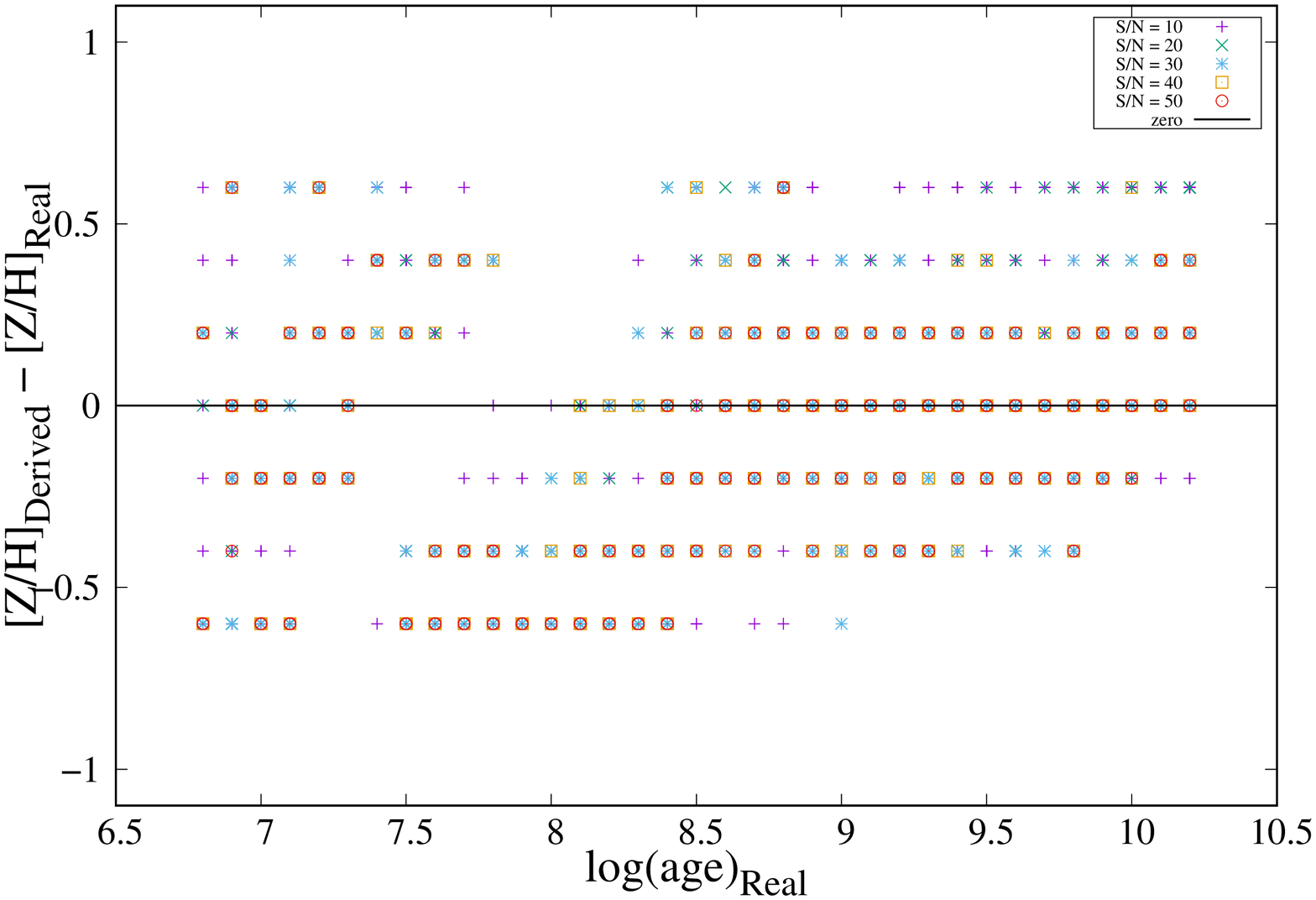}}
\resizebox{75mm}{!}{\includegraphics[width=\columnwidth, angle=270]{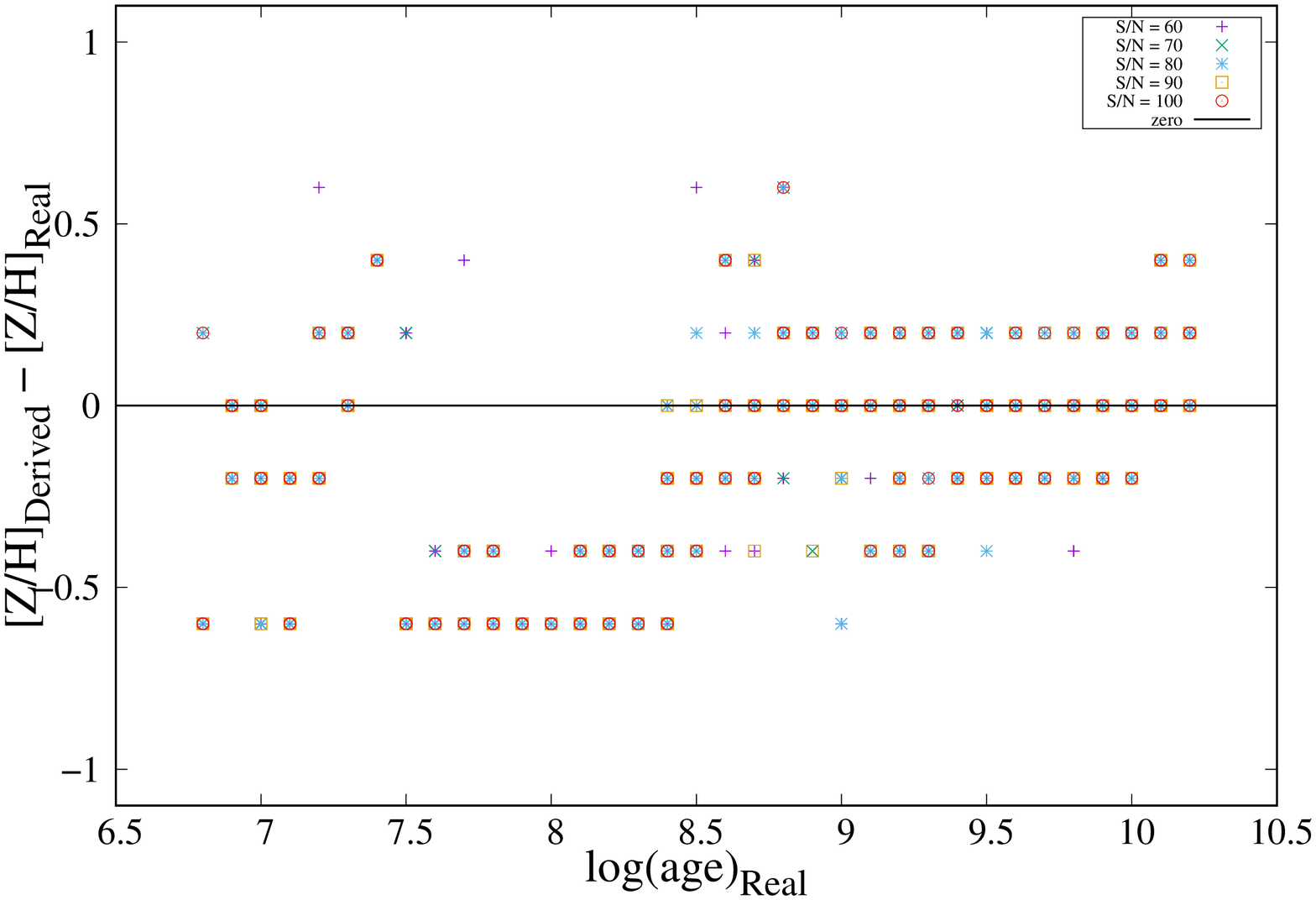}}

\caption{The same as figure \ref{Fig1} for the range $3700 \leqslant \lambda/\mbox{\AA} \leqslant 6200$ using Padova models and MIST mock clusters}
\label{Fig10}
\end{figure*}


\begin{figure*}

\resizebox{75mm}{!}{\includegraphics[width=\columnwidth, angle=270]{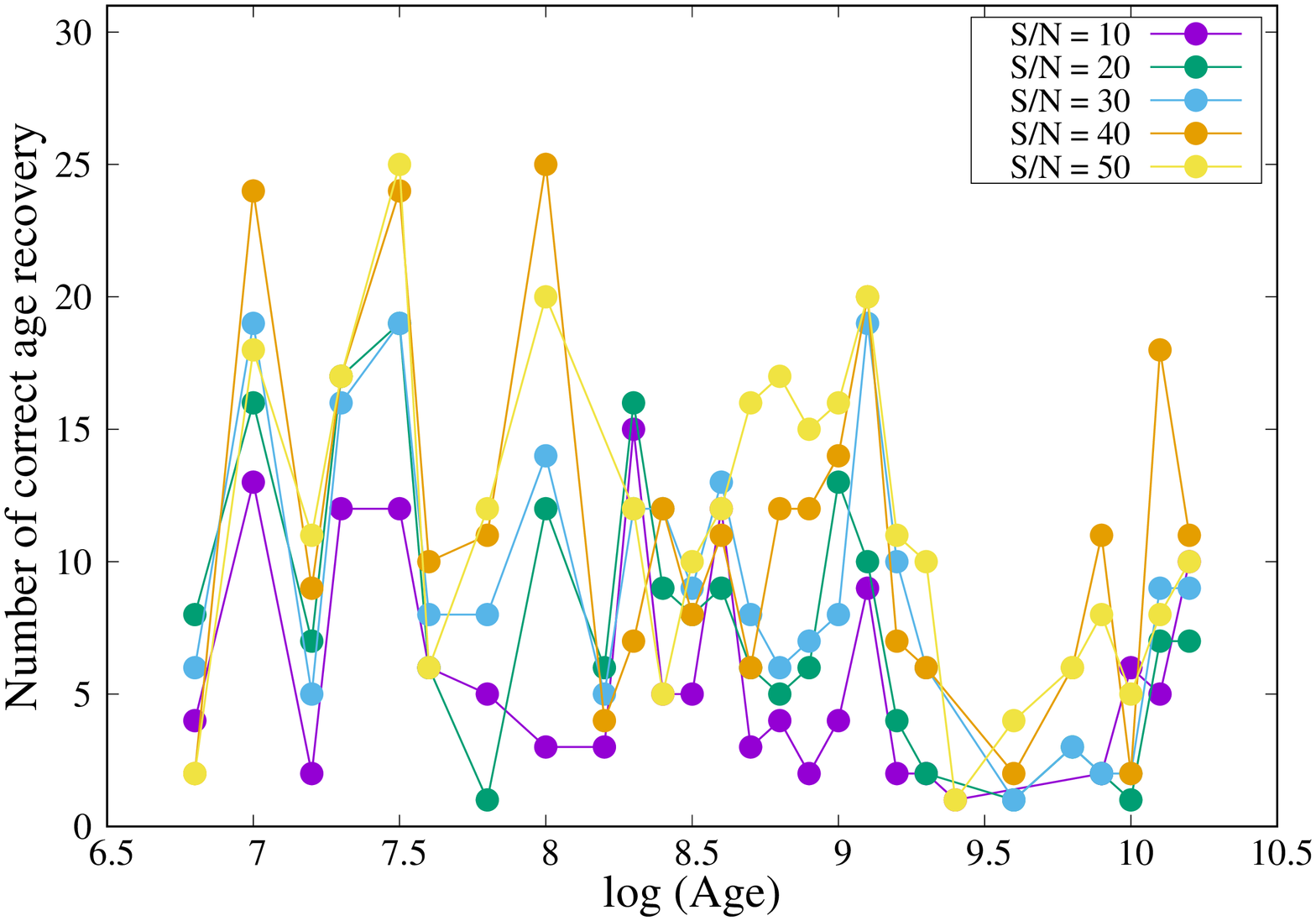}}
\resizebox{75mm}{!}{\includegraphics[width=\columnwidth, angle=270]{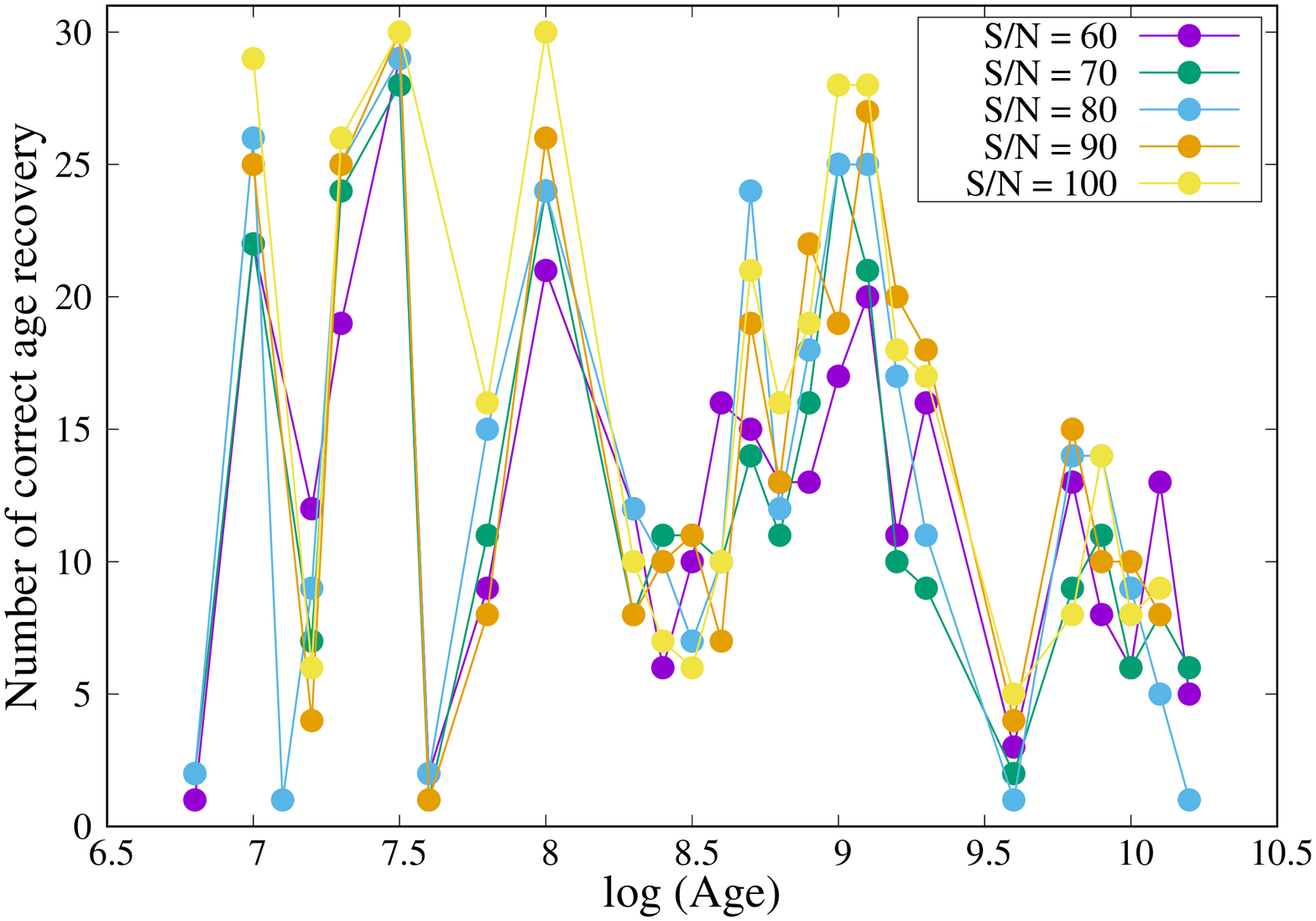}}
\resizebox{75mm}{!}{\includegraphics[width=\columnwidth, angle=270]{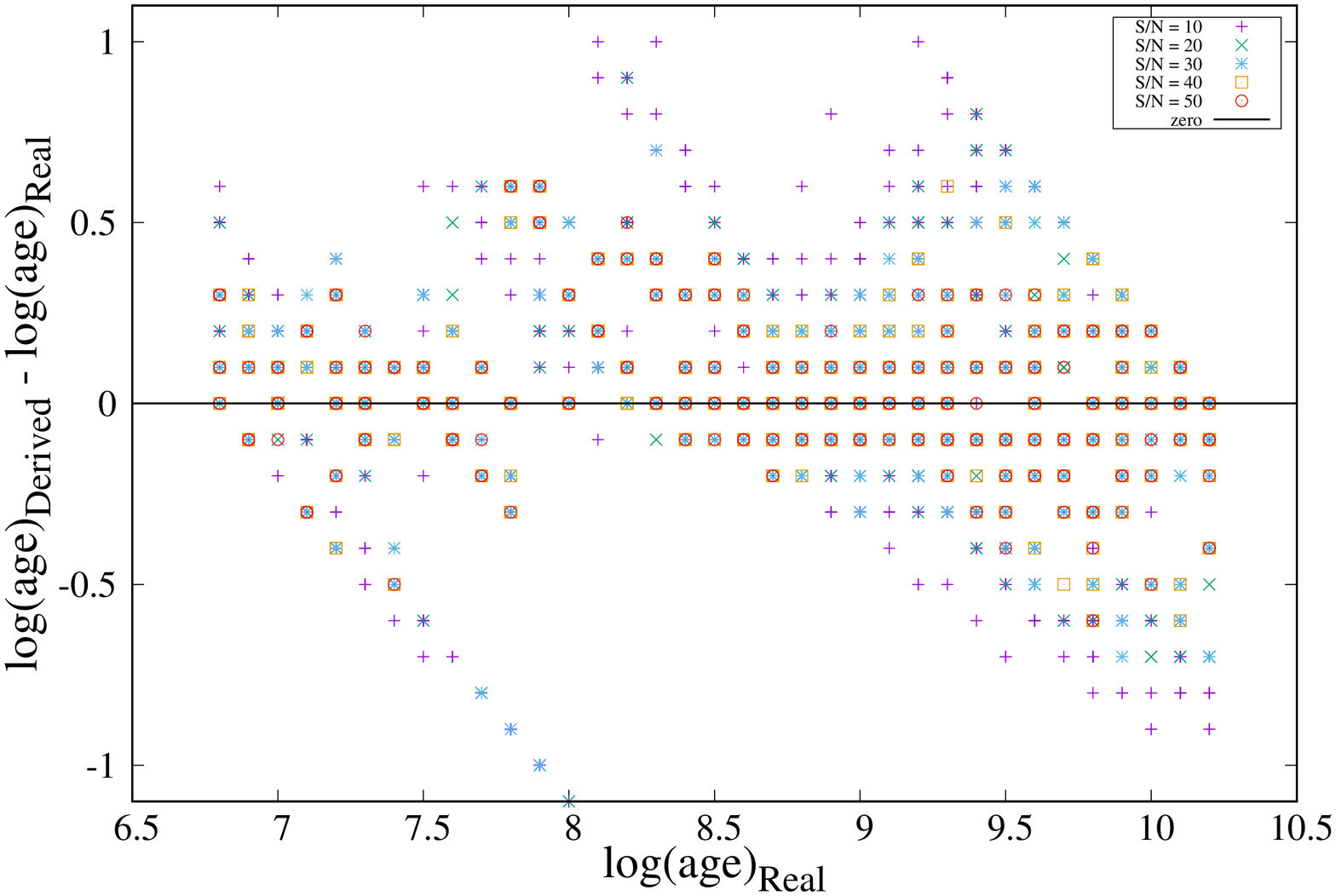}}
\resizebox{75mm}{!}{\includegraphics[width=\columnwidth, angle=270]{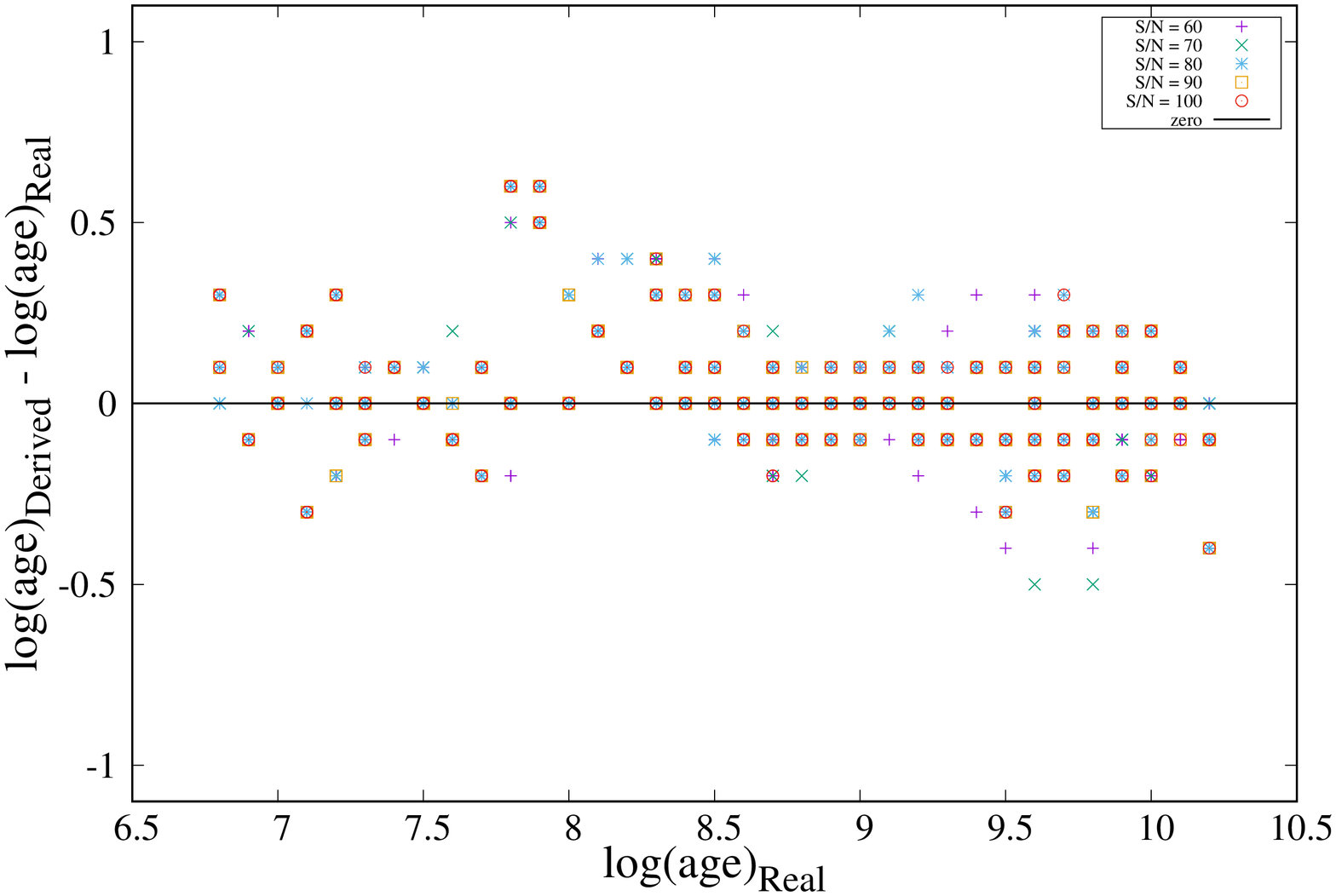}}
\resizebox{75mm}{!}{\includegraphics[width=\columnwidth, angle=270]{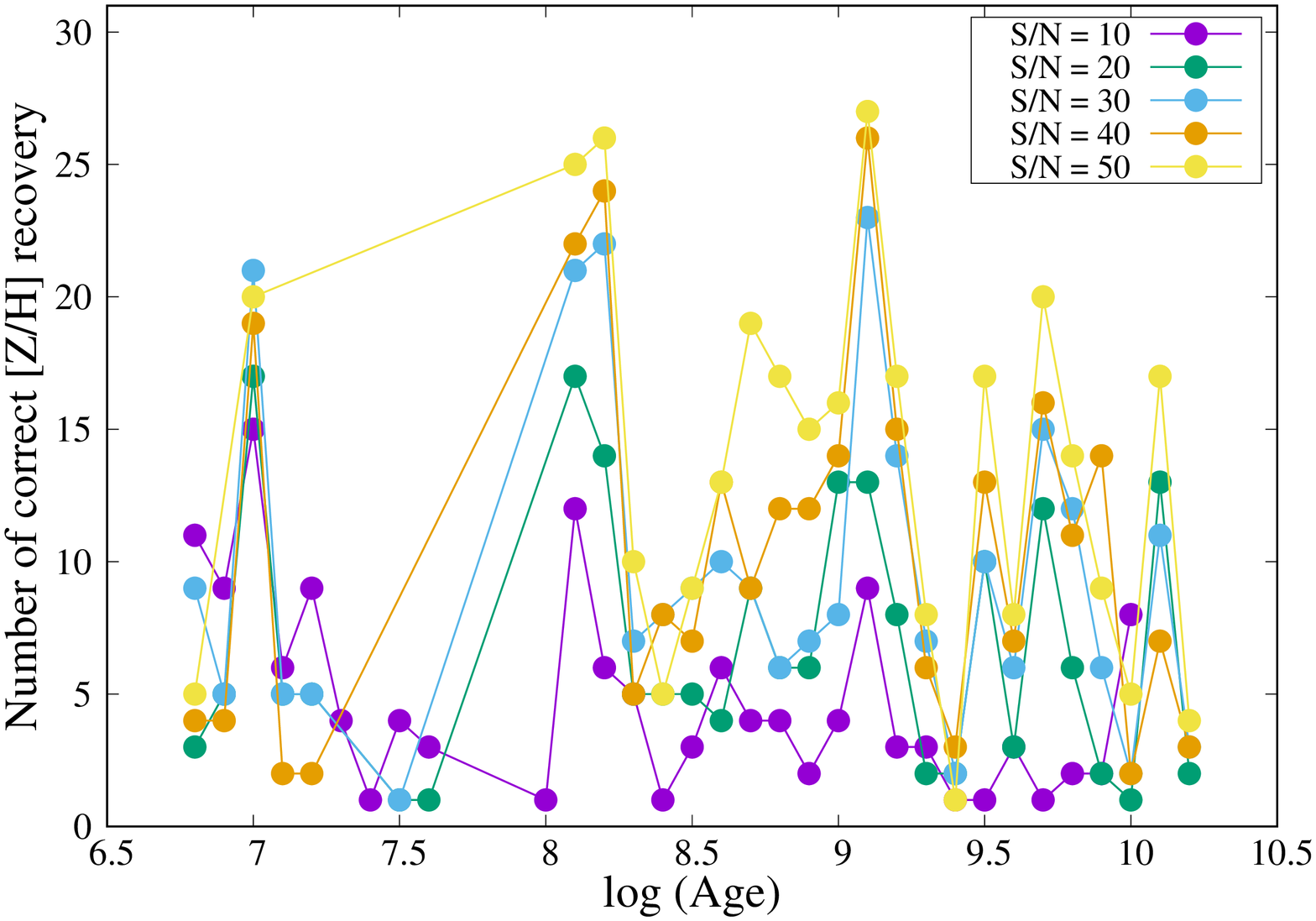}}
\resizebox{75mm}{!}{\includegraphics[width=\columnwidth, angle=270]{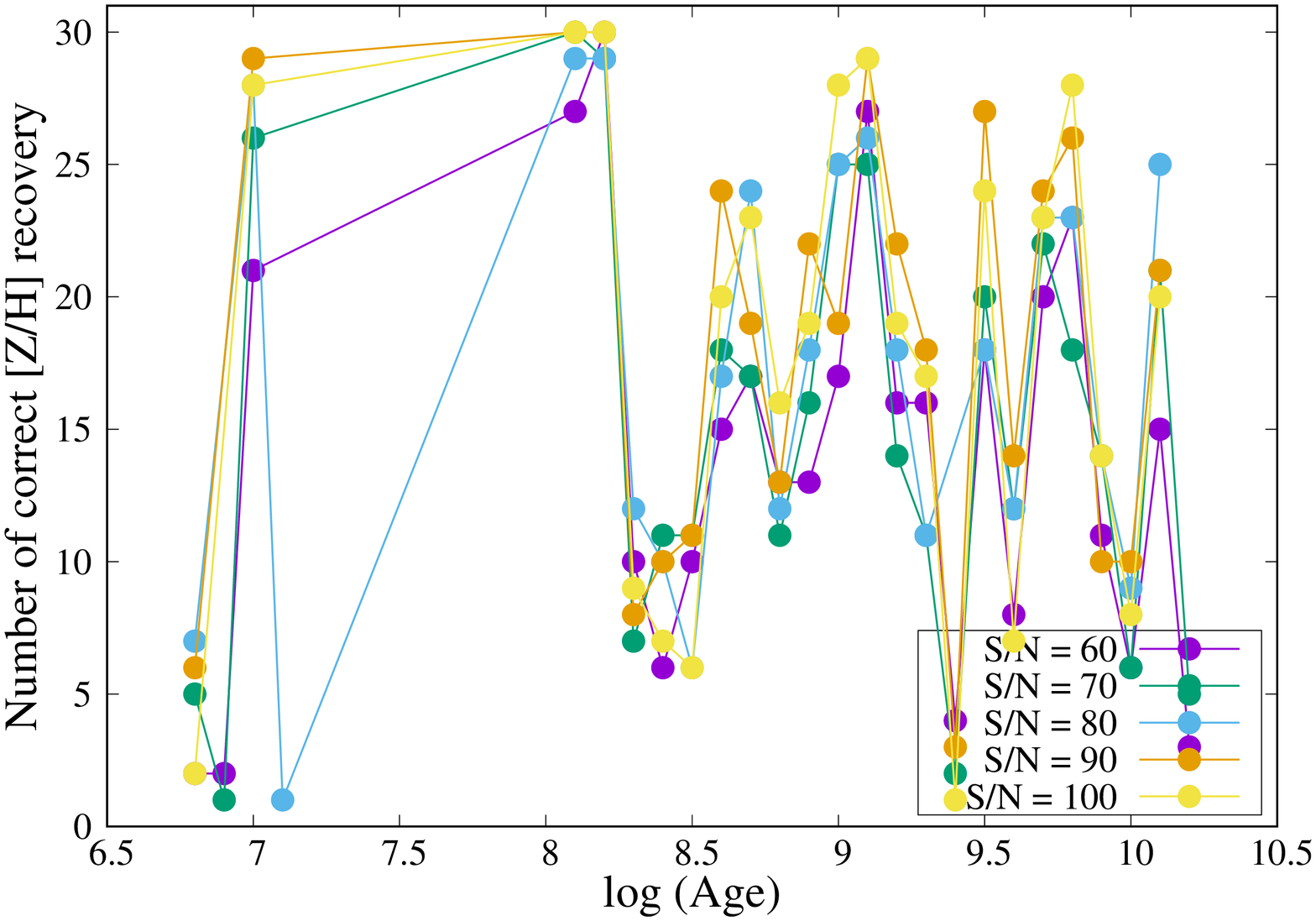}}
\resizebox{75mm}{!}{\includegraphics[width=\columnwidth, angle=270]{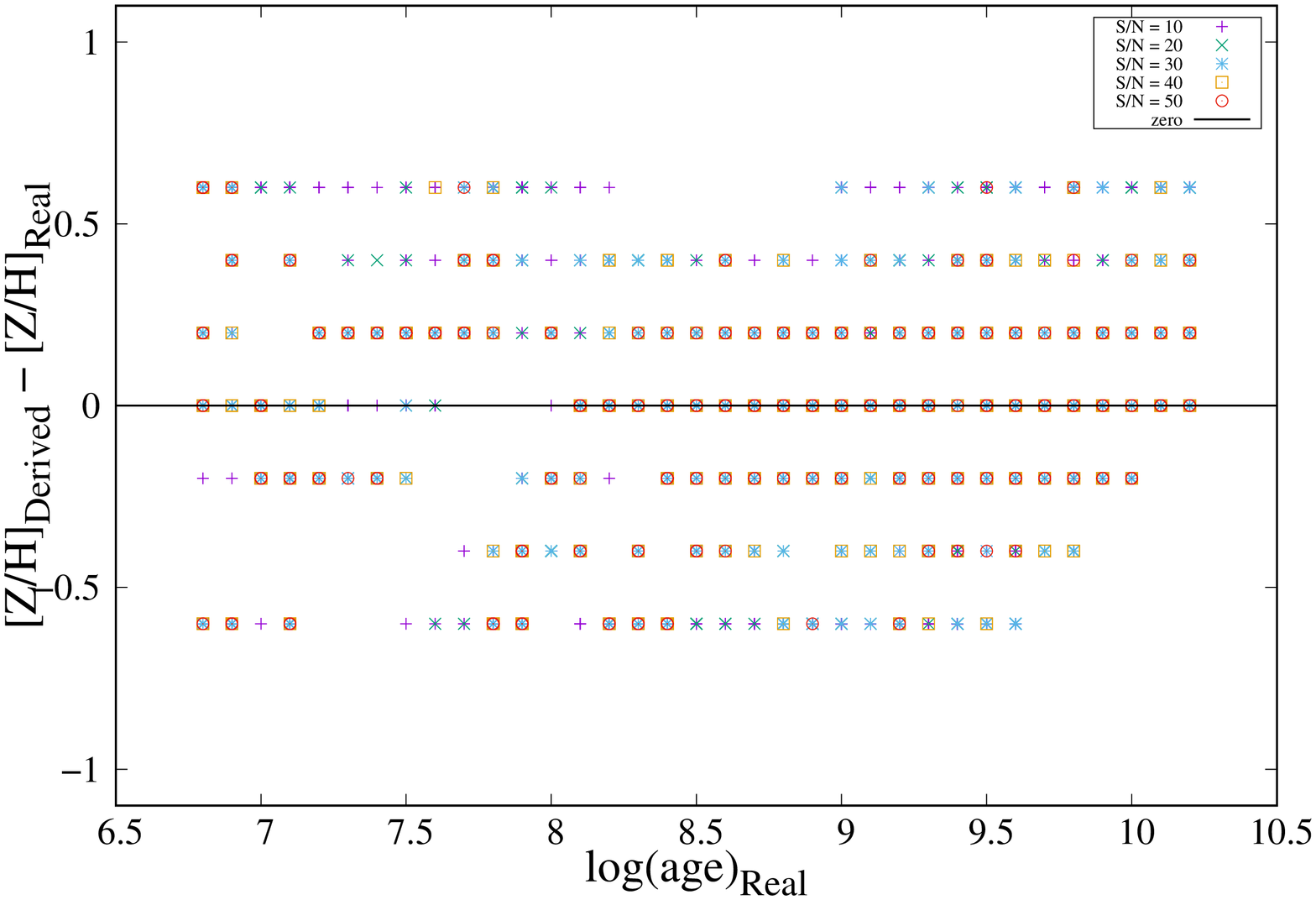}}
\resizebox{75mm}{!}{\includegraphics[width=\columnwidth, angle=270]{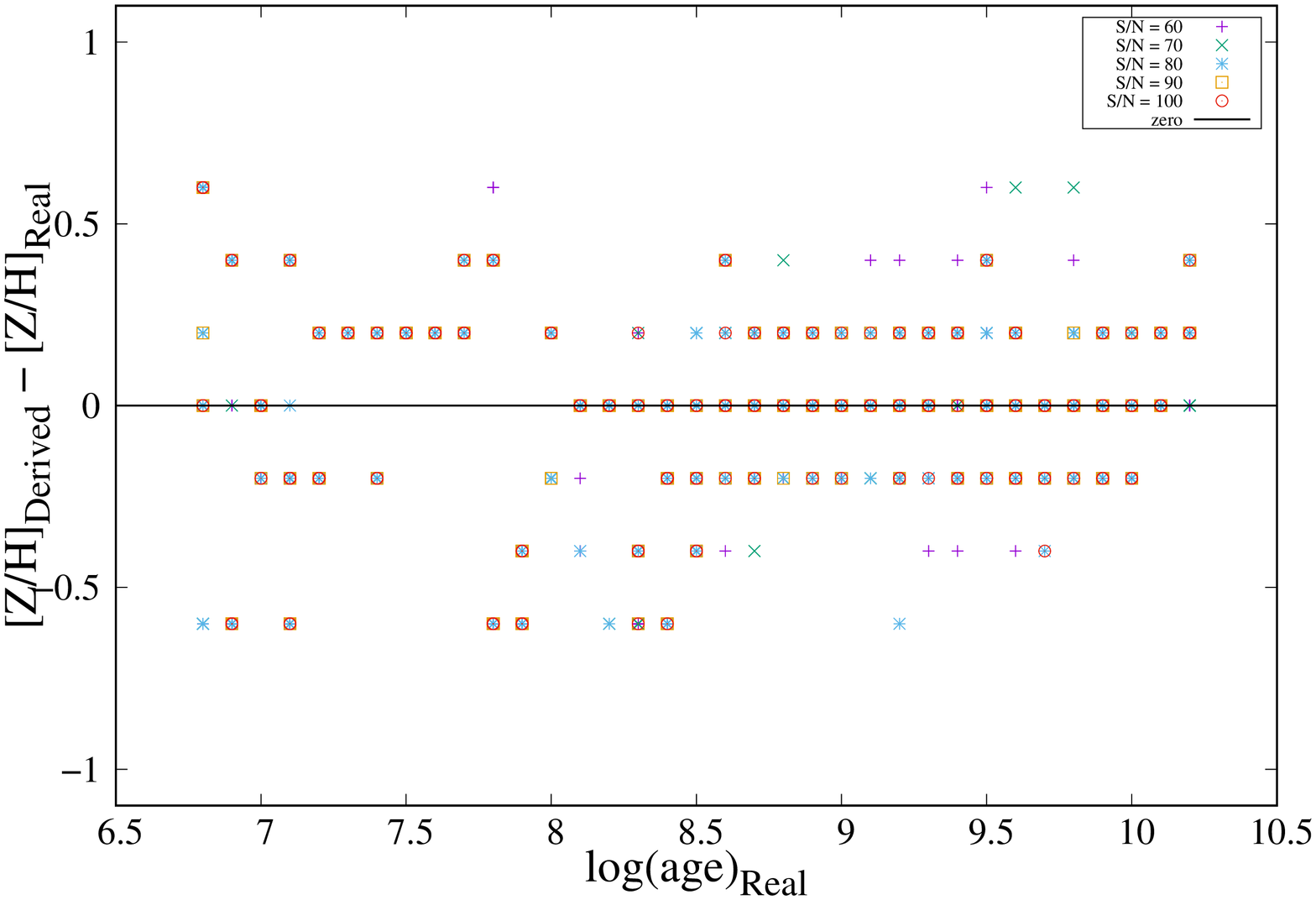}}

\caption{The same as figure \ref{Fig10}, now for the range $3700 \leqslant \lambda/\mbox{\AA} \leqslant 5000$ }
\label{Fig11}
\end{figure*}

\begin{figure*}

\resizebox{75mm}{!}{\includegraphics[width=\columnwidth, angle=270]{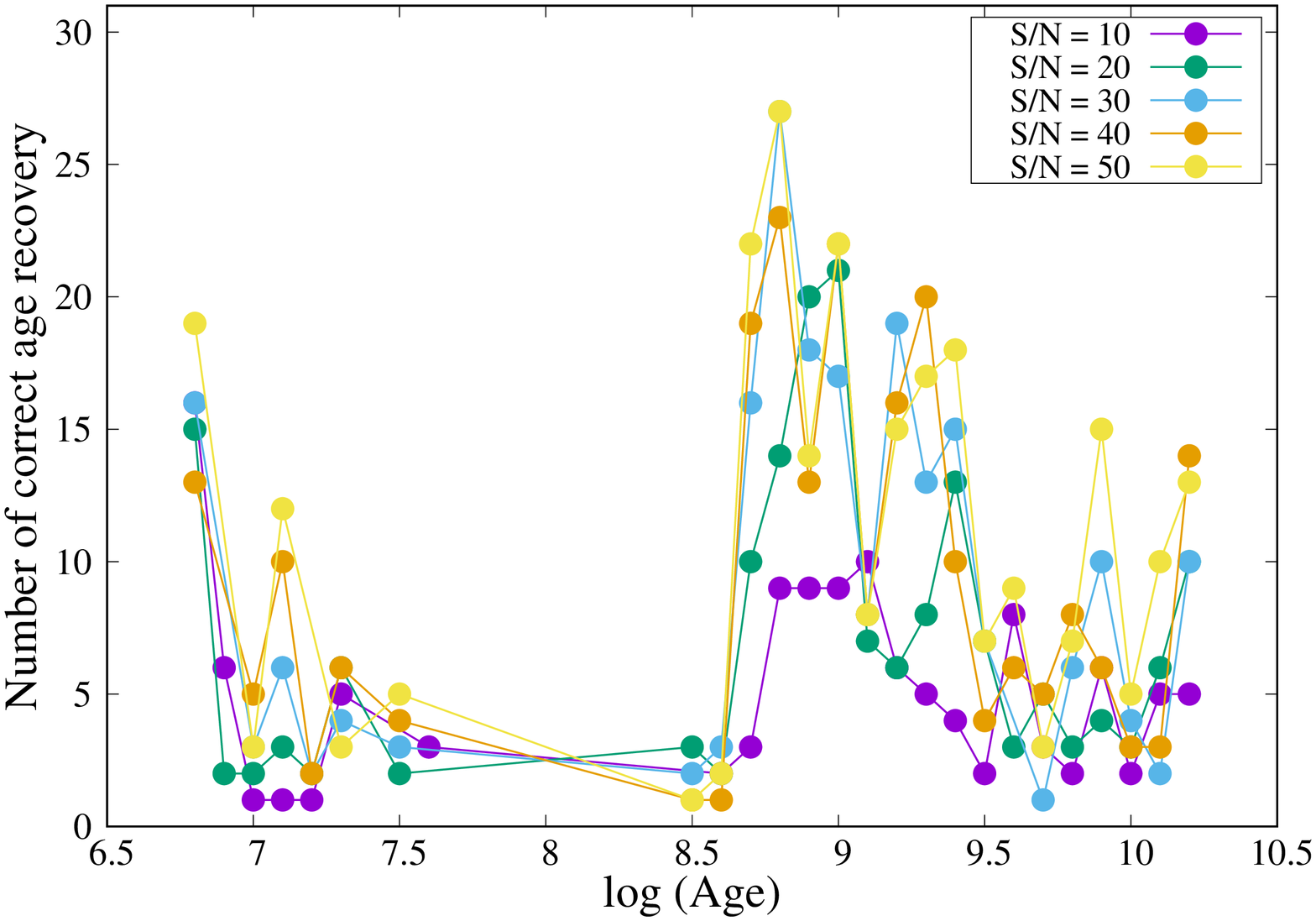}}
\resizebox{75mm}{!}{\includegraphics[width=\columnwidth, angle=270]{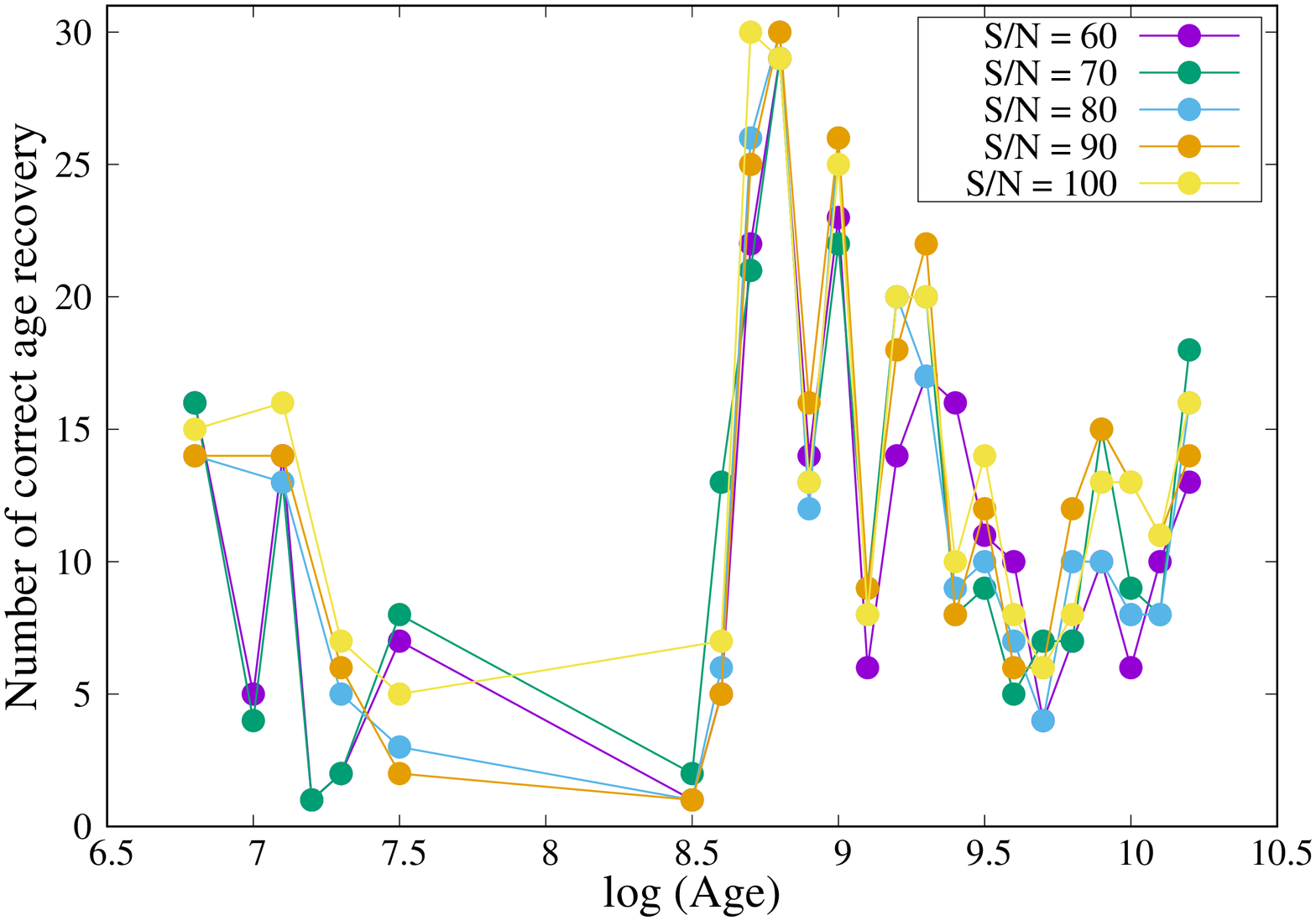}}
\resizebox{75mm}{!}{\includegraphics[width=\columnwidth, angle=270]{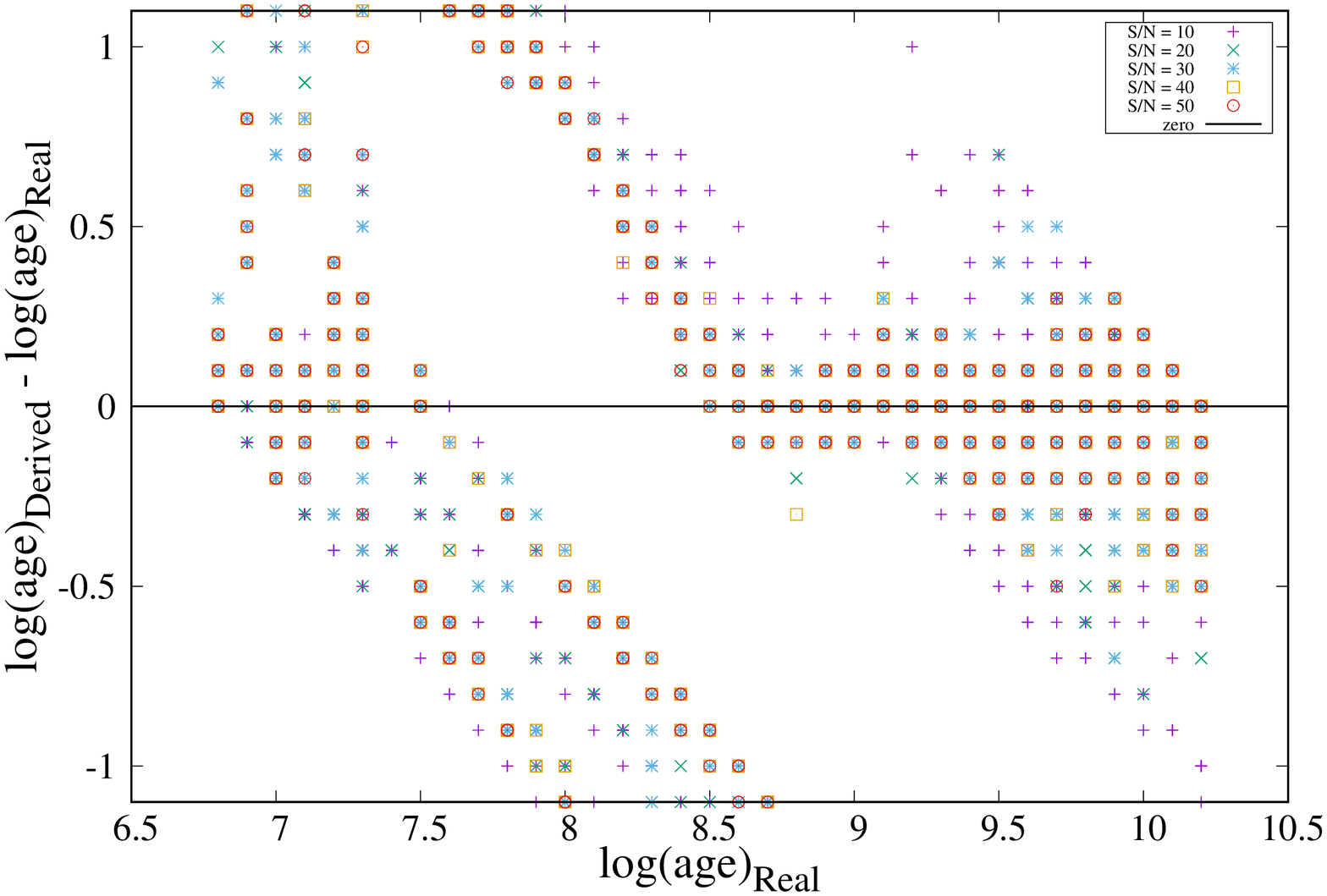}}
\resizebox{75mm}{!}{\includegraphics[width=\columnwidth, angle=270]{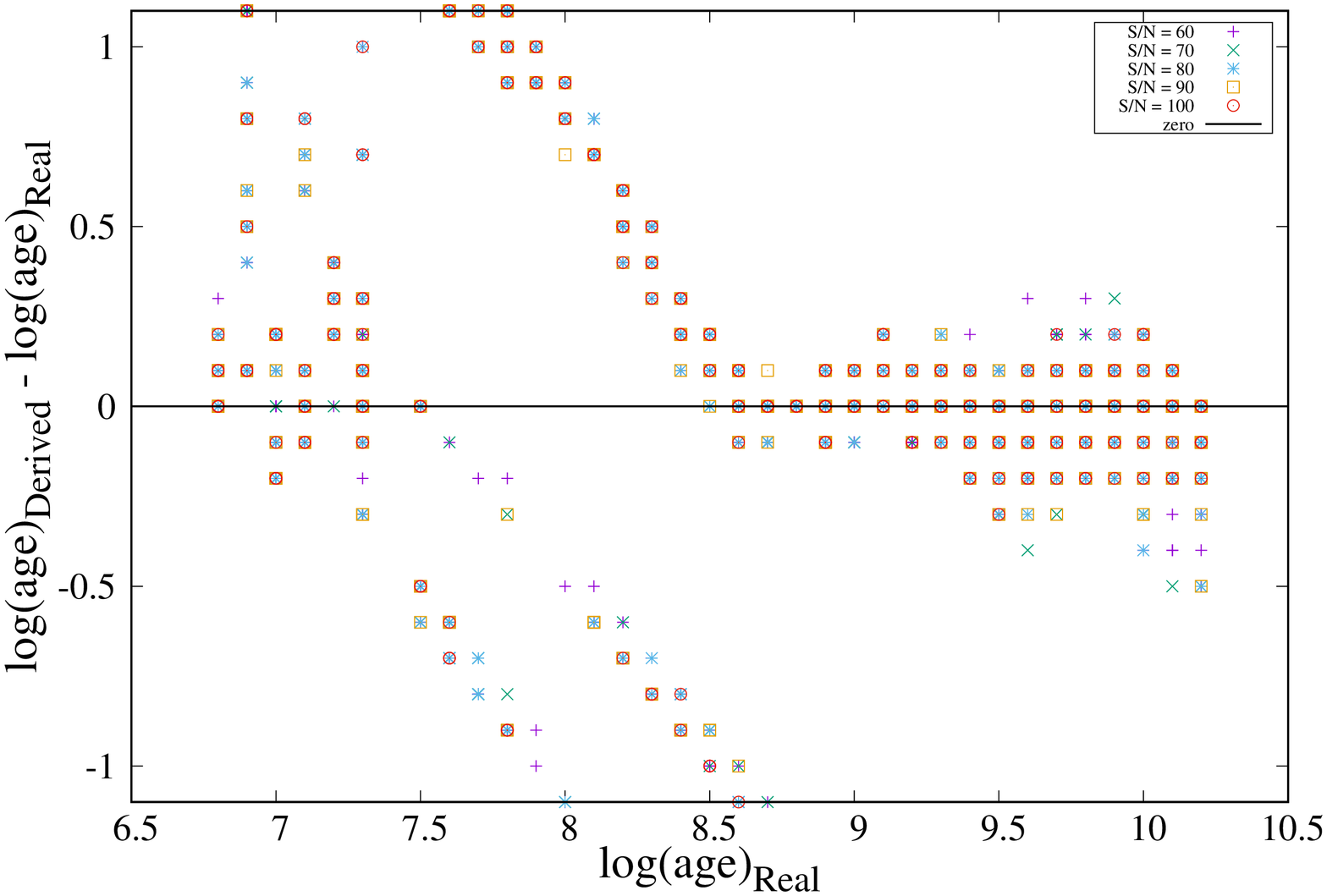}}
\resizebox{75mm}{!}{\includegraphics[width=\columnwidth, angle=270]{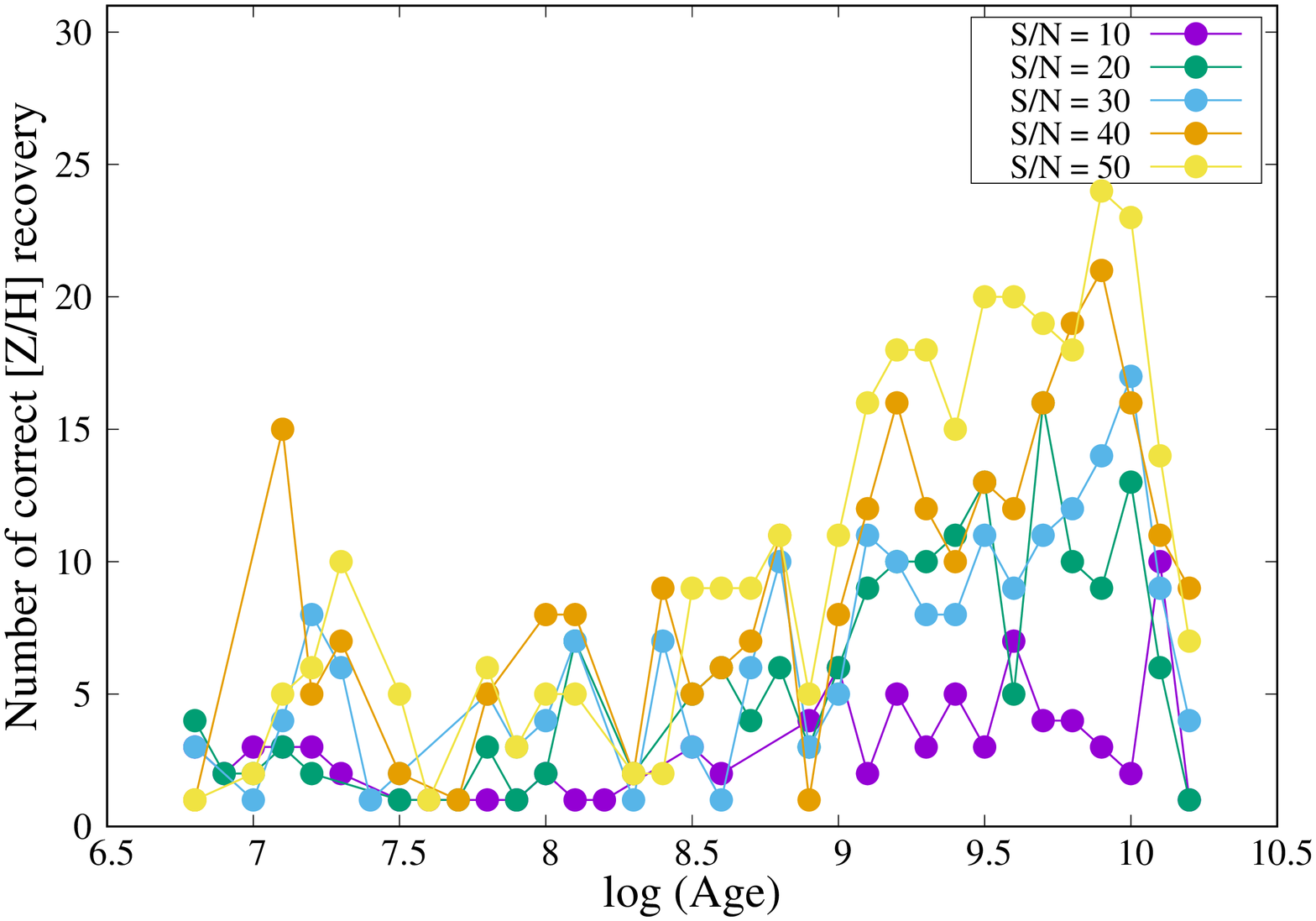}}
\resizebox{75mm}{!}{\includegraphics[width=\columnwidth, angle=270]{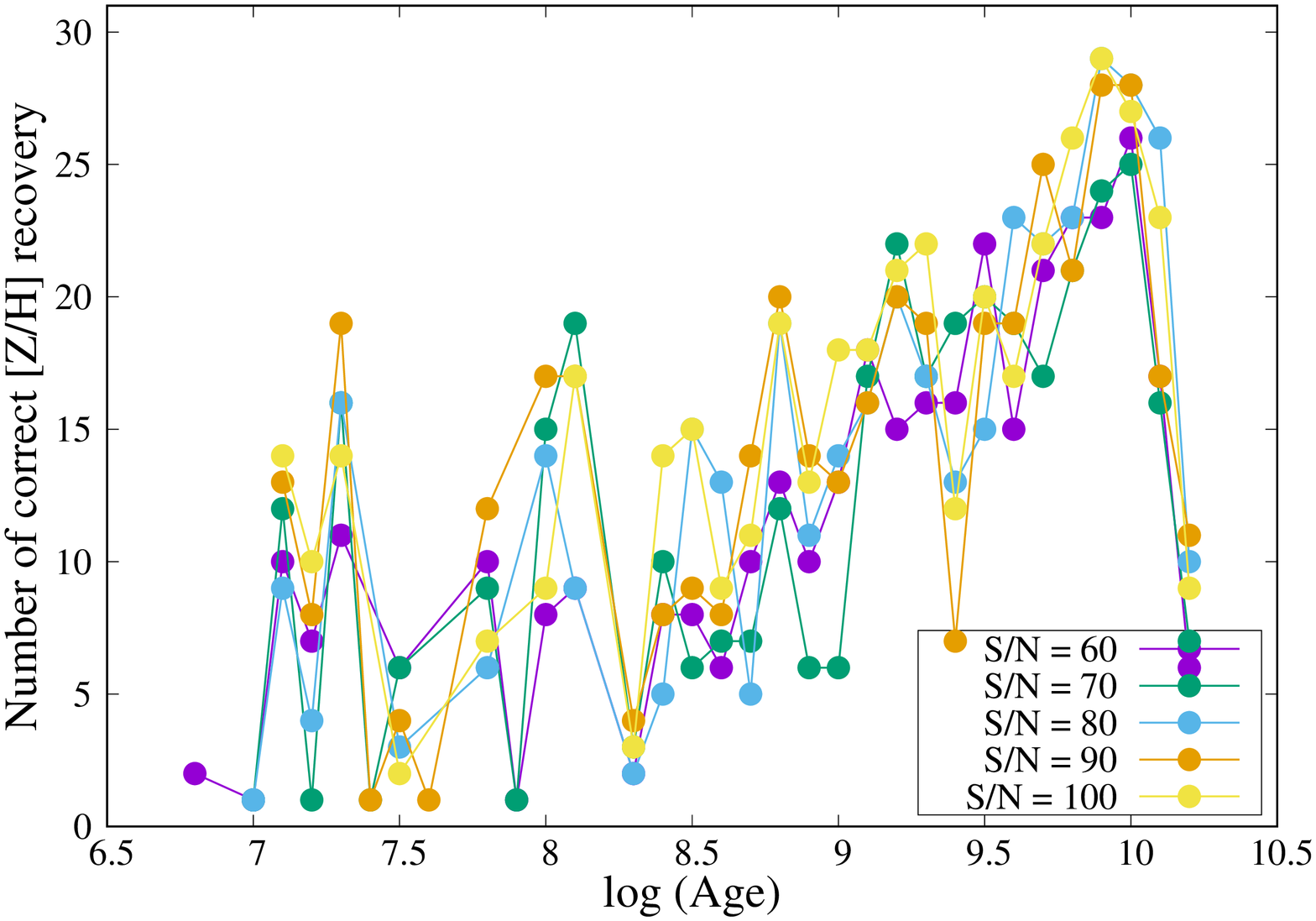}}
\resizebox{75mm}{!}{\includegraphics[width=\columnwidth, angle=270]{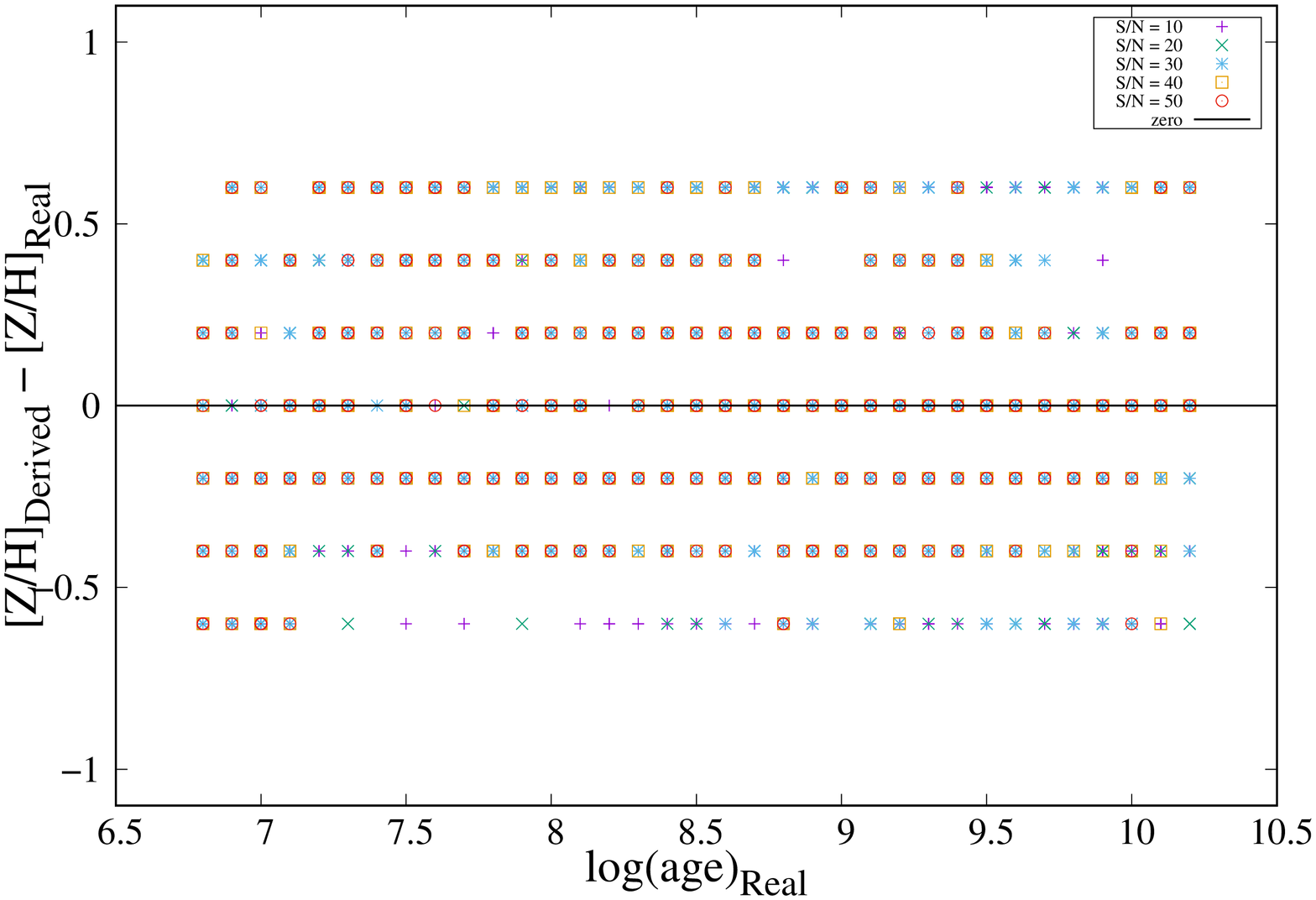}}
\resizebox{75mm}{!}{\includegraphics[width=\columnwidth, angle=270]{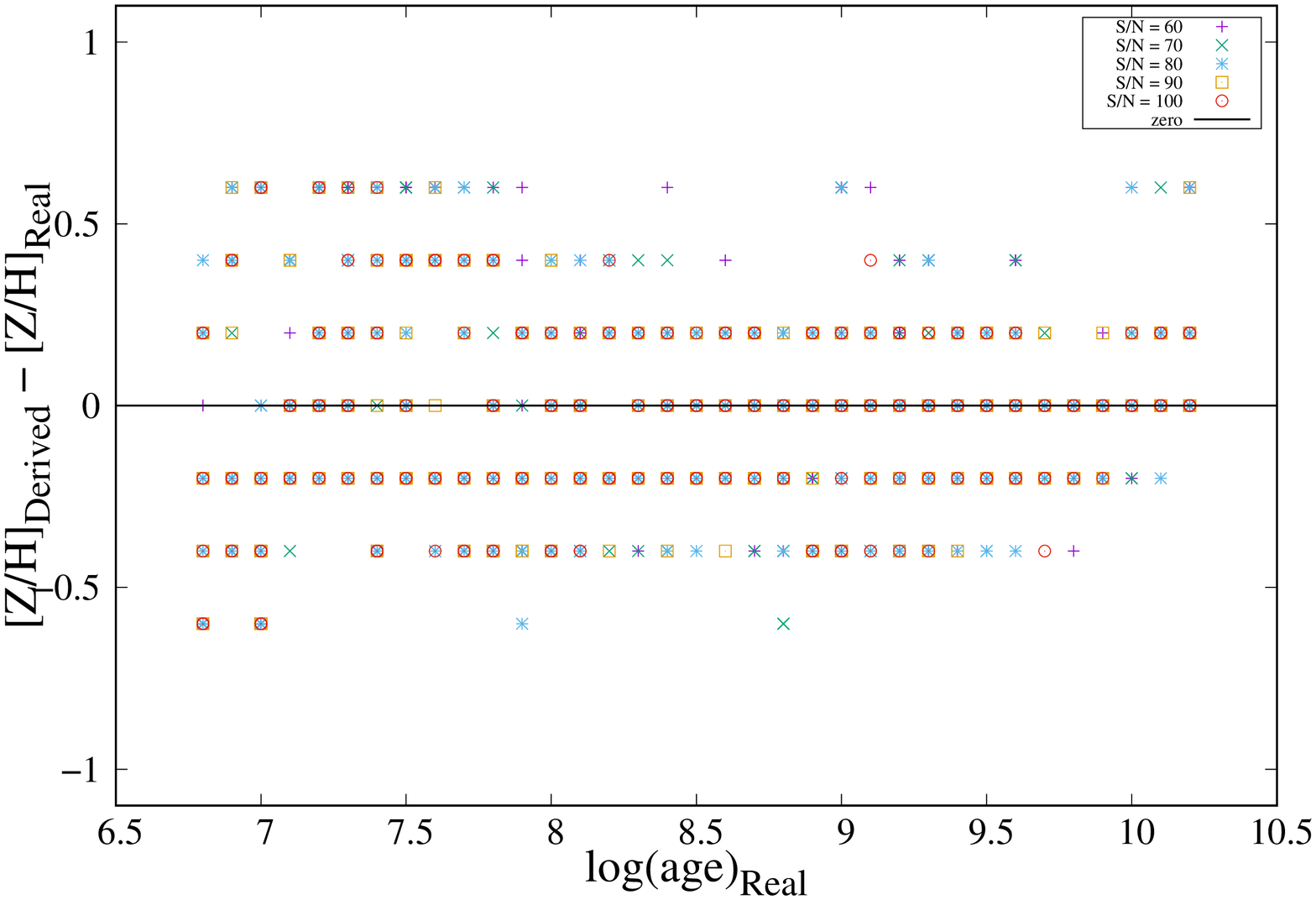}}

\caption{The same as figure \ref{Fig10}, now for the range $5000 \leqslant \lambda/\mbox{\AA} \leqslant 6200$ }
\label{Fig12}
\end{figure*}


\begin{figure*}
\includegraphics[width=14cm,angle=0]{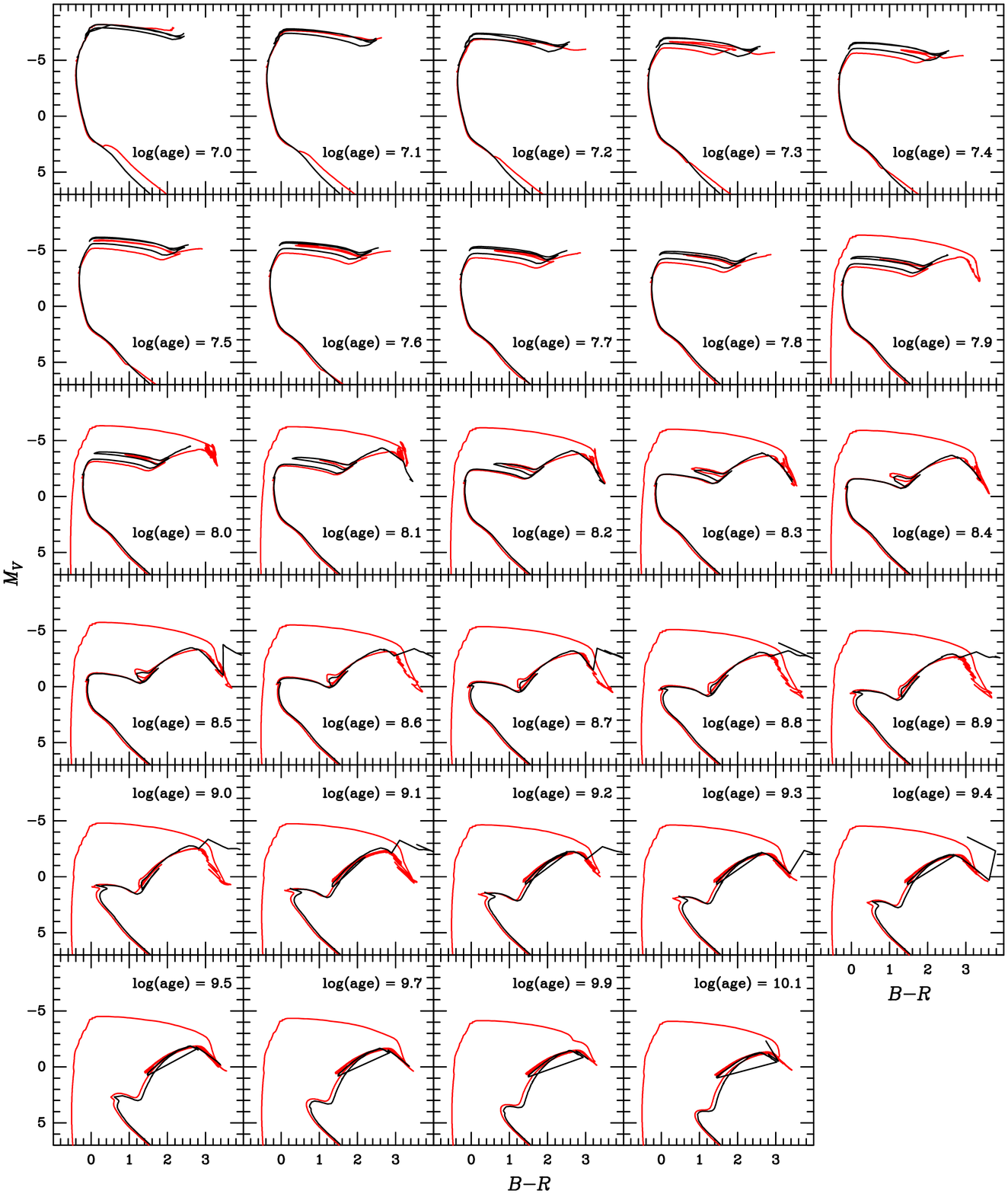}
\caption{$V$ vs.\ $B-R$ CMDs from log\,(age) = 7.0 to 10.1, using [Z/H] = $-$0.4. Black lines represent Padova isochrones while red lines represent MIST isochrones.}
\label{CMD_P_M}
\end{figure*}

\begin{figure*}
\resizebox{75mm}{!}{\includegraphics[angle=270]{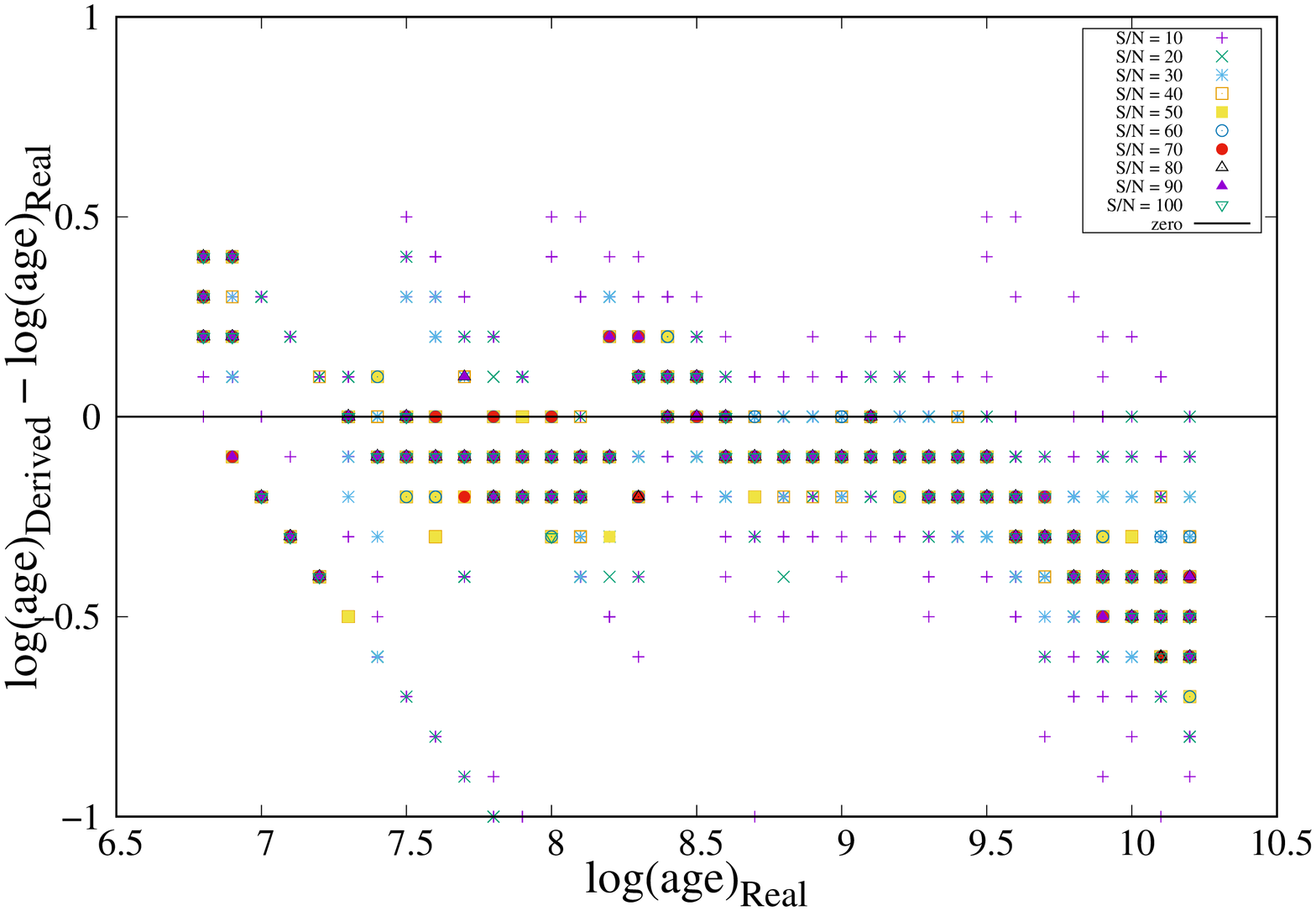}}
\resizebox{75mm}{!}{\includegraphics[angle=270]{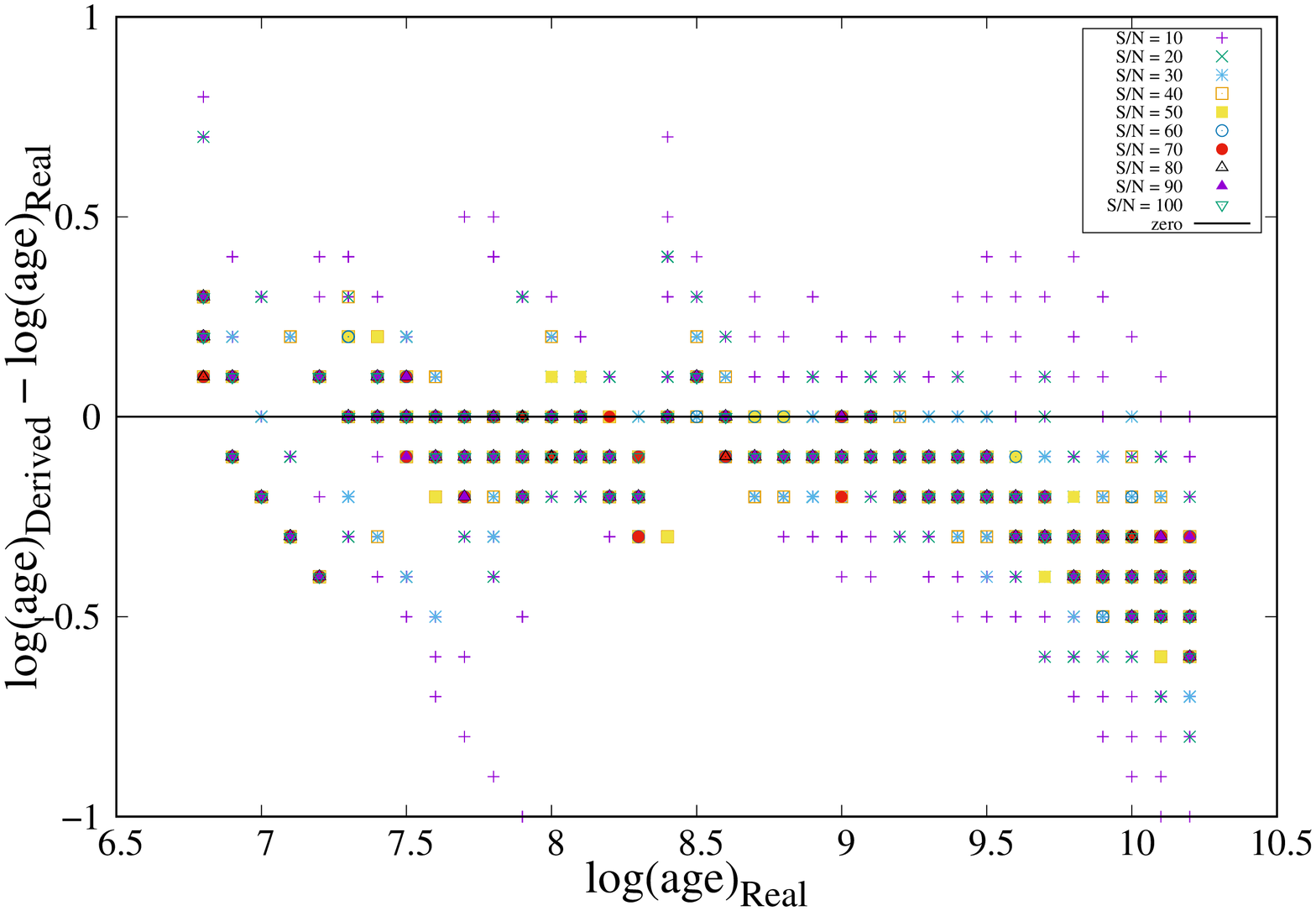}}
\resizebox{75mm}{!}{\includegraphics[angle=270]{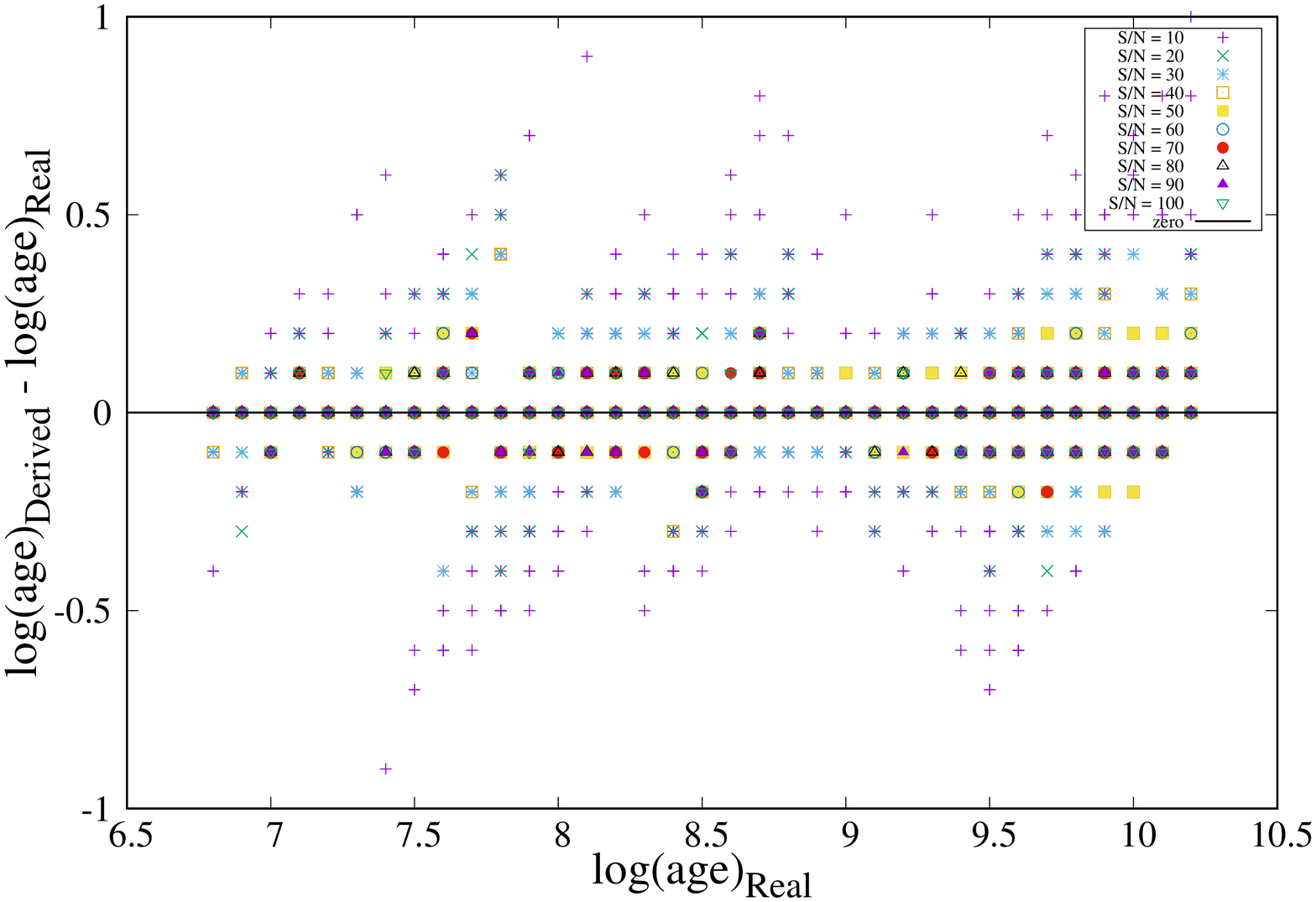}}
\resizebox{75mm}{!}{\includegraphics[width=\columnwidth, angle=270]{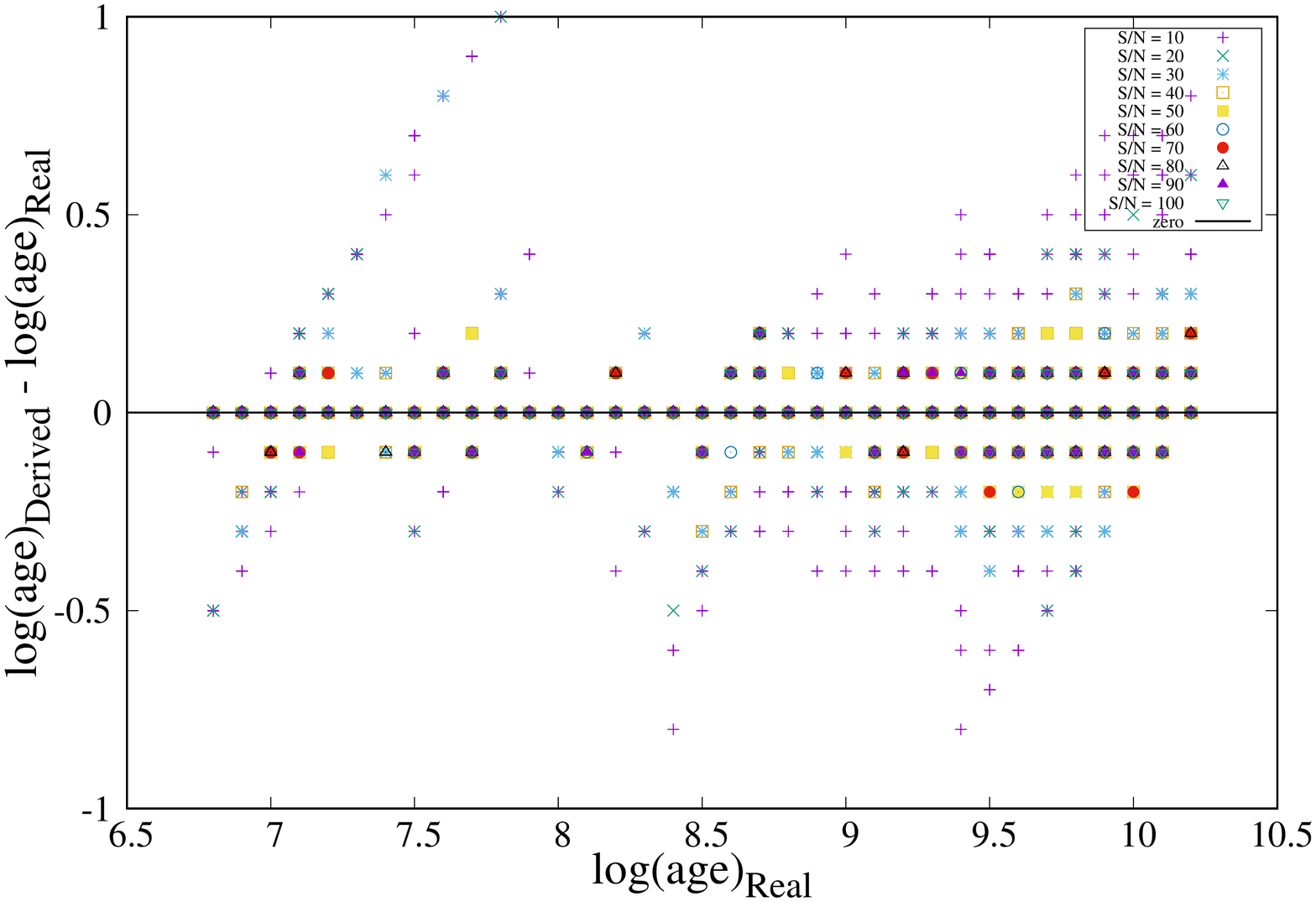}}
\resizebox{75mm}{!}{\includegraphics[angle=270]{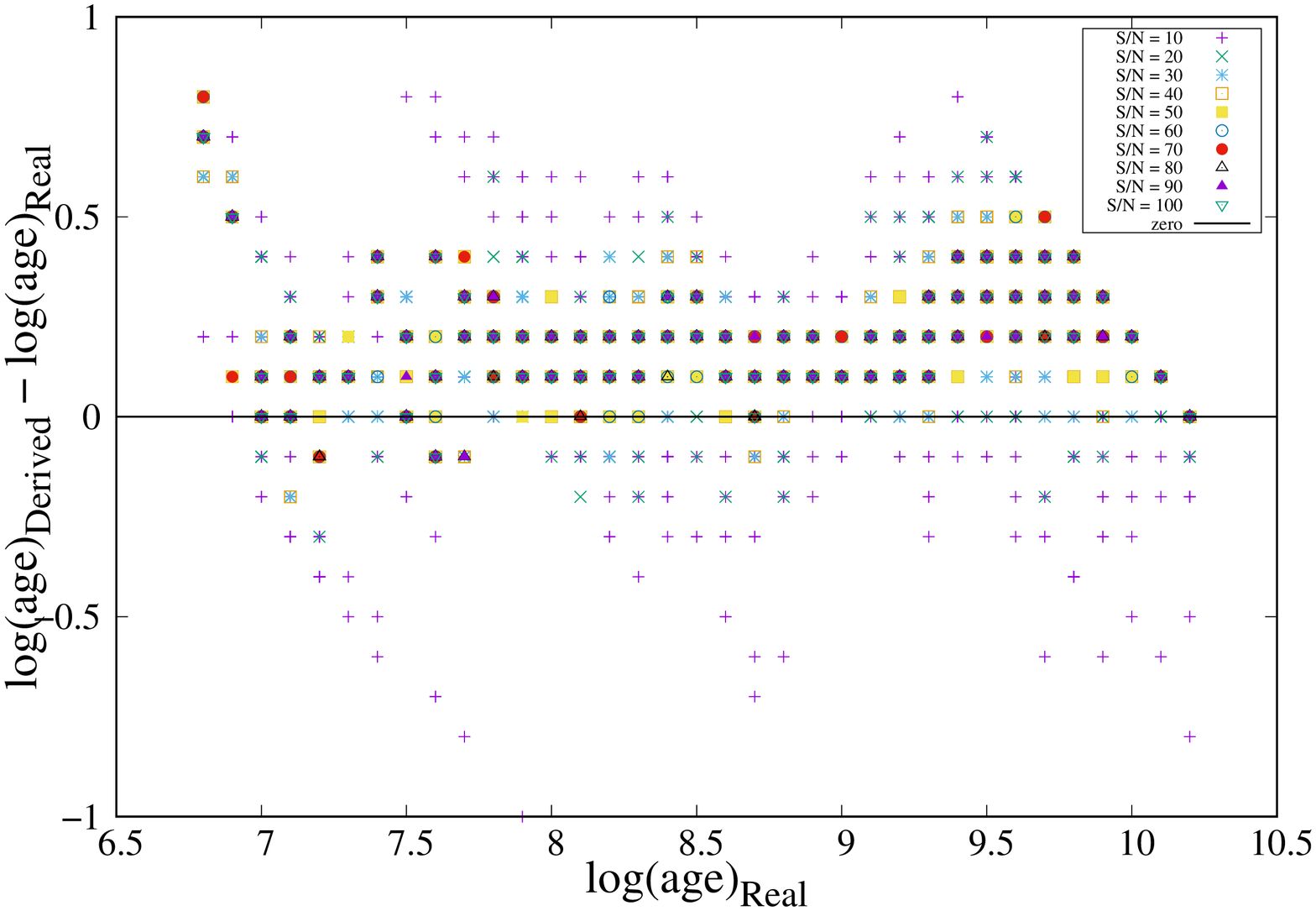}}
\resizebox{75mm}{!}{\includegraphics[width=\columnwidth, angle=270]{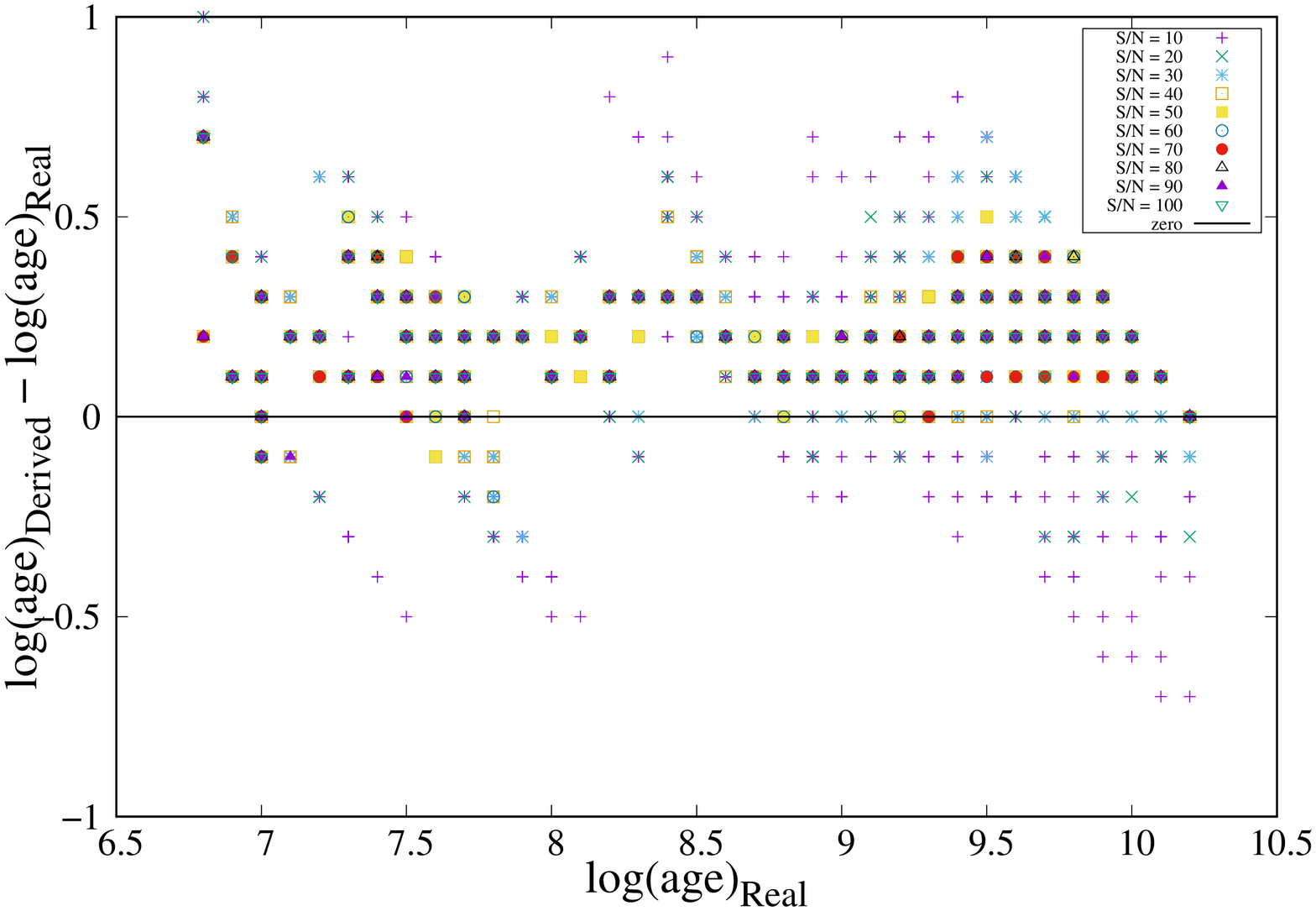}}

\caption{ (Derived\,--\,real) log(age) as a function of log(age), with metallicity fixed. Top panel: [Z/H] = 0.0, middle panel:  [Z/H] = $-$0.4, and bottom panel: [Z/H] = $-$0.8. $3700 - 6200$ \AA\ (left),  $3700 - 5000$ \AA\ (right)}
\label{Z8}
\end{figure*}


\begin{figure*}
\resizebox{75mm}{!}{\includegraphics[angle=270]{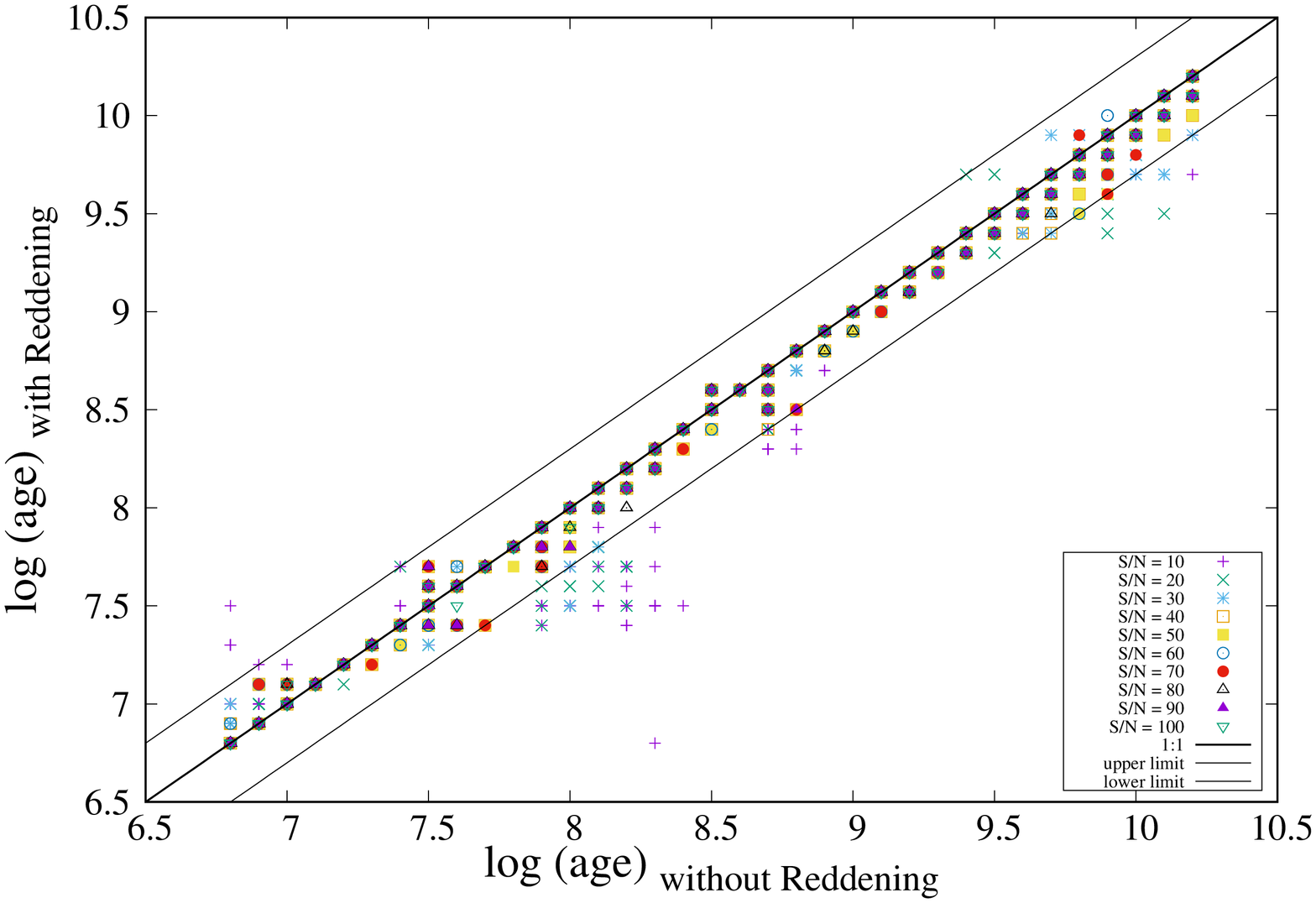}}
\resizebox{75mm}{!}{\includegraphics[angle=270]{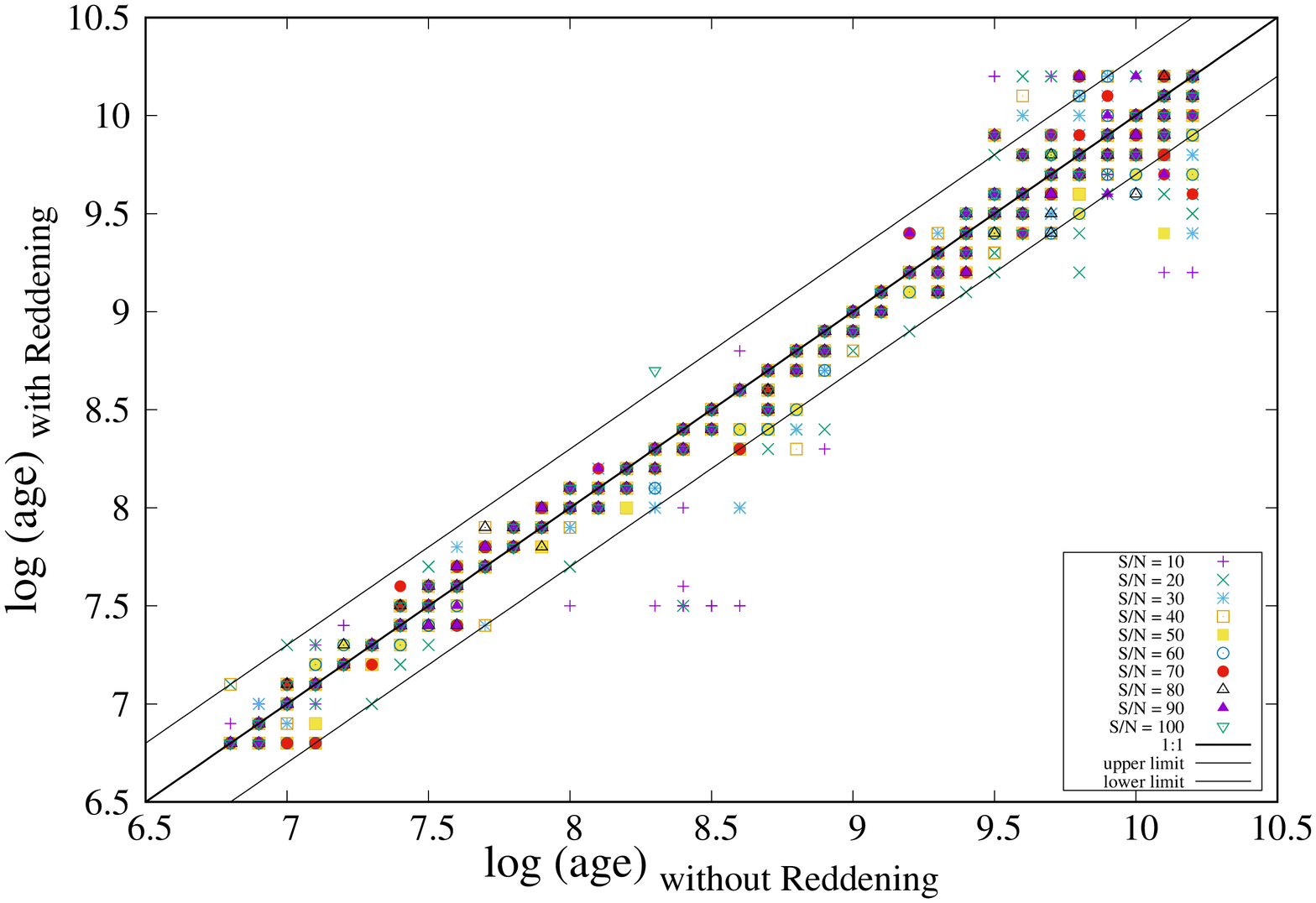}}
\caption{log(age) obtained when accounting for extinction versus 
  log(age) obtained without accounting for extinction} while keeping
  constant metallicity (left panel) and while varying metallicity (right panel) 
\label{red}
\end{figure*}




\begin{table*}
  \caption{Results on age determination - Padova - wavelength: 3700\,--\,6200 \AA}
\makebox[\textwidth][c]{
\begin{tabular}{cccccccccccccccccc}
\toprule

\multicolumn{1}{c}{S/N} &
\multicolumn{1}{c}{6.8} & \multicolumn{1}{c}{6.9} & \multicolumn{1}{c}{7.0} & \multicolumn{1}{c}{7.1} &
\multicolumn{1}{c}{7.2} &
\multicolumn{1}{c}{7.3} & \multicolumn{1}{c}{7.4} & \multicolumn{1}{c}{7.5} & \multicolumn{1}{c}{7.6} &
\multicolumn{1}{c}{7.7}  \\
\cmidrule(lr){1-1}\cmidrule(lr){2-2}\cmidrule(lr){3-3}\cmidrule(lr){4-4}\cmidrule(lr){5-5}\cmidrule(lr){6-6}\cmidrule(lr){7-7}\cmidrule(lr){8-8}\cmidrule(lr){9-9}\cmidrule(lr){10-10}\cmidrule(lr){11-11}

10.0 & 6.9$\pm$0.12 & 7.0$\pm$0.09 & 7.0$\pm$0.11 & 7.0$\pm$0.11 & 7.2$\pm$0.18 & 7.3$\pm$0.16 & 7.35$\pm$0.17 & 7.5$\pm$0.16 & 7.6$\pm$0.22 & 7.65$\pm$0.2\\
20.0 & 6.8$\pm$0.12 & 6.9$\pm$0.12 & 7.0$\pm$0.12 & 7.1$\pm$0.11 & 7.2$\pm$0.05 & 7.3$\pm$0.09 & 7.4$\pm$0.08 & 7.5$\pm$0.11 & 7.6$\pm$0.14 & 7.7$\pm$0.11\\
30.0 & 6.8$\pm$0.11 & 6.9$\pm$0.09 & 7.0$\pm$0.07 & 7.1$\pm$0.1 & 7.2$\pm$0.04 & 7.3$\pm$0.05 & 7.4$\pm$0.06 & 7.5$\pm$0.1 & 7.6$\pm$0.12 & 7.7$\pm$0.06\\
40.0 & 6.8$\pm$0.09 & 6.9$\pm$0.08 & 7.0$\pm$0.05 & 7.1$\pm$0.06 & 7.2$\pm$0.04 & 7.3$\pm$0.02 & 7.4$\pm$0.04 & 7.5$\pm$0.08 & 7.6$\pm$0.12 & 7.7$\pm$0.07\\
50.0 & 6.8$\pm$0.04 & 6.9$\pm$0.05 & 7.0$\pm$0.04 & 7.1$\pm$0.07 & 7.2$\pm$0.03 & 7.3$\pm$0.03 & 7.4$\pm$0.04 & 7.5$\pm$0.07 & 7.6$\pm$0.09 & 7.7$\pm$0.05\\
60.0 & 6.8$\pm$0.06 & 6.9$\pm$0.04 & 7.0$\pm$0.04 & 7.1$\pm$0.04 & 7.2$\pm$0.0 & 7.3$\pm$0.03 & 7.4$\pm$0.03 & 7.5$\pm$0.06 & 7.6$\pm$0.08 & 7.7$\pm$0.04\\
70.0 & 6.8$\pm$0.04 & 6.9$\pm$0.04 & 7.0$\pm$0.04 & 7.1$\pm$0.04 & 7.2$\pm$0.0 & 7.3$\pm$0.0 & 7.4$\pm$0.0 & 7.5$\pm$0.05 & 7.6$\pm$0.08 & 7.7$\pm$0.03\\
\bottomrule
\end{tabular}%
}
\label{T1}%
\end{table*}%

\clearpage
\begin{table*}
  \caption{Results - Padova - wavelength: 3700-5000}
\makebox[\textwidth][c]{
    \begin{tabular}{cccccccccccccccccc}
\toprule

\multicolumn{1}{c}{S/N} &
\multicolumn{1}{c}{6.8} & \multicolumn{1}{c}{6.9} & \multicolumn{1}{c}{7.0} & \multicolumn{1}{c}{7.1} &
\multicolumn{1}{c}{7.2} &
\multicolumn{1}{c}{7.3} & \multicolumn{1}{c}{7.4} & \multicolumn{1}{c}{7.5} & \multicolumn{1}{c}{7.6} &
\multicolumn{1}{c}{7.7}  \\
\cmidrule(lr){1-1}\cmidrule(lr){2-2}\cmidrule(lr){3-3}\cmidrule(lr){4-4}\cmidrule(lr){5-5}\cmidrule(lr){6-6}\cmidrule(lr){7-7}\cmidrule(lr){8-8}\cmidrule(lr){9-9}\cmidrule(lr){10-10}\cmidrule(lr){11-11}

10.0 & 6.9$\pm$0.21 & 7.0$\pm$0.12 & 7.0$\pm$0.1 & 7.05$\pm$0.12 & 7.2$\pm$0.13 & 7.3$\pm$0.16 & 7.3$\pm$0.16 & 7.35$\pm$0.31 & 7.5$\pm$0.3 & 7.7$\pm$0.28\\
20.0 & 6.8$\pm$0.13 & 6.9$\pm$0.08 & 7.0$\pm$0.07 & 7.1$\pm$0.08 & 7.2$\pm$0.17 & 7.3$\pm$0.14 & 7.4$\pm$0.1 & 7.5$\pm$0.14 & 7.5$\pm$0.07 & 7.7$\pm$0.11\\
30.0 & 6.8$\pm$0.09 & 6.9$\pm$0.06 & 7.0$\pm$0.04 & 7.1$\pm$0.08 & 7.2$\pm$0.09 & 7.3$\pm$0.09 & 7.4$\pm$0.08 & 7.5$\pm$0.1 & 7.6$\pm$0.1 & 7.7$\pm$0.11\\
40.0 & 6.8$\pm$0.05 & 6.9$\pm$0.07 & 7.0$\pm$0.04 & 7.1$\pm$0.06 & 7.2$\pm$0.04 & 7.3$\pm$0.0 & 7.4$\pm$0.06 & 7.5$\pm$0.07 & 7.6$\pm$0.05 & 7.7$\pm$0.08\\
50.0 & 6.8$\pm$0.07 & 6.9$\pm$0.04 & 7.0$\pm$0.04 & 7.1$\pm$0.04 & 7.2$\pm$0.04 & 7.3$\pm$0.03 & 7.4$\pm$0.03 & 7.5$\pm$0.06 & 7.6$\pm$0.06 & 7.7$\pm$0.09\\
60.0 & 6.8$\pm$0.08 & 6.9$\pm$0.04 & 7.0$\pm$0.03 & 7.1$\pm$0.03 & 7.2$\pm$0.03 & 7.3$\pm$0.03 & 7.4$\pm$0.0 & 7.5$\pm$0.05 & 7.6$\pm$0.05 & 7.7$\pm$0.04\\
70.0 & 6.8$\pm$0.08 & 6.9$\pm$0.0 & 7.0$\pm$0.03 & 7.1$\pm$0.04 & 7.2$\pm$0.03 & 7.3$\pm$0.0 & 7.4$\pm$0.0 & 7.5$\pm$0.05 & 7.6$\pm$0.05 & 7.7$\pm$0.03\\
\bottomrule
\end{tabular}%
}
\label{T2}%
\end{table*}%

\clearpage
\begin{table*}
  \caption{Results - Padova - wavelength: 5000-6200}
\makebox[\textwidth][c]{
    \begin{tabular}{cccccccccccccccccc}
\toprule

\multicolumn{1}{c}{S/N} &
\multicolumn{1}{c}{6.8} & \multicolumn{1}{c}{6.9} & \multicolumn{1}{c}{7.0} & \multicolumn{1}{c}{7.1} &
\multicolumn{1}{c}{7.2} &
\multicolumn{1}{c}{7.3} & \multicolumn{1}{c}{7.4} & \multicolumn{1}{c}{7.5} & \multicolumn{1}{c}{7.6} &
\multicolumn{1}{c}{7.7}  \\
\cmidrule(lr){1-1}\cmidrule(lr){2-2}\cmidrule(lr){3-3}\cmidrule(lr){4-4}\cmidrule(lr){5-5}\cmidrule(lr){6-6}\cmidrule(lr){7-7}\cmidrule(lr){8-8}\cmidrule(lr){9-9}\cmidrule(lr){10-10}\cmidrule(lr){11-11}

10.0 & 6.9$\pm$0.55 & 6.9$\pm$0.64 & 7.05$\pm$0.67 & 7.0$\pm$0.37 & 7.0$\pm$0.5 & 7.2$\pm$0.66 & 7.2$\pm$0.66 & 7.3$\pm$0.69 & 7.25$\pm$0.54 & 7.2$\pm$0.68\\
20.0 & 7.7$\pm$0.56 & 7.3$\pm$0.58 & 7.2$\pm$0.56 & 7.1$\pm$0.46 & 7.2$\pm$0.56 & 7.25$\pm$0.33 & 7.5$\pm$0.52 & 7.45$\pm$0.58 & 7.35$\pm$0.46 & 7.5$\pm$0.57\\
30.0 & 7.8$\pm$0.52 & 7.3$\pm$0.56 & 7.2$\pm$0.35 & 7.1$\pm$0.4 & 7.2$\pm$0.35 & 7.3$\pm$0.32 & 7.45$\pm$0.28 & 7.5$\pm$0.52 & 7.5$\pm$0.52 & 7.7$\pm$0.45\\
40.0 & 8.0$\pm$0.52 & 7.2$\pm$0.5 & 7.2$\pm$0.4 & 7.2$\pm$0.25 & 7.15$\pm$0.33 & 7.3$\pm$0.14 & 7.4$\pm$0.37 & 7.5$\pm$0.36 & 7.5$\pm$0.38 & 7.7$\pm$0.42\\
50.0 & 7.75$\pm$0.55 & 7.0$\pm$0.44 & 7.1$\pm$0.4 & 7.1$\pm$0.26 & 7.2$\pm$0.23 & 7.3$\pm$0.25 & 7.4$\pm$0.2 & 7.5$\pm$0.29 & 7.5$\pm$0.29 & 7.6$\pm$0.34\\
60.0 & 7.8$\pm$0.58 & 6.9$\pm$0.51 & 7.1$\pm$0.32 & 7.1$\pm$0.13 & 7.2$\pm$0.14 & 7.3$\pm$0.14 & 7.5$\pm$0.18 & 7.5$\pm$0.2 & 7.5$\pm$0.26 & 7.6$\pm$0.32\\
70.0 & 7.1$\pm$0.58 & 7.0$\pm$0.33 & 7.1$\pm$0.37 & 7.1$\pm$0.12 & 7.2$\pm$0.13 & 7.3$\pm$0.14 & 7.4$\pm$0.17 & 7.5$\pm$0.1 & 7.5$\pm$0.1 & 7.65$\pm$0.27\\
\bottomrule
\end{tabular}%
}
\label{T3}%
\end{table*}%

\clearpage
\begin{table*}
  \caption{Results - MIST - wavelength: 3700-6200}
\makebox[\textwidth][c]{
    \begin{tabular}{cccccccccccccccccc}
\toprule

\multicolumn{1}{c}{S/N} &
\multicolumn{1}{c}{6.8} & \multicolumn{1}{c}{6.9} & \multicolumn{1}{c}{7.0} & \multicolumn{1}{c}{7.1} &
\multicolumn{1}{c}{7.2} &
\multicolumn{1}{c}{7.3} & \multicolumn{1}{c}{7.4} & \multicolumn{1}{c}{7.5} & \multicolumn{1}{c}{7.6} &
\multicolumn{1}{c}{7.7}  \\
\cmidrule(lr){1-1}\cmidrule(lr){2-2}\cmidrule(lr){3-3}\cmidrule(lr){4-4}\cmidrule(lr){5-5}\cmidrule(lr){6-6}\cmidrule(lr){7-7}\cmidrule(lr){8-8}\cmidrule(lr){9-9}\cmidrule(lr){10-10}\cmidrule(lr){11-11}

10.0 & 6.8$\pm$0.14 & 6.9$\pm$0.09 & 7.0$\pm$0.08 & 7.0$\pm$0.1 & 7.3$\pm$0.1 & 7.3$\pm$0.09 & 7.4$\pm$0.25 & 7.55$\pm$0.13 & 7.5$\pm$0.2 & 7.8$\pm$0.33\\
20.0 & 6.8$\pm$0.12 & 6.9$\pm$0.08 & 7.0$\pm$0.08 & 7.0$\pm$0.09 & 7.2$\pm$0.1 & 7.3$\pm$0.11 & 7.4$\pm$0.09 & 7.5$\pm$0.08 & 7.6$\pm$0.1 & 7.9$\pm$0.14\\
30.0 & 6.8$\pm$0.09 & 6.9$\pm$0.08 & 7.0$\pm$0.05 & 7.1$\pm$0.06 & 7.2$\pm$0.09 & 7.3$\pm$0.09 & 7.4$\pm$0.09 & 7.5$\pm$0.06 & 7.6$\pm$0.1 & 7.8$\pm$0.09\\
40.0 & 6.8$\pm$0.09 & 6.9$\pm$0.07 & 7.0$\pm$0.09 & 7.1$\pm$0.04 & 7.2$\pm$0.05 & 7.3$\pm$0.07 & 7.4$\pm$0.09 & 7.5$\pm$0.06 & 7.6$\pm$0.09 & 7.75$\pm$0.09\\
50.0 & 6.8$\pm$0.11 & 6.9$\pm$0.05 & 7.0$\pm$0.08 & 7.1$\pm$0.06 & 7.2$\pm$0.05 & 7.3$\pm$0.03 & 7.4$\pm$0.04 & 7.5$\pm$0.04 & 7.6$\pm$0.08 & 7.7$\pm$0.08\\
60.0 & 6.8$\pm$0.1 & 6.9$\pm$0.03 & 7.0$\pm$0.06 & 7.1$\pm$0.03 & 7.2$\pm$0.0 & 7.3$\pm$0.0 & 7.4$\pm$0.04 & 7.5$\pm$0.04 & 7.6$\pm$0.06 & 7.7$\pm$0.07\\
70.0 & 6.8$\pm$0.05 & 6.9$\pm$0.03 & 7.0$\pm$0.05 & 7.1$\pm$0.03 & 7.2$\pm$0.0 & 7.3$\pm$0.0 & 7.4$\pm$0.06 & 7.5$\pm$0.04 & 7.6$\pm$0.07 & 7.7$\pm$0.06\\
\bottomrule
\end{tabular}%
}
\label{T4}%
\end{table*}%

\clearpage
\begin{table*}
  \caption{Results - MIST - wavelength: 3700-5000}
\makebox[\textwidth][c]{
    \begin{tabular}{cccccccccccccccccc}
\toprule

\multicolumn{1}{c}{S/N} &
\multicolumn{1}{c}{6.8} & \multicolumn{1}{c}{6.9} & \multicolumn{1}{c}{7.0} & \multicolumn{1}{c}{7.1} &
\multicolumn{1}{c}{7.2} &
\multicolumn{1}{c}{7.3} & \multicolumn{1}{c}{7.4} & \multicolumn{1}{c}{7.5} & \multicolumn{1}{c}{7.6} &
\multicolumn{1}{c}{7.7}  \\
\cmidrule(lr){1-1}\cmidrule(lr){2-2}\cmidrule(lr){3-3}\cmidrule(lr){4-4}\cmidrule(lr){5-5}\cmidrule(lr){6-6}\cmidrule(lr){7-7}\cmidrule(lr){8-8}\cmidrule(lr){9-9}\cmidrule(lr){10-10}\cmidrule(lr){11-11}

10.0 & 7.15$\pm$0.21 & 6.95$\pm$0.15 & 7.0$\pm$0.22 & 6.95$\pm$0.17 & 7.2$\pm$0.21 & 7.3$\pm$0.21 & 7.3$\pm$0.21 & 7.4$\pm$0.2 & 7.5$\pm$0.23 & 7.7$\pm$0.41\\
20.0 & 6.9$\pm$0.2 & 6.9$\pm$0.11 & 7.0$\pm$0.12 & 7.1$\pm$0.15 & 7.2$\pm$0.16 & 7.3$\pm$0.11 & 7.4$\pm$0.15 & 7.5$\pm$0.14 & 7.5$\pm$0.11 & 7.7$\pm$0.21\\
30.0 & 6.8$\pm$0.13 & 6.9$\pm$0.08 & 7.0$\pm$0.06 & 7.1$\pm$0.13 & 7.2$\pm$0.15 & 7.3$\pm$0.04 & 7.4$\pm$0.11 & 7.5$\pm$0.07 & 7.5$\pm$0.09 & 7.7$\pm$0.18\\
40.0 & 6.8$\pm$0.12 & 6.9$\pm$0.08 & 7.0$\pm$0.05 & 7.1$\pm$0.11 & 7.2$\pm$0.12 & 7.3$\pm$0.03 & 7.4$\pm$0.08 & 7.5$\pm$0.06 & 7.5$\pm$0.1 & 7.7$\pm$0.12\\
50.0 & 6.8$\pm$0.12 & 6.9$\pm$0.06 & 7.0$\pm$0.06 & 7.1$\pm$0.08 & 7.2$\pm$0.04 & 7.3$\pm$0.04 & 7.4$\pm$0.08 & 7.5$\pm$0.0 & 7.6$\pm$0.07 & 7.7$\pm$0.06\\
60.0 & 6.8$\pm$0.1 & 6.9$\pm$0.07 & 7.0$\pm$0.03 & 7.1$\pm$0.05 & 7.2$\pm$0.02 & 7.3$\pm$0.03 & 7.4$\pm$0.04 & 7.5$\pm$0.03 & 7.6$\pm$0.05 & 7.7$\pm$0.06\\
70.0 & 6.8$\pm$0.08 & 6.9$\pm$0.05 & 7.0$\pm$0.05 & 7.1$\pm$0.04 & 7.2$\pm$0.03 & 7.3$\pm$0.02 & 7.4$\pm$0.06 & 7.5$\pm$0.0 & 7.6$\pm$0.04 & 7.7$\pm$0.04\\
\bottomrule
\end{tabular}%
}
\label{T5}%
\end{table*}%

\clearpage
\begin{table*}
  \caption{Results - MIST - wavelength: 5000-6200}
\makebox[\textwidth][c]{
    \begin{tabular}{cccccccccccccccccc}
\toprule

\multicolumn{1}{c}{S/N} &
\multicolumn{1}{c}{6.8} & \multicolumn{1}{c}{6.9} & \multicolumn{1}{c}{7.0} & \multicolumn{1}{c}{7.1} &
\multicolumn{1}{c}{7.2} &
\multicolumn{1}{c}{7.3} & \multicolumn{1}{c}{7.4} & \multicolumn{1}{c}{7.5} & \multicolumn{1}{c}{7.6} &
\multicolumn{1}{c}{7.7}  \\
\cmidrule(lr){1-1}\cmidrule(lr){2-2}\cmidrule(lr){3-3}\cmidrule(lr){4-4}\cmidrule(lr){5-5}\cmidrule(lr){6-6}\cmidrule(lr){7-7}\cmidrule(lr){8-8}\cmidrule(lr){9-9}\cmidrule(lr){10-10}\cmidrule(lr){11-11}

10.0 & 7.0$\pm$0.53 & 7.1$\pm$0.62 & 7.25$\pm$0.59 & 7.3$\pm$0.41 & 7.25$\pm$0.69 & 7.5$\pm$0.69 & 7.4$\pm$1.0 & 7.7$\pm$0.86 & 7.5$\pm$0.85 & 7.5$\pm$0.83\\
20.0 & 7.0$\pm$0.46 & 7.3$\pm$0.45 & 7.1$\pm$0.36 & 7.3$\pm$0.45 & 7.4$\pm$0.59 & 7.7$\pm$0.39 & 7.4$\pm$0.76 & 7.5$\pm$0.67 & 7.65$\pm$0.67 & 8.65$\pm$0.69\\
30.0 & 6.8$\pm$0.26 & 7.4$\pm$0.5 & 7.0$\pm$0.39 & 7.4$\pm$0.39 & 7.4$\pm$0.41 & 7.3$\pm$0.34 & 7.4$\pm$0.71 & 7.5$\pm$0.48 & 7.6$\pm$0.52 & 7.9$\pm$0.49\\
40.0 & 6.8$\pm$0.2 & 7.4$\pm$0.45 & 7.0$\pm$0.15 & 7.1$\pm$0.18 & 7.2$\pm$0.28 & 7.4$\pm$0.25 & 7.4$\pm$0.51 & 7.5$\pm$0.43 & 7.6$\pm$0.5 & 7.9$\pm$0.53\\
50.0 & 6.8$\pm$0.19 & 7.35$\pm$0.57 & 7.0$\pm$0.2 & 7.3$\pm$0.25 & 7.2$\pm$0.26 & 7.55$\pm$0.3 & 7.4$\pm$0.72 & 7.5$\pm$0.15 & 7.6$\pm$0.4 & 7.7$\pm$0.44\\
60.0 & 7.0$\pm$0.19 & 6.9$\pm$0.56 & 7.0$\pm$0.16 & 7.5$\pm$0.21 & 7.2$\pm$0.28 & 7.4$\pm$0.2 & 7.4$\pm$0.42 & 7.5$\pm$0.15 & 7.6$\pm$0.23 & 7.8$\pm$0.42\\
70.0 & 6.8$\pm$0.18 & 7.15$\pm$0.55 & 7.0$\pm$0.16 & 7.3$\pm$0.21 & 7.2$\pm$0.29 & 7.3$\pm$0.21 & 7.4$\pm$0.06 & 7.5$\pm$0.13 & 7.6$\pm$0.33 & 7.7$\pm$0.3\\
\bottomrule
\end{tabular}%
}
\label{T6}%
\end{table*}%

\clearpage
\begin{table*}
  \caption{Results - MIST models Padova clusters - wavelength: 3700-6200}
\makebox[\textwidth][c]{
    \begin{tabular}{cccccccccccccccccc}
\toprule

\multicolumn{1}{c}{S/N} &
\multicolumn{1}{c}{6.8} & \multicolumn{1}{c}{6.9} & \multicolumn{1}{c}{7.0} & \multicolumn{1}{c}{7.1} &
\multicolumn{1}{c}{7.2} &
\multicolumn{1}{c}{7.3} & \multicolumn{1}{c}{7.4} & \multicolumn{1}{c}{7.5} & \multicolumn{1}{c}{7.6} &
\multicolumn{1}{c}{7.7}  \\
\cmidrule(lr){1-1}\cmidrule(lr){2-2}\cmidrule(lr){3-3}\cmidrule(lr){4-4}\cmidrule(lr){5-5}\cmidrule(lr){6-6}\cmidrule(lr){7-7}\cmidrule(lr){8-8}\cmidrule(lr){9-9}\cmidrule(lr){10-10}\cmidrule(lr){11-11}

10.0 & 6.9$\pm$0.15 & 6.9$\pm$0.11 & 6.95$\pm$0.06 & 7.0$\pm$0.07 & 7.05$\pm$0.11 & 7.3$\pm$0.09 & 7.3$\pm$0.12 & 7.5$\pm$0.15 & 7.5$\pm$0.18 & 7.6$\pm$0.15\\
20.0 & 7.1$\pm$0.15 & 7.05$\pm$0.12 & 6.9$\pm$0.06 & 7.0$\pm$0.07 & 7.0$\pm$0.1 & 7.3$\pm$0.04 & 7.3$\pm$0.09 & 7.5$\pm$0.1 & 7.5$\pm$0.14 & 7.6$\pm$0.13\\
30.0 & 7.2$\pm$0.08 & 7.0$\pm$0.1 & 6.9$\pm$0.06 & 7.0$\pm$0.07 & 7.0$\pm$0.06 & 7.3$\pm$0.07 & 7.3$\pm$0.09 & 7.4$\pm$0.13 & 7.5$\pm$0.12 & 7.6$\pm$0.12\\
40.0 & 7.2$\pm$0.05 & 6.9$\pm$0.08 & 6.9$\pm$0.04 & 7.0$\pm$0.04 & 7.0$\pm$0.05 & 7.3$\pm$0.03 & 7.3$\pm$0.08 & 7.4$\pm$0.09 & 7.5$\pm$0.12 & 7.6$\pm$0.13\\
50.0 & 7.2$\pm$0.05 & 6.9$\pm$0.07 & 6.9$\pm$0.03 & 7.0$\pm$0.02 & 7.0$\pm$0.03 & 7.3$\pm$0.03 & 7.3$\pm$0.08 & 7.3$\pm$0.09 & 7.3$\pm$0.1 & 7.6$\pm$0.13\\
60.0 & 7.2$\pm$0.04 & 6.9$\pm$0.05 & 6.9$\pm$0.03 & 7.0$\pm$0.0 & 7.0$\pm$0.03 & 7.3$\pm$0.02 & 7.3$\pm$0.07 & 7.3$\pm$0.09 & 7.4$\pm$0.13 & 7.6$\pm$0.13\\
70.0 & 7.2$\pm$0.04 & 6.9$\pm$0.05 & 6.9$\pm$0.02 & 7.0$\pm$0.0 & 7.0$\pm$0.02 & 7.3$\pm$0.02 & 7.3$\pm$0.08 & 7.3$\pm$0.07 & 7.4$\pm$0.13 & 7.8$\pm$0.11\\
\bottomrule
\end{tabular}%
}
\label{T7}%
\end{table*}%

\clearpage
\begin{table*}
  \caption{Results - MIST models Padova clusters - wavelength: 3700-5000}
\makebox[\textwidth][c]{
    \begin{tabular}{cccccccccccccccccc}
\toprule

\multicolumn{1}{c}{S/N} &
\multicolumn{1}{c}{6.8} & \multicolumn{1}{c}{6.9} & \multicolumn{1}{c}{7.0} & \multicolumn{1}{c}{7.1} &
\multicolumn{1}{c}{7.2} &
\multicolumn{1}{c}{7.3} & \multicolumn{1}{c}{7.4} & \multicolumn{1}{c}{7.5} & \multicolumn{1}{c}{7.6} &
\multicolumn{1}{c}{7.7}  \\
\cmidrule(lr){1-1}\cmidrule(lr){2-2}\cmidrule(lr){3-3}\cmidrule(lr){4-4}\cmidrule(lr){5-5}\cmidrule(lr){6-6}\cmidrule(lr){7-7}\cmidrule(lr){8-8}\cmidrule(lr){9-9}\cmidrule(lr){10-10}\cmidrule(lr){11-11}

10.0 & 6.9$\pm$0.21 & 6.9$\pm$0.17 & 7.0$\pm$0.23 & 7.0$\pm$0.2 & 7.3$\pm$0.13 & 7.3$\pm$0.18 & 7.3$\pm$0.16 & 7.3$\pm$0.25 & 7.3$\pm$0.24 & 7.5$\pm$0.33\\
20.0 & 6.9$\pm$0.15 & 6.9$\pm$0.1 & 7.0$\pm$0.14 & 7.0$\pm$0.14 & 7.3$\pm$0.14 & 7.3$\pm$0.11 & 7.3$\pm$0.09 & 7.45$\pm$0.1 & 7.5$\pm$0.1 & 7.5$\pm$0.22\\
30.0 & 6.9$\pm$0.13 & 6.9$\pm$0.06 & 7.0$\pm$0.04 & 7.1$\pm$0.11 & 7.1$\pm$0.11 & 7.3$\pm$0.05 & 7.3$\pm$0.09 & 7.5$\pm$0.08 & 7.5$\pm$0.13 & 7.5$\pm$0.19\\
40.0 & 6.8$\pm$0.08 & 6.9$\pm$0.07 & 7.0$\pm$0.09 & 7.0$\pm$0.1 & 7.2$\pm$0.11 & 7.3$\pm$0.0 & 7.3$\pm$0.06 & 7.5$\pm$0.08 & 7.5$\pm$0.1 & 7.5$\pm$0.15\\
50.0 & 6.8$\pm$0.1 & 6.9$\pm$0.05 & 7.0$\pm$0.03 & 7.0$\pm$0.05 & 7.3$\pm$0.1 & 7.3$\pm$0.0 & 7.3$\pm$0.09 & 7.5$\pm$0.07 & 7.5$\pm$0.09 & 7.5$\pm$0.17\\
60.0 & 6.8$\pm$0.08 & 6.9$\pm$0.05 & 7.0$\pm$0.03 & 7.0$\pm$0.05 & 7.3$\pm$0.1 & 7.3$\pm$0.0 & 7.3$\pm$0.08 & 7.5$\pm$0.05 & 7.5$\pm$0.04 & 7.5$\pm$0.16\\
70.0 & 6.8$\pm$0.04 & 6.9$\pm$0.04 & 7.0$\pm$0.03 & 7.0$\pm$0.04 & 7.1$\pm$0.09 & 7.3$\pm$0.0 & 7.3$\pm$0.07 & 7.5$\pm$0.04 & 7.5$\pm$0.06 & 7.5$\pm$0.07\\
\bottomrule
\end{tabular}%
}
\label{T8}%
\end{table*}%

\clearpage
\begin{table*}
  \caption{Results - MIST models Padova clusters - wavelength: 5000-6200}
\makebox[\textwidth][c]{
    \begin{tabular}{cccccccccccccccccc}
\toprule

\multicolumn{1}{c}{S/N} &
\multicolumn{1}{c}{6.8} & \multicolumn{1}{c}{6.9} & \multicolumn{1}{c}{7.0} & \multicolumn{1}{c}{7.1} &
\multicolumn{1}{c}{7.2} &
\multicolumn{1}{c}{7.3} & \multicolumn{1}{c}{7.4} & \multicolumn{1}{c}{7.5} & \multicolumn{1}{c}{7.6} &
\multicolumn{1}{c}{7.7}  \\
\cmidrule(lr){1-1}\cmidrule(lr){2-2}\cmidrule(lr){3-3}\cmidrule(lr){4-4}\cmidrule(lr){5-5}\cmidrule(lr){6-6}\cmidrule(lr){7-7}\cmidrule(lr){8-8}\cmidrule(lr){9-9}\cmidrule(lr){10-10}\cmidrule(lr){11-11}

10.0 & 7.15$\pm$0.72 & 7.1$\pm$0.59 & 7.3$\pm$0.62 & 7.25$\pm$0.52 & 7.1$\pm$0.44 & 7.5$\pm$0.66 & 7.35$\pm$0.64 & 7.7$\pm$0.53 & 7.45$\pm$0.58 & 7.55$\pm$0.61\\
20.0 & 7.7$\pm$0.53 & 7.4$\pm$0.56 & 7.4$\pm$0.42 & 7.3$\pm$0.43 & 7.3$\pm$0.38 & 7.3$\pm$0.43 & 7.6$\pm$0.29 & 7.7$\pm$0.36 & 7.65$\pm$0.4 & 7.6$\pm$0.46\\
30.0 & 8.0$\pm$0.5 & 7.6$\pm$0.49 & 7.3$\pm$0.3 & 7.3$\pm$0.33 & 7.3$\pm$0.25 & 7.35$\pm$0.26 & 7.7$\pm$0.2 & 7.75$\pm$0.42 & 7.75$\pm$0.34 & 7.6$\pm$0.4\\
40.0 & 8.1$\pm$0.35 & 7.5$\pm$0.42 & 7.3$\pm$0.27 & 7.15$\pm$0.25 & 7.4$\pm$0.27 & 7.4$\pm$0.21 & 7.65$\pm$0.25 & 7.8$\pm$0.28 & 7.65$\pm$0.29 & 7.5$\pm$0.32\\
50.0 & 7.9$\pm$0.32 & 7.6$\pm$0.45 & 7.3$\pm$0.32 & 7.3$\pm$0.24 & 7.2$\pm$0.25 & 7.3$\pm$0.21 & 7.6$\pm$0.21 & 7.8$\pm$0.21 & 7.7$\pm$0.19 & 7.55$\pm$0.22\\
60.0 & 8.0$\pm$0.28 & 7.6$\pm$0.38 & 7.3$\pm$0.2 & 7.4$\pm$0.23 & 7.1$\pm$0.22 & 7.3$\pm$0.19 & 7.4$\pm$0.23 & 7.8$\pm$0.2 & 7.6$\pm$0.2 & 7.5$\pm$0.21\\
70.0 & 8.0$\pm$0.28 & 7.6$\pm$0.36 & 7.5$\pm$0.21 & 7.2$\pm$0.2 & 7.2$\pm$0.23 & 7.3$\pm$0.19 & 7.6$\pm$0.18 & 7.8$\pm$0.28 & 7.7$\pm$0.22 & 7.6$\pm$0.17\\
\bottomrule
\end{tabular}%
}
\label{T9}%
\end{table*}%

\clearpage
\begin{table*}
  \caption{Results - Padova models MIST clusters - wavelength: 3700-6200}
\makebox[\textwidth][c]{
    \begin{tabular}{cccccccccccccccccc}
\toprule

\multicolumn{1}{c}{S/N} &
\multicolumn{1}{c}{6.8} & \multicolumn{1}{c}{6.9} & \multicolumn{1}{c}{7.0} & \multicolumn{1}{c}{7.1} &
\multicolumn{1}{c}{7.2} &
\multicolumn{1}{c}{7.3} & \multicolumn{1}{c}{7.4} & \multicolumn{1}{c}{7.5} & \multicolumn{1}{c}{7.6} &
\multicolumn{1}{c}{7.7}  \\
\cmidrule(lr){1-1}\cmidrule(lr){2-2}\cmidrule(lr){3-3}\cmidrule(lr){4-4}\cmidrule(lr){5-5}\cmidrule(lr){6-6}\cmidrule(lr){7-7}\cmidrule(lr){8-8}\cmidrule(lr){9-9}\cmidrule(lr){10-10}\cmidrule(lr){11-11}

10.0 & 6.85$\pm$0.12 & 6.9$\pm$0.11 & 7.0$\pm$0.13 & 6.9$\pm$0.19 & 7.3$\pm$0.28 & 7.3$\pm$0.18 & 6.9$\pm$0.42 & 7.5$\pm$0.54 & 8.05$\pm$0.72 & 8.4$\pm$0.59\\
20.0 & 6.8$\pm$0.14 & 6.95$\pm$0.13 & 7.1$\pm$0.11 & 6.9$\pm$0.22 & 7.5$\pm$0.26 & 7.4$\pm$0.12 & 6.9$\pm$0.0 & 8.3$\pm$0.56 & 8.4$\pm$0.52 & 8.5$\pm$0.79\\
30.0 & 6.8$\pm$0.17 & 6.9$\pm$0.08 & 7.1$\pm$0.1 & 7.3$\pm$0.25 & 7.5$\pm$0.19 & 7.3$\pm$0.11 & 6.9$\pm$0.0 & 8.4$\pm$0.46 & 8.4$\pm$0.47 & 8.5$\pm$0.42\\
40.0 & 6.8$\pm$0.09 & 6.9$\pm$0.06 & 7.1$\pm$0.15 & 7.3$\pm$0.23 & 7.5$\pm$0.24 & 7.3$\pm$0.1 & 6.9$\pm$0.0 & 8.4$\pm$0.5 & 8.45$\pm$0.43 & 8.5$\pm$0.57\\
50.0 & 6.8$\pm$0.14 & 6.9$\pm$0.05 & 7.1$\pm$0.12 & 7.3$\pm$0.23 & 7.5$\pm$0.2 & 7.4$\pm$0.1 & 6.9$\pm$0.0 & 8.4$\pm$0.23 & 8.45$\pm$0.29 & 8.5$\pm$0.3\\
60.0 & 6.8$\pm$0.14 & 6.9$\pm$0.03 & 7.1$\pm$0.1 & 7.3$\pm$0.21 & 7.5$\pm$0.18 & 7.4$\pm$0.07 & 6.9$\pm$0.0 & 8.4$\pm$0.31 & 8.5$\pm$0.06 & 8.55$\pm$0.31\\
70.0 & 6.8$\pm$0.05 & 6.9$\pm$0.04 & 7.1$\pm$0.11 & 7.3$\pm$0.15 & 7.5$\pm$0.24 & 7.4$\pm$0.09 & 6.9$\pm$0.0 & 8.4$\pm$0.27 & 8.5$\pm$0.06 & 8.55$\pm$0.05\\
\bottomrule
\end{tabular}%
}
\label{T10}%
\end{table*}%

\clearpage
\begin{table*}
  \caption{Results - Padova models MIST clusters - wavelength: 3700-5000}
\makebox[\textwidth][c]{
    \begin{tabular}{cccccccccccccccccc}
\toprule

\multicolumn{1}{c}{S/N} &
\multicolumn{1}{c}{6.8} & \multicolumn{1}{c}{6.9} & \multicolumn{1}{c}{7.0} & \multicolumn{1}{c}{7.1} &
\multicolumn{1}{c}{7.2} &
\multicolumn{1}{c}{7.3} & \multicolumn{1}{c}{7.4} & \multicolumn{1}{c}{7.5} & \multicolumn{1}{c}{7.6} &
\multicolumn{1}{c}{7.7}  \\
\cmidrule(lr){1-1}\cmidrule(lr){2-2}\cmidrule(lr){3-3}\cmidrule(lr){4-4}\cmidrule(lr){5-5}\cmidrule(lr){6-6}\cmidrule(lr){7-7}\cmidrule(lr){8-8}\cmidrule(lr){9-9}\cmidrule(lr){10-10}\cmidrule(lr){11-11}

10.0 & 7.1$\pm$0.19 & 7.0$\pm$0.17 & 7.05$\pm$0.13 & 7.0$\pm$0.21 & 7.2$\pm$0.26 & 7.3$\pm$0.22 & 7.3$\pm$0.26 & 7.5$\pm$0.33 & 7.6$\pm$0.31 & 7.6$\pm$0.49\\
20.0 & 7.1$\pm$0.14 & 6.9$\pm$0.14 & 7.0$\pm$0.07 & 7.1$\pm$0.24 & 7.25$\pm$0.24 & 7.3$\pm$0.1 & 7.4$\pm$0.22 & 7.5$\pm$0.16 & 7.5$\pm$0.16 & 7.8$\pm$0.27\\
30.0 & 7.0$\pm$0.13 & 6.8$\pm$0.13 & 7.0$\pm$0.06 & 7.2$\pm$0.24 & 7.25$\pm$0.28 & 7.3$\pm$0.09 & 7.5$\pm$0.23 & 7.5$\pm$0.08 & 7.5$\pm$0.09 & 7.5$\pm$0.2\\
40.0 & 7.1$\pm$0.1 & 6.8$\pm$0.12 & 7.0$\pm$0.04 & 6.8$\pm$0.25 & 7.3$\pm$0.2 & 7.3$\pm$0.07 & 7.5$\pm$0.16 & 7.5$\pm$0.04 & 7.5$\pm$0.07 & 7.65$\pm$0.15\\
50.0 & 7.1$\pm$0.1 & 6.8$\pm$0.05 & 7.0$\pm$0.05 & 7.05$\pm$0.25 & 7.25$\pm$0.18 & 7.3$\pm$0.07 & 7.5$\pm$0.11 & 7.5$\pm$0.04 & 7.5$\pm$0.04 & 7.5$\pm$0.15\\
60.0 & 7.1$\pm$0.07 & 6.8$\pm$0.08 & 7.0$\pm$0.04 & 7.05$\pm$0.25 & 7.2$\pm$0.2 & 7.3$\pm$0.06 & 7.5$\pm$0.04 & 7.5$\pm$0.02 & 7.5$\pm$0.03 & 7.65$\pm$0.15\\
70.0 & 7.1$\pm$0.1 & 6.8$\pm$0.05 & 7.0$\pm$0.04 & 7.3$\pm$0.25 & 7.5$\pm$0.19 & 7.3$\pm$0.05 & 7.5$\pm$0.0 & 7.5$\pm$0.03 & 7.5$\pm$0.06 & 7.65$\pm$0.15\\
\bottomrule
\end{tabular}%
}
\label{T11}%
\end{table*}%

\clearpage
\begin{table*}
  \caption{Results - Padova models MIST clusters - wavelength: 5000-6200}
\makebox[\textwidth][c]{
    \begin{tabular}{cccccccccccccccccc}
\toprule

\multicolumn{1}{c}{S/N} &
\multicolumn{1}{c}{6.8} & \multicolumn{1}{c}{6.9} & \multicolumn{1}{c}{7.0} & \multicolumn{1}{c}{7.1} &
\multicolumn{1}{c}{7.2} &
\multicolumn{1}{c}{7.3} & \multicolumn{1}{c}{7.4} & \multicolumn{1}{c}{7.5} & \multicolumn{1}{c}{7.6} &
\multicolumn{1}{c}{7.7}  \\
\cmidrule(lr){1-1}\cmidrule(lr){2-2}\cmidrule(lr){3-3}\cmidrule(lr){4-4}\cmidrule(lr){5-5}\cmidrule(lr){6-6}\cmidrule(lr){7-7}\cmidrule(lr){8-8}\cmidrule(lr){9-9}\cmidrule(lr){10-10}\cmidrule(lr){11-11}

10.0 & 6.8$\pm$0.53 & 7.0$\pm$0.69 & 6.95$\pm$0.66 & 7.05$\pm$0.6 & 7.3$\pm$0.6 & 7.25$\pm$0.75 & 9.1$\pm$0.76 & 7.3$\pm$0.88 & 7.6$\pm$0.94 & 7.55$\pm$0.89\\
20.0 & 6.85$\pm$0.44 & 7.3$\pm$0.5 & 7.05$\pm$0.38 & 7.2$\pm$0.53 & 7.5$\pm$0.6 & 7.3$\pm$0.53 & 9.0$\pm$0.61 & 7.25$\pm$0.77 & 7.2$\pm$0.91 & 8.8$\pm$0.85\\
30.0 & 6.8$\pm$0.17 & 7.4$\pm$0.46 & 7.2$\pm$0.36 & 7.2$\pm$0.49 & 7.5$\pm$0.59 & 7.4$\pm$0.41 & 9.0$\pm$0.07 & 7.0$\pm$0.8 & 7.25$\pm$0.92 & 8.8$\pm$0.84\\
40.0 & 6.9$\pm$0.07 & 7.6$\pm$0.38 & 7.0$\pm$0.17 & 7.1$\pm$0.29 & 7.5$\pm$0.42 & 7.45$\pm$0.38 & 9.0$\pm$0.07 & 7.55$\pm$0.83 & 7.1$\pm$0.91 & 8.8$\pm$0.86\\
50.0 & 6.8$\pm$0.08 & 7.7$\pm$0.48 & 7.0$\pm$0.16 & 7.1$\pm$0.43 & 7.5$\pm$0.44 & 7.4$\pm$0.37 & 9.05$\pm$0.08 & 8.7$\pm$0.81 & 7.85$\pm$0.94 & 8.9$\pm$0.58\\
60.0 & 6.8$\pm$0.08 & 7.7$\pm$0.4 & 7.0$\pm$0.15 & 7.1$\pm$0.27 & 7.5$\pm$0.39 & 7.4$\pm$0.29 & 9.1$\pm$0.09 & 8.7$\pm$0.76 & 8.7$\pm$0.91 & 8.9$\pm$0.76\\
70.0 & 6.8$\pm$0.06 & 7.7$\pm$0.38 & 7.05$\pm$0.17 & 7.1$\pm$0.31 & 7.5$\pm$0.39 & 7.4$\pm$0.26 & 9.1$\pm$0.08 & 7.5$\pm$0.8 & 8.7$\pm$0.91 & 8.9$\pm$0.67\\
\bottomrule
\end{tabular}%
}
\label{T12}%
\end{table*}%



\clearpage
\begin{table*}
  \caption{Metallicity Results - Padova - wavelength: 3700-6200}
\makebox[\textwidth][c]{
    \begin{tabular}{cccccccccccccccccc}
\toprule

\multicolumn{1}{c}{S/N} &
\multicolumn{1}{c}{6.8} & \multicolumn{1}{c}{6.9} & \multicolumn{1}{c}{7.0} & \multicolumn{1}{c}{7.1} &
\multicolumn{1}{c}{7.2} &
\multicolumn{1}{c}{7.3} & \multicolumn{1}{c}{7.4} & \multicolumn{1}{c}{7.5} & \multicolumn{1}{c}{7.6} &
\multicolumn{1}{c}{7.7}  \\
\cmidrule(lr){1-1}\cmidrule(lr){2-2}\cmidrule(lr){3-3}\cmidrule(lr){4-4}\cmidrule(lr){5-5}\cmidrule(lr){6-6}\cmidrule(lr){7-7}\cmidrule(lr){8-8}\cmidrule(lr){9-9}\cmidrule(lr){10-10}\cmidrule(lr){11-11}

10.0 & -0.2$\pm$0.28 & -0.2$\pm$0.32 & -0.6$\pm$0.28 & -0.6$\pm$0.29 & -0.6$\pm$0.24 & -0.4$\pm$0.23 & -0.4$\pm$0.19 & -0.4$\pm$0.27 & -0.4$\pm$0.33 & -0.4$\pm$0.37\\
20.0 & -0.4$\pm$0.23 & -0.4$\pm$0.41 & -0.6$\pm$0.1 & -0.6$\pm$0.25 & -0.4$\pm$0.11 & -0.4$\pm$0.09 & -0.4$\pm$0.11 & -0.4$\pm$0.22 & -0.5$\pm$0.33 & -0.4$\pm$0.33\\
30.0 & -0.4$\pm$0.21 & -0.4$\pm$0.28 & -0.4$\pm$0.1 & -0.4$\pm$0.13 & -0.4$\pm$0.08 & -0.4$\pm$0.12 & -0.4$\pm$0.1 & -0.4$\pm$0.16 & -0.4$\pm$0.23 & -0.4$\pm$0.2\\
40.0 & -0.4$\pm$0.13 & -0.4$\pm$0.13 & -0.4$\pm$0.08 & -0.4$\pm$0.07 & -0.4$\pm$0.07 & -0.4$\pm$0.06 & -0.4$\pm$0.0 & -0.4$\pm$0.1 & -0.4$\pm$0.23 & -0.4$\pm$0.14\\
50.0 & -0.4$\pm$0.04 & -0.4$\pm$0.16 & -0.4$\pm$0.08 & -0.4$\pm$0.08 & -0.4$\pm$0.05 & -0.4$\pm$0.05 & -0.4$\pm$0.0 & -0.4$\pm$0.1 & -0.4$\pm$0.09 & -0.4$\pm$0.12\\
60.0 & -0.4$\pm$0.11 & -0.4$\pm$0.04 & -0.4$\pm$0.07 & -0.4$\pm$0.04 & -0.4$\pm$0.0 & -0.4$\pm$0.06 & -0.4$\pm$0.0 & -0.4$\pm$0.06 & -0.4$\pm$0.08 & -0.4$\pm$0.1\\
70.0 & -0.4$\pm$0.04 & -0.4$\pm$0.04 & -0.4$\pm$0.08 & -0.4$\pm$0.04 & -0.4$\pm$0.0 & -0.4$\pm$0.05 & -0.4$\pm$0.0 & -0.4$\pm$0.04 & -0.4$\pm$0.06 & -0.4$\pm$0.06\\
\bottomrule
\end{tabular}%
}
\label{T13}%
\end{table*}%

\clearpage
\begin{table*}
  \caption{Metallicity Results - Padova - wavelength: 3700-5000}
\makebox[\textwidth][c]{
    \begin{tabular}{cccccccccccccccccc}
\toprule

\multicolumn{1}{c}{S/N} &
\multicolumn{1}{c}{6.8} & \multicolumn{1}{c}{6.9} & \multicolumn{1}{c}{7.0} & \multicolumn{1}{c}{7.1} &
\multicolumn{1}{c}{7.2} &
\multicolumn{1}{c}{7.3} & \multicolumn{1}{c}{7.4} & \multicolumn{1}{c}{7.5} & \multicolumn{1}{c}{7.6} &
\multicolumn{1}{c}{7.7}  \\
\cmidrule(lr){1-1}\cmidrule(lr){2-2}\cmidrule(lr){3-3}\cmidrule(lr){4-4}\cmidrule(lr){5-5}\cmidrule(lr){6-6}\cmidrule(lr){7-7}\cmidrule(lr){8-8}\cmidrule(lr){9-9}\cmidrule(lr){10-10}\cmidrule(lr){11-11}

10.0 & -0.2$\pm$0.23 & -0.2$\pm$0.26 & -0.4$\pm$0.23 & -0.4$\pm$0.32 & -0.4$\pm$0.13 & -0.4$\pm$0.22 & -0.2$\pm$0.29 & -0.4$\pm$0.32 & -0.4$\pm$0.29 & -0.4$\pm$0.35\\
20.0 & -0.4$\pm$0.14 & -0.4$\pm$0.26 & -0.4$\pm$0.15 & -0.4$\pm$0.18 & -0.4$\pm$0.23 & -0.4$\pm$0.18 & -0.4$\pm$0.15 & -0.4$\pm$0.17 & -0.4$\pm$0.17 & -0.4$\pm$0.15\\
30.0 & -0.4$\pm$0.11 & -0.4$\pm$0.23 & -0.4$\pm$0.08 & -0.4$\pm$0.08 & -0.4$\pm$0.11 & -0.4$\pm$0.14 & -0.4$\pm$0.13 & -0.4$\pm$0.16 & -0.4$\pm$0.15 & -0.4$\pm$0.17\\
40.0 & -0.4$\pm$0.05 & -0.4$\pm$0.16 & -0.4$\pm$0.08 & -0.4$\pm$0.04 & -0.4$\pm$0.0 & -0.4$\pm$0.0 & -0.4$\pm$0.13 & -0.4$\pm$0.14 & -0.4$\pm$0.1 & -0.4$\pm$0.12\\
50.0 & -0.4$\pm$0.07 & -0.4$\pm$0.24 & -0.4$\pm$0.05 & -0.4$\pm$0.0 & -0.4$\pm$0.0 & -0.4$\pm$0.05 & -0.4$\pm$0.09 & -0.4$\pm$0.1 & -0.4$\pm$0.05 & -0.4$\pm$0.14\\
60.0 & -0.4$\pm$0.08 & -0.4$\pm$0.11 & -0.4$\pm$0.07 & -0.4$\pm$0.0 & -0.4$\pm$0.0 & -0.4$\pm$0.05 & -0.4$\pm$0.0 & -0.4$\pm$0.07 & -0.4$\pm$0.0 & -0.4$\pm$0.0\\
70.0 & -0.4$\pm$0.08 & -0.4$\pm$0.0 & -0.4$\pm$0.06 & -0.4$\pm$0.0 & -0.4$\pm$0.0 & -0.4$\pm$0.0 & -0.4$\pm$0.0 & -0.4$\pm$0.1 & -0.4$\pm$0.0 & -0.4$\pm$0.0\\
\bottomrule
\end{tabular}%
}
\label{T14}%
\end{table*}%

\clearpage
\begin{table*}
  \caption{Metallicity Results - Padova - wavelength: 5000-6200}
\makebox[\textwidth][c]{
    \begin{tabular}{cccccccccccccccccc}
\toprule

\multicolumn{1}{c}{S/N} &
\multicolumn{1}{c}{6.8} & \multicolumn{1}{c}{6.9} & \multicolumn{1}{c}{7.0} & \multicolumn{1}{c}{7.1} &
\multicolumn{1}{c}{7.2} &
\multicolumn{1}{c}{7.3} & \multicolumn{1}{c}{7.4} & \multicolumn{1}{c}{7.5} & \multicolumn{1}{c}{7.6} &
\multicolumn{1}{c}{7.7}  \\
\cmidrule(lr){1-1}\cmidrule(lr){2-2}\cmidrule(lr){3-3}\cmidrule(lr){4-4}\cmidrule(lr){5-5}\cmidrule(lr){6-6}\cmidrule(lr){7-7}\cmidrule(lr){8-8}\cmidrule(lr){9-9}\cmidrule(lr){10-10}\cmidrule(lr){11-11}

10.0 & -0.4$\pm$0.39 & -0.3$\pm$0.36 & -0.4$\pm$0.41 & -0.4$\pm$0.34 & -0.4$\pm$0.3 & -0.2$\pm$0.34 & -0.2$\pm$0.4 & -0.2$\pm$0.36 & -0.2$\pm$0.46 & -0.3$\pm$0.34\\
20.0 & -0.4$\pm$0.4 & -0.4$\pm$0.41 & -0.4$\pm$0.34 & -0.5$\pm$0.33 & -0.3$\pm$0.25 & -0.4$\pm$0.24 & -0.2$\pm$0.33 & -0.2$\pm$0.29 & -0.2$\pm$0.32 & -0.4$\pm$0.29\\
30.0 & -0.3$\pm$0.39 & -0.4$\pm$0.4 & -0.4$\pm$0.26 & -0.4$\pm$0.33 & -0.4$\pm$0.27 & -0.2$\pm$0.16 & -0.2$\pm$0.27 & -0.2$\pm$0.29 & -0.2$\pm$0.28 & -0.3$\pm$0.23\\
40.0 & -0.3$\pm$0.37 & -0.4$\pm$0.34 & -0.4$\pm$0.24 & -0.5$\pm$0.25 & -0.2$\pm$0.24 & -0.2$\pm$0.24 & -0.2$\pm$0.19 & -0.2$\pm$0.25 & -0.2$\pm$0.28 & -0.2$\pm$0.28\\
50.0 & -0.4$\pm$0.36 & -0.4$\pm$0.32 & -0.4$\pm$0.18 & -0.4$\pm$0.3 & -0.4$\pm$0.22 & -0.3$\pm$0.16 & -0.4$\pm$0.18 & -0.2$\pm$0.22 & -0.2$\pm$0.21 & -0.3$\pm$0.24\\
60.0 & -0.4$\pm$0.31 & -0.4$\pm$0.3 & -0.4$\pm$0.18 & -0.4$\pm$0.2 & -0.4$\pm$0.16 & -0.2$\pm$0.19 & -0.4$\pm$0.23 & -0.4$\pm$0.24 & -0.4$\pm$0.25 & -0.4$\pm$0.29\\
70.0 & -0.4$\pm$0.29 & -0.4$\pm$0.27 & -0.4$\pm$0.18 & -0.4$\pm$0.17 & -0.4$\pm$0.17 & -0.2$\pm$0.1 & -0.4$\pm$0.23 & -0.4$\pm$0.26 & -0.3$\pm$0.19 & -0.4$\pm$0.2\\
\bottomrule
\end{tabular}%
}
\label{T15}%
\end{table*}%


\clearpage
\begin{table*}
  \caption{Metallicity Results - MIST - wavelength: 3700-6200}
\makebox[\textwidth][c]{
    \begin{tabular}{cccccccccccccccccc}
\toprule

\multicolumn{1}{c}{S/N} &
\multicolumn{1}{c}{6.8} & \multicolumn{1}{c}{6.9} & \multicolumn{1}{c}{7.0} & \multicolumn{1}{c}{7.1} &
\multicolumn{1}{c}{7.2} &
\multicolumn{1}{c}{7.3} & \multicolumn{1}{c}{7.4} & \multicolumn{1}{c}{7.5} & \multicolumn{1}{c}{7.6} &
\multicolumn{1}{c}{7.7}  \\
\cmidrule(lr){1-1}\cmidrule(lr){2-2}\cmidrule(lr){3-3}\cmidrule(lr){4-4}\cmidrule(lr){5-5}\cmidrule(lr){6-6}\cmidrule(lr){7-7}\cmidrule(lr){8-8}\cmidrule(lr){9-9}\cmidrule(lr){10-10}\cmidrule(lr){11-11}

10.0 & -0.4$\pm$0.27 & -0.3$\pm$0.2 & -0.4$\pm$0.22 & -0.2$\pm$0.27 & -0.4$\pm$0.29 & -0.3$\pm$0.29 & -0.4$\pm$0.25 & -0.3$\pm$0.26 & -0.4$\pm$0.24 & -0.4$\pm$0.35\\
20.0 & -0.4$\pm$0.13 & -0.4$\pm$0.19 & -0.4$\pm$0.22 & -0.4$\pm$0.2 & -0.4$\pm$0.23 & -0.4$\pm$0.27 & -0.4$\pm$0.1 & -0.4$\pm$0.16 & -0.4$\pm$0.16 & -0.4$\pm$0.14\\
30.0 & -0.4$\pm$0.08 & -0.4$\pm$0.16 & -0.4$\pm$0.14 & -0.4$\pm$0.11 & -0.4$\pm$0.17 & -0.4$\pm$0.23 & -0.4$\pm$0.09 & -0.4$\pm$0.09 & -0.4$\pm$0.07 & -0.4$\pm$0.1\\
40.0 & -0.4$\pm$0.0 & -0.4$\pm$0.19 & -0.4$\pm$0.15 & -0.4$\pm$0.08 & -0.4$\pm$0.04 & -0.4$\pm$0.18 & -0.4$\pm$0.09 & -0.4$\pm$0.1 & -0.4$\pm$0.07 & -0.4$\pm$0.08\\
50.0 & -0.4$\pm$0.04 & -0.4$\pm$0.07 & -0.4$\pm$0.13 & -0.4$\pm$0.17 & -0.4$\pm$0.05 & -0.4$\pm$0.08 & -0.4$\pm$0.04 & -0.4$\pm$0.06 & -0.4$\pm$0.05 & -0.4$\pm$0.08\\
60.0 & -0.4$\pm$0.05 & -0.4$\pm$0.12 & -0.4$\pm$0.09 & -0.4$\pm$0.11 & -0.4$\pm$0.0 & -0.4$\pm$0.09 & -0.4$\pm$0.04 & -0.4$\pm$0.08 & -0.4$\pm$0.06 & -0.4$\pm$0.06\\
70.0 & -0.4$\pm$0.0 & -0.4$\pm$0.12 & -0.4$\pm$0.1 & -0.4$\pm$0.08 & -0.4$\pm$0.0 & -0.4$\pm$0.08 & -0.4$\pm$0.06 & -0.4$\pm$0.06 & -0.4$\pm$0.05 & -0.4$\pm$0.05\\
\bottomrule
\end{tabular}%
}
\label{T16}%
\end{table*}%

\clearpage
\begin{table*}
  \caption{Metallicity Results - MIST - wavelength: 3700-5000}
\makebox[\textwidth][c]{
    \begin{tabular}{cccccccccccccccccc}
\toprule

\multicolumn{1}{c}{S/N} &
\multicolumn{1}{c}{6.8} & \multicolumn{1}{c}{6.9} & \multicolumn{1}{c}{7.0} & \multicolumn{1}{c}{7.1} &
\multicolumn{1}{c}{7.2} &
\multicolumn{1}{c}{7.3} & \multicolumn{1}{c}{7.4} & \multicolumn{1}{c}{7.5} & \multicolumn{1}{c}{7.6} &
\multicolumn{1}{c}{7.7}  \\
\cmidrule(lr){1-1}\cmidrule(lr){2-2}\cmidrule(lr){3-3}\cmidrule(lr){4-4}\cmidrule(lr){5-5}\cmidrule(lr){6-6}\cmidrule(lr){7-7}\cmidrule(lr){8-8}\cmidrule(lr){9-9}\cmidrule(lr){10-10}\cmidrule(lr){11-11}

10.0 & -0.2$\pm$0.22 & -0.4$\pm$0.18 & -0.4$\pm$0.43 & -0.2$\pm$0.17 & -0.2$\pm$0.16 & -0.2$\pm$0.17 & -0.2$\pm$0.17 & -0.2$\pm$0.15 & -0.2$\pm$0.16 & -0.2$\pm$0.36\\
20.0 & -0.4$\pm$0.17 & -0.4$\pm$0.13 & -0.4$\pm$0.34 & -0.4$\pm$0.11 & -0.4$\pm$0.12 & -0.3$\pm$0.16 & -0.4$\pm$0.14 & -0.2$\pm$0.14 & -0.3$\pm$0.1 & -0.4$\pm$0.16\\
30.0 & -0.4$\pm$0.11 & -0.4$\pm$0.11 & -0.4$\pm$0.22 & -0.4$\pm$0.11 & -0.4$\pm$0.1 & -0.4$\pm$0.1 & -0.4$\pm$0.1 & -0.4$\pm$0.11 & -0.4$\pm$0.1 & -0.4$\pm$0.15\\
40.0 & -0.4$\pm$0.09 & -0.4$\pm$0.13 & -0.4$\pm$0.09 & -0.4$\pm$0.07 & -0.4$\pm$0.08 & -0.4$\pm$0.1 & -0.4$\pm$0.09 & -0.4$\pm$0.11 & -0.4$\pm$0.11 & -0.4$\pm$0.1\\
50.0 & -0.4$\pm$0.09 & -0.4$\pm$0.1 & -0.4$\pm$0.1 & -0.4$\pm$0.06 & -0.4$\pm$0.0 & -0.4$\pm$0.11 & -0.4$\pm$0.1 & -0.4$\pm$0.1 & -0.4$\pm$0.08 & -0.4$\pm$0.04\\
60.0 & -0.4$\pm$0.08 & -0.4$\pm$0.12 & -0.4$\pm$0.06 & -0.4$\pm$0.09 & -0.4$\pm$0.0 & -0.4$\pm$0.08 & -0.4$\pm$0.04 & -0.4$\pm$0.07 & -0.4$\pm$0.07 & -0.4$\pm$0.04\\
70.0 & -0.4$\pm$0.06 & -0.4$\pm$0.04 & -0.4$\pm$0.09 & -0.4$\pm$0.0 & -0.4$\pm$0.0 & -0.4$\pm$0.09 & -0.4$\pm$0.06 & -0.4$\pm$0.09 & -0.4$\pm$0.07 & -0.4$\pm$0.04\\
\bottomrule
\end{tabular}%
}
\label{T17}%
\end{table*}%

\clearpage
\begin{table*}
  \caption{Metallicity Results - MIST - wavelength: 5000-6200}
\makebox[\textwidth][c]{
    \begin{tabular}{cccccccccccccccccc}
\toprule

\multicolumn{1}{c}{S/N} &
\multicolumn{1}{c}{6.8} & \multicolumn{1}{c}{6.9} & \multicolumn{1}{c}{7.0} & \multicolumn{1}{c}{7.1} &
\multicolumn{1}{c}{7.2} &
\multicolumn{1}{c}{7.3} & \multicolumn{1}{c}{7.4} & \multicolumn{1}{c}{7.5} & \multicolumn{1}{c}{7.6} &
\multicolumn{1}{c}{7.7}  \\
\cmidrule(lr){1-1}\cmidrule(lr){2-2}\cmidrule(lr){3-3}\cmidrule(lr){4-4}\cmidrule(lr){5-5}\cmidrule(lr){6-6}\cmidrule(lr){7-7}\cmidrule(lr){8-8}\cmidrule(lr){9-9}\cmidrule(lr){10-10}\cmidrule(lr){11-11}

10.0 & -0.4$\pm$0.47 & 0.0$\pm$0.4 & 0.0$\pm$0.36 & 0.0$\pm$0.41 & 0.2$\pm$0.26 & 0.2$\pm$0.23 & -0.2$\pm$0.28 & 0.2$\pm$0.29 & -0.2$\pm$0.31 & -0.2$\pm$0.29\\
20.0 & -0.2$\pm$0.36 & 0.2$\pm$0.28 & -0.3$\pm$0.38 & 0.1$\pm$0.24 & -0.2$\pm$0.22 & 0.1$\pm$0.13 & -0.4$\pm$0.27 & -0.2$\pm$0.21 & -0.4$\pm$0.27 & -0.2$\pm$0.35\\
30.0 & -0.4$\pm$0.35 & 0.0$\pm$0.23 & -0.4$\pm$0.31 & 0.0$\pm$0.24 & -0.2$\pm$0.22 & 0.0$\pm$0.28 & -0.4$\pm$0.27 & -0.4$\pm$0.21 & -0.4$\pm$0.22 & -0.4$\pm$0.18\\
40.0 & -0.4$\pm$0.28 & 0.0$\pm$0.28 & -0.4$\pm$0.29 & -0.4$\pm$0.28 & -0.2$\pm$0.21 & 0.0$\pm$0.28 & -0.4$\pm$0.21 & -0.4$\pm$0.23 & -0.4$\pm$0.21 & -0.4$\pm$0.28\\
50.0 & -0.4$\pm$0.23 & 0.0$\pm$0.26 & -0.4$\pm$0.33 & 0.0$\pm$0.27 & -0.4$\pm$0.14 & 0.0$\pm$0.24 & -0.4$\pm$0.22 & -0.3$\pm$0.13 & -0.4$\pm$0.2 & -0.4$\pm$0.16\\
60.0 & -0.6$\pm$0.26 & -0.4$\pm$0.26 & -0.5$\pm$0.26 & 0.0$\pm$0.26 & -0.3$\pm$0.16 & 0.0$\pm$0.27 & -0.4$\pm$0.16 & -0.4$\pm$0.15 & -0.4$\pm$0.11 & -0.4$\pm$0.16\\
70.0 & -0.4$\pm$0.25 & -0.4$\pm$0.28 & -0.4$\pm$0.22 & 0.0$\pm$0.29 & -0.2$\pm$0.11 & -0.4$\pm$0.24 & -0.4$\pm$0.06 & -0.2$\pm$0.14 & -0.4$\pm$0.12 & -0.4$\pm$0.07\\
\bottomrule
\end{tabular}%
}
\label{T18}%
\end{table*}%

\clearpage
\begin{table*}
  \caption{Metallicity Results - MIST models Padova clusters - wavelength: 3700-6200}
\makebox[\textwidth][c]{
    \begin{tabular}{cccccccccccccccccc}
\toprule

\multicolumn{1}{c}{S/N} &
\multicolumn{1}{c}{6.8} & \multicolumn{1}{c}{6.9} & \multicolumn{1}{c}{7.0} & \multicolumn{1}{c}{7.1} &
\multicolumn{1}{c}{7.2} &
\multicolumn{1}{c}{7.3} & \multicolumn{1}{c}{7.4} & \multicolumn{1}{c}{7.5} & \multicolumn{1}{c}{7.6} &
\multicolumn{1}{c}{7.7}  \\
\cmidrule(lr){1-1}\cmidrule(lr){2-2}\cmidrule(lr){3-3}\cmidrule(lr){4-4}\cmidrule(lr){5-5}\cmidrule(lr){6-6}\cmidrule(lr){7-7}\cmidrule(lr){8-8}\cmidrule(lr){9-9}\cmidrule(lr){10-10}\cmidrule(lr){11-11}

10.0 & -0.2$\pm$0.13 & -0.2$\pm$0.2 & -0.4$\pm$0.29 & -0.3$\pm$0.31 & -0.2$\pm$0.3 & -0.4$\pm$0.31 & -0.1$\pm$0.34 & -0.6$\pm$0.38 & -0.3$\pm$0.45 & -0.6$\pm$0.4\\
20.0 & -0.2$\pm$0.1 & -0.4$\pm$0.11 & -0.4$\pm$0.26 & -0.3$\pm$0.25 & -0.4$\pm$0.12 & -0.6$\pm$0.35 & -0.2$\pm$0.26 & -0.4$\pm$0.31 & -0.4$\pm$0.37 & -0.6$\pm$0.37\\
30.0 & -0.4$\pm$0.1 & -0.4$\pm$0.15 & -0.6$\pm$0.29 & -0.4$\pm$0.16 & -0.4$\pm$0.11 & -0.6$\pm$0.29 & -0.2$\pm$0.24 & -0.4$\pm$0.28 & -0.4$\pm$0.31 & -0.6$\pm$0.39\\
40.0 & -0.4$\pm$0.09 & -0.4$\pm$0.18 & -0.6$\pm$0.27 & -0.4$\pm$0.11 & -0.2$\pm$0.11 & -0.6$\pm$0.23 & -0.2$\pm$0.16 & -0.4$\pm$0.3 & 0.0$\pm$0.29 & 0.0$\pm$0.39\\
50.0 & -0.4$\pm$0.09 & -0.4$\pm$0.16 & -0.6$\pm$0.29 & -0.4$\pm$0.11 & -0.2$\pm$0.09 & -0.6$\pm$0.17 & -0.2$\pm$0.2 & -0.4$\pm$0.24 & -0.4$\pm$0.26 & 0.0$\pm$0.37\\
60.0 & -0.4$\pm$0.08 & -0.4$\pm$0.16 & -0.6$\pm$0.21 & -0.2$\pm$0.09 & -0.2$\pm$0.08 & -0.6$\pm$0.12 & -0.2$\pm$0.23 & -0.4$\pm$0.24 & 0.0$\pm$0.33 & 0.0$\pm$0.39\\
70.0 & -0.4$\pm$0.08 & -0.4$\pm$0.15 & -0.8$\pm$0.14 & -0.2$\pm$0.08 & -0.2$\pm$0.06 & -0.6$\pm$0.1 & -0.2$\pm$0.23 & -0.4$\pm$0.27 & -0.4$\pm$0.33 & -0.8$\pm$0.39\\
\bottomrule
\end{tabular}%
}
\label{T19}%
\end{table*}%

\clearpage
\begin{table*}
  \caption{Metallicity Results - MIST models Padova clusters - wavelength: 3700-5000}
\makebox[\textwidth][c]{
    \begin{tabular}{cccccccccccccccccc}
\toprule

\multicolumn{1}{c}{S/N} &
\multicolumn{1}{c}{6.8} & \multicolumn{1}{c}{6.9} & \multicolumn{1}{c}{7.0} & \multicolumn{1}{c}{7.1} &
\multicolumn{1}{c}{7.2} &
\multicolumn{1}{c}{7.3} & \multicolumn{1}{c}{7.4} & \multicolumn{1}{c}{7.5} & \multicolumn{1}{c}{7.6} &
\multicolumn{1}{c}{7.7}  \\
\cmidrule(lr){1-1}\cmidrule(lr){2-2}\cmidrule(lr){3-3}\cmidrule(lr){4-4}\cmidrule(lr){5-5}\cmidrule(lr){6-6}\cmidrule(lr){7-7}\cmidrule(lr){8-8}\cmidrule(lr){9-9}\cmidrule(lr){10-10}\cmidrule(lr){11-11}

10.0 & -0.2$\pm$0.15 & -0.2$\pm$0.21 & 0.2$\pm$0.51 & -0.1$\pm$0.43 & -0.2$\pm$0.18 & -0.2$\pm$0.16 & -0.2$\pm$0.21 & -0.2$\pm$0.17 & -0.2$\pm$0.22 & -0.2$\pm$0.42\\
20.0 & -0.2$\pm$0.13 & -0.2$\pm$0.12 & -0.4$\pm$0.37 & -0.4$\pm$0.35 & -0.4$\pm$0.22 & -0.2$\pm$0.17 & -0.2$\pm$0.18 & -0.2$\pm$0.21 & -0.2$\pm$0.24 & 0.0$\pm$0.44\\
30.0 & -0.1$\pm$0.14 & -0.2$\pm$0.1 & -0.4$\pm$0.29 & -0.3$\pm$0.34 & -0.4$\pm$0.23 & -0.2$\pm$0.18 & -0.4$\pm$0.15 & -0.4$\pm$0.27 & -0.2$\pm$0.24 & -0.3$\pm$0.41\\
40.0 & -0.2$\pm$0.1 & -0.2$\pm$0.1 & -0.4$\pm$0.32 & -0.2$\pm$0.31 & -0.5$\pm$0.28 & -0.2$\pm$0.18 & -0.2$\pm$0.16 & -0.2$\pm$0.28 & -0.1$\pm$0.24 & 0.0$\pm$0.4\\
50.0 & -0.2$\pm$0.11 & -0.2$\pm$0.1 & -0.4$\pm$0.26 & -0.2$\pm$0.23 & -0.6$\pm$0.24 & -0.4$\pm$0.16 & -0.4$\pm$0.15 & 0.0$\pm$0.27 & -0.1$\pm$0.26 & 0.0$\pm$0.42\\
60.0 & -0.2$\pm$0.07 & -0.2$\pm$0.12 & -0.4$\pm$0.24 & -0.2$\pm$0.21 & -0.6$\pm$0.26 & -0.2$\pm$0.17 & -0.4$\pm$0.13 & 0.0$\pm$0.29 & 0.0$\pm$0.23 & 0.0$\pm$0.39\\
70.0 & -0.2$\pm$0.08 & -0.2$\pm$0.1 & -0.4$\pm$0.21 & -0.2$\pm$0.17 & -0.2$\pm$0.27 & -0.2$\pm$0.16 & -0.4$\pm$0.11 & 0.0$\pm$0.29 & 0.0$\pm$0.23 & 0.0$\pm$0.18\\
\bottomrule
\end{tabular}%
}
\label{T20}%
\end{table*}%

\clearpage
\begin{table*}
  \caption{Metallicity Results - MIST models Padova clusters - wavelength: 5000-6200}
\makebox[\textwidth][c]{
    \begin{tabular}{cccccccccccccccccc}
\toprule

\multicolumn{1}{c}{S/N} &
\multicolumn{1}{c}{6.8} & \multicolumn{1}{c}{6.9} & \multicolumn{1}{c}{7.0} & \multicolumn{1}{c}{7.1} &
\multicolumn{1}{c}{7.2} &
\multicolumn{1}{c}{7.3} & \multicolumn{1}{c}{7.4} & \multicolumn{1}{c}{7.5} & \multicolumn{1}{c}{7.6} &
\multicolumn{1}{c}{7.7}  \\
\cmidrule(lr){1-1}\cmidrule(lr){2-2}\cmidrule(lr){3-3}\cmidrule(lr){4-4}\cmidrule(lr){5-5}\cmidrule(lr){6-6}\cmidrule(lr){7-7}\cmidrule(lr){8-8}\cmidrule(lr){9-9}\cmidrule(lr){10-10}\cmidrule(lr){11-11}

10.0 & 0.2$\pm$0.14 & 0.0$\pm$0.49 & 0.2$\pm$0.16 & 0.0$\pm$0.34 & 0.0$\pm$0.36 & 0.2$\pm$0.21 & 0.2$\pm$0.24 & 0.2$\pm$0.19 & 0.2$\pm$0.13 & 0.1$\pm$0.26\\
20.0 & 0.0$\pm$0.38 & 0.0$\pm$0.48 & 0.0$\pm$0.22 & 0.0$\pm$0.29 & 0.0$\pm$0.21 & 0.0$\pm$0.22 & 0.0$\pm$0.12 & 0.1$\pm$0.2 & 0.1$\pm$0.16 & 0.0$\pm$0.23\\
30.0 & 0.0$\pm$0.42 & -0.8$\pm$0.49 & 0.0$\pm$0.3 & 0.0$\pm$0.23 & 0.0$\pm$0.25 & 0.0$\pm$0.19 & 0.0$\pm$0.15 & 0.0$\pm$0.16 & 0.0$\pm$0.13 & 0.0$\pm$0.25\\
40.0 & 0.0$\pm$0.4 & -0.8$\pm$0.37 & 0.0$\pm$0.34 & -0.2$\pm$0.23 & 0.0$\pm$0.21 & 0.0$\pm$0.21 & 0.0$\pm$0.23 & 0.0$\pm$0.14 & 0.0$\pm$0.13 & 0.0$\pm$0.29\\
50.0 & 0.0$\pm$0.46 & -0.8$\pm$0.35 & 0.0$\pm$0.32 & 0.0$\pm$0.25 & -0.2$\pm$0.23 & -0.2$\pm$0.19 & 0.0$\pm$0.23 & 0.0$\pm$0.12 & 0.0$\pm$0.14 & 0.0$\pm$0.23\\
60.0 & 0.0$\pm$0.46 & -0.8$\pm$0.24 & -0.4$\pm$0.33 & 0.0$\pm$0.22 & -0.2$\pm$0.22 & -0.2$\pm$0.18 & 0.0$\pm$0.29 & 0.1$\pm$0.11 & 0.0$\pm$0.21 & 0.0$\pm$0.31\\
70.0 & 0.0$\pm$0.47 & -0.8$\pm$0.11 & 0.0$\pm$0.31 & -0.2$\pm$0.26 & -0.2$\pm$0.21 & -0.2$\pm$0.19 & 0.0$\pm$0.26 & 0.0$\pm$0.14 & 0.0$\pm$0.2 & 0.0$\pm$0.27\\
\bottomrule
\end{tabular}%
}
\label{T21}%
\end{table*}%

\clearpage
\begin{table*}
  \caption{Metallicity Results - Padova models MIST clusters - wavelength: 3700-6200}
\makebox[\textwidth][c]{
    \begin{tabular}{cccccccccccccccccc}
\toprule

\multicolumn{1}{c}{S/N} &
\multicolumn{1}{c}{6.8} & \multicolumn{1}{c}{6.9} & \multicolumn{1}{c}{7.0} & \multicolumn{1}{c}{7.1} &
\multicolumn{1}{c}{7.2} &
\multicolumn{1}{c}{7.3} & \multicolumn{1}{c}{7.4} & \multicolumn{1}{c}{7.5} & \multicolumn{1}{c}{7.6} &
\multicolumn{1}{c}{7.7}  \\
\cmidrule(lr){1-1}\cmidrule(lr){2-2}\cmidrule(lr){3-3}\cmidrule(lr){4-4}\cmidrule(lr){5-5}\cmidrule(lr){6-6}\cmidrule(lr){7-7}\cmidrule(lr){8-8}\cmidrule(lr){9-9}\cmidrule(lr){10-10}\cmidrule(lr){11-11}

10.0 & -0.4$\pm$0.39 & -0.2$\pm$0.37 & -0.6$\pm$0.21 & -0.6$\pm$0.39 & -0.2$\pm$0.28 & -0.2$\pm$0.13 & 0.0$\pm$0.24 & -0.2$\pm$0.35 & -0.9$\pm$0.46 & -1.0$\pm$0.37\\
20.0 & -0.3$\pm$0.4 & -0.6$\pm$0.41 & -0.6$\pm$0.24 & -1.0$\pm$0.36 & -0.2$\pm$0.31 & -0.2$\pm$0.11 & 0.0$\pm$0.1 & -0.8$\pm$0.41 & -0.8$\pm$0.36 & -0.8$\pm$0.46\\
30.0 & -1.0$\pm$0.38 & -0.4$\pm$0.29 & -0.4$\pm$0.18 & -0.6$\pm$0.37 & -0.6$\pm$0.29 & -0.2$\pm$0.11 & 0.0$\pm$0.08 & -1.0$\pm$0.36 & -1.0$\pm$0.3 & -1.0$\pm$0.25\\
40.0 & -1.0$\pm$0.36 & -0.4$\pm$0.2 & -0.6$\pm$0.21 & -0.6$\pm$0.23 & -0.6$\pm$0.25 & -0.2$\pm$0.11 & 0.0$\pm$0.05 & -1.0$\pm$0.37 & -1.0$\pm$0.29 & -0.8$\pm$0.33\\
50.0 & -1.0$\pm$0.3 & -0.4$\pm$0.14 & -0.6$\pm$0.23 & -1.0$\pm$0.27 & -0.6$\pm$0.21 & -0.2$\pm$0.11 & 0.0$\pm$0.0 & -1.0$\pm$0.2 & -1.0$\pm$0.19 & -0.8$\pm$0.19\\
60.0 & -1.0$\pm$0.38 & -0.4$\pm$0.06 & -0.6$\pm$0.19 & -1.0$\pm$0.2 & -0.6$\pm$0.19 & -0.2$\pm$0.04 & 0.0$\pm$0.0 & -1.0$\pm$0.28 & -1.0$\pm$0.05 & -1.0$\pm$0.19\\
70.0 & -1.0$\pm$0.24 & -0.4$\pm$0.08 & -0.6$\pm$0.22 & -0.8$\pm$0.2 & -0.6$\pm$0.19 & -0.2$\pm$0.06 & 0.0$\pm$0.0 & -1.0$\pm$0.24 & -1.0$\pm$0.06 & -1.0$\pm$0.1\\
\bottomrule
\end{tabular}%
}
\label{T22}%
\end{table*}%

\clearpage
\begin{table*}
  \caption{Metallicity Results - Padova models MIST clusters - wavelength: 3700-5000}
\makebox[\textwidth][c]{
    \begin{tabular}{cccccccccccccccccc}
\toprule

\multicolumn{1}{c}{S/N} &
\multicolumn{1}{c}{6.8} & \multicolumn{1}{c}{6.9} & \multicolumn{1}{c}{7.0} & \multicolumn{1}{c}{7.1} &
\multicolumn{1}{c}{7.2} &
\multicolumn{1}{c}{7.3} & \multicolumn{1}{c}{7.4} & \multicolumn{1}{c}{7.5} & \multicolumn{1}{c}{7.6} &
\multicolumn{1}{c}{7.7}  \\
\cmidrule(lr){1-1}\cmidrule(lr){2-2}\cmidrule(lr){3-3}\cmidrule(lr){4-4}\cmidrule(lr){5-5}\cmidrule(lr){6-6}\cmidrule(lr){7-7}\cmidrule(lr){8-8}\cmidrule(lr){9-9}\cmidrule(lr){10-10}\cmidrule(lr){11-11}

10.0 & -0.4$\pm$0.37 & -0.2$\pm$0.39 & -0.4$\pm$0.28 & -0.4$\pm$0.5 & -0.2$\pm$0.2 & -0.2$\pm$0.2 & -0.2$\pm$0.15 & -0.2$\pm$0.22 & -0.2$\pm$0.21 & -0.1$\pm$0.34\\
20.0 & -0.2$\pm$0.46 & 0.0$\pm$0.47 & -0.4$\pm$0.2 & -0.6$\pm$0.3 & -0.2$\pm$0.17 & -0.2$\pm$0.17 & -0.2$\pm$0.18 & -0.2$\pm$0.14 & -0.2$\pm$0.15 & -0.2$\pm$0.19\\
30.0 & -0.3$\pm$0.44 & -0.4$\pm$0.49 & -0.4$\pm$0.09 & -0.6$\pm$0.33 & -0.4$\pm$0.19 & -0.2$\pm$0.15 & -0.2$\pm$0.19 & -0.2$\pm$0.08 & -0.2$\pm$0.0 & -0.2$\pm$0.08\\
40.0 & 0.2$\pm$0.35 & -0.3$\pm$0.48 & -0.4$\pm$0.1 & -0.6$\pm$0.36 & -0.2$\pm$0.19 & -0.2$\pm$0.0 & -0.2$\pm$0.19 & -0.2$\pm$0.07 & -0.2$\pm$0.07 & -0.2$\pm$0.06\\
50.0 & 0.2$\pm$0.35 & 0.0$\pm$0.52 & -0.4$\pm$0.1 & -0.6$\pm$0.35 & -0.2$\pm$0.2 & -0.2$\pm$0.07 & -0.2$\pm$0.2 & -0.2$\pm$0.0 & -0.2$\pm$0.0 & -0.2$\pm$0.1\\
60.0 & 0.2$\pm$0.28 & -0.2$\pm$0.49 & -0.4$\pm$0.09 & -0.6$\pm$0.35 & -0.2$\pm$0.2 & -0.2$\pm$0.0 & -0.2$\pm$0.18 & -0.2$\pm$0.0 & -0.2$\pm$0.0 & -0.2$\pm$0.06\\
70.0 & 0.2$\pm$0.36 & -0.2$\pm$0.5 & -0.4$\pm$0.07 & -0.6$\pm$0.31 & -0.6$\pm$0.2 & -0.2$\pm$0.0 & -0.2$\pm$0.2 & -0.2$\pm$0.0 & -0.2$\pm$0.0 & -0.2$\pm$0.08\\
\bottomrule
\end{tabular}%
}
\label{T23}%
\end{table*}%

\clearpage
\begin{table*}
  \caption{Metallicity Results - Padova models MIST clusters - wavelength: 5000-6200}
\makebox[\textwidth][c]{
    \begin{tabular}{cccccccccccccccccc}
\toprule

\multicolumn{1}{c}{S/N} &
\multicolumn{1}{c}{6.8} & \multicolumn{1}{c}{6.9} & \multicolumn{1}{c}{7.0} & \multicolumn{1}{c}{7.1} &
\multicolumn{1}{c}{7.2} &
\multicolumn{1}{c}{7.3} & \multicolumn{1}{c}{7.4} & \multicolumn{1}{c}{7.5} & \multicolumn{1}{c}{7.6} &
\multicolumn{1}{c}{7.7}  \\
\cmidrule(lr){1-1}\cmidrule(lr){2-2}\cmidrule(lr){3-3}\cmidrule(lr){4-4}\cmidrule(lr){5-5}\cmidrule(lr){6-6}\cmidrule(lr){7-7}\cmidrule(lr){8-8}\cmidrule(lr){9-9}\cmidrule(lr){10-10}\cmidrule(lr){11-11}

10.0 & -0.6$\pm$0.34 & -0.4$\pm$0.41 & -0.4$\pm$0.39 & -0.8$\pm$0.32 & 0.2$\pm$0.35 & -0.2$\pm$0.39 & 0.0$\pm$0.44 & 0.2$\pm$0.34 & 0.2$\pm$0.38 & 0.2$\pm$0.43\\
20.0 & -0.6$\pm$0.36 & -0.4$\pm$0.48 & -0.7$\pm$0.34 & -0.6$\pm$0.34 & -0.1$\pm$0.28 & -0.2$\pm$0.32 & 0.2$\pm$0.43 & 0.2$\pm$0.25 & 0.0$\pm$0.34 & 0.1$\pm$0.36\\
30.0 & -0.8$\pm$0.31 & 0.0$\pm$0.43 & -0.8$\pm$0.37 & -0.4$\pm$0.32 & -0.2$\pm$0.26 & -0.2$\pm$0.23 & -0.2$\pm$0.45 & 0.0$\pm$0.28 & 0.0$\pm$0.3 & -0.2$\pm$0.33\\
40.0 & -0.8$\pm$0.29 & -0.1$\pm$0.4 & -0.8$\pm$0.31 & -0.4$\pm$0.21 & -0.2$\pm$0.33 & -0.2$\pm$0.25 & -0.6$\pm$0.4 & 0.0$\pm$0.32 & 0.0$\pm$0.28 & -0.2$\pm$0.36\\
50.0 & -0.8$\pm$0.35 & -0.6$\pm$0.33 & -0.8$\pm$0.26 & -0.6$\pm$0.28 & -0.2$\pm$0.23 & -0.2$\pm$0.19 & -0.6$\pm$0.41 & -0.1$\pm$0.3 & 0.0$\pm$0.32 & -0.6$\pm$0.33\\
60.0 & -0.6$\pm$0.24 & -0.6$\pm$0.33 & -0.8$\pm$0.31 & -0.4$\pm$0.23 & -0.2$\pm$0.25 & -0.2$\pm$0.18 & -0.6$\pm$0.43 & -0.4$\pm$0.3 & 0.0$\pm$0.33 & -0.6$\pm$0.31\\
70.0 & -0.7$\pm$0.35 & -0.6$\pm$0.36 & -0.8$\pm$0.25 & -0.4$\pm$0.23 & -0.2$\pm$0.27 & -0.4$\pm$0.18 & -0.7$\pm$0.35 & -0.4$\pm$0.3 & 0.0$\pm$0.29 & -0.6$\pm$0.31\\
\bottomrule
\end{tabular}%
}
\label{T24}%
\end{table*}%



\label{lastpage}
\end{document}